# UNIVERSIDAD NACIONAL DE INGENIERÍA
## FACULTAD DE ELECTROTECNIA Y COMPUTACIÓN

TRABAJO MONOGRÁFICO

**Propuesta de sistema GeoInformático con representación de escenarios para auxiliar en la nueva metodología propuesta por funcionarios de INETER y la UNI para el estudio a gran escala de la vulnerabilidad y daños debido a sismos en las edificaciones**

**AUTORES:**

**Federico Vladimir Gutiérrez Corea**
**Adolfo Javier Urrutia Zambrana**

**PARA OPTAR AL TITULO DE:**

**Ingeniero en Computación**

**TUTOR:**

**Profa. Dra. Marisela Quintana**

**ASESORES:**

**Prof. Dr. Armando Ugarte (UNI-FARQ)**
**Msc. Álvaro Amador (INETER)**

**Managua, Julio de 2007**

# RESUMEN

En el estudio es presentada una herramienta software basado en un Sistema de Información Geográfico (SIG). La cual permite la estimación de la vulnerabilidad sísmica y la presentación de los resultados para cada casa, grupo de edificaciones, cuadras o a nivel de todo el proyecto o barrio a través de mapas digitales.

Nicaragua es un país con una alta vulnerabilidad sísmica, lo que ha significado que en los últimos años se hayan realizado numerosos estudios de vulnerabilidad y daños debido a sismos, basados en varias metodologías y por distintos grupos científicos.

La estimación de la vulnerabilidad sísmica requiere de la ejecución de distintas tareas, por ejemplo, la recolección de datos en el campo, la integración de datos del catastro municipal, reprocesamiento o pruebas en pantalla de confiabilidad de los datos, la definición de las funciones para el cálculo de la vulnerabilidad para las edificaciones, el cálculo a nivel de cuadras o barrios y finalmente la presentación de estos resultados sobre mapas.

Observando el flujo de trabajo de los proyectos llevados a cabo con anterioridad en Nicaragua, se identificó que la preparación de los datos y la presentación de los resultados tomaba demasiado tiempo, principalmente debido a la necesidad de distintas herramientas de software para las variadas tareas y cálculos científicos, además de tener que trasladar los datos de un software a otro hasta llegar a la aplicación que finalmente presenta los resultados. Es común que estos procesos se tuvieran que realizar nuevamente para poder observar los resultados a través de otros parámetros, como daños o vulnerabilidad.

Con el propósito de reducir el tiempo requerido en esos procedimientos y en la utilización de distintas herramientas de software no especializadas, se ha creado un sistema de software integral, donde el usuario no tiene que preocuparse por usar distintos instrumentos informáticos para cada parte del proceso. El mayor beneficio de este software se obtiene por la combinación del SIG con la lógica específica de la metodología del índice de vulnerabilidad sísmica, índice de daños y la presentación de los resultados. Ese beneficio es debido a la explotación de la habilidad de personalización que presenta el software SIG (ArcGIS, ESRI), el cual permite conectar distintas bibliotecas de software con funcionalidades especializadas y lógica de programación basada en los requerimientos del usuario.

Este nuevo software usa una base de datos empresarial externa al SIG, donde se almacena toda la información de entrada y de resultados, que automáticamente es sincronizada con el Sistema de información geográfico para la presentación de los resultados sobre mapas.

Con el nuevo requerimiento de agrupación de las edificaciones en tipologías, presentado por la nueva metodología. La subida de información catastral de la municipalidad al sistema es algo indispensable, esta información se vuelve una base para el proyecto en ejecución. Los datos catastrales contienen información sobre el tipo constructivo de las viviendas; Tipo de pared, tipo de techo, tipo de uso, número de habitantes, dimensiones, etc. Esa información es luego procesada para definir grupos de viviendas que por tener características similares constituyen una misma clase o tipología.

Por cada tipología se selecciona un número de edificaciones que van a ser visitadas para recoger la información utilizada en el cálculo de la vulnerabilidad sísmica. Las otras casas no seleccionadas toman el valor promedio de la tipología a la que pertenecen. La visualización de los resultados sobre un "layout" de mapas es automática, pudiendo cambiarse su visualización dependiendo de distintos parámetros seleccionados, por ejemplo, los daños que ocasionó algún sismo definido en el proyecto, o la vulnerabilidad, ya sean a nivel de viviendas, cuadras o todo el proyecto.

Otro resultado importante de esta tesis es su documentación, la cual puede servir como guía para estudiantes que trabajen con ingeniería de software orientada a objetos, UML y arquitectura lógica de software en capas (3-tier).

Este trabajo fue llevado a cabo en la dirección de geofísica, del Instituto Nicaragüense de Estudios Territoriales (INETER) y la Universidad Nacional de Ingeniería (UNI), ambas de Managua, Nicaragua.

# ABSTRACT

A GIS based software is presented which permits the estimation of seismic vulnerability and the presentation of results in digital maps for single houses, groups of buildings, parts of settlements or even complete towns.

Nicaragua is a country with a high seismic activity. Thus, seismic vulnerability and risk studies have been carried out in recent years by several scientific groups applying different methodologies.

The assessment of seismic vulnerability requires the execution of distinct tasks, e.g. recollection of field data, integration of data from the municipal cadastre, reprocessing or screening to test the reliability of the data, definition of calculation of vulnerability functions, calculation of vulnerability for single objects as houses or buildings, calculation of mean vulnerability for certain areas as barrios or squares. Finally, the results are normally presented on maps.

Observing the work flows of the projects carried out in Nicaragua we can see that the preparation of data and the presentation of the results take much time, mainly because several separate software tools are used for distinct tasks and scientific calculations. Then the results have to be passed to other software used for the graphical presentation of results. It is common, that all the process has to be done repeatedly to observe the influence of different parameters to the final result.

In order to reduce time and effort to be spent with several unspecialized tools and procedures, an integrated software system was created, the user of which has not to care about separate software tools for each part of the process. The main advantage of the software is the combination of Geographical Information System (GIS) with the logics that surrounds the specific methodologies of seismic vulnerability index, index of damages and presentation of results. This advantage is accomplished by exploiting the ability of customizing the GIS software (ArcGIS, ESRI) trough plug-in´s of its several software libraries with the specialized functionalities and the logical programming with the requested user requirements.

The new software uses a connection with an external centralized Enterprise Data Base which stores all the input information and calculation results and which is automatically synchronized for the presentation of results using GIS.

When new requirements of typologies for houses are introduced, cadastral data from the municipality is uploaded to the system. This information becomes the base data set of the current project. The cadastral information contains data on the constructive type of the house, dimensions, year of construction, type of walls, roof, number of inhabitants, etc.. These data are then processed to define groups of houses belonging to certain types or classes. For each type a number of houses is selected, which will be visited to confirm the correctness of the classification.

The other not selected houses get the average value of its type. This process can be run repeatedly to generate distinct results for several sets of input parameters visualizing the results automatically over a map layout.

The system also allows to present damage scenarios for specific seismic events with given hypocenter and magnitude. Moreover, there is another and just as important byproduct. The documentation of the software serves as a guide for students working on object oriented software engineering by using unified modeling language (UML) and software logic architecture (3-tier).

This work was carried out at the Geophysical Department of the Instituto Nicaragüense de Estudios Territoriales (INETER) and the Universidad Nacional de Ingeniería (UNI), Managua, Nicaragua.

## ÍNDICE DE CONTENIDO

## Lista de figuras

## Lista de Tablas

# *Dedicatoria*

*Todo permanecería tal como está si nadie lo moviese.*

*Si los sueños sólo se tienen flotando en mar*
*de la contemplación y el pensamiento,*
*nunca se van a realizar.*

*El mañana no importaría mucho después de nuestra muerte,*
*si supiéramos que nadie igual de parecido a nosotros*
*jamás volverá a poner pie en este suelo,*
*o crear sombra ante ese mismo sol.*

*Pero no es necesario buscar en el fondo,*
*porque todos sabemos que al menos*
*uno de los que vienen después, esos herederos del futuro,*
*repetirán nuestras distintivas palabras,*
*y sobre todos nuestros más tontos errores.*

*Por esos que no tiene que ser nuestros hijos, ni compartir*
*nuestro apellido para llamarlos mis hermanos.*
*Por ellos que se reirán de los mismos chistes,*
*que preguntaran por los mismos enigmas, que lloraran por*
*nuestra compartida existencia.*

*Hoy, he decido hacer algo.*

*Ajuz Adolfo*

# Agradecimientos

Primero que nada quiero dar gracias a Dios, por darme la oportunidad de formar esta vida tan entretenida que llevo.

A toda mi familia especialmente a mi papá, por aconsejarme desde muy pequeño sobre como ser estudiante (todavía recuerdo aquellos días cuando apenas estando en 3er nivel de preescolar me enseño las tablas de multiplicar) también por darme la dicha de poderme dedicar a los estudios sin preocuparme por la parte económica, agradezco a mi abuelita, por estar siempre al lado de mis hermanos y mió atenta a que siguiéramos un buen camino, también deseo agradecer a mi esposa Elia Miranda sin la que toda esta felicidad no existiría.

A todo el pueblo Nicaragüense, porque gracias a sus aportes la Universidad Nacional de Ingeniería continua siendo publica y formadora de profesionales.

A todos aquellos excelentes docentes, debido a que en ellos descansa parte fundamental de la calidad académica que posee la UNI, especialmente a los que me impartieron clases, como nuestra tutora Marisela Quintana, los profesores José Díaz Chow, Rommel Briones, Alfonso Boza, Carla Reyes, Flor de Maria, Irene González, Glenda Barrios, Magda Luna, Enrique Silva, Narciso Aguilera, Marlon Ramírez, Carlos Sánchez, Svetlana Herrera y a nuestro asesor en la UNI Armando Ugarte de la FARQ.

En INETER a Guillermo Chávez, Norwing Acosta, Alex Castellón, Steffen Schillenger, Wilfried Strauch y a Nuestro asesor Álvaro Amador, gracias a ellos se nos facilito el camino desde el punto de vista técnico donde se mezcla la "programación SIG", cuando en ese tema nos sentimos solos en un mar de ideas e igualmente de dudas, aparecieron ellos y nos apoyaron mediante el compartimiento de sus experiencias, enseñanzas, materiales, ideas y tiempo.

A Ajuz, mi amigo y compañero de monografía, con el que compartí este maravilloso éxito.

Vladimir Gutiérrez

*Entre todas las personas que directamente o indirectamente colaboraron para que esta tesis monográfica se llevara a cabo me gustaría darles las gracias a mi mamá, por haberme impulsado a continuar con mi educación superior en un momento de mi vida en el cual no estaba muy seguro del camino a seguir.*

*A mi esposa Karla Vanesa por su constante y desinteresada ayuda, sin la cual hubiera sido más difícil lograr esta meta en tiempo y forma. A mi abuela materna (mayi) por su interés en mi bienestar personal y profesional a lo largo de mis años de vida universitaria.*

*A mi compañero de tesis Vladimir Gutiérrez, quien me enseñó que dos cabezas piensa mucho más que una y que me acompañó durante estos dos años y medio de sueños y mucho aprendizaje.*

*A mi tutora Marisela Quintana, y a nuestros asesores Armando Ugarte y Álvaro Amador, además del profesor José Díaz Chow por compartir con nosotros sus conocimientos y puntos de vista.*

*Y a todas esas personas que han sabido esperarme mientras realizaba este trabajo, que supieron ver a largo plazo el potencial que poseo. A todos ustedes* $9.99 \times 10^{9,999,999,999}$ *¡Gracias!*

*Ajuz Adolfo*

# CAPÍTULO 1 INTRODUCCIÓN

La necesidad de minimizar los daños causados en viviendas, edificaciones y otros tipos de elementos en riesgo debido a fenómenos naturales, ha llevado a la humanidad a realizar estudios de vulnerabilidad de estos elementos en riesgo y estudios de la amenaza que los fenómenos naturales representan. Estos tipos de investigaciones por lo general se presentan sobre mapas para analizar sus resultados de forma espacial.

Con el avance de las tecnologías de la información y comunicación (TIC), se han desarrollado numerosas herramientas informáticas que ayudan parcialmente a la realización de estos tipos de análisis. Debido a la gran diversidad de metodologías con las que se pueden llevar a cabo estos estudios de vulnerabilidad, daños, amenazas, entre otros, provoca que, a diferencia de sistemas más comerciales (pej: contabilidad, inventarios), no exista un sistema informático que supla a cabalidad las necesidades de cada una de las investigaciones que se lleven a cabo con una metodología en particular, teniendo entonces los especialistas que recurrir a herramientas de software que resultan más aptas en algunas áreas y débiles en otras.

Sin embargo, dentro del desarrollo de las TIC se encuentran las tecnologías de componentes que permite reutilizar partes que empaquetan lógicas y funcionalidades de otros sistemas para ser nuevamente usadas en programas especializados en los que el diseñador y programador se abstraen del funcionamiento interno de este componente y sólo les interesa su resultado.

Como uno de los resultados de este trabajo, se ha desarrollado un sistema informático que hace uso de la tecnología de componentes para reutilizar las funcionalidades del sistema de información geográfico ArcGIS y para generar sus propios componentes que extiendan al Sistema de información Geográfico (SIG), mediante la implementación de la lógica que envuelve el estudio de la vulnerabilidad y daños debido a sismos en las edificaciones, y así presentar su resultado de forma espacial sobre mapas utilizando una metodología en particular propuesta por funcionarios del Instituto Nicaragüense de Estudios Territoriales (INETER) y la Universidad Nacional de Ingeniería (UNI). Metodología que mezcla la información catastral, la agrupación de las viviendas en tipologías, la metodología Índice de Vulnerabilidad sísmica, Índice de Daño y el SIG para generar resultados más amplios desde el punto de vista de estos expertos.

Actualmente se realizan este tipo de estudios utilizando varias herramientas informáticas y procedimientos manuales para ligar los múltiples resultados que generan estas distintas herramientas, lo que conlleva una cantidad considerable de tiempo en la generación de los resultados.

Como otra consecuencia del trabajo y no menos importante, es que su documentación podrá servir como guía para estudiantes que deseen investigar sobre la Ingeniería de Software Orientada a Objetos utilizando UML y la Arquitectura de Software en capas lógicas 3-tier, que son las tecnologías importantes en su ramo y a las que están apostando los desarrolladores e investigadores para el futuro de la informática en lo que corresponde al desarrollo de software.

## 1.1 ANTECEDENTES

Es la primera vez que en Nicaragua se desarrolla un sistema para auxiliar en el estudio de la vulnerabilidad y daños debido a sismos (VDS) que implemente y acople de forma automática en un mismo programa los siguientes elementos:

1. Información sobre datos catastrales
2. Agrupación de las viviendas en tipologías y generación automática de estas últimas.
3. Selección de las viviendas a encuestar de forma automática, aleatoria y espacialmente distribuidas.
4. El método Italiano Índice de Vulnerabilidad Sísmica de Benedetti-Pretini
5. El método Índice de daños
6. El Sistema de Información Geográfico (SIG)

Además es novedoso que el sistema resultante haya sido en base a las siguientes características:

a. Completamente construido bajo la Ingeniería de Software Orientada a Objetos
b. Basado en Arquitectura de Software en Capas (3-Tier)
c. Programación Orientada a Objetos POO
d. Modelado con UML
e. Uso de componentes ArcObjects del SIG y creación de propios componentes.

Existen antecedentes de proyectos donde se utilizan alguno de los elementos enumerados primeramente, mientras que los ítems de nuestra lista que no son utilizados los reemplazan con diferentes técnicas, a continuación enumeramos algunos de dichos estudios:

1. Elementos para la evaluación de la vulnerabilidad sísmica en un centro urbano. Aplicaciones y propuestas para la ciudad de Granada (Nicaragua); Ing. Carmelo Mario Attubato; Diciembre de 2005; Trabajo realizado dentro del marco “Maestría en Evaluación y Prevención de Riesgos ambientales en el entorno Centroamericano”; Universidad Autónoma de Barcelona – Univiversitat de Girona.
   Utilizo el Índice de Vulnerabilidad Sísmica, Índice de Daños y el SIG.

2. Estudio de la vulnerabilidad sísmica en el barrio Monseñor Lezcano, Septiembre de 2005; por estudiantes de Ing. Civil UNI-IES; Rodrigo Ibarra González, Francisco Centeno Saravia; Tutor: Dr. Armando Ugarte.
   Utilizaron el Índice de Vulnerabilidad Sísmica e Índice de Daños.

3. Proyecto de reducción de la vulnerabilidad ante desastres naturales; Estudio de la vulnerabilidad sísmica de Managua; Mayo 2005; INETER - SINAPRED Sistema Nacional para la Prevención de Desastres. Utilizaron Datos catastrales y el SIG.

4. Metodología para la determinación de la vulnerabilidad sísmica en edificaciones, Abril 2003, Tesis Monográfica Ing. Civil UNI-RUPAP.
   Utilizaron el Índice de Vulnerabilidad Sísmica y el SIG.

5. Estudio de la vulnerabilidad y riesgo sísmico en Posoltega y Quezalguaque, Noviembre de 2001, MOVIMONDO – UNI.
   Utilizaron el Índice de Vulnerabilidad Sísmica, Índice de Daños y el SIG.

## 1.2 OBJETIVOS

### 1.2.1 General:

Brindar a INETER y la UNI una herramienta alternativa que auxiliará a la nueva metodología para el estudio a gran escala de la vulnerabilidad y daño debido a sismos en las edificaciones, mediante la reducción del tiempo y los pasos necesarios para la realización de los cálculos y procedimientos que permitirán presentar los resultados de dicho estudio para su posterior análisis.

### 1.2.2 Específicos:

1. Investigar sobre metodologías de ingeniería de software orientadas a objeto para realizar el análisis y diseño del sistema.

2. Indagar y profundizar sobre arquitectura de software en capas para aplicarlo al diseño e implementación del sistema.

3. Investigar sobre las funcionalidades brindadas por el ArcGIS para reutilizar funciones del SIG mediante la tecnología de componentes para expandir sus funcionalidades.

4. Desarrollar un sistema en capas lógicas orientado a objetos, que mejore nuestra técnica de desarrollo de software y permita darle mantenimiento menos complicado al sistema resultante.

## 1.3 HIPÓTESIS:

Una herramienta informática que implemente la nueva metodología propuesta por funcionarios de INETER y la UNI para el estudio a gran escala de la vulnerabilidad y daños debido a sismos en las edificaciones, reducirá los pasos y el tiempo necesario para la realización de los cálculos y procedimientos que permitirán presentar los resultados de dicho estudio para su posterior análisis.

## 1.4 VARIABLES:

X1: Cantidad de pasos a utilizar cuando se use la herramienta Geoinformática en auxilio de estudios de la vulnerabilidad y daños debido a sismos.

X2: Cantidad de pasos resultantes cuando no se utilizó la herramienta Geoinformática y se emplean los recursos actuales (Diciembre de 2006), en estudios de la vulnerabilidad y daños debido a sismos.

X3: Tiempo a invertir en el caso de usar la herramienta Geoinformática en auxilio del estudio en cuestión

X4: Tiempo a invertir en el caso de no usar la herramienta Geoinformática y emplear recursos actuales, en auxilio del estudio en cuestión.

## 1.5 JUSTIFICACIÓN

El método de Índice de Vulnerabilidad Sísmica, el Índice de Daños, el Análisis de Datos Catastrales para agrupar las viviendas en tipologías, en combinación con el Sistema de Información Geográfico (SIG) estándar, poseen una serie de pasos de tediosa repetición. Entre sus desventajas resalta el hecho que el SIG estándar no contempla funciones para calcular la Vulnerabilidad y Daños debido a Sismos (VDS), ni tampoco rutinas para analizar datos catastrales con el objetivo de generar tipologías, por lo que se usan otras herramientas en estas operaciones. Sin embargo, esa desventaja está lejos de ser la única, basta con tomar en cuenta la necesidad de crear un mapa diferente cada vez que se desea ver los datos desde otra perspectiva, o los procesos inherentes a mover datos entre dos o más aplicaciones que no fueron diseñadas para comunicarse directamente entre si.

A la incapacidad actual de reutilización directa de productos obtenidos en proyectos anteriores, se le debe sumar el grado de error en el que se incurre al obtener índices de vulnerabilidad de viviendas de distintas categorías, además que en un proyecto sólo se pueda observar el resultado de las pocas viviendas a las que directamente se les realizó trabajo de campo, obviando de esta manera las otras viviendas, que por lo general son la mayoría. Por eso, la nueva metodología también contempla el uso de datos catastrales, para obtener así mayor exactitud en el cálculo del índice de vulnerabilidad sísmica de las viviendas y aminorar costos, gracias a la sinergia de utilizar datos que fueron recuperados con el fin de cobrar impuestos, para ayudarse a calcular vulnerabilidad en las viviendas. Esta mejora en el cálculo del índice de vulnerabilidad, se logra gracias a la agrupación de las viviendas en tipologías y a que se realizará un proceso para seleccionar muestras de las edificaciones de cada tipología de forma aleatoria y distribuida espacialmente. El resultado de este muestreo serán las edificaciones que se levanten en el trabajo de campo, información para calcular el índice de vulnerabilidad sísmica.

Producto de la mayor amplitud en las estimaciones que implica la nueva metodología propuesta por funcionarios de INETER y la UNI para el estudio de la vulnerabilidad y daños debido a sismos a gran escala. Se da un aumento en la cantidad de pasos y en el tiempo utilizado en los cálculos y presentación de los resultados, con respecto a la metodología anteriormente utilizada. La necesidad de reducir este número de pasos y el tiempo empleado además de la buena definición del procedimiento y la oportunidad de aplicar un sistema informático que integre, implemente y acople todas las funcionalidades requeridas en dicho estudio, tales como información sobre datos catastrales, el método italiano Índice de Vulnerabilidad Sísmica de Benedetti-Pretini y el SIG en la presentación de los resultados de forma espacial para el análisis, por mencionar algunas; nos han servido como base para proponer una herramienta alternativa para el estudio VDS.

Esta permitirá una reducción de los pasos, tiempo y como consecuencia una mayor concentración en el análisis de los resultados. Resultados que servirán para la planeación de medidas predesastres y posdesastres, como la preparación del personal médico, paramédico y grupos de emergencia, determinación anticipada de las necesidades de la población después del evento (alojamiento, alimentación, medicamentos y otros), y así contribuir a una atención oportuna y eficiente de la población afectada en casos de desastres sísmicos.

# CAPÍTULO 2 MARCO TEÓRICO

## 2.1 INGENIERÍA DE SOFTWARE ORIENTADA A OBJETOS

La Ingeniería de Software es una disciplina o área de la informática o ciencias de la computación que ofrece métodos y técnicas para desarrollar y mantener software de calidad que resuelva problemas de todo tipo. Hoy en día es cada vez más frecuente la consideración de la Ingeniería de Software como una nueva área de la ingeniería, y el ingeniero de software comienza a ser una profesión implantada en el mundo laboral internacional, con derechos, deberes y responsabilidades que cumplir, junto a una ya reconocida consideración social en el mundo empresarial y, por suerte para el ingeniero de software, con brillante futuro.

La Ingeniería de Software Orientada a Objetos IngSOO, es un enfoque que se centra en el problema basado en una visión orientada a objetos, donde el dominio del problema se caracteriza mediante un conjunto de estos. El software Orientado a Objetos es fácil de mantener debido a que su estructura es inherentemente poco acoplada. Esto lleva a menores efectos colaterales cuando se deben hacer cambios y provoca menos frustración en el Ingeniero de Software y en el cliente.[1]

En sus comienzos la ingeniería de software separó la naturaleza de los datos de sus procesos asociados. Una herencia de este principio son las Bases de Datos Relacionales, verdaderos repositorios donde se mantienen esquemas modelados de datos y un lenguaje propio para manipularlos.[2]

### 2.1.1 Conceptos de la Orientación a Objetos

Cualquier discusión sobre ingeniería de software orientada a objetos debe comenzar por el término *orientado a objetos.*[3] El paradigma OO se basa en el concepto de objeto.

| |
|---|
| Un objeto es aquello que tiene estado (propiedades más valores), comportamiento (acciones y reacciones a mensajes) e identidad (propiedad que lo distingue de los demás). |

La estructura y comportamiento de objetos similares están definidos en su clase común; los términos instancia y objeto son intercambiables. Una clase es un conjunto de objetos que comparten una estructura y comportamiento común. La diferencia entre un objeto y una clase es que un objeto es una entidad concreta que existe en tiempo y espacio, mientras que una clase representa una abstracción, la "esencia" de un objeto, tal como es. De aquí que un objeto no es una clase, sin embargo, una clase puede ser un objeto[4].

1 [PRESSMAN02] página 343
2 [PEMOBD] página 5
3 [PRESSMAN02] página 345
4 [ZAVALA00]

De forma resumida podemos decir que un objeto es una entidad que incluye datos y el código que lo maneja[5] y la clase es una descripción generalizada (una plantilla, un patrón o un prototipo) que describe una colección de objetos similares. Las clases "viven" como código en archivos, este código controla la forma en la que el objeto debe trabajar, mientras los objetos "viven" en la memoria y son creados como instancias de clases.[6]

En el enfoque OO las propiedades del objeto son claves.

Los principios del modelo OO son fundamentalmente: abstracción, encapsulación, modularidad y jerarquía, y en menor grado tipificación (typing), concurrencia, persistencia.

Se dice que si un modelo que se considera OO no contiene alguno de los primeros cuatro elementos, entonces no es OO.

**Abstracción**. Es una descripción simplificada o especificación de un sistema que enfatiza algunos de los detalles o propiedades del sistema, mientras suprime otros.

**Encapsulación**. Es el proceso de ocultar todos los detalles de un objeto que no contribuyen a sus características esenciales.

**Modularidad**. Es la propiedad de un sistema por estar formado de un conjunto de módulos coherentes e independientes.

**Jerarquía o herencia**. Es el orden de las abstracciones organizado por niveles."[7]

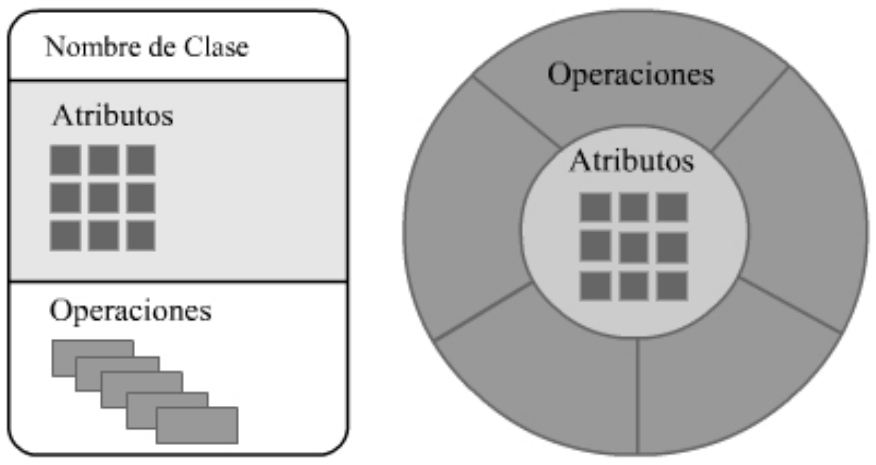

Figura 1. Representación alternativa de una clase orientada a objetos.[8]

En la figura de la clase con forma redonda podemos observar como las operaciones (métodos) envuelven a las los atributos, evitando así el acceso directo a los datos.

### 2.1.2 Elementos claves para la Ingeniería de Software

La ingeniería de software nos brinda una metodología para realizar el análisis y diseño de un sistema de software mediante el modelado, el cual es una forma de pensar acerca de problemas usando modelos organizados de ideas del mundo real. El modelado permite visualizar el sistema construido o a construir.

[5] [BALENA00] página 246
[6] [BURKE04] página 5
[7] [ZAVALA00]
[8] [PRESSMAN02] página 346

A menudo se utiliza el triángulo del éxito, para explicar los componentes claves para un proyecto de software exitoso:

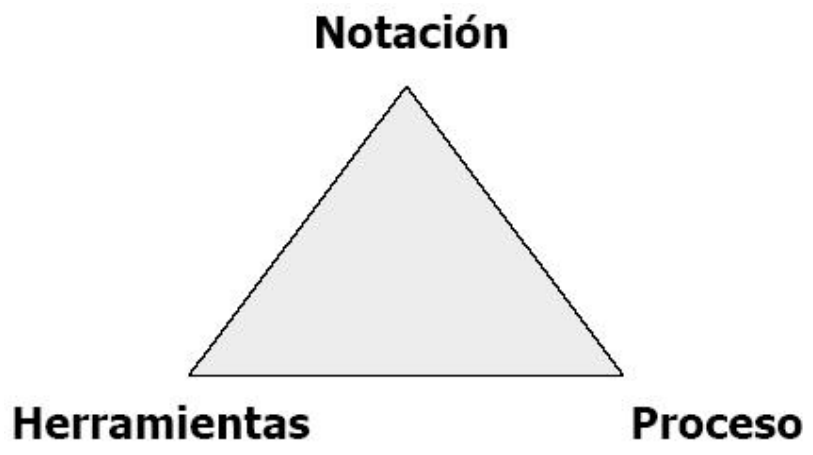

Figura 2. Triángulo del éxito

Se necesitan tres elementos para desarrollar ingeniería de software: la notación, el proceso (o metodología) y la herramienta.

Se puede aprender la notación pero si no se conoce como usarla (proceso) probablemente se fallará; se puede tener un gran proceso, pero si no se sabe como comunicarlo (notación) probablemente también se fallará; por último, si no se documenta el artefacto del trabajo (herramienta), probablemente se malogrará.

Se crean modelos de sistemas complejos porque no se puede comprender el sistema en su totalidad. Existen límites de la capacidad humana para entender la complejidad. Este concepto se ve en el mundo de la arquitectura. Si uno quiere construir una casa para la mascota en el patio de su casa sólo se tiene que empezar a construirla, si se desea construir una nueva casa probablemente se necesitará un plano, si se va a construir un rascacielos definitivamente se necesitarán planos.[9]

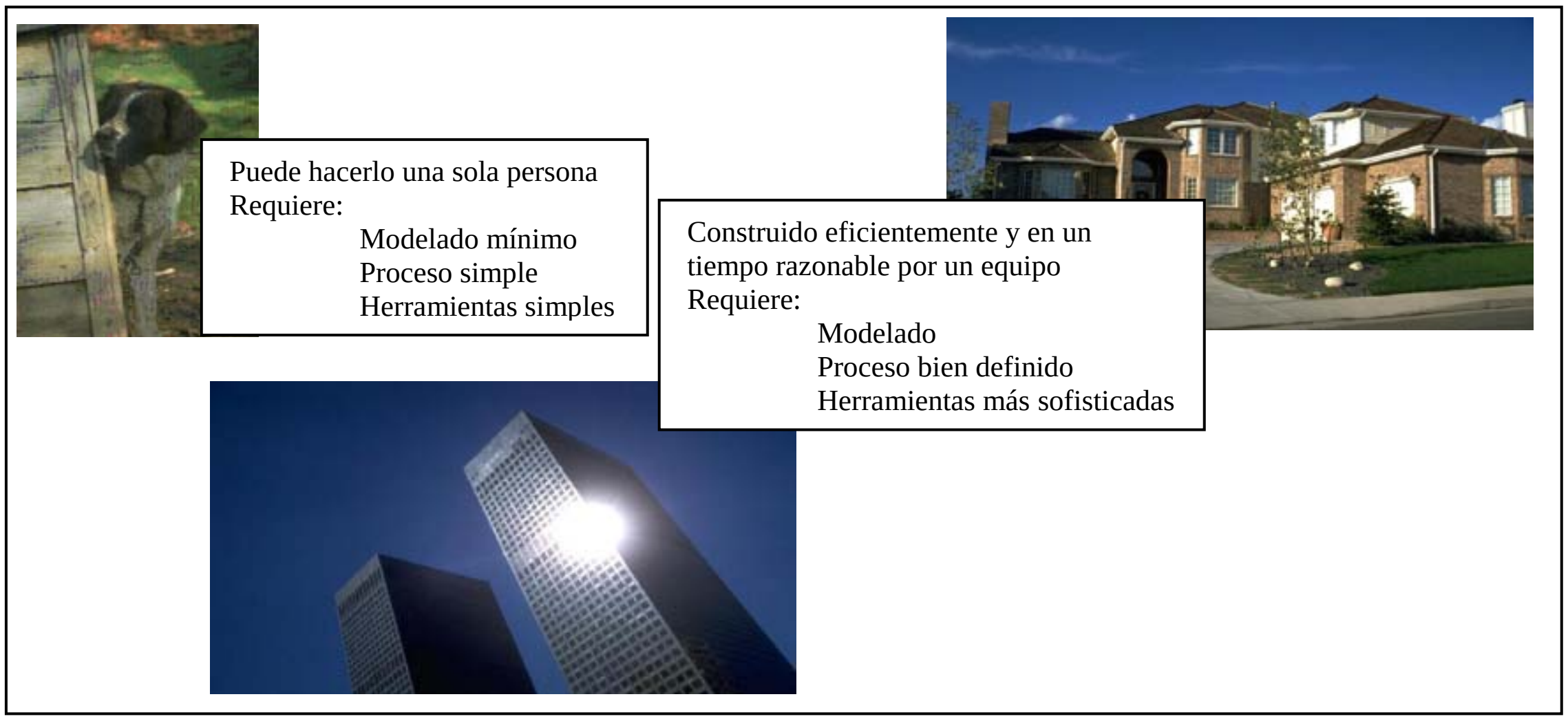

Figura 3. Construcción de una casa de mascota, una casa de hogar y un rascacielos[10]

[9] [QUATRANI00] páginas 13 y 14
[10] [LETELIER]

### 2.1.3 Notación en la Ingeniería Software OO

La notación es un juego de modelos estándares usados para diseñar un proyecto, no describe la forma de implementar esos modelos.[11]

La notación juega un papel muy importante en cualquier modelo, es la pega que mantiene a todos los procesos juntos. La notación juega tres roles:

1. Sirve como un lenguaje para la comunicación de decisiones que no son obvias o que no pueden ser inferidas por el mismo código.

2. Provee semántica que es lo suficientemente rica para capturar todas las estrategias importantes y decisiones tácticas.

3. Ofrece una forma concreta, suficiente para el razonamiento humano y las herramientas de manipulación.

#### 2.1.3.1 Múltiples notaciones

Desde finales de los ochentas y durante los noventas se generaron docenas de metodologías para el análisis y diseño orientado a objetos con sus notaciones.[12] Durante los noventas diferentes metodologías para la ingeniería de software orientada a objetos, cada una con su propio juego de notación, estaban en el mercado.[13]

Las metodologías (con sus respectivas notaciones) destacadas fueron:

1. El método de Booch
2. Técnica de modelado de objetos OMT (por sus siglas en inglés) de Rumbaugh
3. Ingeniería de Software Orientada a Objetos OOSE (por sus siglas en inglés) una versión simplificada de Objectory, ambos de Jacobson.
4. El método y notación de modelado de Coad Yourdon.[14]

Esta cantidad de métodos y notaciones generaron la guerra de los métodos. Pero a medida que el tiempo avanzaba los autores incorporaban en sus propios métodos mejoras de los métodos de los otros autores. Esto concluyó en que estas principales metodologías comenzaron a converger, pero los autores continuaron manteniendo sus propias y únicas notaciones. El uso de distintas notaciones trajo confusiones al mercado, debido a que un símbolo significaba diferentes cosas para diferentes personas.

---

[11] [STRUM99] página 8
[12] [PRESSMAN02] página 262
[13] [QUATRANI00] página 15
[14] [STRUM99] página 8

### 2.1.3.2 Lenguaje de Modelado Unificado (UML)

El fin de la guerra de las metodologías hasta donde concierne a la notación vino con la adopción del Lenguaje de Modelado Unificado (UML) por sus siglas en inglés Unified Model Language.

UML es un lenguaje para especificar, visualizar y documentar los artefactos de un sistema orientado a objeto bajo desarrollo.

Este representa la unificación de las notaciones de Booch, OMT y Objectory así como las mejores ideas de otro sinnúmero de metodologías como se muestra en las siguientes figuras.[13]

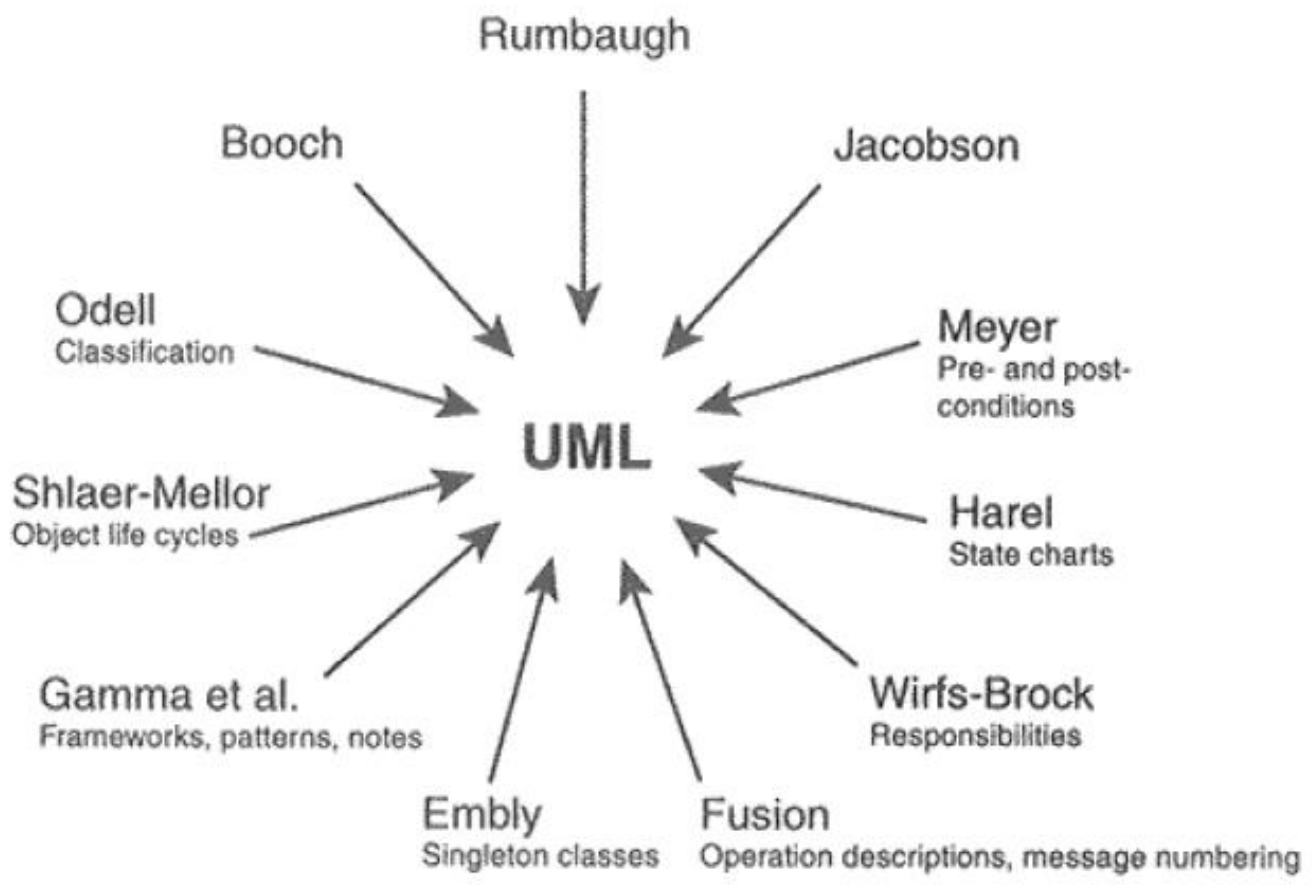

Figura 4. Unificación de notaciones generan UML

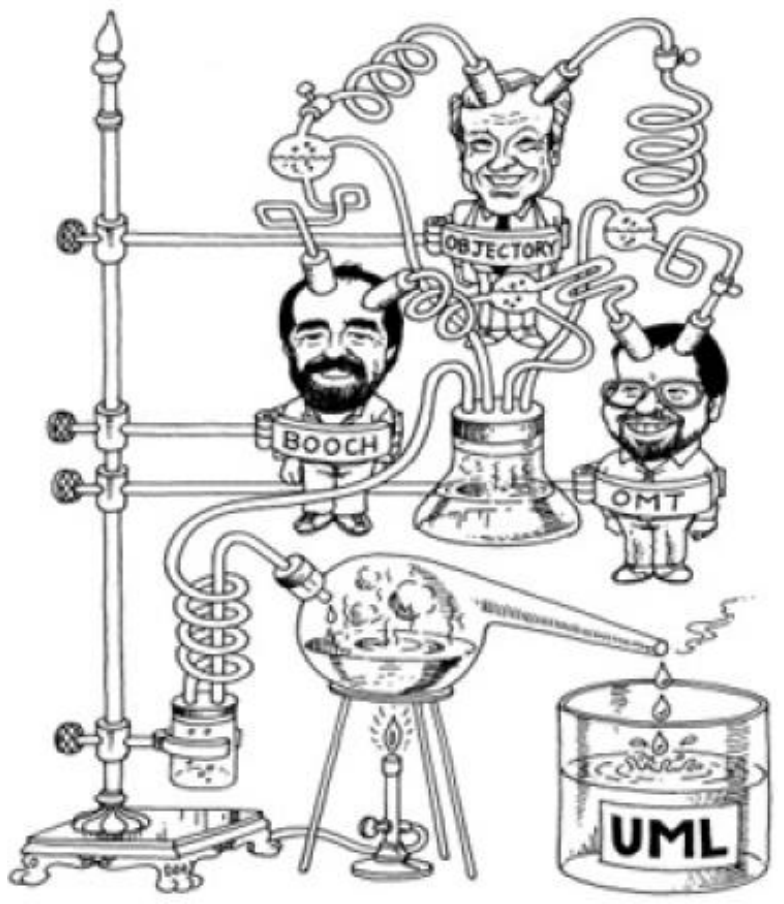

Figura 5. Como empezó todo

La especificación de UML, por ser una notación, no define un proceso de desarrollo de software, pero pretende ser útil con una metodología de desarrollo iterativo. Intenta soportar la mayoría de los procesos de desarrollo orientados a objeto.

UML captura información acerca de la estructura estática y comportamiento dinámico de un sistema. Un sistema es modelado mediante una colección de objetos discretos que interactúan para realizar un trabajo que beneficia a un usuario externo. La estructura estática define los tipos de objetos importantes para un sistema y su implementación, así como la relación entre los objetos. El comportamiento dinámico define la historia de los objetos sobre el tiempo y la comunicación con otros objetos para lograr metas. Modelar un sistema desde distintos y separados puntos de vistas pero a la vez relacionados permite que este sea interpretado para diferentes propósitos.[15]

[15] [JRIJGB99] página 21

### 2.1.3.3 Diagramas de UML

A continuación se indicarán los diagramas que componen al UML, su explicación y los elementos que los componen.

Un modelo captura una vista de un sistema del mundo real. Es una abstracción de dicho sistema, considerando un cierto propósito. Así el modelo describe completamente aquellos aspectos del sistema que son relevantes al propósito del modelo y a un apropiado nivel de detalle. El código fuente del sistema es el modelo más detallado (y además ejecutable). Sin embargo, se requieren otros modelos. Cada modelo es completo desde el punto de vista del sistema, no obstante, existen relaciones de trazabilidad entre los diferentes modelos.

El diagrama es una representación gráfica de una colección de elementos de modelado, a menudo dibujada como un grafo con vértices conectados por arcos. A continuación se enumeran los diagramas de UML:

1. Diagrama de Casos de Uso
2. Diagrama de Clase
3. Diagrama de Objetos (en realidad no se provee como un tipo de diagrama separado)

Diagramas de comportamiento

4. Diagrama de estado
5. Diagrama de actividad

Diagramas de Interacción

6. Diagrama de secuencia
7. Diagrama de colaboración

Diagramas de implementación

8. Diagrama de componentes
9. Diagrama de despliegue

En los diagramas de secuencia, colaboración y actividad se modelan objetos. Existe bibliografía que también se refiere a diagramas de paquetes, de subsistemas y de modelos; sin embargo, estos corresponden a casos particulares de los diagramas arriba mencionados.[16]

[16] [LETELIER] páginas 23-26

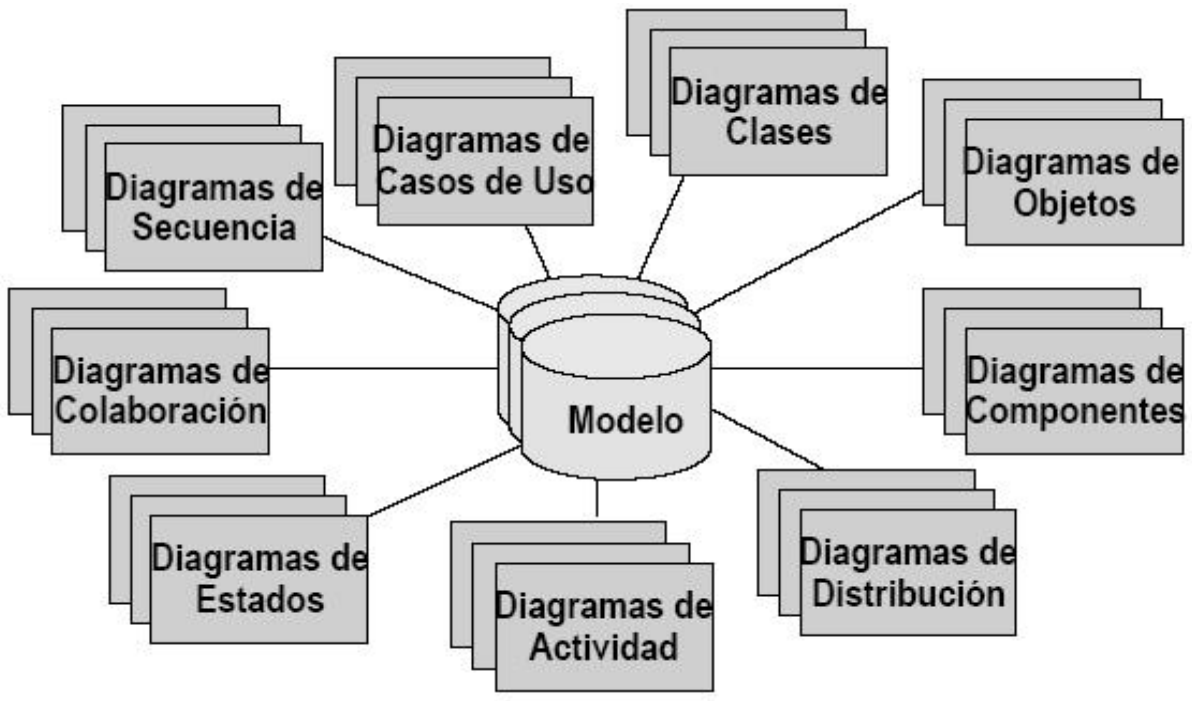

Figura 6. Diagramas de UML[17]

### 2.1.3.3.1 Diagrama de casos de uso

El comportamiento del sistema bajo desarrollo (qué funcionalidades deben de ser proveídas por el sistema) es documentada mediante un modelo de caso de uso, que indica las funcionalidades pretendidas por el sistema (casos de uso), su entorno (actores) y la relación entre los casos de uso y los actores (diagrama de casos de uso). El rol más importante del modelo de casos de uso es el de la comunicación.

Este provee un medio usado por el cliente, el usuario final y el desarrollador para discutir las funcionalidades y comportamiento del sistema.[18]

Casos de uso es una técnica para capturar información de cómo un sistema o negocio trabaja, o de cómo se desea que trabaje.

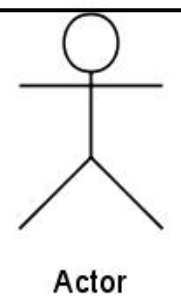

Figura x. UML: Notación para un Actor

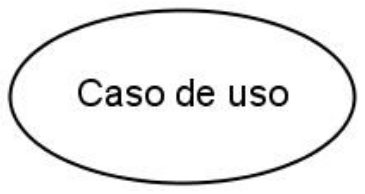

Figura x. UML: Notación para un Caso de Uso

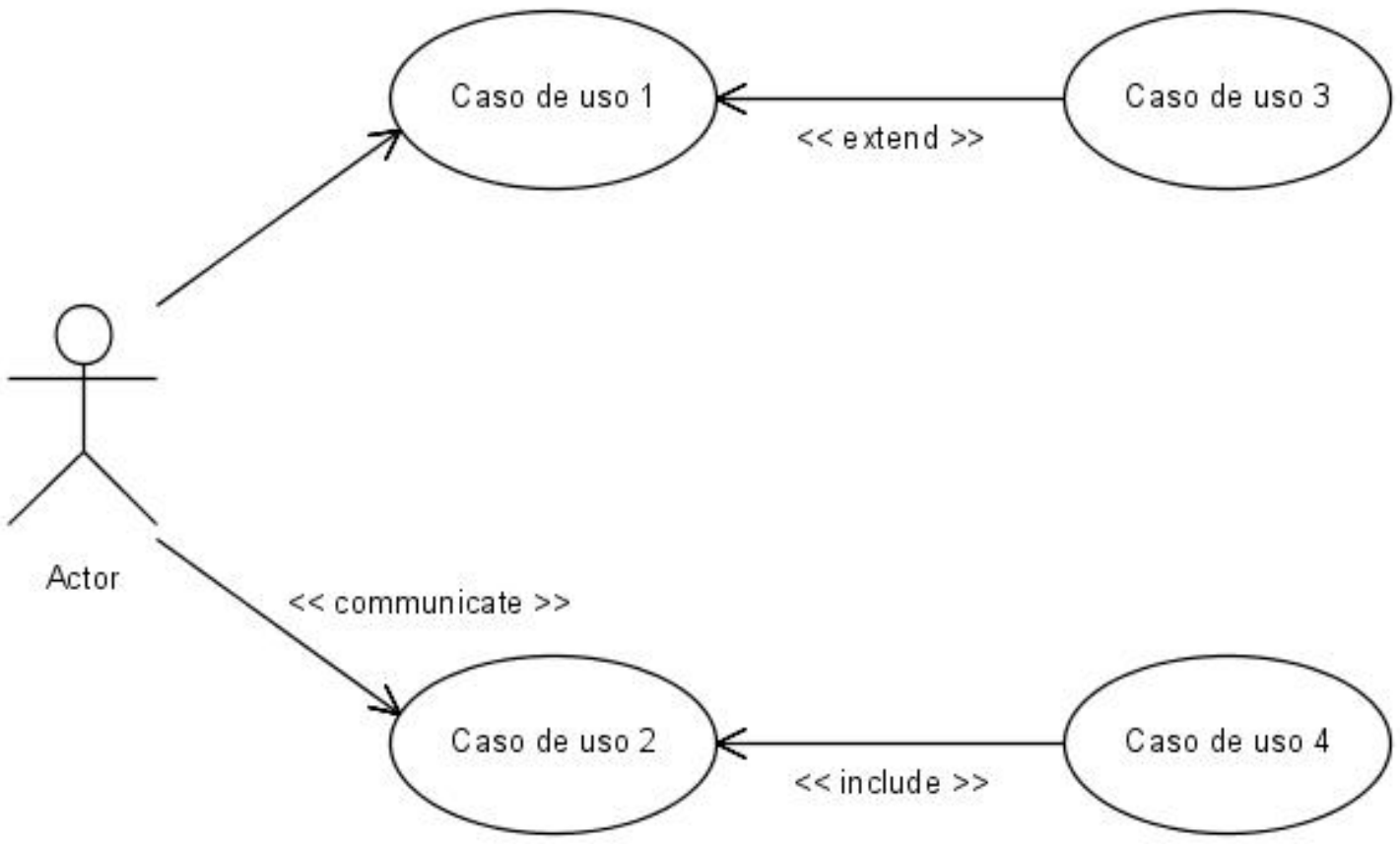

Figura 7. Diagrama de Casos de Uso

[17] [LETELIER]
[18] [QUATRANI00] página 23

Cada caso de uso puede estar definido por:

1. Texto que lo describe
2. Secuencia de pasos (**Flujo de Eventos**) ejecutados dentro del caso de uso
3. Precondiciones y poscondiciones para que el caso de uso comience o termine
4. Combinación de las anteriores

#### 2.1.3.3.2 Diagrama de secuencia

Son usados para describir gráficamente un caso de uso o un escenario. Un diagrama de secuencia muestra los objetos de un escenario mediante líneas verticales y los mensajes entre objetos como flechas conectando objetos. Los mensajes son dibujados cronológicamente desde arriba hacia abajo.

Este diagrama contendrá una vista de las clases participantes en un caso de uso y es más adecuado para observar la perspectiva cronológica de las interacciones.

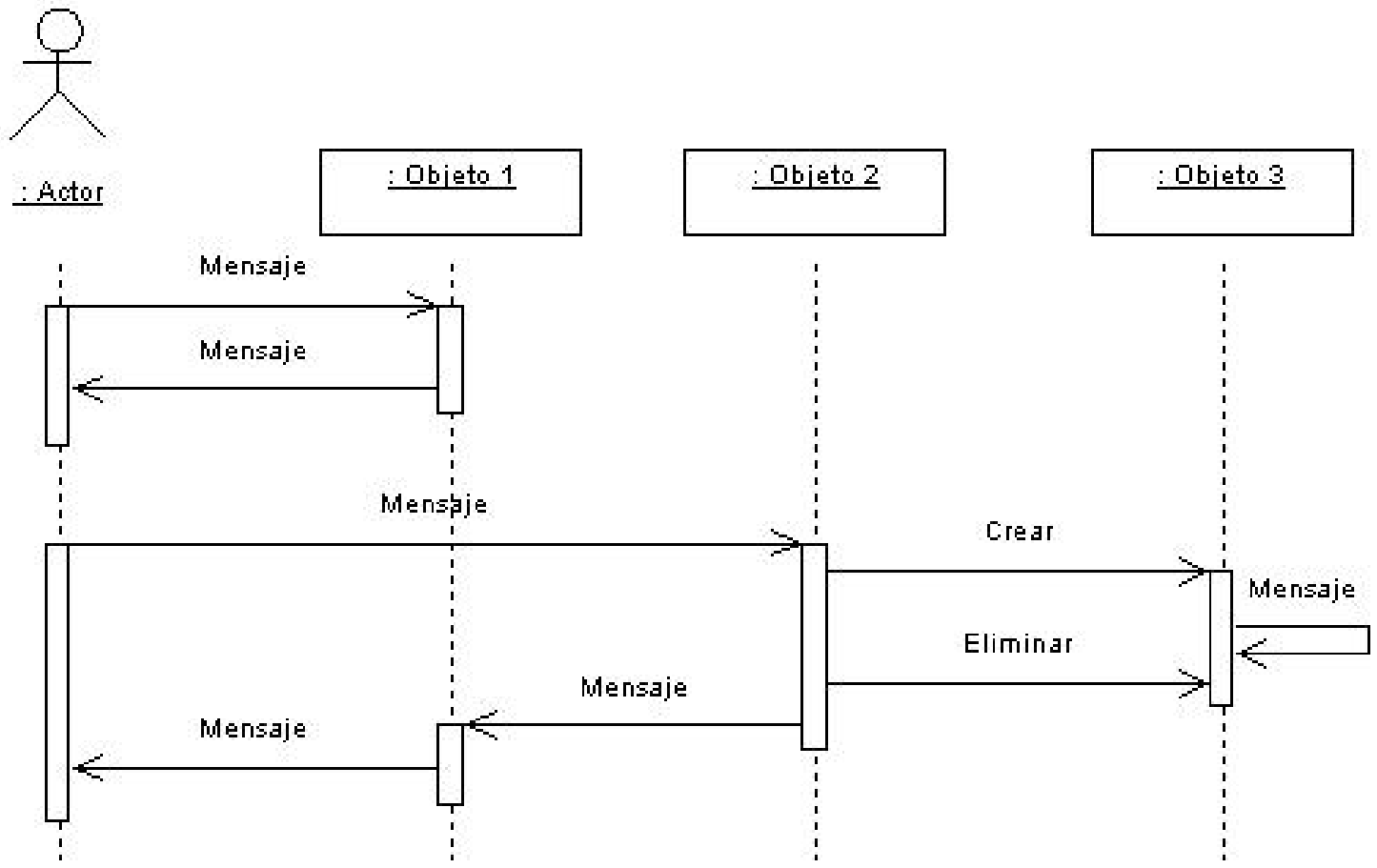

Figura 8. Diagrama de Secuencia

#### 2.1.3.3.3 Diagrama de colaboración

Este tipo de diagrama también es usado para describir gráficamente un caso de uso o escenario, pero este diagrama ofrece una mejor visión del escenario cuando el analista está intentando comprender la participación de un objeto en el sistema.

El diagrama de colaboración modela la interacción entre los objetos de un caso de uso, los objetos están conectados por enlaces en los cuales se representan los mensajes enviados acompañados de una flecha que indica la dirección.

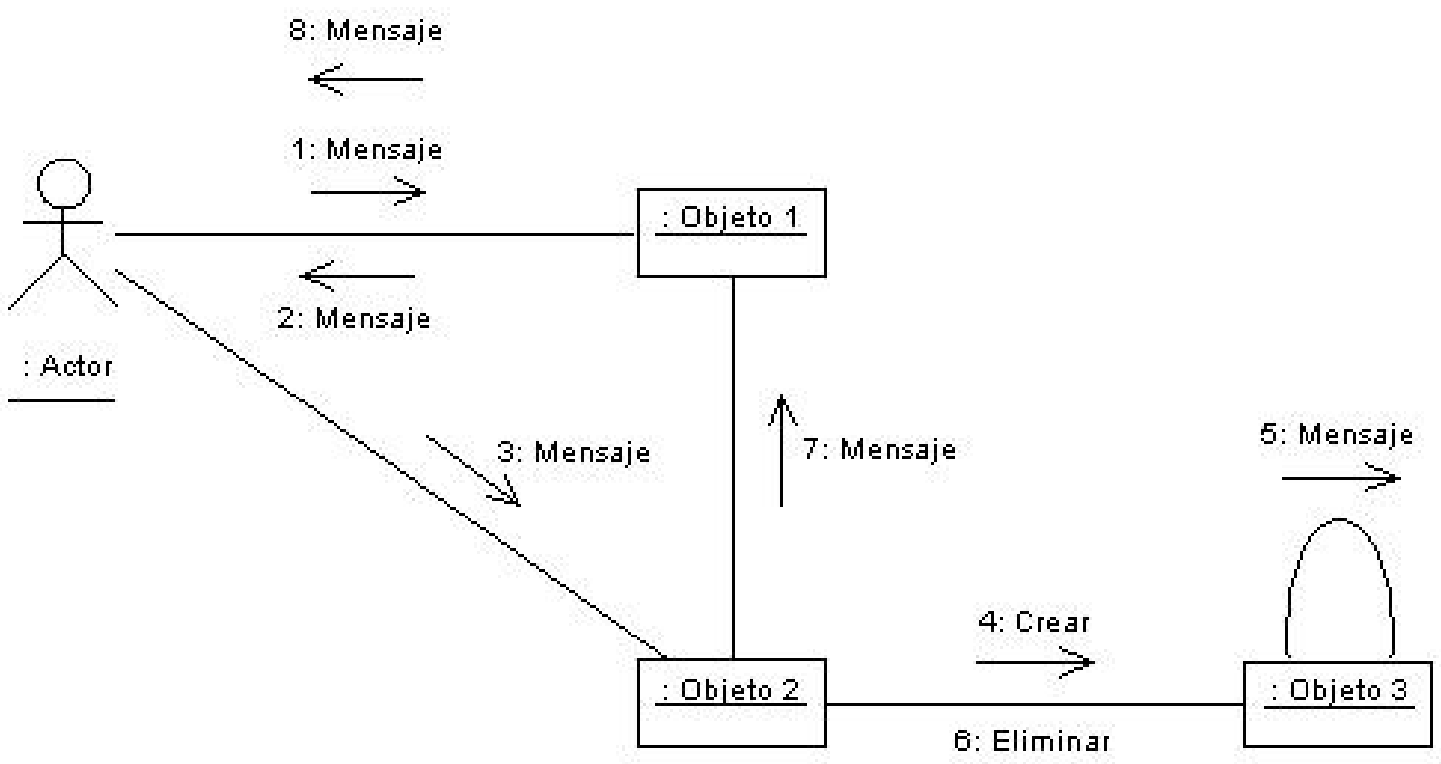

Figura 9. Diagrama de Colaboración

### 2.1.3.3.4 Diagrama de clases

Un diagrama de clases presenta las clases del sistema con sus relaciones estructurales y de herencia. La definición de clase incluye definiciones para atributos y operaciones. El diagrama de clases es el principal para el análisis y diseño. [19]

Un diagrama de clases está compuesto por los siguientes elementos:

1. Clase: atributos, métodos y visibilidad.
2. Relaciones: Asociación, Herencia, Agregación simple y Agregación compuesta.[20]

**Clase:**

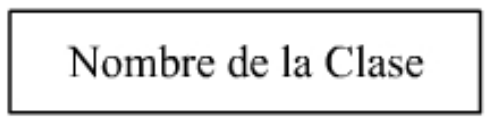

Figura 10. Representación más simple de una clase

La representación estándar de una clase muestra más información y está divida en tres secciones:

1. La parte de arriba contiene el nombre de la clase
2. En la parte media se ubican las propiedades
3. La sección de abajo contiene a los métodos

[19] [LETELIER] paginas 33,38-39, 96-97
[20] [STRUM99] páginas 134-135

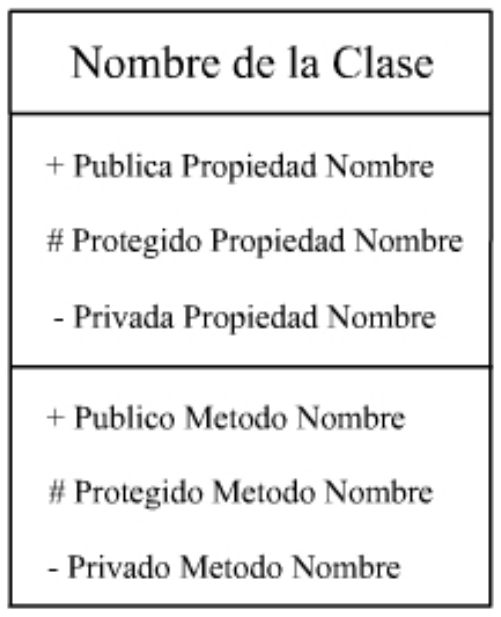

Figura 11. Representación típica de una clase

El símbolo ubicado al lado de cada propiedad y método indica el nivel de visibilidad. El símbolo de más (+) indica público, el negativo (-) indica privado, el numeral (#) indica protegido, que sólo se puede acceder desde otros métodos de clases del mismo proyecto.

**Relaciones:**

1. Asociación (“usa”): En esta relación un objeto emplea los servicios de otro objeto, es la relación entre dos clases donde una clase no requiere a la otra clase para existir. Cada clase tiene igual responsabilidad. Es representada por una línea conectando a las clases.

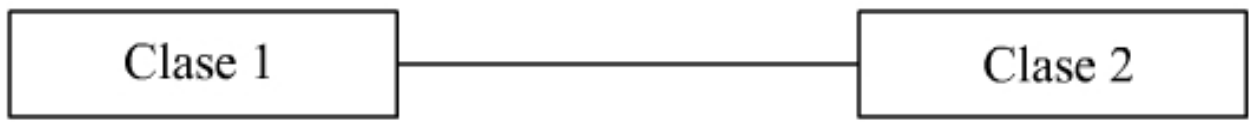

Figura 12. Relación de Asociación

2. Herencia (“es una”): Existe cuando una clase es una versión especializada de otra clase. Es representada mediante un triángulo del lado de la clase base.

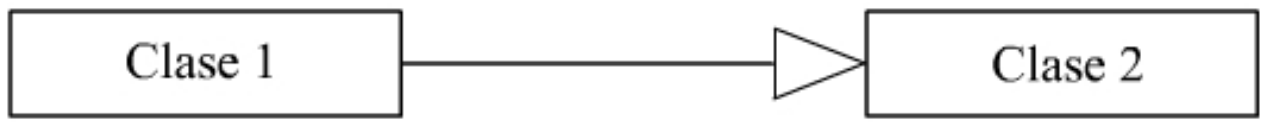

Figura 13. Relación de Herencia

3. Agregación Simple (“contiene, hecho de”): En esta relación un objeto está hecho de otros objetos (subordinados), es un tipo de relación más general donde una clase simplemente tiene más responsabilidad que otras clases. Este tipo de relación se representa con un diamante sin relleno.

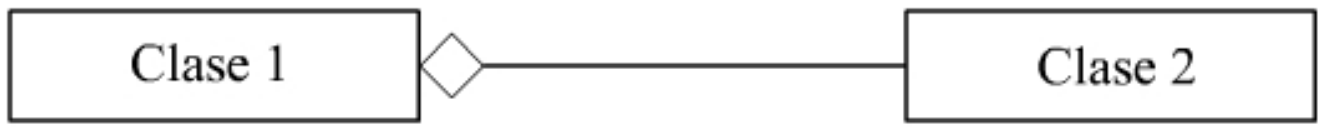

Figura 14. Relación de Agregación Simple

4. Agregación compuesta ("posee, tiene"): Es una relación padre hijo, una relación entre dos clases donde una clase no puede existir sin la otra clase. Es representada mediante un diamante relleno.

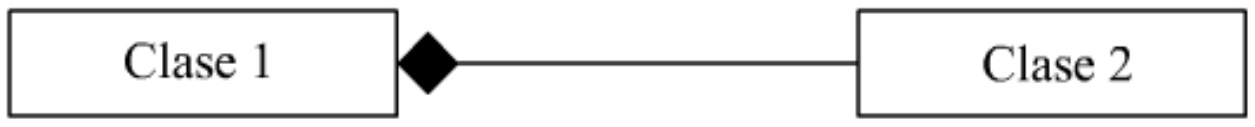

Figura 15. Relación de Agregación Compuesta

Para los tipos de asociaciones 1, 3 y 4 existen indicadores de cardinalidades, para hacer referencia a cuantos elementos de una clase se relaciona la otra clase. [21]

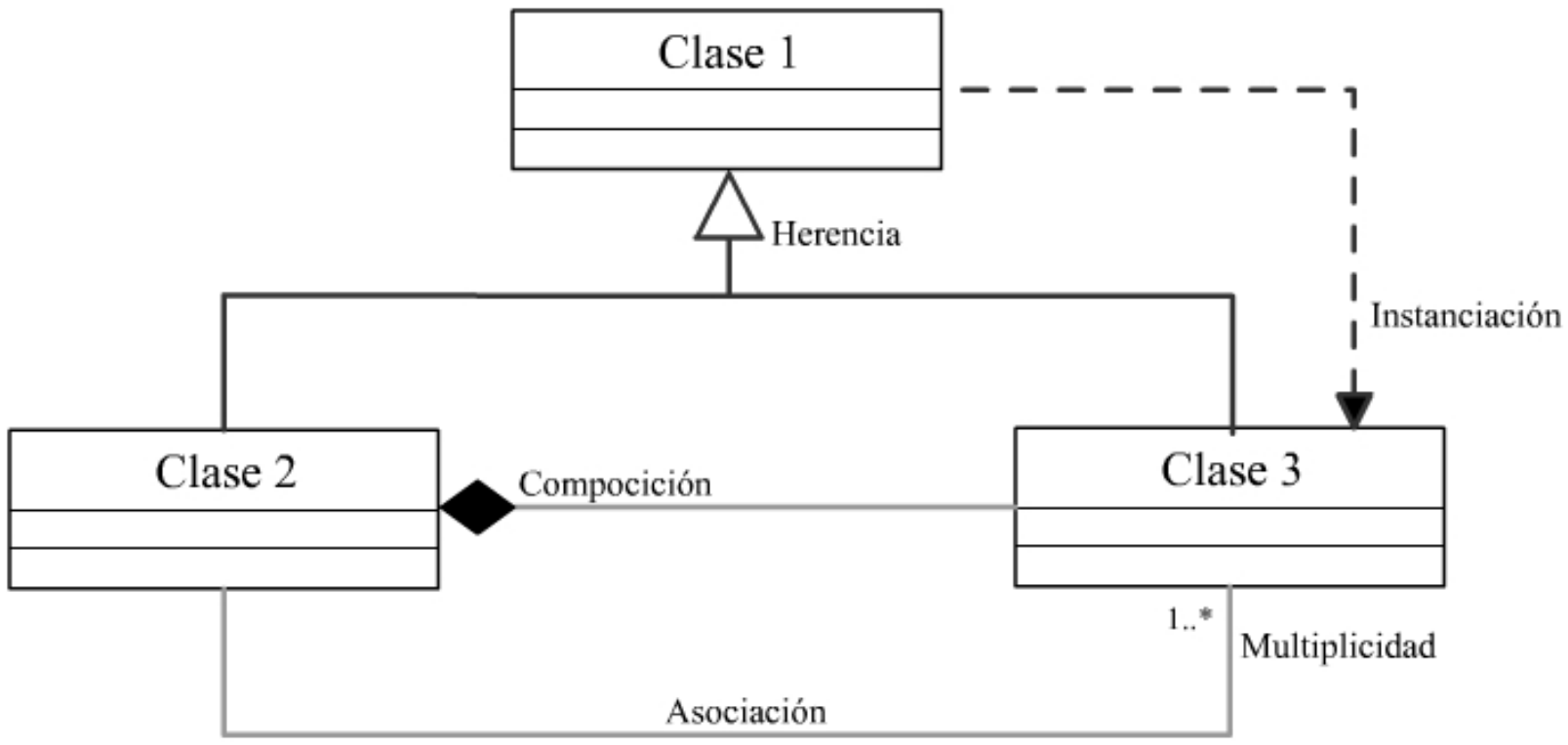

Figura 16. Diagrama de distintas relaciones

### 2.1.3.3.5 Diagrama de Estados

Un diagrama de estado muestra los estados de un simple objeto, los eventos o mensajes que causan una transición de un estado a otro y las acciones que resultan de un cambio de estado.

No es necesario crear un diagrama de estado por cada clase del sistema, sólo para clases con comportamiento dinámico significativo.

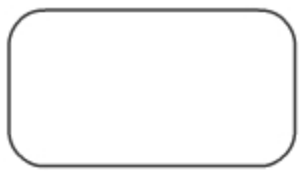

Figura 17. Notación UML para un estado

Existen dos tipos de estados especiales: el estado de inicio y el de fin, cada diagrama sólo puede tener un estado de inicio y uno de fin.

Figura 18. Estado Inicial

Figura 19. Estado Final

[21] [STRUM99] páginas 138-139 y [LHOTKA98] página 93

Las transiciones entre estados representan un cambio de un estado original a otro estado sucesor (que puede ser el mismo estado original). Una transición puede estar acompañada de una acción.

Existen dos formas de transición de un estado: automático y no automático. Un estado de transición automático ocurre cuando se completa una actividad de un estado original, para este caso no hay nombre asociado al evento con la transición del estado. Un estado de transición no automático es causado por el nombre de un evento, ambos tipos de transiciones de estados son considerados de tiempo cero y no pueden ser interrumpidos.
Un estado de transición es representado por una flecha que va desde el estado original hacia el estado sucesor.

Una transición de estado puede tener asociado una acción y una condición de guarda y también puede disparar un evento. Una acción es una actividad que ocurre cuando hay una transición de estado. Un evento es un mensaje que es enviado a otro objeto en el sistema. Una condición de guarda es una expresión booleana de un valor de un atributo, que permite la transición de estado sólo si la condición es verdadera. Tanto las acciones como las condiciones de guarda son comportamientos del objeto y típicamente se vuelven operaciones o métodos, a menudo esas operaciones son privadas.

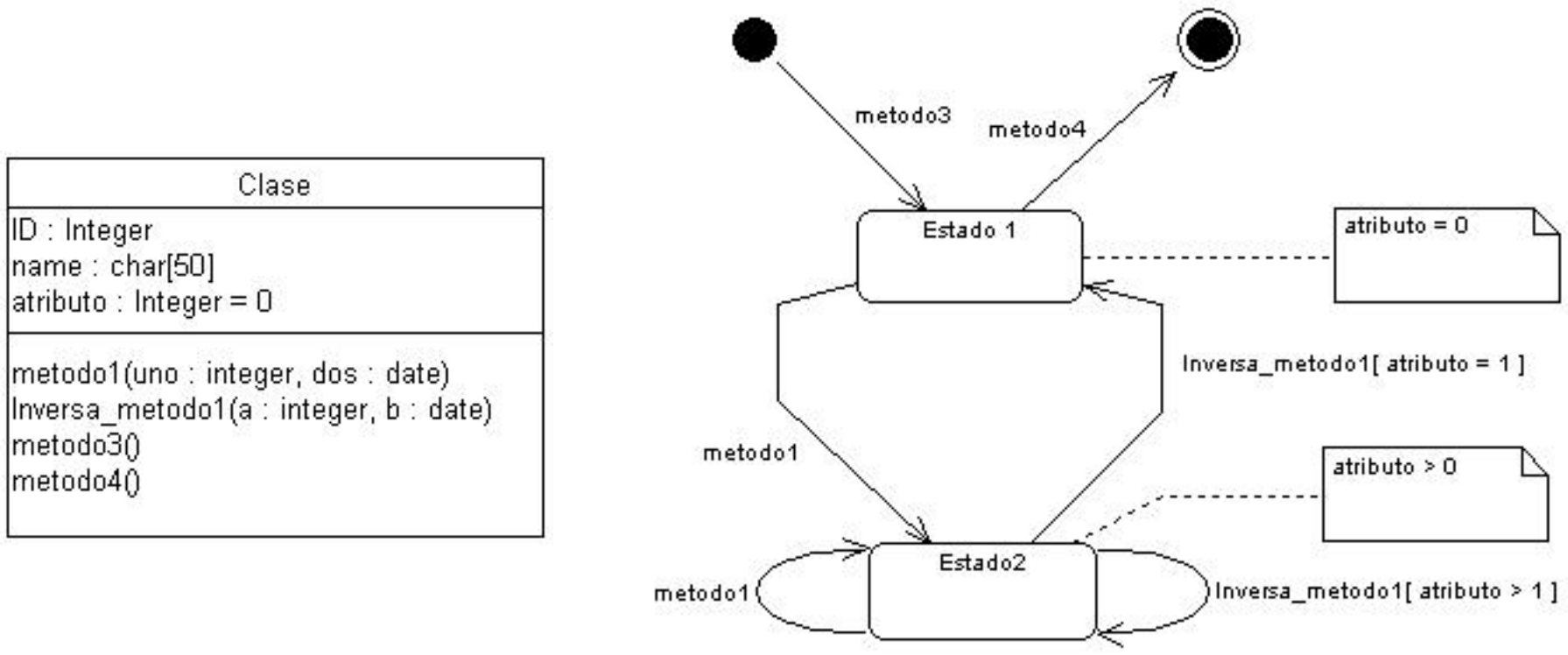

Figura 20. Diagrama de estado para la Clase

Las acciones que acompañan toda la transición de un estado a otro, deben de ubicarse en acciones de entrada (entry action) del estado, así mismo todas las acciones que acompañan la transición de salida de un estado deben de ubicarse en acciones de salida del estado.

Se puede ejecutar una acción como consecuencia de entrar, salir, estar en un estado o por la ocurrencia de un evento. [22]

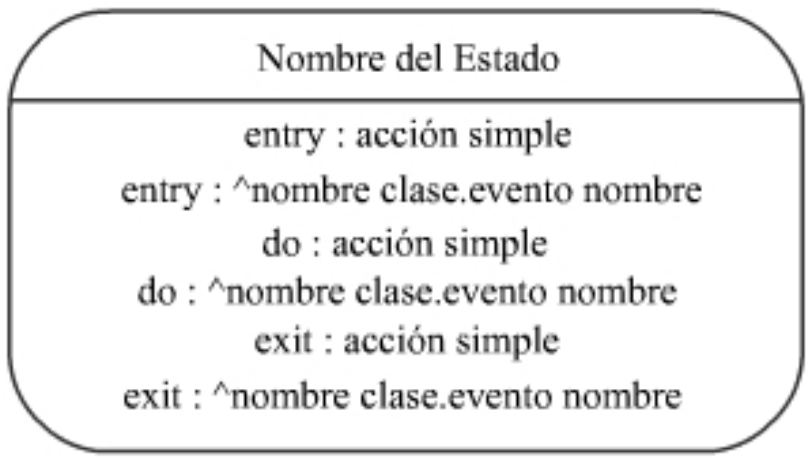

Figura 21. Estado detallado

[22] [QUATRANI00] páginas 150, 108, 111, 113

### 2.1.3.3.6 Diagrama de Actividad

El diagrama de actividad ha sido creado para mostrar una visión simplificada de lo que ocurre durante una operación o proceso.

El diagrama de actividades de UML es muy parecido a los viejos diagramas de flujos, que son de los primeros modelos visuales que se aplicaron a la computación y que siempre se enseñan en los primeros cursos de programación. El diagrama UML muestra los pasos (conocidos como actividades), puntos de decisión y bifurcaciones.[23]

Los diagramas de actividades sirven para especificar:

1. Modelado de flujo de eventos de un caso de uso
2. El comportamiento de los objetos de una clase
3. La lógica de una operación o método
4. La descripción de un flujo de trabajo [24]

Cada actividad se representa por un rectángulo con las esquinas redondeadas (más angosto y ovalado que la representación del estado). El procesamiento dentro de una actividad se lleva a cabo, y al realizarse, se continúa con la siguiente actividad. Una flecha representa la transición de una actividad a otra. Al igual que el diagrama de estados, el de actividad cuenta con un punto inicial (representado por un círculo relleno) y uno final (representado por una diana).

Las decisiones se representan mediante un rombo desde donde parten lar rutas de decisión.

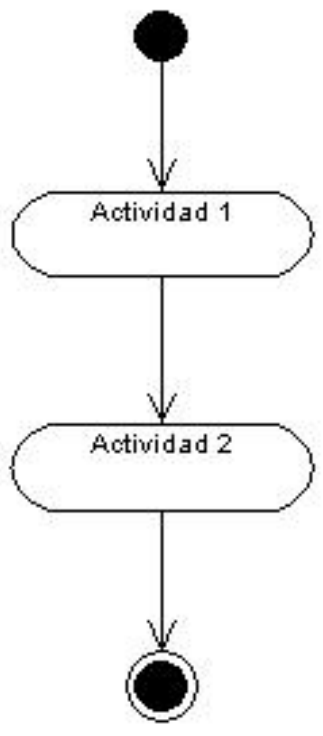

Figura 22. Transición de una actividad a otra

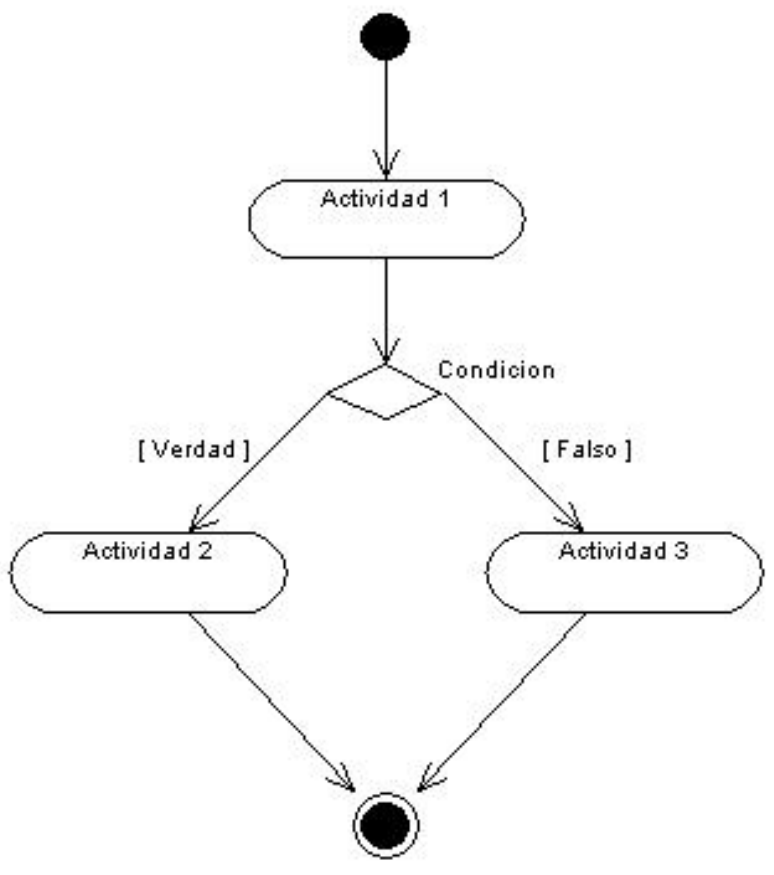

Figura 23. Diagrama de actividad con condición

[23] [SCHMULLER] páginas 152-153
[24] [LETELIER] página 47

Es muy probable que se necesite modelar dos rutas de actividades que se ejecuten al mismo tiempo y luego se reúnan en un punto en común, por eso el diagrama de actividad permite modelar rutas concurrentes. Para representar esta división se utiliza una línea gruesa perpendicular a la transición y las rutas partirán de esta línea. Para representar la reincorporación ambas rutas apuntarán a otra línea gruesa.

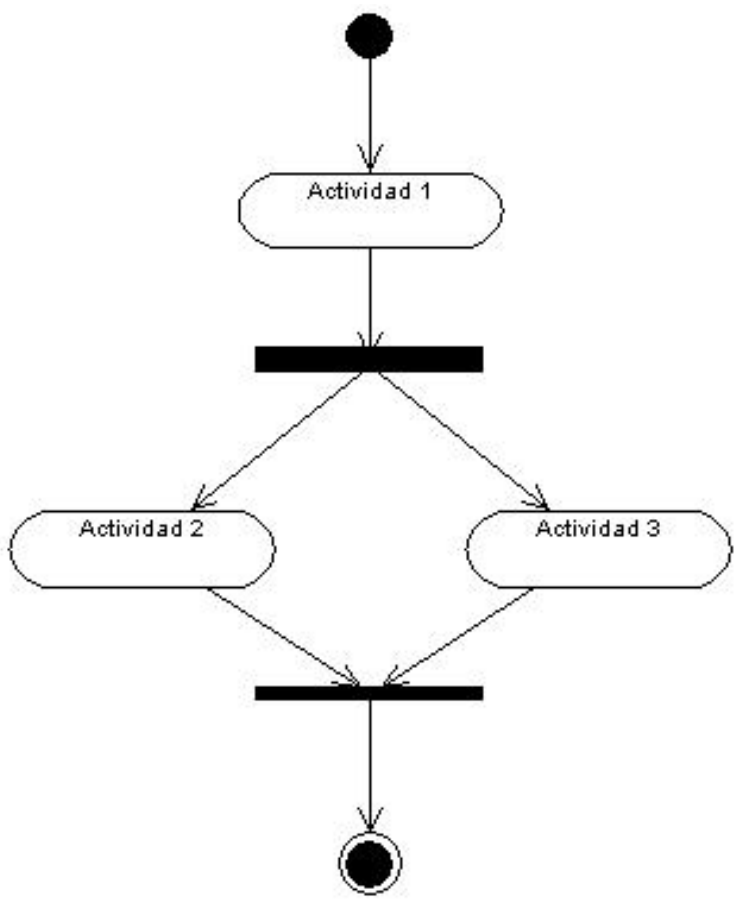

Figura 24. Representación de una transición que se bifurca en dos rutas y se ejecutan concurrentemente para luego reincorporarse

### 2.1.3.3.7 Diagrama de Componentes

El diagrama de componente permite modelar la estructura del software y la dependencia entre componentes Los componentes representan todos los tipos de elementos de software que entran en la fabricación de aplicaciones informáticas. Pueden ser simples archivos, paquetes de software, bibliotecas cargadas dinámicamente, etc.

Las relaciones de dependencia se utilizan en los diagramas de componentes para indicar que un componente utiliza los servicios ofrecidos de otro componente.

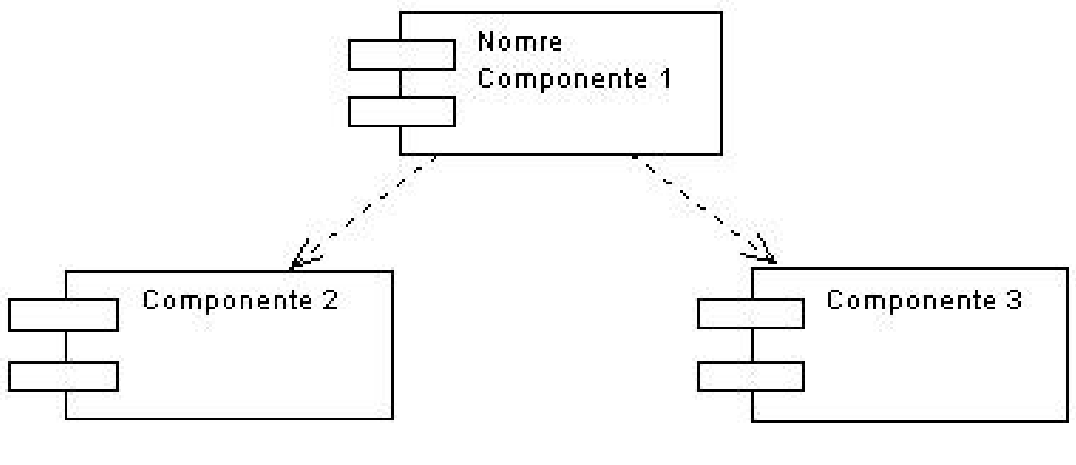

Figura 25. Diagrama de componentes

### 2.1.3.3.8 Diagrama de Despliegue

Los diagramas de despliegue muestran la disposición física de los distintos nodos que componen un sistema y el reparto de los componentes sobre dicho nodo.

Los estereotipos permiten precisar la naturaleza del equipo:

1. Dispositivos
2. Procesador
3. Memoria

Los nodos se conectan mediante soportes bidireccionales que también se pueden estereotipar.[25]

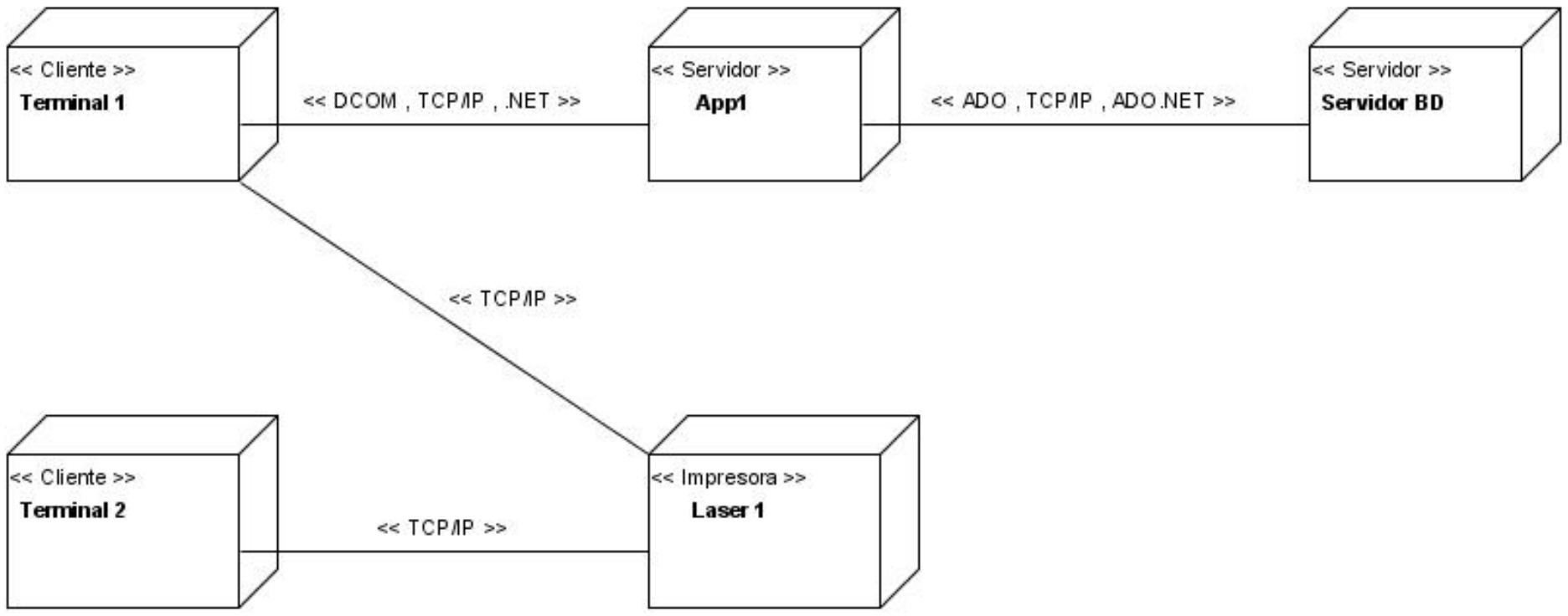

Figura 26. Diagrama de Despliegue

[25] [LETELIER] páginas 175 y 179

### 2.1.4 El Proceso en la Ingeniería de Software OO

Para desarrollar un proyecto de forma satisfactoria que alcance o exceda las expectaciones, debe ser trabajado en tiempo, de forma económica y adaptable al cambio. El ciclo de desarrollo debe promover creatividad e innovación. Al mismo tiempo el proceso debe ser controlable y medido para asegurar que sea completado.[26]

El proceso de desarrollo define **Quién** debe hacer **Qué**, **Cuándo** y **Cómo** debe hacerlo. No existe un proceso de desarrollo de software universal. Las características de cada proyecto (equipo de desarrollo, recursos, etc.) exigen que el proceso sea configurable.[27]

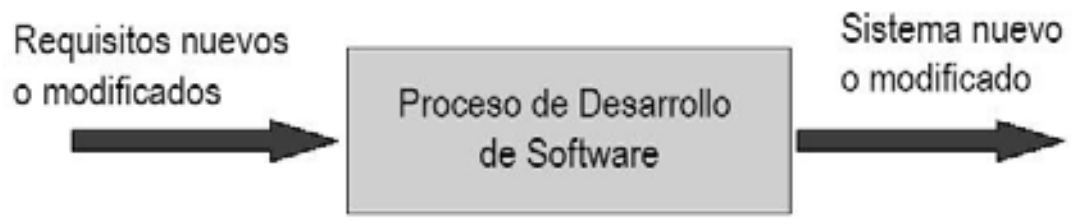

Figura 27. Proceso de desarrollo de software

De manera formal el proceso de desarrollo de software presenta cuatro reglas:
1. Proveer una guía sobre el orden de actividades para el equipo de trabajo.
2. Especificar que artefactos (documentación resultante) deben de ser desarrollados.
3. Dirigir las tareas de desarrolladores individuales y del equipo como un todo.
4. Ofrecer criterios de medición y monitoreo de los productos y actividades del proyecto.[28]

No es bueno que en un proyecto de software los desarrolladores comiencen a codificar sin más un sistema desde su inicio, porque no se sabría que codifican.
Se debe proceder de manera más estructurada y metódica; la estructura y naturaleza de los pasos de un esfuerzo de desarrollo es lo que se conoce como metodología.[29]

Las metodologías para el desarrollo de software se pueden agrupar según la cantidad de componentes de proceso que realizan en un intervalo de tiempo:
1. Lineales o en cascada
2. Evolutivos (iterativo, incremental)

#### 2.1.4.1 Metodologías en cascada

La influencia de estas metodologías trascendió por varios años, y establece que el análisis, diseño, codificación y distribución va uno después del otro; solamente cuando se haya completado uno se puede iniciar el otro. Además tiende a enmascarar el riesgo real del proyecto hasta que es muy tarde hacer algo manejable sobre este. [30]

---

[26] [QUATRINI00] página 16

[27] [LETELIER] página 182

[28] [CONALLEN00] página 61

[29] [KROLL, KRUCHTEN03] Capítulo 3 tema 1

[30] [SCHMULLER] páginas 207-208, [KRUCHTEN00] capítulo 1 tema 1

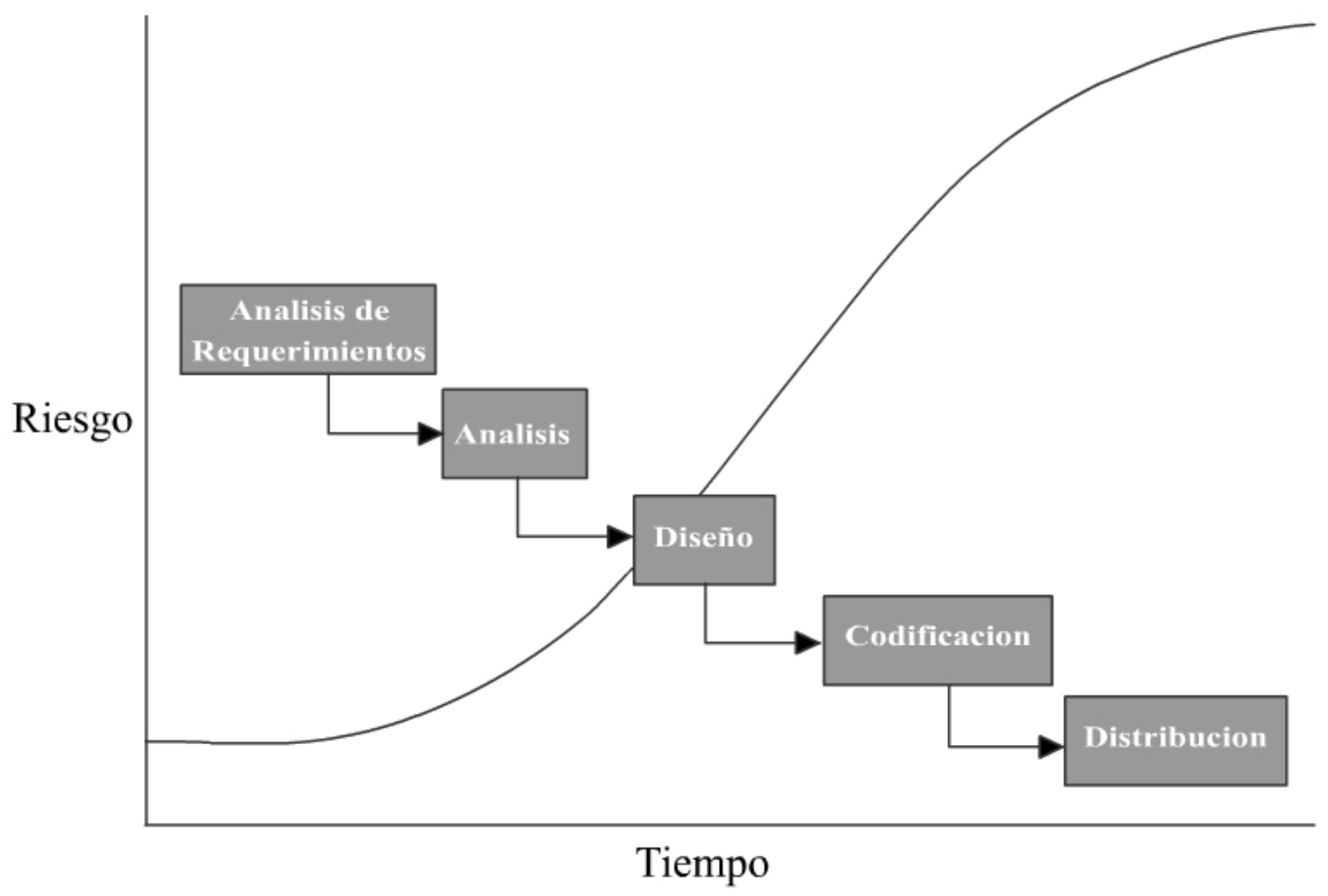

Figura 28. El ciclo de vida en cascada

Esta metodología tiende a la realización de tareas individuales y reduce el impacto de compresión en el proyecto. Si el analista no tiene contactos con el diseñador y éste a su vez no tiene contacto con el desarrollador, existe la posibilidad de que los miembros rara vez trabajen juntos para compartir puntos de vistas importantes. Por otro lado, si el proceso no puede retroceder y volver a los primeros estados es posible que las ideas desarrolladas no sean utilizadas. La revisión del análisis y diseño en la incorporación de una idea desarrollada establece una mayor oportunidad de éxito.[31]

#### 2.1.4.2 Proceso iterativo incremental

Esta metodología fuerza la identificación del riesgo del proyecto temprano en el ciclo de vida, cuando es posible reaccionar y atacar el riesgo a tiempo y de una manera eficiente. Este proceso tiene mucho en común con el propuesto en 1988 por Barry Bohem “El modelo espiral”[32]

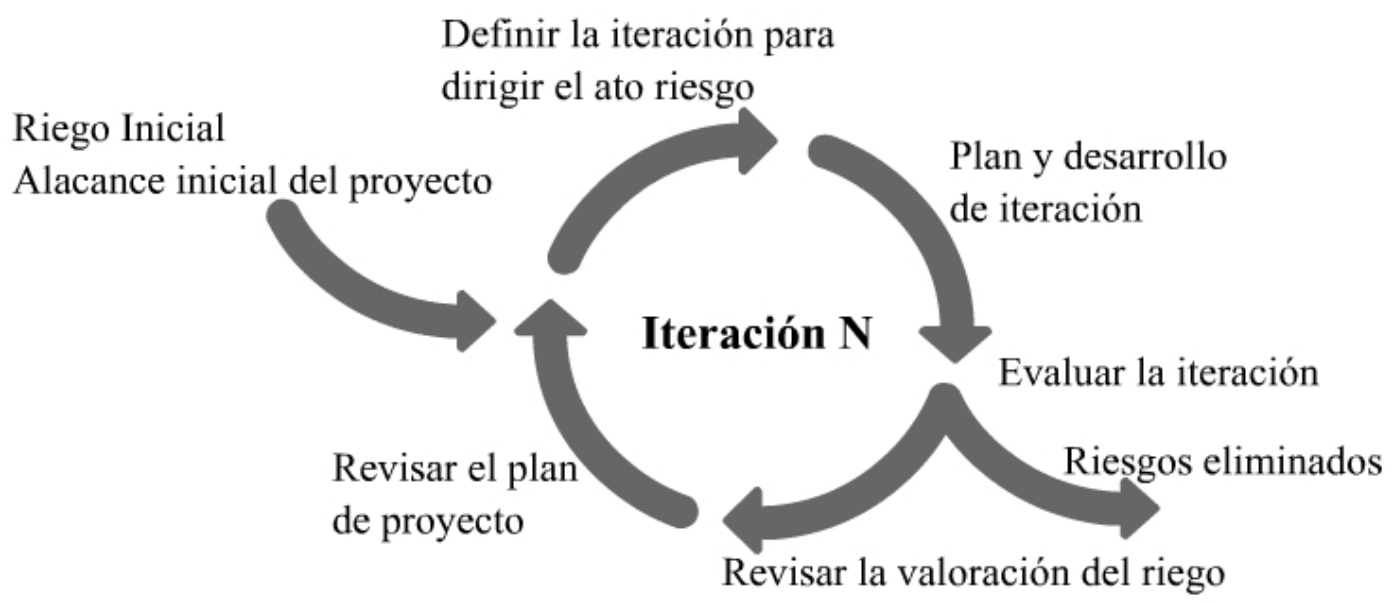

Figura 29. Un proceso iterativo incremental

[31] [SCHMULLER] páginas 207-208
[32] [KRUCHTEN00] capítulo 1 tema 1, [SCHMULLER] página 196

En el proceso iterativo incremental el ciclo de vida de desarrollo procede como una serie de iteraciones que consiste de uno o más de los siguientes componentes de proceso: modelado del negocio, requerimientos, análisis, diseño, implementación, prueba y distribución.[33]

Las metodologías para el desarrollo de software también se pueden agrupar según el formalismo en el proceso de trabajo y la cantidad de documentación a generar en:

A. Metodologías ágiles (Baja ceremonia)
B. Metodologías pesadas (Alta ceremonia)

#### 2.1.4.3 Metodologías ágiles

Las metodologías ágiles hacen algunos sacrificios en términos de la ceremonia y el rigor a favor de la flexibilidad y la habilidad de adaptar y envolver los ambientes de negocios. Éstas enfocan gran valor en producir el software trabajando más que en crear una extensiva y comprensiva documentación. En vez de hacer un plan rígido de trabajo, una de las principales doctrinas de las metodologías ágiles es responder al cambio que ocurre durante el proceso. Esto no significa que la implantación y la documentación no son importantes, es sólo que son tan importantes como asegurarse que el software trabaje o la adaptación del plan de trabajo a la realidad del cambio. Estas metodologías han incrementado su popularidad en desarrolladores individuales y grupos de trabajo pequeños. [34]

Asumen que el factor más importante en el éxito de un proyecto es la calidad de las personas y lo bien que puedan trabajar juntas en términos humanos. Estas metodologías tienden a usar intervalos cortos de tiempo en sus iteraciones, porque no hacen mucho énfasis en documentación pesada. Estas metodologías también se denominan de peso ligero (lightweight) o de baja ceremonia porque el proceso no tiene gran documentación y puntos de control, al considerar que éstos entorpecen los cambios y trabajos que se pueden desarrollar con el talento de las personas.[35]

Las características de estas metodologías las hacen ideales para sistemas pequeños y proyectos menos complejos, aunque algunas de estas metodologías tienen lineamientos adicionales para proyectos más complejos.

Algunas de las metodologías ágiles son: XP, Scrum, Adaptive development (AD), Crystal y GRAPPLE, las que están construidas sobre las buenas prácticas de desarrollo de sistemas" [34]

[33] [QUATRANI00] página 17
[34] [KROLL, KRUCHTEN03] capítulo 3 tema 2
[35] [FOWLER3E] páginas 29-30

#### 2.1.4.4 Metodologías pesadas

Las metodologías pesadas o de alta ceremonia, son procesos con abundante documentación y puntos de control en el proyecto.[36] Crecientemente las compañías están viendo al software como una inversión en las estrategias de negocio y como algo que ofrece resultados predecibles en términos de costos y asuntos de calidad. Como resultado de ello muchas compañías se han vuelto conscientes de la importancia de usar un proceso de desarrollo de software bien definido y bien documentado que facilite el éxito en sus proyectos de software.

Las características de metodologías pesadas son:

1. Las planificaciones son cuidadosamente documentadas.
2. Son producidos muchos artefactos relacionados al manejo, requerimientos, diseños, pruebas y procedimientos.
3. La administración de artefactos de requerimientos, diseños y pruebas son detalladas, documentadas y puestas bajo el control de versiones.
4. Es creado y mantenido un rastreo de enlaces entre los requerimientos, elementos de diseño y artefactos de pruebas.
5. Para los cambios es requerido un tablero de aprobación y control de cambios.
6. Los resultados de inspecciones son cuidadosamente seguidos y registrados.

Algunas de las metodologías pesadas son: El Rational Unified Proces (RUP), Rational Objectory Process (ROP), Coad Yourdon, CMM, CMMI, los desarrollados por el departamento de defensa de los Estados Unidos (DOD), DOD-STD-2167, DOD-STD-2167, MIL-STD-1521B, MIL-STD-498.[37]

[36] [FOWLER3E] páginas 29-30

[37] [KROLL, KRUCHTEN03] capítulo 3 tema 3

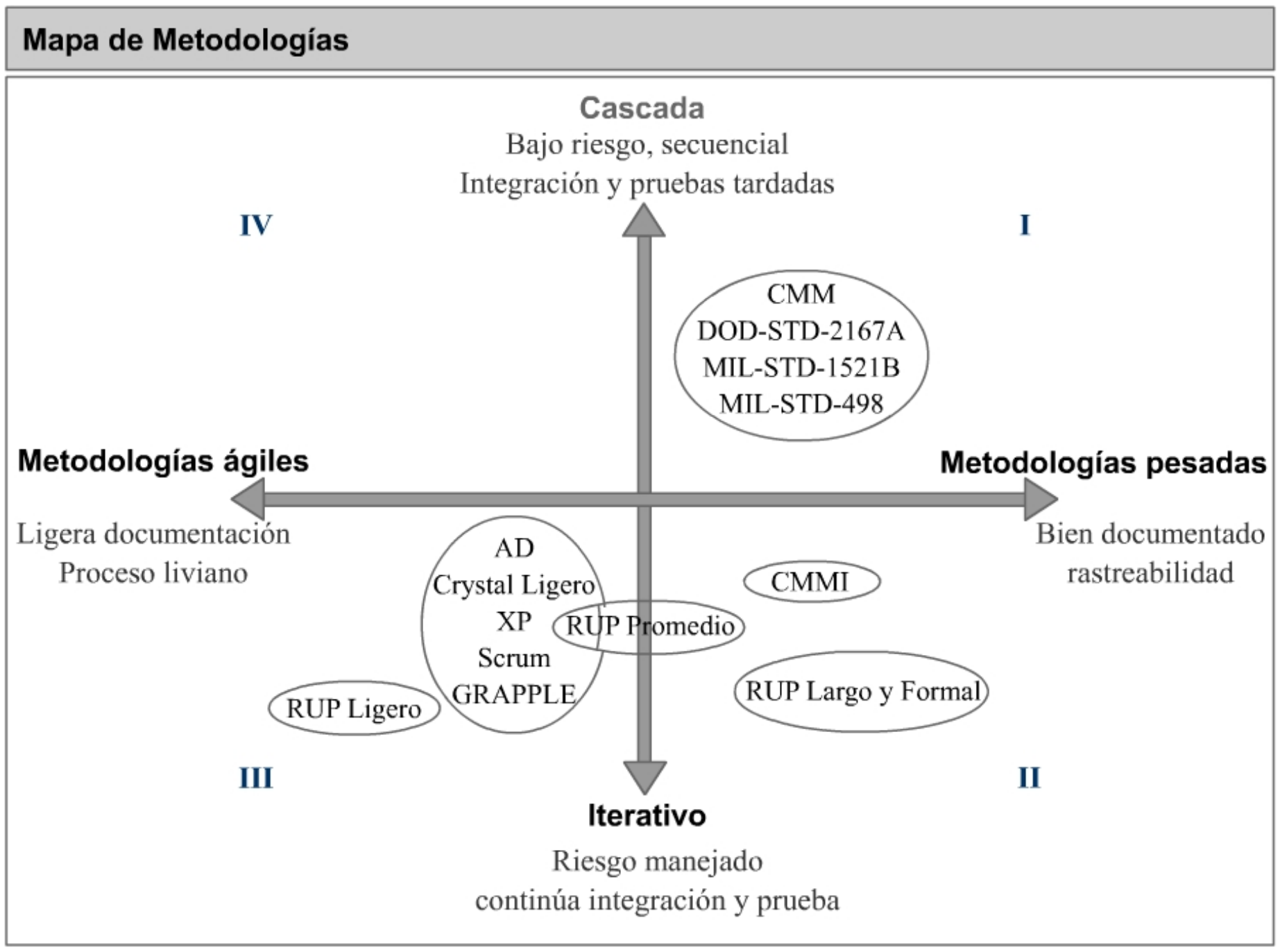

Figura 30. Mapa de Procesos

### 2.1.5 Herramientas para la Ingeniería de Software OO

Cualquier método de desarrollo de software es mejor soportado por una herramienta. Al igual que las herramientas de ingeniería y diseño que utilizan ingenieros de otras disciplinas, las herramientas CASE ayudan a garantizar que la calidad se diseñe antes de llegar a construir el producto. La ingeniería de software asistida por computador (CASE) es el conjunto de herramientas que proporcionan al ingeniero la posibilidad de automatizar actividades manuales como la generación de diagramas, modelos y cierta generación de códigos, además de mejorar su visión general de la ingeniería. [38]

Actualmente existen muchas herramientas CASE, desde simples herramientas de dibujo hasta sofisticadas herramientas de modelado.

Sin embargo, una herramienta que nos genere la sintaxis deseada siguiendo cierta metodología a utilizar en un proyecto de software, pudiera ser tan sencilla como un lápiz y un papel, pero dejaría mucho que desear, un ejemplo sería la problemática que se presentaría a la hora que se necesite cambiar de disposición los elementos de un diagrama ya modelado.

[38] [QUATRANI00] página 20, [PRESSMAN02] página 560

### 2.1.5.1 Características deseables en CASE con UML

Características a encontrar en una herramienta de modelado con sintaxis UML:

1. Una paleta de elementos UML, la cual permitirá la creación de diagramas mediante la selección de elementos para luego ser arrastrados sobre la página de dibujo.

2. Diagramación de tipo banda elástica que permitirá crear conexión entre elementos y que estos elementos se ajusten cuando se arrastran sobre la página.

3. Cajas de diálogo para la edición. Esto es porque los modelos consisten en más que diagramas debido a que los elementos poseen información de sus atributos y características.

4. Diccionario de elementos en el que se grabarán los elementos que se van agregando al diagrama y permitirá reutilizar elementos de un diagrama a otro.

5. Una característica final que presentan algunas herramientas debería ser la generación de códigos desde los modelos.[39]

### 2.1.5.2 Algunas herramientas CASE que soportan la Sintaxis UML

1. Rational Rose
2. Poseidon
3. Microsoft Visual Modeler
4. Microsoft Visio
5. SELECT Enterprise
6. Select Component Architect
7. Visual UML

### 2.1.5.3 Comentarios sobre algunas herramientas CASE

Todas la herramientas tienen interfaz de usuario similar y cumplen con la mayoría de las características deseables para las herramientas case.

**Rational Rose:**
Es el líder en el mercado de las herramientas de modelado. Esta herramienta comparte un estándar común universal, haciendo modelos accesibles a no programadores que desean modelar procesos de negocios, así como la modelación lógica de aplicaciones para los programadores.[40]

[39] [SCHMULLER04] página 511
[40] [SCHMULLER] página 389

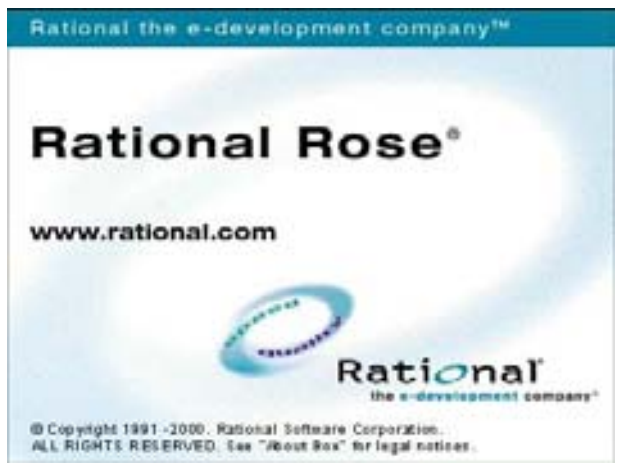

Figura 31. Pantalla de presentación de Rational Rose

**Poseidon:**

Es una herramienta CASE que implementa la sintaxis UML de muy buena calidad. Una ventaja que presenta es que se puede conseguir una versión comunitaria (Free). Los desarrolladores de poseidon dan continuo mantenimiento y mejoras a esta herramienta, como la implementación de las nuevas variantes del UML.

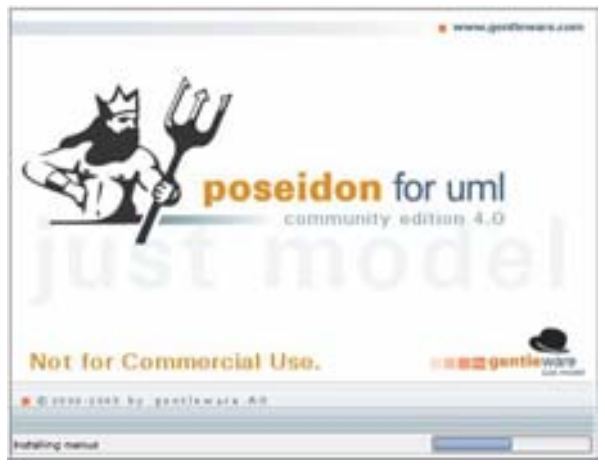

Figura 32. Pantalla de presentación de Poseido 4.0 community Edition

**Microsoft Visual Modeler:**

Es un producto de nivel de entrada, diseñado especialmente para modeladores principiantes. Esta herramienta implementa un subconjunto de los diagramas que dispone el rational rose. Es una derivada del rational rose y producida por la misma compañía que produce el rational rose. Esta herramienta viene integrada en el Visual Studio 6.[41]

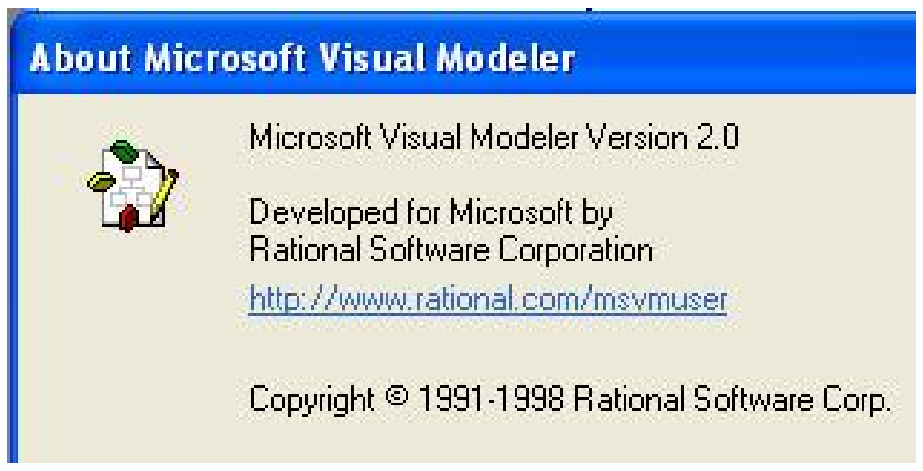

Figura 33. Pantalla de Información de Visual Modeler 2.0

.

[41] [QUATRANI00] página 20

## Microsoft Visio:

Es una de las mejores herramientas de diagramado, Visio versión profesional agrega un número de capacidades UML que lo convierten en una sorprendente herramienta de modelado. UML es sólo una de sus capacidades.[42]

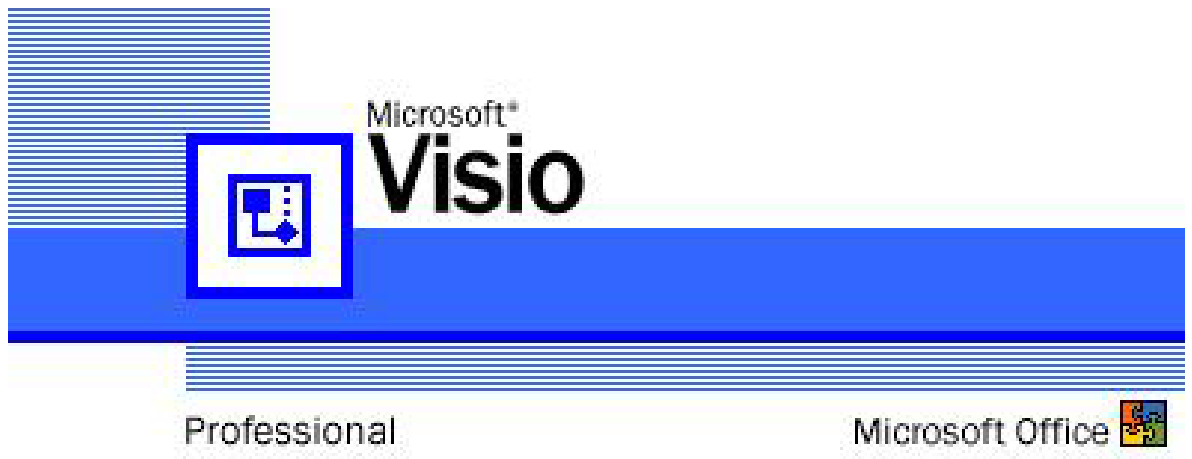

Figura 34. Pantalla de presentación de Microsoft Visio Professional

## SELECT Enterprise:

Ha sido diseñada para ser una herramienta de modelado para toda una empresa. En este sentido automáticamente se conecta con un depósito en red en donde es accesible para los modeladores y los usuarios de los modelos de toda una empresa. Las ventajas de usar este depósito es que se lleva un control de las personas que acceden a los modelos y permite llevar un control de versiones.[43]

## Select Component Architect:

Esta herramienta es la versión extendida y actualizada de Select Enterprise. Select Component Architect. Está engranada hacia el desarrollo vía software de componentes reutilizables y provee extensiones UML para ese propósito, también incluye capacidades para el diseño de base de datos vía diagramas entidad relación.

## Visual UML:

Es una herramienta de ventana abierta, es muy fácil de usar, con esta herramienta se estará hacienda diagramas UML tan pronto como se finalicé la instalación. [44]

[42] [SCHMULLER04] página.512
[43] [SCHMULLER] página 391
[44] [SCHMULLER04] página 536

### 2.1.6 Tecnología de componentes

Los componentes son creados sobre la tecnología de objetos para proveer una mejor vía de reutilización de los mismos.

Algunas veces el término objeto y componente se usa de forma intercambiable, pero existe una importante diferencia entre ellos.

Los objetos son creados por clases. Las clases son hechas de código fuente que define los datos y rutinas que necesita para manipular los datos. Por ser hechas de código fuente, las clases son específicas del lenguaje en que se crearon, y por lo tanto no pueden ser usadas por programas hechos en otros lenguajes.

Los componentes son código binario precompilado. Por ser precompilado son independientes del lenguaje de programación en el que fueron creados.

Esencialmente un componente provee de una serie de servicios que son utilizados por los clientes del componente a través de una o más interfaces. Un componente simple puede consistir de un objeto simple o puede estar hecho de múltiples objetos.

La tecnología orientada a objeto no es tan nueva como la de componentes, pero sólo hasta hace unos años se volvió popular, siendo una de las razones que la técnica de desarrollo facilita la reutilización de código. Desafortunadamente sólo crear objetos en código no fue suficiente para maximizar la reutilización. Es ahí donde entra el diseño orientado a componentes.

Otras técnicas de reutilización de código utilizadas, como la copia de código o las librerías de código reutilizables, pueden ayudar, pero estas técnicas siguen dejando mucho que desear.

La copia de código consiste en recrear una y otra vez muchos de los mismos procesos y procedimientos usados en programas anteriores. Algunos programadores copian el código de un programa existente a un nuevo programa y luego modifican la copia para que el código encaje en los nuevos requerimientos.

Esto implica copiar el código en muchos lugares y si se necesitara cambiar algo en la lógica del código original, esto llevaría a encontrar y cambiar el mismo código en todos lados donde ha sido usado.

La reutilización de código a través de rutinas precompiladas también ayuda, pero es común crear un programa donde se necesite un cambio a una de estas bibliotecas. Desafortunadamente, un cambio en la biblioteca implicaría reconstruir toda la aplicación relacionada a ésta, peor aún, algunas veces la nueva funcionalidad implica ir a cambiar la forma en que utilizan la biblioteca en todos los programas que la emplean y aumenta increíblemente el riesgo de meter fallas (bugs).

Con componentes binarios que empaqueten clases se obtiene un nuevo nivel de reutilización. Se pone el código dentro de un objeto que a su vez está dentro del componente y muchas aplicaciones pueden usar al objeto directamente del componente, beneficiándose así de la facilidad de mantenimiento y de compatibilidad que ofrece la orientación a objeto. Al evolucionar un objeto de negocio con el tiempo, puede ganar nuevas propiedades (datos) o más funcionalidades (métodos, eventos); el punto está en que las propiedades y métodos previos pueden y deben mantenerse iguales para que las aplicaciones existentes no necesiten ser cambiadas. [45]

Sin embargo, existe la limitación de no lograr integración total de componentes producidos por terceras partes. En este sentido y buscando satisfacer esa necesidad de mecanismos estándar e interfaces abiertas, son tres los esfuerzos que más han sobresalido.

1. Microsoft, con sus tecnologías .NET que actualmente se encuentran en su versión 2 y están desarrollando la versión 3; la anterior tecnología de componentes de Microsoft (COM) “Component Object Model” traducido al español “Modelos de Objetos Componente” de la que evoluciono .NET, Distributed Component Object Model (DCOM), Microsoft Transaction Server (MTS)/COM+.

2. Sun Microsystems, que ha presentado Java Beans.

3. Object Management Group, un consorcio integrado por varias industrias importantes, que ha desarrollado CORBA (Common Request Broker Architecture).[46]

[45] [LHOTKA98] páginas 16-20
[46] [RAMIREZ]

## 2.2 ARQUITECTURA LÓGICA EN CAPAS

### 2.1.1 Arquitectura de una aplicación

Arquitectura es un término bastante amplio, que actualmente se está utilizando en una variedad de áreas del conocimiento humano. Desde el concepto más clásico de lo que es arquitectura, tal como sucedía en tiempos de Roma y Grecia, hasta el uso de esta idea en ámbito de los negocios y las finanzas.

Sin embargo, esta tendencia no se limita a ramas sin correlación directa; sino que dentro de la Ingeniería de software el término Arquitectura conlleva en si mismo distintos enfoques. Entres los principales tenemos la Arquitectura física y la Arquitectura lógica. Aunque entre nuestro gremio la división entre lo físico y lógico resulta en algunas ocasiones algo turbio.

Inspirados en la Arquitectura lógica escalable basada en componentes (CSLA) de Rockford Lhotka www.lhotka.net, hacemos eco de la propuesta de separar en dos visiones la Arquitectura de software, la parte lógica y la física. La arquitectura lógica de software trata sobre los tipos de servicios lógicos que provee una aplicación y la forma en que estos se pueden agrupar. La parte física entonces, hace referencia a dónde se encontrarán espacialmente las partes de la aplicación, es decir los servicios lógicos de software como elementos físicos (discos duros, estaciones clientes, servidores y ejecutables).. [47]

Por experiencia laboral y académica, hemos notado que por lo general cuando se habla de arquitectura de una aplicación, automáticamente se piensa en arquitecturas físicas en 3 capas (en el caso de aplicaciones Web) o en cliente servidor (cuando son aplicaciones de escritorio), de tal forma que se ve la arquitectura lógica del software como un todo. Todos los servicios del software revueltos de forma monolítica, lo que lleva al código con lógica directamente detrás de la interfaz del usuario.

En una aplicación común se pueden identificar las siguientes funcionalidades o servicios:

1. Presentación (UI Interfaz de usuario)
2. Procesamiento de la lógica de negocio
3. Procesamiento de los datos

Para comprender mejor la arquitectura lógica de software en capas es necesario primero hablar de la arquitectura monolítica, que además es el tipo de arquitectura de software más utilizada.

### 2.2.2 Arquitectura lógica monolítica

Las aplicaciones tradicionales emplean la arquitectura de software monolítica. Se mezclan los tres elementos anteriormente mencionados como un solo código, teniendo rutinas para manejar la interfaz del usuario, la lógica de procesamiento de la información y el código para hablar con la base de datos (almacenar y recuperar información), como una sola cosa.[48]

[47] [MONOGRA04] página 26, [LHOTKA98] página.51
[48] [LHOTKA98] página 51

Otra técnica que dará igual resultado es centralizar las subrutinas en lugares separados como librerías o módulos de funciones, por el hecho de que el código es directamente invocado desde un evento atrás de la interfaz de usuario. No fomenta la reutilización ya que para usar dichas rutinas en otro proyecto se deben copiar y luego adaptar, lo que vendría a dar problemas de mantenimiento por el simple hecho de tener la misma rutina de variadas funcionalidades en distintos proyectos. En caso de usar el mismo archivo para distintos proyectos y luego modificarlo para un proyecto específico, se corre el riesgo de que el código modificado quiebre otro programa que confiaba en el código anterior. Sin embargo, si la lógica tanto de proceso como de datos estuviera implementada en objetos, se podrían no cambiar las propiedades y métodos ya existentes, sino agregar nuevos, con menor riesgo de quebrar un programa existente.

Según lo abordado autores versados en estos temas, las siguientes tecnologías fomentan el desarrollo de arquitectura lógica de software monolítica:

**(1) La ingeniería de software orientada a datos** y **(2) orientada a procesos,** por el hecho de hacer énfasis en el modelado de los datos como entidades separadas de los procesos, e igualmente modelar los procesos como entidades separadas de sus datos. En ambas metodologías se modelan a que datos van a acceder los procesos luego de haberse identificado ambos (procesos y datos) de forma separada.

La **(3) programación manejada por eventos** cuando típicamente está compuesta de subrutinas que responden a un evento externo. Es diferente cuando el evento llama a un objeto, en este caso trabaja bien porque un evento es básicamente un tipo de mensaje.[49]

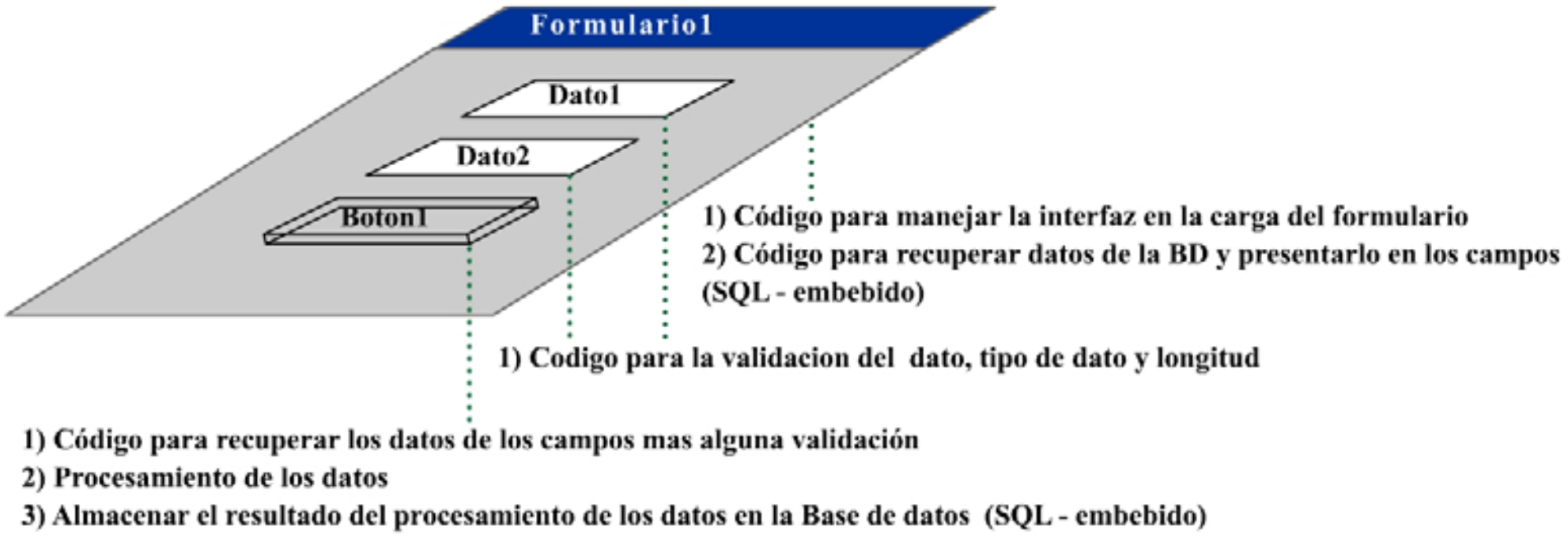

Figura 35. Código Monolítico (de presentación, lógica y persistencia) detrás de la

En la figuran anterior se ilustra como detrás de la interfaz se escribe código monolítico donde se mezclan los servicios de software para el manejo de la interfaz, lógica de negocio y persistencia. Esta apreciación es válida tanto para las aplicaciones Web como de escritorio.

[49] [LHOTKA98] páginas 44-47

La diferencia con respecto a la lógica detrás de la interfaz entre las aplicaciones Web y de escritorio, está en que así como en ambos se embebe SLQ para hablar con la base de datos, en las aplicaciones Web existen lenguajes apartes para crear la interfaz, donde el código de la interfaz es ejecutado en el cliente pero generado desde el servidor; esto significa que el cliente sólo se encarga de interpretarlo y el servidor se encarga de enviarlo ya con su lógica, por lo tanto, aunque pueda existir una separación física de ejecución, la lógica es enviada desde un lugar centralizado.

Algunos lenguajes para generar la interfaz de usuario en aplicaciones Web:

- HTML
- Lenguaje para la generación de HTML Dinámico del lado del servidor (JSP, ASP, ASP.NET o PHP) los que también se utilizan para el procesamiento de la lógica de la aplicación.
- Lenguajes de script, para el manejo dinámico de la interfaz desde el lado del cliente como, java scrtipt, vb script y DHTML

Mientras no se separe lógicamente los servicios del software no importa que existan distintos lenguajes para generar la interfaz, ni importa donde se ejecute dicho código de manejo de interfaz. El código de manejo de interfaz, procesamiento y persistencia de la aplicación seguirá siendo de arquitectura lógica monolítica, donde los tres elementos anteriores se encuentran mezclados. Estos tipos de lenguajes embebidos se pueden apreciar en la figura anterior, el SQL – embebido para hablar con la base de datos.

Con lo ya mencionado no se pretende decir que todas las aplicaciones web sean monolíticas. Siempre y cuando se separe la lógica de los servicios, se tendrá una aplicación web en Capas.

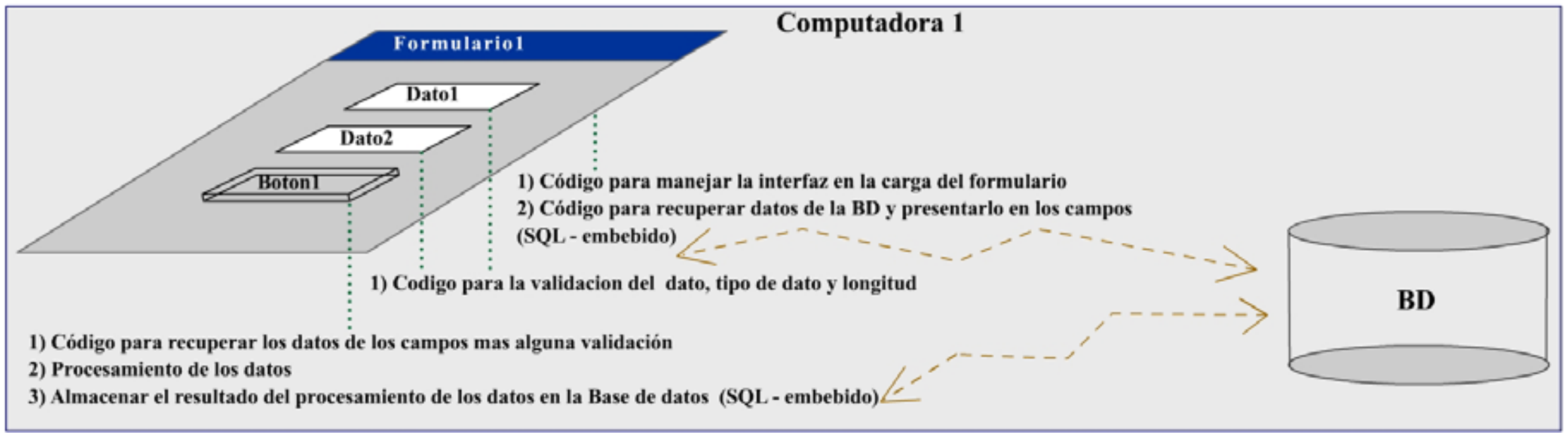

Figura 36. Código Monolítico en una capa física

En la figura anterior se muestra una aplicación mono usuario con código y arquitectura lógica monolítica, donde tanto el software como la base de datos se encuentran en una misma computadora.

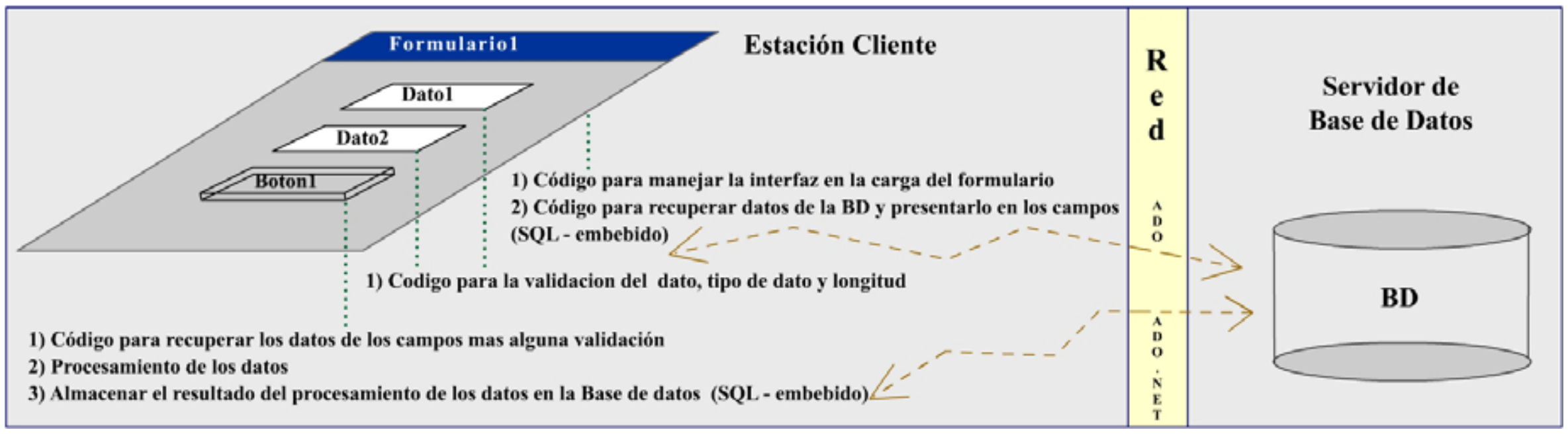

Figura 37. Código Monolítico con la arquitectura Cliente/Servidor más común

En la figura anterior se ilustra la misma aplicación con código y arquitectura monolítica de la figura 36, pero ahora multiusuario corriendo en una arquitectura física de 2 capas Cliente/Servidor, donde se han centralizados los datos, convirtiéndose en un recurso compartido entre todos los usuarios. Como se puede observar, la arquitectura física no hará cambiar la arquitectura lógica sin importar la cantidad de capas físicas que agreguemos; mientras no se separe las funcionalidades del software en capas lógicas, el software seguirá siendo monolítico, mezclado y fuertemente acoplado.

Para evitar tener un código y una arquitectura lógica monolítica, se deben aplicar la tecnología de objetos, IngSOO, POO y así encapsular entidades de negocio que manejen sus propios datos mediante sus propias rutinas.

### 2.2.3 Objetos de negocios y arquitectura lógica en capas

Cuando se realice una separación entre los servicios de software correspondientes al manejo de la interfaz de usuario de la lógica de negocio de la aplicación, se dejará de tener una arquitectura monolítica y se pasará a una arquitectura lógica en capas.

Viendo el mundo real, uno puede darse cuenta que existen objetos que reflejan o representan los negocios. La orientación a objeto permite construir el software como representación del mundo real y los negocios, los que están hechos de objetos físicos y conceptos abstractos. La idea principal de los objetos de negocios es crear representaciones de objetos del mundo real dentro de las aplicaciones.

Un buen diseño de un objeto de negocio contendrá todos los datos y rutinas necesarias para representar una entidad de negocio y podrá ser usado dentro de toda la aplicación. Por supuesto con el tiempo un objeto de negocio puede evolucionar y ganar nuevas propiedades (datos) o más funcionalidades (métodos, eventos), el punto está en que las propiedades y métodos previos pueden y deben mantenerse iguales para que las aplicaciones existentes no necesiten ser cambiadas.

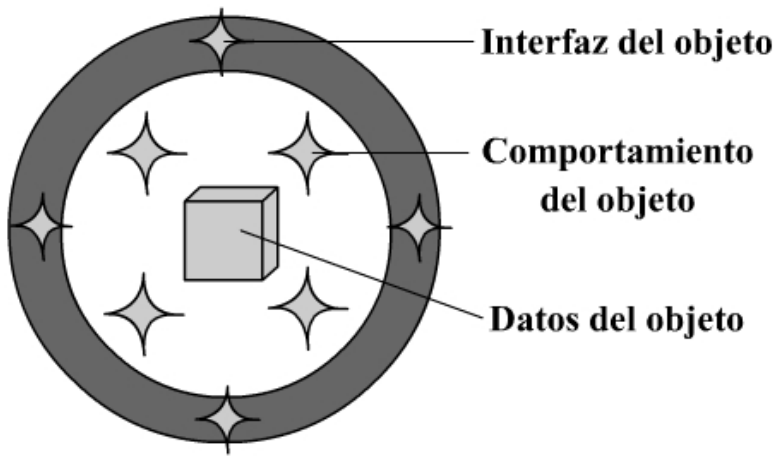

Figura 38. **Objeto de Negocio** (Representación alternativa de una clase orientada a objetos)

Como se observa en la figura anterior, un objeto de negocio es una abstracción basada en código de una regla o entidad del mundo real; es el mismo concepto de una clase u objeto en orientación a objeto, la diferencia radica en el tipo de entidad que se abstrae.

Se explicó anteriormente que una aplicación típica generalmente utiliza arquitectura lógica monolítica, donde casi todo el código y por lo tanto todo el procesamiento y lógica de la aplicación, está mezclada en la interfaz del usuario. También se explicó que otra forma de programar las rutinas de la lógica de la aplicación es poniendo el código en librerías o módulos de rutinas a ser llamadas por la interfaz, pero realmente el procesamiento sigue centrado alrededor de la interfaz.

Esta es la técnica utilizada cuando se trabaja con una programación manejada por eventos. Debido a que los eventos vienen de la interfaz del usuario, tiene sentido que todo el código este ligado a ésta.

Desafortunadamente esa aproximación no llega lejos cuando se desarrolla un modelo orientado a objetos para los programas, donde el foco del software es la interacción entre objetos que modelan entidades y reglas de negocio, no la interacción con eventos provocados por la interfaz.

La clave está en tener en mente que la interfaz del usuario no es la aplicación. La aplicación es la implementación de las reglas y lógicas del negocio, que se abstraerán en objetos de negocio; la interfaz es sólo un medio por el cual el usuario interactúa con los objetos de negocio (aplicación). La meta es generar aplicaciones independientes de la presentación del usuario, lo que significará que la aplicación podrá tener diferentes interfaces de usuario implementadas en diferentes tecnologías.

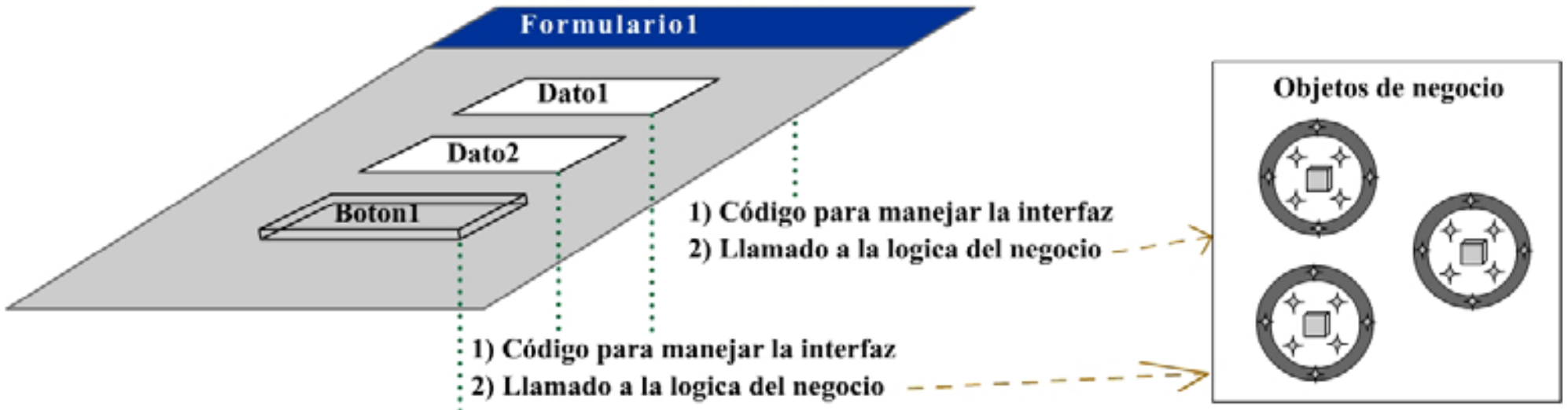

Figura 39. La interfaz del usuario como cliente de la lógica del negocio

En la figura anterior se puede apreciar la interfaz del usuario actuando como cliente de los objetos del negocio. La lógica de procesamiento es implementada por estos objetos y todo trabaja junto para implementar las funcionalidades requeridas para la aplicación.

La interfaz del usuario es la parte de la aplicación que típicamente más cambia. Utilizando objetos de negocio, un cambio en la interfaz es más simple y seguro porque la lógica del negocio es independiente del código de manejo de la interfaz de usuario. De esta forma se podrán hacer cambios a la interfaz del usuario (ya sea mover un control de lugar o cambiar radio buttons por un combo box) sin caer en la problemática que trae escribir lógica de procesamiento directamente detrás de la interfaz del usuario, con lo que probablemente se tendría que arreglar esta lógica de procesamiento para compensar cambios de la interfaz.[50]

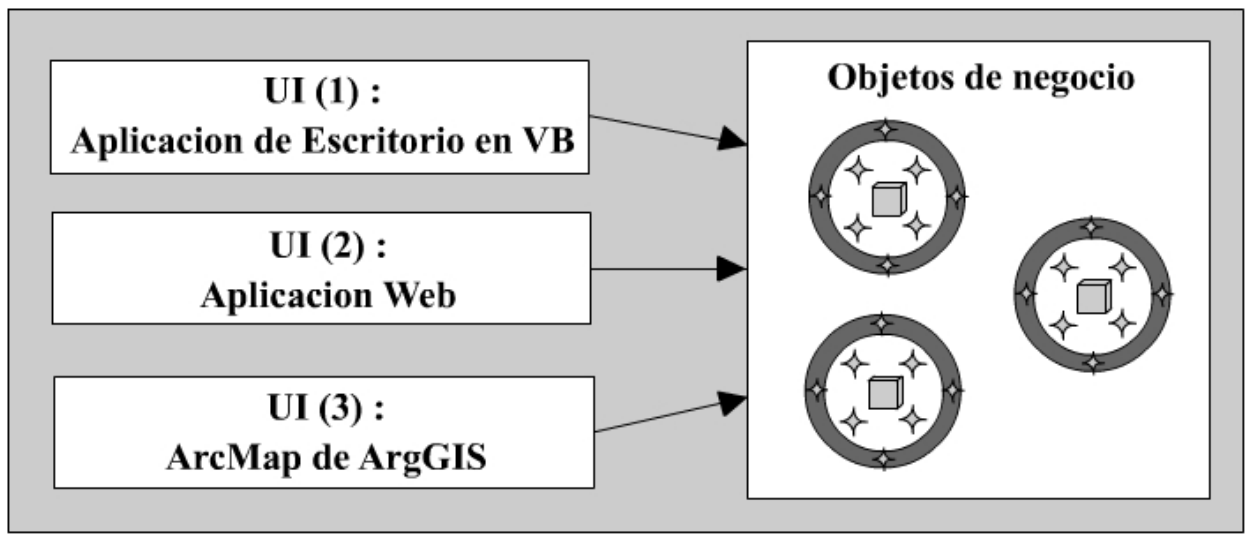

Figura 40. Clara diferenciación de la aplicación en capas lógicas
Tres interfaces de usuario como cliente de la misma lógica del negocio

#### 2.2.3.1 Arquitectura lógica en 2 capas

En esta arquitectura lógica en 2 capas, la lógica del negocio también maneja la persistencia o habla con la Base de datos, lo que es transparente para el programador de la UI (Interfaz de usuario).

De esta forma el programador de la UI sólo habla orientación a objeto con la capa de la lógica del negocio. Por ejemplo, si quisiera salvar los datos de un objeto1 sólo tendría que llamar al método correspondiente **objeto1.salvar** sin tener que escribir nada de SQL. Otro ejemplo pudiera ser el caso en que se presenta mucha información sobre una matriz; en aplicaciones convencionales el programador escribiría SQL, que recuperaría en un objeto particular (recordset en VB) y luego recorrería cada elemento de este objeto para irlo agregando uno a uno a la matriz. Sin embargo, con esta arquitectura en capas, la capa de lógica de negocio implementaría un objeto colección que recuperaría el criterio deseado por el programador de UI, abstrayendo así a éste del contacto con la BD, luego sólo le tocaría recorrer la colección para ir agregando uno a uno los elementos a la matriz.

En las siguientes figuras se puede apreciar la independencia entre las capas lógicas y físicas. En una capa física puede haber una o más capas lógicas.

[50] [LHOTKA98] páginas.8, 16, 51, 63-4

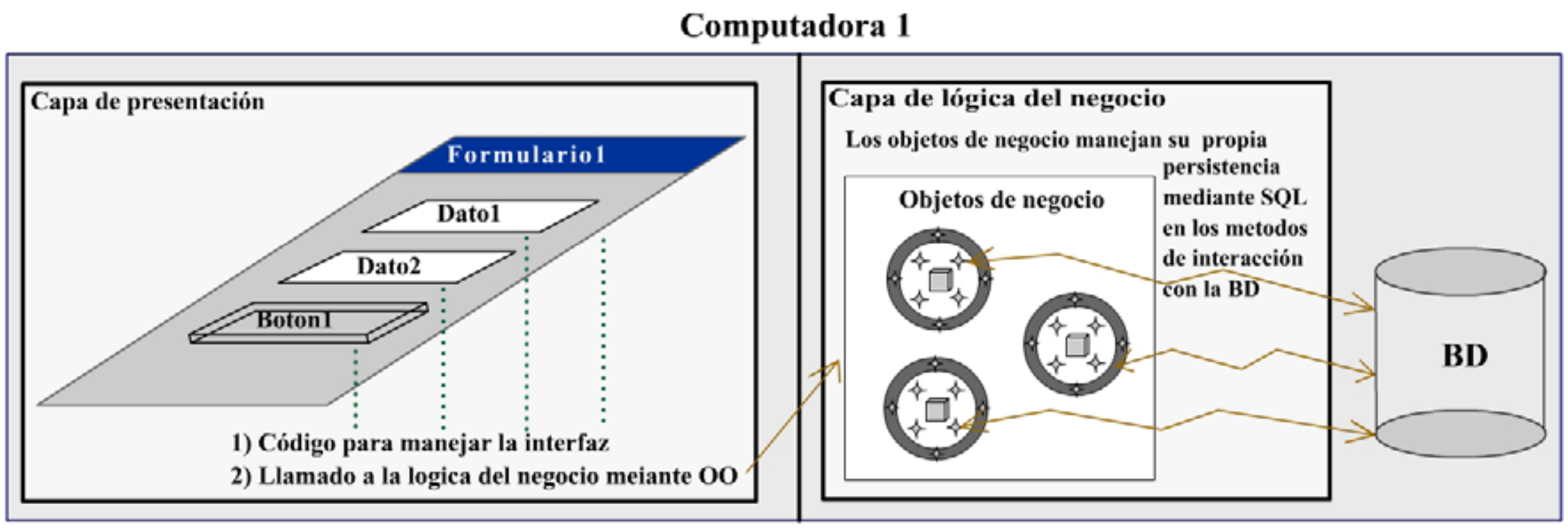

Figura 41. Dos capas lógicas y una capa física (mono usuario)

En la figura anterior se muestra la configuración de una aplicación monousuario con arquitectura lógica en 2 capas, donde ambas capas y la base de datos se encuentran en la misma computadora.

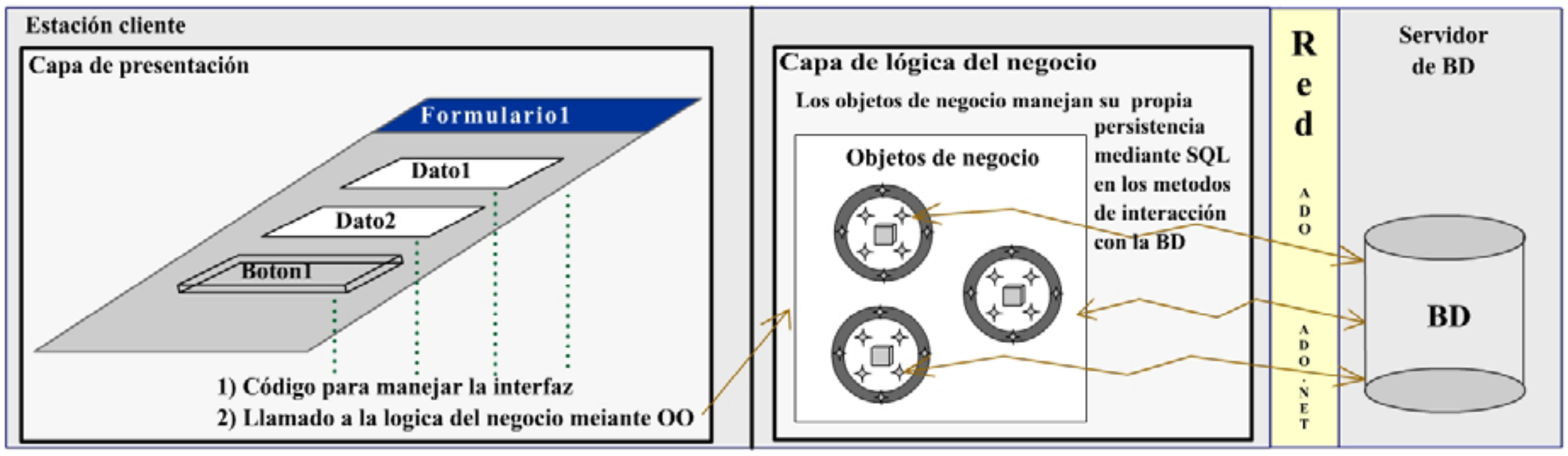

Figura 42. Dos capas lógicas y dos capas físicas (Cliente/Servidor con 2 Capas lógicas)

En la figura anterior se ilustra la misma aplicación con arquitectura lógica en 2 capas de la figura 40, pero ahora multiusuario corriendo en una arquitectura física de 2 capas Cliente/Servidor, donde se han centralizados los datos convirtiéndose en un recurso compartido entre todos los usuarios.

Con el tiempo se volvió aparente que dos niveles físicos (cliente/servidor) simplemente no eran suficientemente poderosos en el manejo de aplicaciones grandes, debido a que cada cliente en las estaciones de trabajo mantiene un diálogo constante con el servidor de base de datos, con lo que el tráfico de la red se vuelve muy alto, convirtiendo el servidor de base de datos en un cuello de botella debido al intento de muchos usuarios de acceder al mismo recurso compartido al mismo tiempo.

Las tres capas físicas ayudan a direccionar el asunto anterior mediante la agregación de otro servidor físico entre los usuarios y la BD. El **servidor de aplicaciones** central puede manejar más eficientemente el tráfico de la red y la carga del servidor de BD, y si además se agrega una conexión de red rápida entre el servidor de aplicaciones y el servidor de BD, se hará a la aplicación aún más eficiente. También es más fácil agregar otros servidores de aplicaciones, lo que requerirá cambios menores en partes de la aplicación, en lugar de regar más servidores de BD, lo que requeriría una división de los datos entre estos últimos.[51]

---
[51] [LHOTKA98] página 64

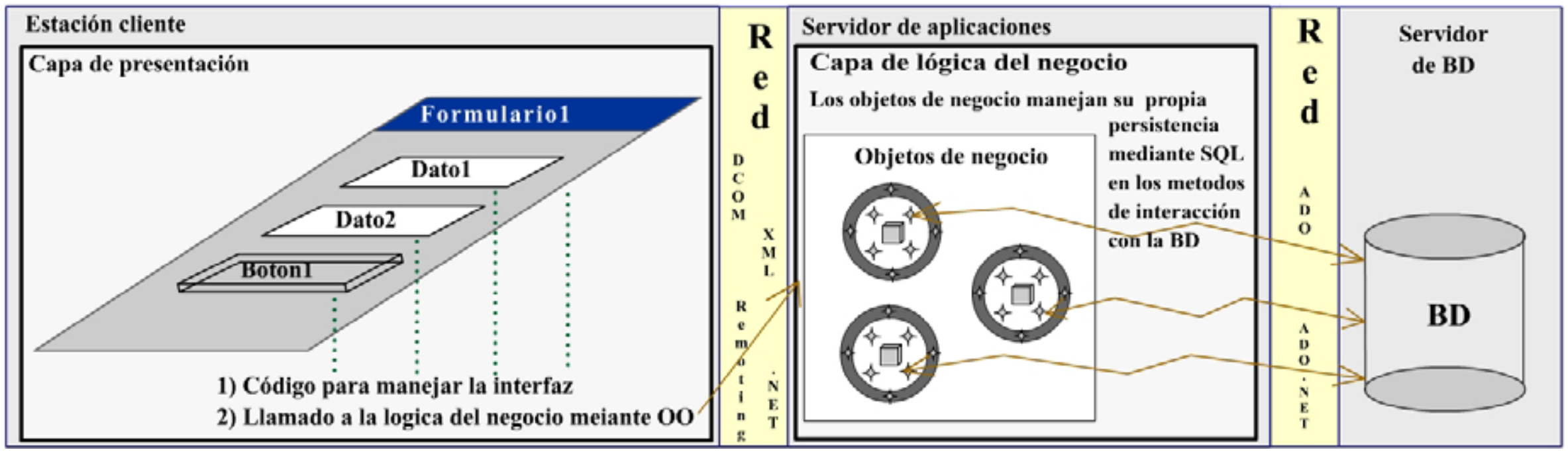

Figura 43. Dos capas lógicas y tres capas físicas

En la figura anterior se muestra la misma aplicación con arquitectura lógica en 2 capas de las figuras 41 y 42, sólo que ahora se ha agregado un nivel físico más, teniendo un total de 3 capas físicas.

### 2.2.3.2 Arquitectura lógica en 3 capas basada en CSLA

La teoría y las ilustraciones del tema 2.2.3.1 (Arquitectura lógica en 2 capas) han servido para aclarar la diferencia entre las capas lógicas y físicas; sin embargo, también hay que tener en cuenta otros aspectos como la persistencia de los objetos, el rendimiento y la escalabilidad.

Recordemos que el punto de clave para la arquitectura lógica en capas, esta en que la UI es solamente un cliente de la lógica del negocio. La lógica del negocio es el corazón de la aplicación y la UI es secundaria.

Esto hace fácil cambiar la UI sin arriesgar la lógica del negocio almacenada en objetos de negocio y permite generar aplicaciones menos acopladas.

Siguiendo la Arquitectura lógica escalable basada en componentes CSLA por sus siglas en inglés (Component-Based Scalable Logical Architecture), si desde una visión lógica se identifica el comportamiento de la persistencia en los objetos de negocio y se separa del resto del comportamiento de la lógica del negocio pura, se obtienen dos perspectivas distintas de cada objeto de negocio, una para servir a la Interfaz de usuario compuesta solamente de la pura lógica de la aplicación y la otra para manejar la persistencia del objeto y hacer que sobreviva en el tiempo.

Lo que realmente se pretende es dividir los objetos de negocios simples en dos objetos conceptuales, ambos con los mismos datos internos, pero cada uno con sus elementos de interfaz y comportamiento, sin olvidar que ambas partes son conceptualmente dos mitades del mismo objeto.[52]

[52] [LHOTKA98] páginas 138, 438-440

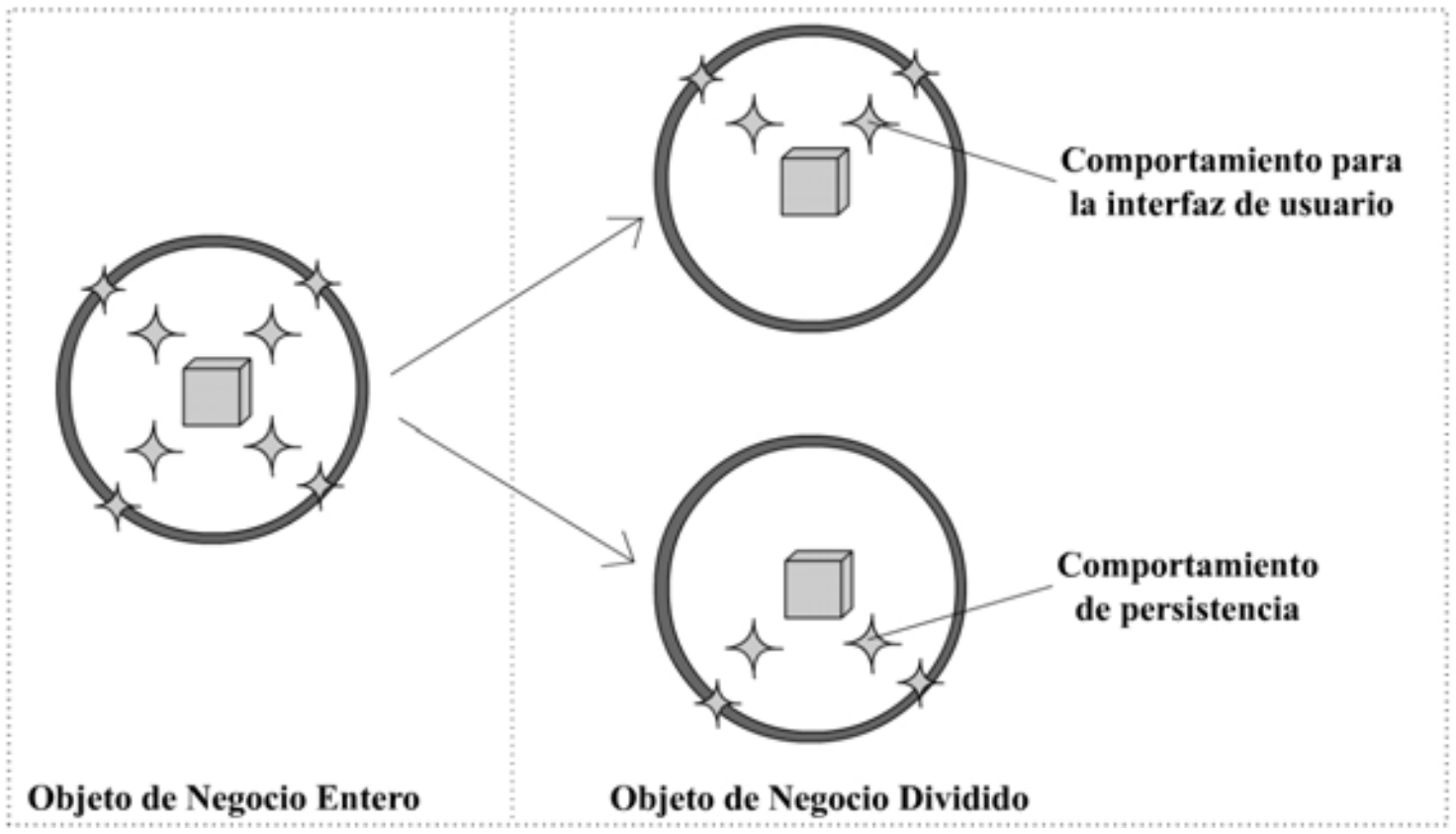

Figura 44. Objeto de Negocio Divido en dos objetos físicos continua siendo un objeto conceptual

Esta separación lógica del objeto de negocio permitirá de forma alternativa ejecutar en computadoras diferentes con carga de red aceptable las mitades generadas (lógica del negocio y persistencia).

Mediante la ejecución de la persistencia lo más cerca posible al repositorio de datos, cargando y salvando de una sola vez el estado del objeto, se evita aumentar el tráfico de la red. Y en caso de aplicaciones de muchos usuarios en línea se abre la posibilidad de tener varias computadoras o servidores de aplicaciones que sirvan como muros de contención para balancear la carga hacia la BD. Una vez ejecutados los procesos de persistencias el objeto puede ser enviado de forma serializada a la computadora que corre la lógica del negocio lo mas contiguo a la UI, para tener una interfaz rica e interactiva y evitar que sea de tipo Batch.

Las tecnologías de componentes anteriores, como DCOM y COM+, al ser utilizadas de forma tradicional se repercute en el rendimiento de la aplicación debido a que los objetos siempre son pasados por referencia. Lo que significa que al pasar un objeto de una máquina o proceso a otro, el objeto siempre se mantiene en su proceso o computadora original (remota), mientras que la computadora solicitante lo que recibe es un puntero o referencia de dicho objeto. Lo anterior implica, que cualquier llamada a propiedades o métodos del objeto remoto debe ser enviada a través de la red y procesada remotamente, luego el resultado es devuelto al objeto que hizo la solicitud. Para evitar lo anterior las dos capas lógica del objeto de negocio (persistencia y lógica del negocio pura) se deben ejecutar en dos objetos físicos y realizar su comunicación de forma serializada una vez por cada solicitud de persistencia.

Cabe recalcar que en **CSLA.NET** no es necesario realizar dicha separación física de un mismo objeto lógico, debido a que en ese framework se utilizan las potencialidades del remoting que ofrece el .NET framework de Microsoft, para hacer objetos móviles y enviar todo el objeto de una máquina a otra y luego simplemente ejecutar ciertos procesos en el equipo remoto deseado para así reenviarse a su computadora origen. También en .NET es posible instanciar objetos remotos por referencia y existen casos donde lógicamente esto es deseado.

Según Lhotka, la ganancia de tener el mismo objeto lógico en un solo objeto físico, tiene el inconveniente de confundir el límite entre la lógica del negocio pura y la lógica del manejo de la persistencia o acceso a los datos, principalmente para los que están aprendiendo a construir sus aplicaciones en 3 o más capas. [53]

Los siguientes pasos sirven al proceso de identificación y división lógica de los objetos en la capa de la persistencia y la capa de la lógica del negocio pura:

**Separar las rutinas de orientación a datos:**
Realizar la división de las rutinas en los objetos de negocio no es difícil, lo único que se debe hacer es agrupar las sentencias y rutinas de orientación a datos en métodos específicos para el propósito de la persistencia del objeto.

**Manejar los datos compartidos:**
Lo siguiente solo aplica en el caso de utilizar tecnologías DCOM y COM+, donde la división lógica del objeto de negocio también debe caer en una división física. Cuando ambas partes del objeto se encuentran en un simple módulo de clase se puede compartir el estado de las variables. Sin embargo, al dividir el objeto en dos partes es necesario comunicar el estado relevante de los datos en el nuevo objeto, tendremos 2 partes del procesamiento que manejan su comunicación internamente.

Aunque se haya divido la lógica del negocio en dos partes lógicas, estas continúan siendo integrales. De hecho, deben de trabajar eficientemente juntas, dado que por si solas ninguna de las dos partes proveerá toda la lógica de aplicación requerida. Esencialmente se debe aparentar como si es una única y entera capa de software, de tal forma que ni la capa de presentación ni la BD (con sus Servicios de datos, procedimiento almacenado y “Triggers”) están conscientes que la capa de lógica del negocio ha sido dividida.

La división de la lógica del negocio en dos partes genera dos nuevas capas:

1. Objetos de Negocio Centrados en la UI (**centrados-en-UI objetos negocio**)
2. Objetos de Negocio Centrados en los Datos (**centrados-en-Datos objetos negocio**)

Para generar nuestro nuevo gráfico que represente esta arquitectura en capas, se procederá a llamar a la interfaz de usuario como **Capa de Presentación** y con el propósito de ilustrar la comunicación que existe entre la capa de lógica del negocio y la Base de Datos (**Base de Datos**), se procederá a mostrar a la BD con todos sus servicios de Procedimientos almacenados y disparadores debajo de todas las capas lógicas y en color blanco[54].

[53] [LHOTKA05] páginas xx, xx
[54] [LHOTKA98] páginas 61, 62, 439

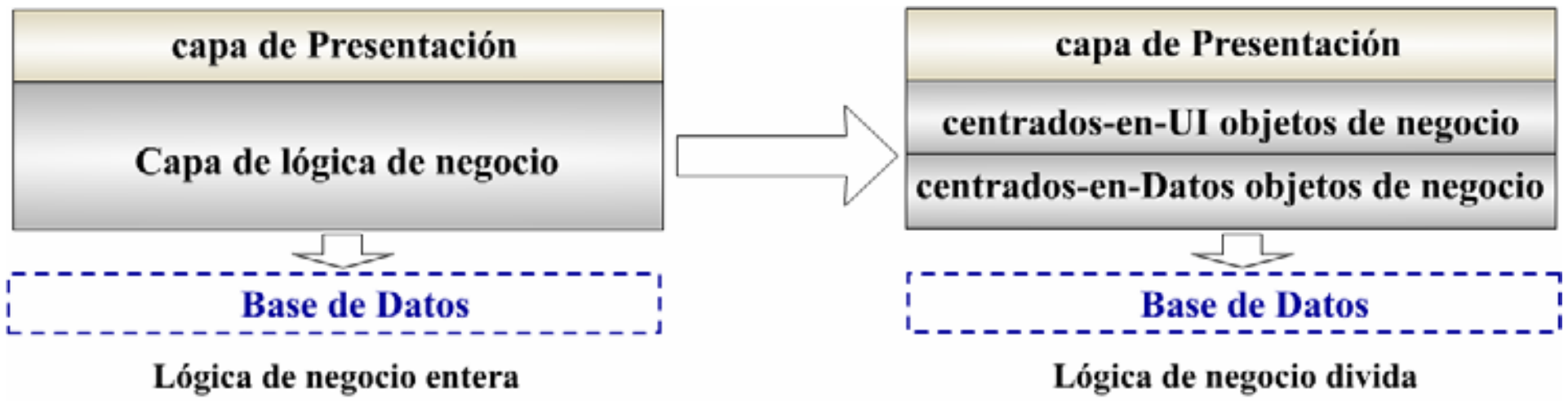

Figura 45. Particionamiento interno de la lógica de negocio

#### 2.2.3.2.1 Capa de Presentación

Esta es la parte más visible de toda la aplicación: la interfaz de usuario. Tradicionalmente está compuesta de formularios, menús y controles que los usuarios usan para interactuar con el programa. Por eso es razonable expandir esta definición para que alcance en los ambientes actuales. La presentación generada para un tipo de aplicación con arquitectura lógica de capas no sólo podrá estar formada de interfaz de escritorio, sino también de paginas html, dentro de un mapa ArcMap de ArcGIS, un procesador de palabra u otro tipo de programa que soporte comunicación con los paquetes donde se almacena la lógica del negocio.

#### 2.2.3.2.2 Objetos de negocio centrados en la UI

En vez de poner todo el papel importante del código de la aplicación en la interfaz del usuario, se mueven hacia los objetos de negocio. Al hacer esto se podrán aplicar técnicas de diseño orientado a objeto para construir un modelo de objetos y poner las reglas de negocio y de procesamiento en los objetos de este modelo.

Los objetos de negocio centrados en la UI, deberían ser diseñados para reflejar de forma acertada las entidades de negocio del mundo real que las aplicaciones necesitan modelar. Esto significa que cada objeto debería representar alguna entidad junto con sus atributos, métodos y reglas de negocio que afectan dicha entidad.

El conjunto de modelo de objeto también debería forzar las reglas de negocio con respecto a cómo los objetos interactúan.

Cualquier interfaz de usuario interactuará exclusivamente con los objetos de negocios centrados en la UI, y estos objetos no sólo implementarán y forzarán todas las reglas del negocio, también proveerán al programador de la UI todas las capacidades requeridas para crear una rica e interactiva interfaz de usuario.

#### 2.2.3.2.3 Objetos de negocio centrados en los Datos

La capa de objetos de negocios centrados en la UI provee al desarrollador de ésta con un modelo intuitivo y confiable de los objetos del mundo real de nuestro negocio. Los objetos centrados en los datos necesitan proveer a los objetos centrados en la UI con un mecanismo robusto y poderoso con el que los datos podrán ser salvados y recuperados como se necesita que sea. Esto significa que cualquier interacción con la fuente de datos será manejada por los objetos de negocios centrados en los datos; los objetos de negocio del lado del cliente no tendrán idea de cómo los datos son almacenados o recuperados.

Mediante la división del procesamiento de los datos fuera de los objetos de negocio, se proveerá con un gran convenio de flexibilidad en términos de cómo desarrollar la aplicación. Se puede escoger implementar los objetos de negocio centrados en los datos como un componente que corre en la máquina de trabajo del usuario o poner el componente en otra máquina en la red.

Aparte de todo eso, los objetos de negocio centrados en los datos permitirán escudar a la aplicación sobre las fuentes de datos, las que pueden ser base de datos relacionales o múltiples bases de datos y de distintos tipos. Siguiendo la misma línea, si luego se necesita migrar de un tipo de BD a otro, entonces sólo se necesitará adaptar o cambiar los objetos de negocio centrados en los datos para ahora hablar con la nueva BD, sin ningún impacto sobre los objetos de negocios centrados en la UI o en la interfaz del usuario.

#### 2.2.3.2.4 Base de Datos

Como mínimo esta capa es responsable de la creación, actualización, recuperación y borrado de los datos de la fuente de datos. En algunos casos también incluye características más complejas, como forzar la integridad referencial y otras reglas relacionadas a los datos.

La BD aparte de almacenar físicamente a los datos, brinda **servicios de datos,** como Procedimientos almacenados (Stored Procedures SP) y Disparadores (Triggers). Esto permite ubicar un poco de procesamiento en la BD, lo que a menudo provee beneficios en el rendimiento, gracias a que muchos servidores de bases de datos SQL optimizan los procedimientos almacenados para correr eficientemente.

Sin embargo, es importante notar que es más apropiado (salvo los casos en que el rendimiento se vea muy beneficiado de tener lógica dentro de la BD) poner la mayoría del trabajo que el servidor de base de datos pueda hacer en los objetos de negocio centrados en los datos, debido a que los lenguajes complementarios al SQL que implementan las BDs no se pueden comparar con las flexibilidades, como la depuración que puede ofrecer un lenguaje de programación; además desde dentro de ellos (lenguajes complementarios al SQL que implementan las BDs) es difícil implementar lógica de negocio como entidades de negocio y más difícil aún hacerla independiente del ambiente.[55]

[55] [LHOTKA98] página 62, 64-68

#### 2.2.3.2.5 Distribuciones de las capas lógicas basadas en CSLA en capas físicas

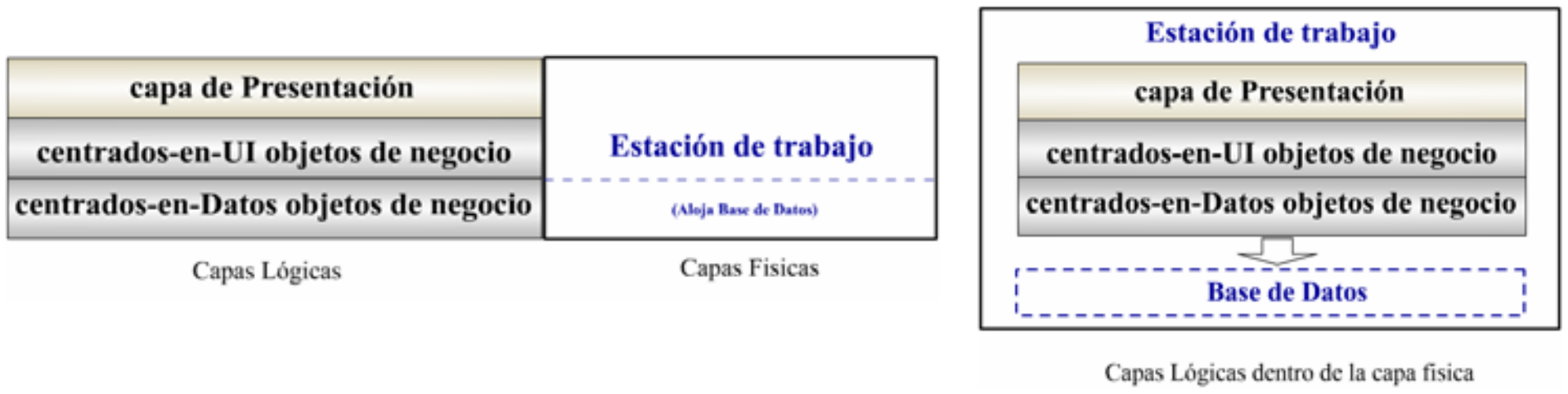

Figura 46. Dos vistas de la misma arquitectura de Tres capas Lógicas y Una capa Física - alojando a la Base de Datos (Aplicación mono usuario en tres capas lógicas)

En la figura anterior con una capa física y tres capas lógicas, se muestra como independientemente de la arquitectura física de una aplicación, se puede generar una arquitectura lógica avanzada. De esta forma se logra tener un aplicación más desacoplada y fácil de mantener, con la posibilidad de hacerla trabajar con distintas interfaces y distintos tipos o múltiples orígenes de datos, además con la facilidad de poder distribuir sus capas lógicas a distintos computadores en la red y así aumentar su escalabilidad; todo esto sin necesidad de volver a escribir la aplicación y con mínimos o ningún cambio.

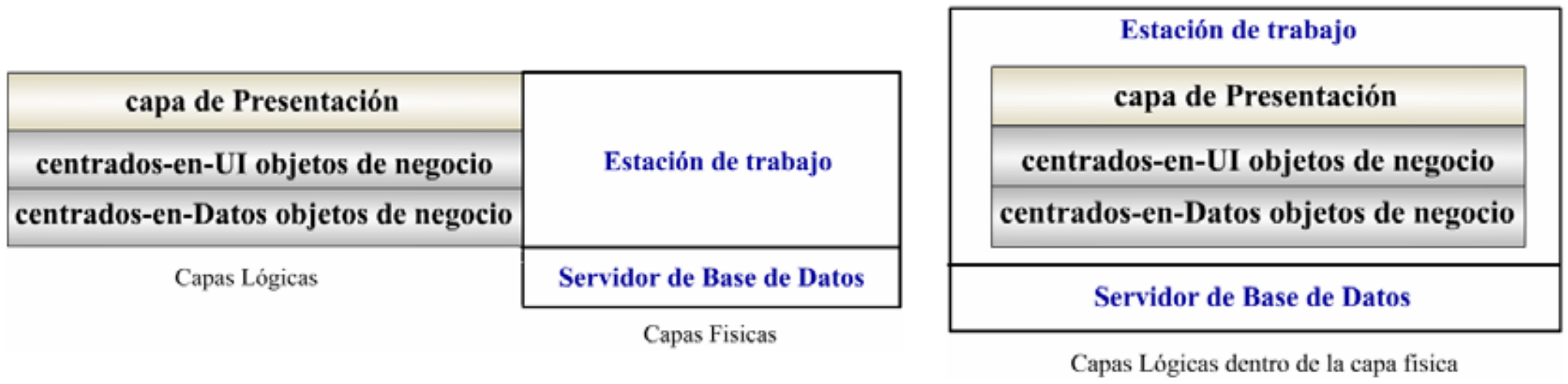

Figura 47. Dos vistas de la misma arquitectura de Tres capas lógicas y Dos capas físicas tradicionales (Cliente

La figura anterior muestra la más común arquitectura física cliente/servidor, donde se pone todo el procesamiento en el cliente, excepto la BD en si. Sin embargo, con esta configuración el servidor de base de datos se puede volver un cuello de botella cuando muchos clientes están intentando hacer uso de los servicios de datos. Esto ocurrirá muy comúnmente con aplicaciones de muchos usuarios.

El punto está en que mientras más procesamiento se delegue al cliente más datos son traídos desde el servidor para ser procesados. De igual forma, si no se pone procesamiento en el cliente, se tiene que enviar mucho material sobre la red por ser el servidor, donde el servidor se convierte en encargado de crear cada pantalla vista por el usuario y las que necesitan ser enviadas por la red.

Idealmente se debe buscar en balance entre el procesamiento en el servidor, el procesamiento en el cliente y la carga de la red.

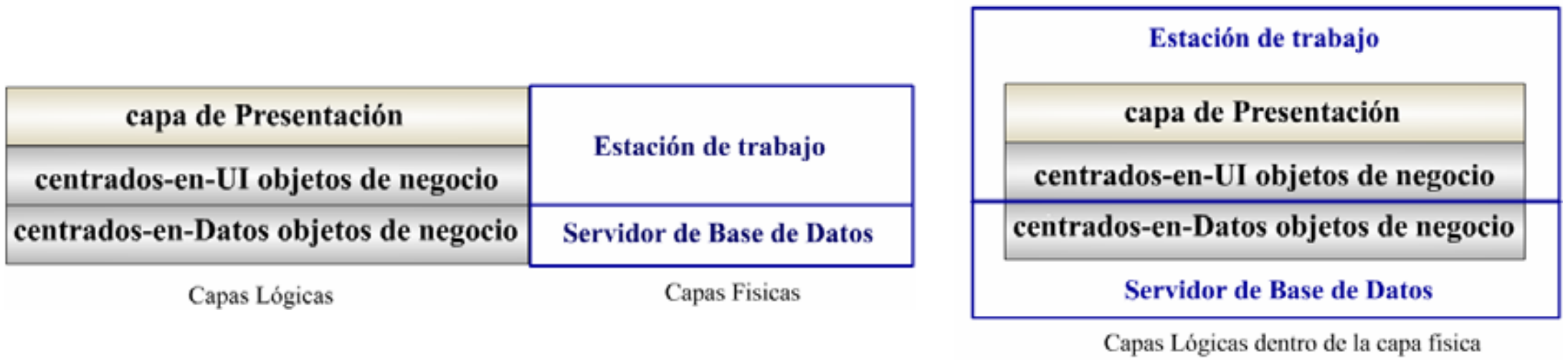

Figura 48. Dos vistas de la misma arquitectura de Tres capas lógicas y Dos capas físicas (procesamiento

En la figura anterior se presenta una configuración física de dos capas alternativas, donde en caso que el servidor de base de datos soporte extra procesamiento, pudiera funcionar bien poner algunos servicios en el servidor de base de datos para proveer procesamiento centralizado o compartido.

Este modelo pone mucho procesamiento extra en el servidor de base de datos. Antes de saltar a esta configuración como la solución perfecta, es necesario evaluar si la máquina podrá manejar el extra procesamiento sin convertirse en un cuello de botella.

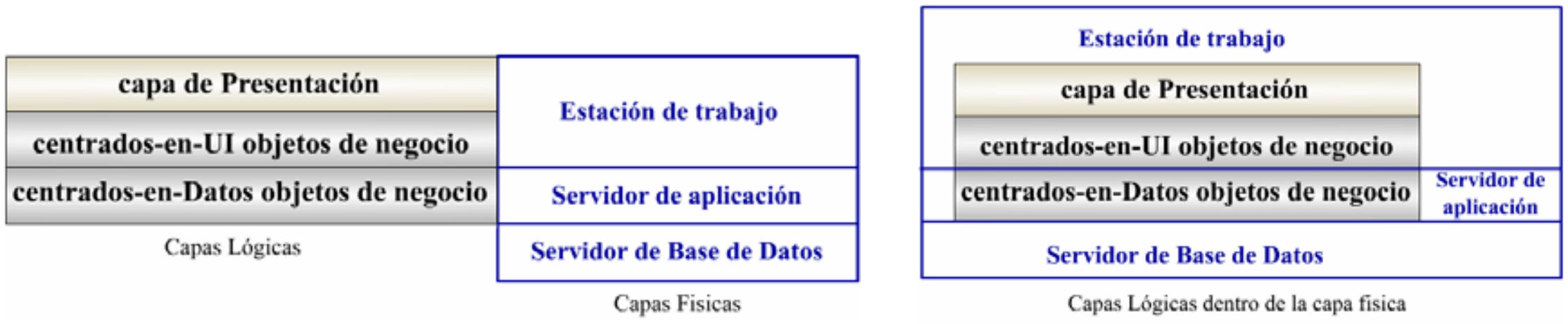

Figura 49. Dos vistas de la misma arquitectura de Tres capas lógicas y Tres capas físicas

En la configuración mostrada en la figura de arriba la estación de trabajo cliente se comunica con el servidor de aplicaciones, y el servidor de aplicaciones se comunica con el servidor de base de datos que soporta la base de datos.

Esta es una configuración muy poderosa y flexible, porque si se necesita agregar otra base de datos se puede hacer sin cambiar los clientes; si se necesita más rendimiento sólo se debe agregar otro servidor de aplicaciones con poco o ningún cambio en los clientes.

Temas de rendimiento juegan un rol importante en la decisión de donde ubicar las partes de la aplicación. En un mundo ideal tendría sentido poner todas las reglas del negocio en un lugar centralizado desde donde pueda ser fácil darle mantenimiento, pero esto llevaría a una interfaz de usuario orientada a batch debido a que ninguna parte de la lógica del negocio de la aplicación se encuentra cerca del cliente. Esto no es la rica interfaz que los usuarios desean. Además de eso, al mover todo el procesamiento fuera de las estaciones clientes se estaría desperdiciando el poder de las computadoras de los usuarios, sería contra productivo ignorar dicho poder.

Otra forma sería poner los objetos en ambas partes, en el cliente y el servidor. Esta es la principal promesa de la CSLA, con la división de los objetos de negocios entre Objetos centrados en la UI y objetos centrados en los Datos; poner una mitad de los objetos en el cliente y la otra mitad en el servidor. Esto lleva a tomar consideraciones para una correcta comunicación entre objetos de distintas máquinas en la red y minimizarle el tráfico.

Independientemente de los argumentos del rendimiento, siempre se debería de mantener físicamente los objetos cerca de donde interactúan más con su entorno: la interfaz del usuario cerca del usuario (en la estación cliente) y el procesamiento de los datos cerca de los datos (los servicios de datos cerca del servidor de datos). Manteniendo los objetos en el lugar correcto se puede evitar comunicación en la red y ganar rendimiento. [56]

[56] [LHOTKA98] páginas .114-117

## 2.3 MAPEO OBJETO-RELACIONAL

El mapeo OBJETO-RELACIONAL trata sobre la integración de tecnologías orientadas a objeto con base de datos relacionales, específicamente sobre las transformaciones para convertir un diagrama de clases de UML a un modelo conceptual de una base de datos relacional. La tendencia sobre la programación y técnicas de diseño de software es hacia la orientación a objeto, pero se siguen utilizando manejadores de base de datos relacionales RDBMS. De hecho las RDBMS son la base en la que residen los datos de la mayoría de las aplicaciones.[57]

Al seguir el paradigma orientado a objeto para las aplicaciones, se pueden generar mayores niveles de abstracción en los modelos de datos. Con este paradigma se generan una serie de modelos que permiten plasmar los requerimientos de una aplicación, generar su diseño e implementarlo. El modelo de clases generado de dicho proceso es el que se relaciona con los datos y a partir de este diagrama y los requerimientos del sistema se puede lograr el modelo conceptual en una RDBMS aplicando el mapeo objeto-relacional.

Esta idea representa un cambio importante en la forma de modelar los datos. El paradigma orientado a datos y las bases de datos relacionales corrientes usualmente son modelos altamente normalizados con poca abstracción. Cada “cosa de interés” es instanciada como una tabla relacional. Como resultado, los sistemas frecuentemente requieren de cientos de tablas de base de datos e igual número de módulos de pantallas y reportes. Los módulos de programas están basados directamente en estas tablas, con el flujo de trabajo del usuario instanciado, solamente a través de la forma en que los cientos de módulos interactúan. Sin embargo, la incorporación del pensamiento objeto-relacional, proveyendo una serie de abstracciones distintas, potencian sus capacidades.[58]

Se ha probado estadísticamente que el tiempo de vida de los datos es mucho mayor que el tiempo promedio de vida de la aplicación que los manipula. Así, la mayoría de las aplicaciones desarrolladas tienen que convivir con algún esquema de datos presente, aumentando la importancia de la tecnología utilizada para modelarlo y persistirlo.

Las bases de datos orientadas a objetos son quizás la forma más sencilla de persistir un modelo de objetos, aunque el mercado de las tecnologías de bases de datos orientadas a objetos es aún pequeño e inestable comparado con el mercado de las bases de datos relacionales. Sólo las bases de datos relacionales han demostrado ser escalables, robustas y lo suficientemente estándares para las aplicaciones empresariales.[59]

El mapeo objeto relacional encaja perfectamente en la capa de **objetos de negocio centrada en los Datos**, que se desarrolla siguiendo el modelo de arquitectura lógica de software en tres capas, debido a que es esta capa a la que se le ha delegado la persistencia de los objetos y ella es la encargada de abstraer el tipo de fuente de datos que utiliza la aplicación.

---

[57] [VB2THEMAX]
[58] [GOLOBISKY03]
[59] [PEMOBD] página 9

### 2.3.1 Mapeo de una clase simple

Por cada clase se crea una tabla, cada fila guarda los datos de un objeto de dicha clase y cada columna es una propiedad del objeto. Para hacer referencia al objeto de forma unívoca, ya sea que esté cargado en la memoria o no lo esté, se utiliza la clave primaria definida en la BD. Esta clave primaria puede ser una característica que identifique al objeto en el mundo real de forma unívoca, o en el caso de ser un objeto que en el mundo real no cuente con esta característica, se puede usar un valor que no se repita generado por la BD.

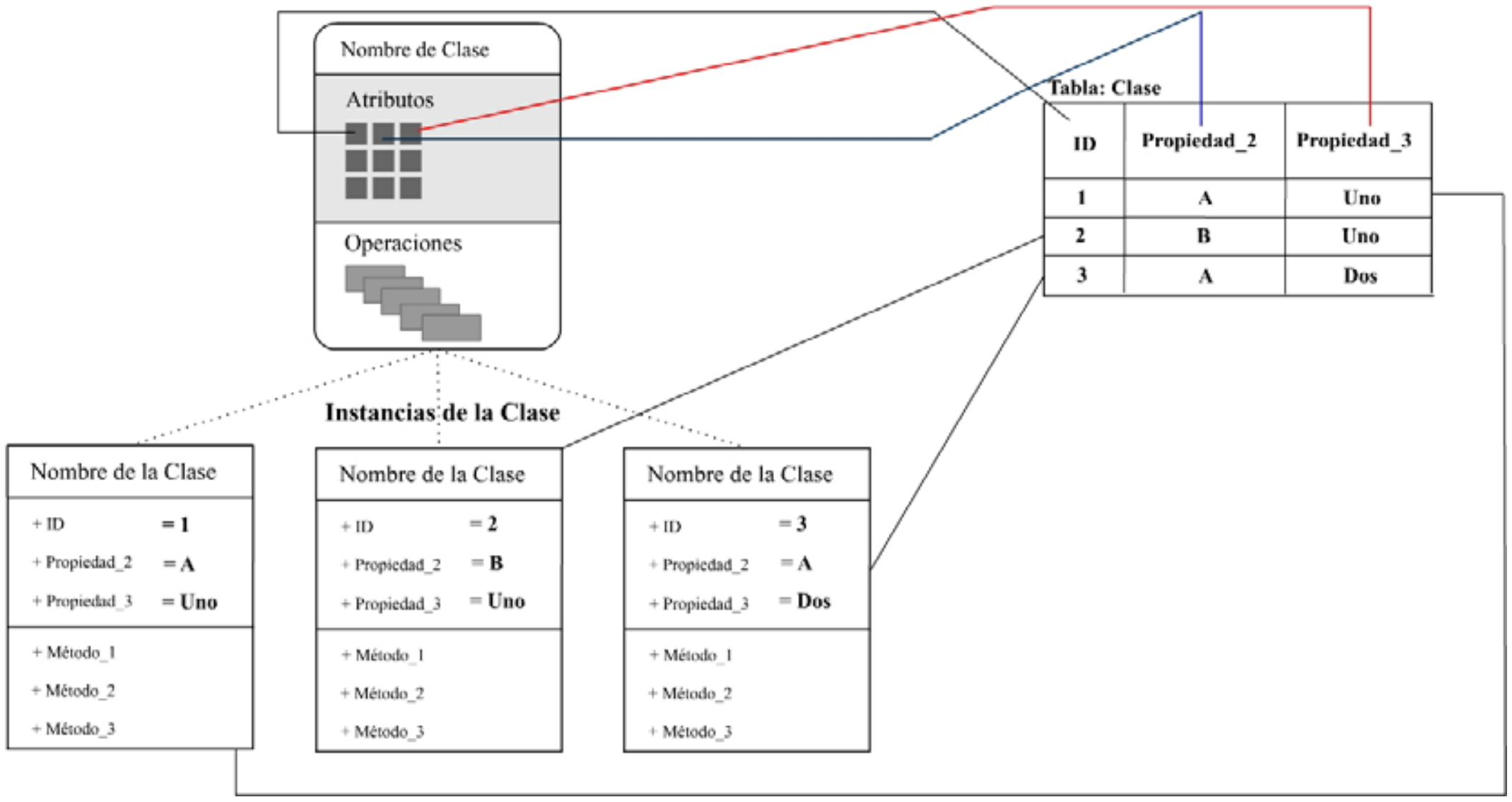

Figura 50. Mapeo de una clase contra una tabla de la BD

Este es el más simple de los mapeos. Se puede ver como una relación de tipo cliente servidor, donde el cliente es el programador que consume directamente dicho objeto. El cliente simplemente declara una variable para crear el objeto, el que se recupera desde una tabla de la BD. Luego se llaman a las propiedades y métodos de dicho objeto; cuando el usuario ha terminado de trabajar con el objeto simplemente lo libera. Los objetos que poseen este comportamiento vienen de clase de tipo CoClass, clases que se pueden instanciar desde cualquier lugar con la simple creación de un nuevo objeto de este tipo.

En la figura anterior se muestra el mapeo objeto relacional más sencillo, cuando se mapea una clase contra una tabla de la BD. Sin embargo, por lo general un diagrama de clases nunca está compuesto de clases aisladas sino de clases relacionadas. También ocurre que un objeto puede tomar sus datos de varias tablas cuando posee un estado y de otras tablas cuando posee otro estado; en este tipo de situaciones es donde se debe extender el concepto anterior.

### 2.3.2 Mapeo de clases relacionadas

Este tipo de mapeo se lleva a cabo cuando se presentan relaciones de asociación simple (relación "usa") de una o dos vías, agregación simple (relación "contiene" o "hecha de") o agregación compuesta (relación "posee" o "tiene"), con sus distintos indicadores de cardinalidades entre las distintas clases relacionadas.

#### 2.3.2.1 Cardinalidad 1 a 1

**Con relación de asociación simple**

Este tipo de mapeo está presente en las asociaciones simples (relaciones de tipo "uses") y también en las relaciones de agregaciones simples (donde un objeto está hecho de otros objetos, aquí una clase simplemente tiene más responsabilidad que otra). En este tipo de asociación el cliente es otro objeto, el cual tiene declarada una propiedad que le retorna el objeto de asociación o agregación simple, que también se recupera desde una tabla de la BD usando la misma lógica (2.31) "mapeo de una clase simple". Luego se llaman a las propiedades y métodos de dicho objeto; cuando se ha dejado de utilizar simplemente se libera. Los objetos que poseen este comportamiento vienen de clase de tipo Class, clases u objetos que sólo se pueden instanciar desde otro objeto.

Desde el enfoque del DOO simplemente se debe tener un propiedad que retorne el objeto asociado; desde el enfoque de RDBMS cada clase tendrá su propia tabla y la clase relacionada deberá tener una clave externa en su tabla que haga referencia a la clave primaria que identifica al objeto relacionador. En caso de que la relación entre ambas clases sean bidireccional significa que ambas clases relacionadas deben de poseer una propiedad que les retorne su objeto relacionado, igualmente ambas tablas en la RDBMS deben de tener una clave externa que haga referencia a la clave primaria de la tabla relacionadora.

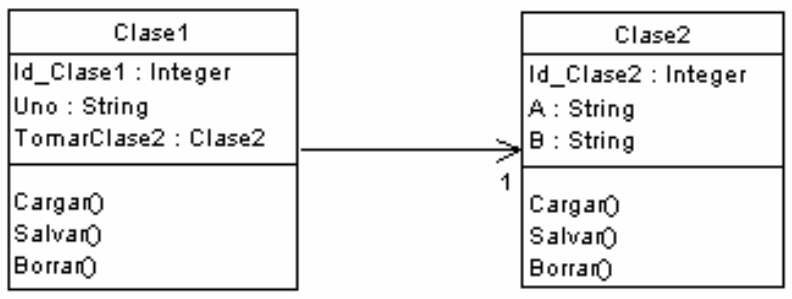

Figura 51. Clases con relación simple unidireccional uno a uno

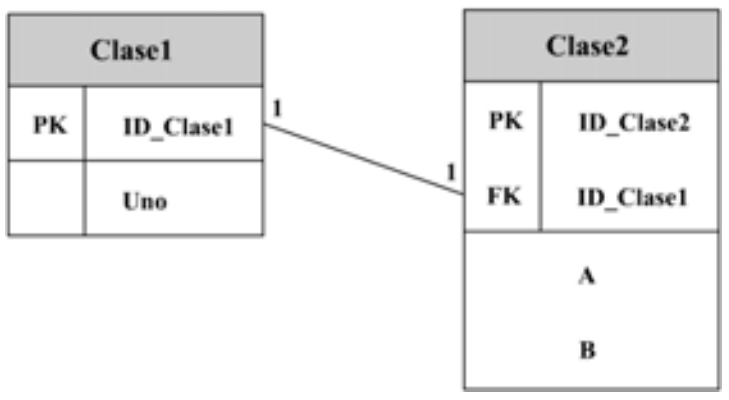

Figura 52. Mapeo correspondiente en tablas de un RDBMS del diagrama de clases de la figura 50

Desde el enfoque de programación, la clase 1 presentará una propiedad que retorna un objeto de tipo clase 2. Dicha propiedad envía un mensaje al método cargar de la clase 2 y le pasa como primer parámetro vacío y como segundo su propio ID. De esta forma la Clase 2 le retornará su objeto correspondiente al que relaciona Clase 1. A continuación se muestran los pseudocódigos correspondientes a la propiedad TomarClase2 de la Clase1.

```
Public Property Get TomarClase2() As Clase2

Static objClase2 As Clase2
 If objClase2 Is Nothing Then
     Set objClase2 = New Clase2
     objClase2.Cargar , ID_Clase1
 End If
 Set TomarClase2 = objClase2

End Property
```

Figura 53. pseudocódigo del la propiedad TomarClase2 de la Clase1

A continuación se muestra el algoritmo de la capa de acceso a los datos del método Cargar de la clase 2 (método que es llamado desde la propiedad TomarClase2 de la Clase1 en la figura 52), el que recibe como parámetro el ID del objeto, luego va a la BD a recuperarlo para así retornar el objeto correspondiente.

```
Public Function Cargar(ByVal ID_Clase2 As Integer, ByVal ID_Clase1 As Integer) As Clase2
' El objeto se puede cargar pasándole por parámetro el ID de la Clase1 y/o el ID de la Clase2

    Dim rs As Recordset
    Dim strSQL As String
    Dim strWHERE As String
    Static objClase2 As Clase2

    strSQL = "SELECT * FROM CLASE2 WHERE "

    If Len(Trim(ID_Clase2)) > 0 Then _
    ' Si se quiere cargar mediante su propio ID
            strWHERE = "ID_Clase2 = " & ID_Clase2

    If Len(Trim(ID_Clase1)) > 0 Then
    ' Si se quiere cargar mediante el ID del objeto que lo asocia
            strWHERE = "ID_Clase1 = " & ID_Clase1
        End If
    End If

    ' Se agregan los parámetros a la consulta
    strSQL = strSQL & strWHERE

    Set rs = New Recordset
    rs.Open strSQL, cn

    If Not rs.EOF Then
    ' Si se retorno algún registro, entones es que existe objeto asociado

        Set objClase2 = New Clase2
        ' Trasladar los valores de la BD a los atributos del Objeto
        objClase2.ID_Clase2 = rs("ID_Clase2"): objClase2.A = rs("A"): objClase2.B = rs("B")

    End If
    rs.Close

    ' Retornar el Objeto con sus atributos cargados
    Cargar = objClase2
    ' Retorna Nothing, cuando no se entro al ultimo if donde se hace la instanciación

End Function
```

Figura 54. pseudocódigo del método Cargar de la Clase2

### 2.3.2.2 Cardinalidad 1 a M

Este tipo de relación por lo general se presenta en la agregación compuesta, o padre hijo, donde el hijo no puede vivir sin su padre (relación "tiene"). También se da en las asociaciones simples con cardinalidad 1 a M (relación "usa") donde el objeto del lado (1) no es propietario de los objetos del lado (M).

Enfocados en el DOO, se debe tener una propiedad en la clase del lado (1) que retorne una colección de objetos de la clase del lado (M) correspondientes al objeto del lado (1).

Por el lado del RDBMS se debe tener una tabla por cada clase, con una relación conceptual de uno a muchos. La tabla del lado (M) deberá tener una clave externa en su tabla que haga referencia a la clave primaria que identifica al objeto relacionador. En caso de que la relación entre ambas clases sea bidireccional, significa que ambas clases relacionadas deben poseer una propiedad que les retorne su objeto relacionado. Igualmente ambas tablas en la RDBMS deben de tener una clave externa que haga referencia a la clave primaria de la tabla relacionadora.

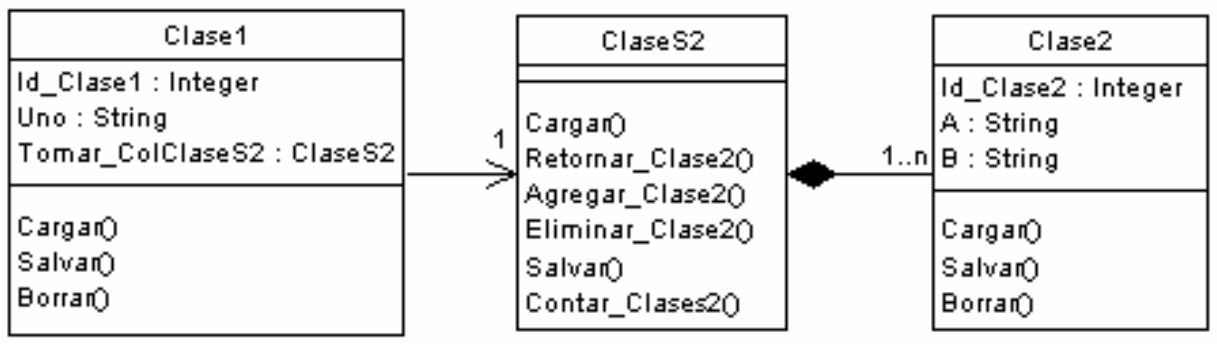

Figura 55. Clases con relación de 1 a M, utilizando una clase de colección intermedia

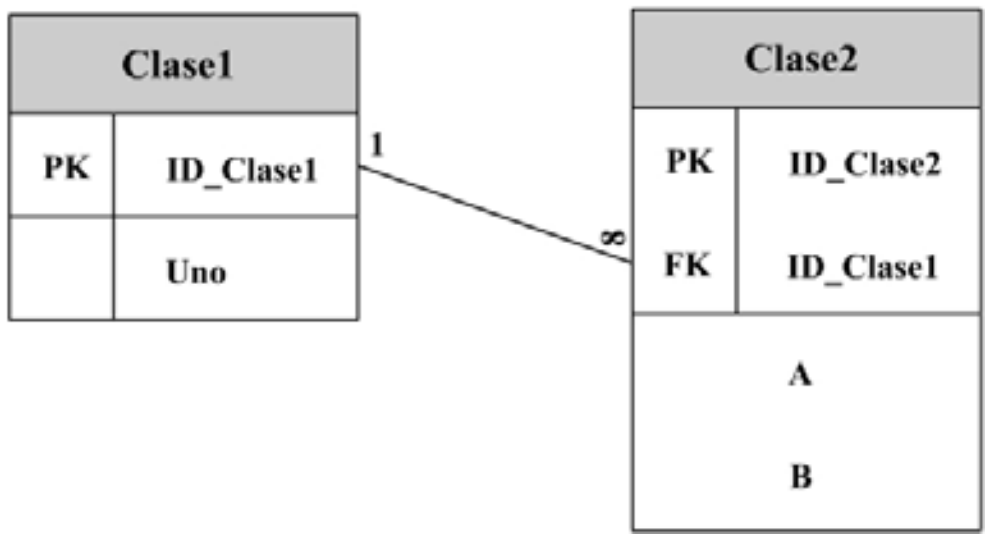

Figura 56. Mapeo correspondiente en tablas de un RDBMS del diagrama de clases de la figura 55

```
Public Property Get Tomar_ColClaseS2() As ClaseS2

Static objClaseS2 As ClaseS2
 If objClaseS2 Is Nothing Then
     Set objClase2 = New ClaseS2
     objClaseS2.Cargar ID_Clase1
 End If
 Set Tomar_ColClaseS2 = objClaseS2

End Property
```

Figura 57. pseudocódigo del la propiedad Tomar_ColClaseS2 de la Clase1

El proceso seguido en el algoritmo mostrado en la figura anterior es el siguiente: se muestra el pseudocódigo de la propiedad Tomar_ColClaseS2 de la clase1. La propiedad manda un mensaje al método cargar de la claseS2 (en plural) y le pasa como parámetro su propio ID. Dicho método retorna un objeto de tipo ClaseS2, el cual es un objeto de tipo colección que contiene sólo objetos del tipo clase2, asociados al objeto de tipo clase1 correspondiente. Luego este objeto colección es retornado por la propiedad.

A continuación se muestra el algoritmo del método Cargar de la claseS2 (método que es llamado desde la propiedad Tomar_ColClaseS2 de la Clase1 en la figura 57). Recibe como parámetro el ID del objeto de la clase1, luego va a la BD a recuperar los datos relacionados a los objetos asociados al pasado por parámetro, para luego convertirlos a objetos y subirlos a la colección de objetos de tipo clase2 que retorna dicho método.

```
Public Function Cargar(ByVal ID_Clase1 As Integer) As ClaseS2
' Este objeto colección se cargar pasándole por parámetro el ID de la Clase1

    Dim rs As Recordset
    Dim strSQL As String
    mColClaseS2 As ClaseS2
    objClase2 As Clase2

    strSQL = "SELECT * FROM CLASE2 WHERE ID_Clase1 = " & ID_Clase1

    Set rs = New Recordset
    rs.Open strSQL, cn

    Do Until rs.EOF

        Set objClase2 = New Clase2
        ' Trasladar los valores de la BD a los atributos del Objeto
        objClase2.ID_Clase2 = rs("ID_Clase2"): objClase2.A = rs("A"): objClase2.B = rs("B")
        'Agregar el objeto a la Coleccion
        mColClaseS2.Add objClase2

        'Moviendo al siguiente registro retornado desde la BD
        rs.MoveNext
    Loop

    rs.Close
    ' Retornar el Objeto colección con elementos cargados
    Cargar = mColClaseS2
    'Retorna Nothing, cuando no se entro ni una vez al Do donde se hace la instanciación

End Function
```

Figura 58. pseudocódigo del método Cargar de la ClaseS2

### 2.3.2.3 Cardinalidad N a M:

Este tipo de relación por lo general se presenta en la asociación simple con cardinalidad N a M (relación tipo “usa”), lo que significa que un objeto no es propietario del otro, sólo utiliza sus servicios.

Desde el enfoque de la OO este tipo de relación es equivalente a tener dos relaciones 1 a M, uno por cada clase (ambos sentidos) y con la misma lógica que se explicó en (2.3.2.2) para la cardinalidad 1 a M. Sin embargo, desde el punto de vista RDBMS se le debe agregar una tabla de relación en la que se ubicarán las claves primarias de cada una de las tablas que representa una clase; dicha tabla creará el muchos a muchos entre ambas tablas relacionadoras.
Desde el punto de vista de programación, los algoritmos son igual a los explicados en las ilustraciones 57 y 56, con la diferencia que ahora ambas clases principales deben poseer la propiedad que le retorne el objeto de tipo colección que desea que se le retorne.

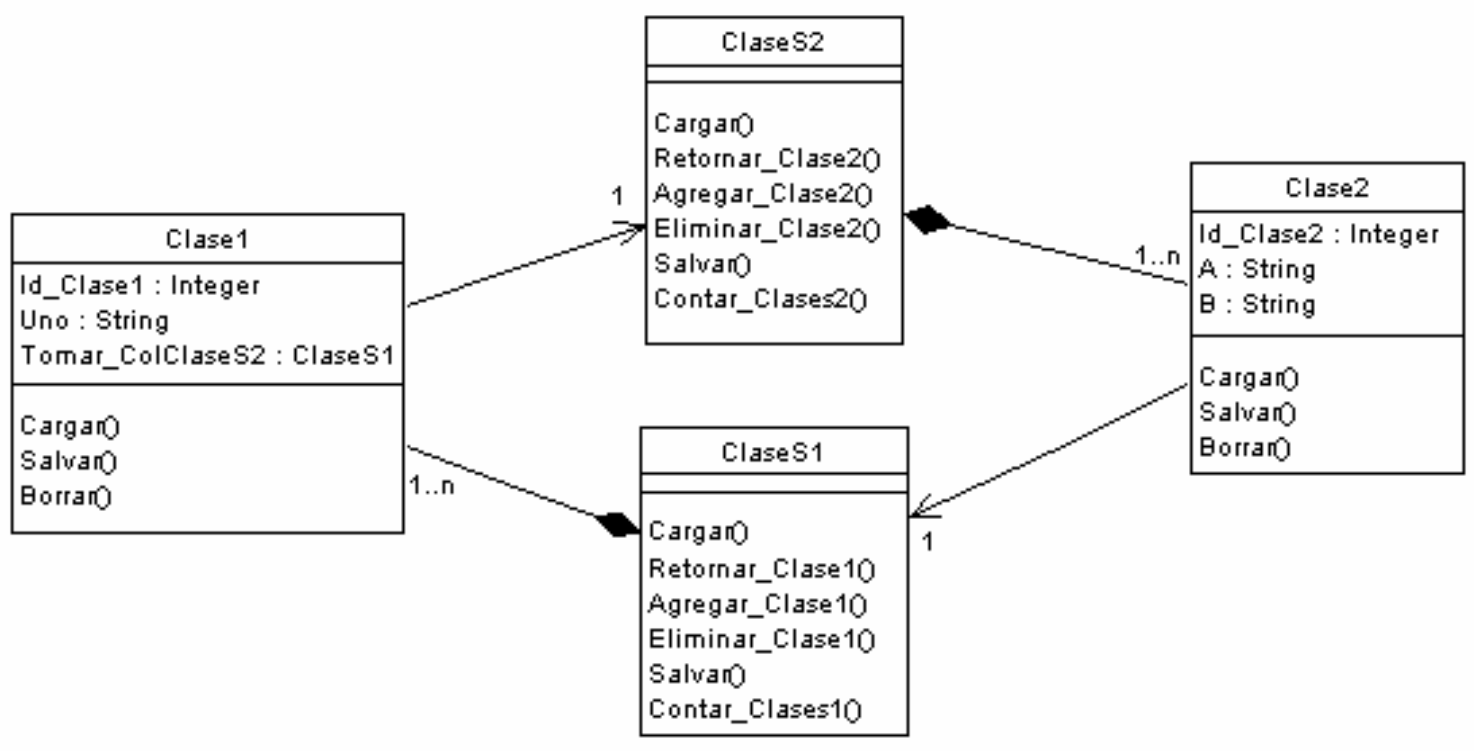

Figura 59. Clases con relación de N a M, utilizando clases de colección intermedias

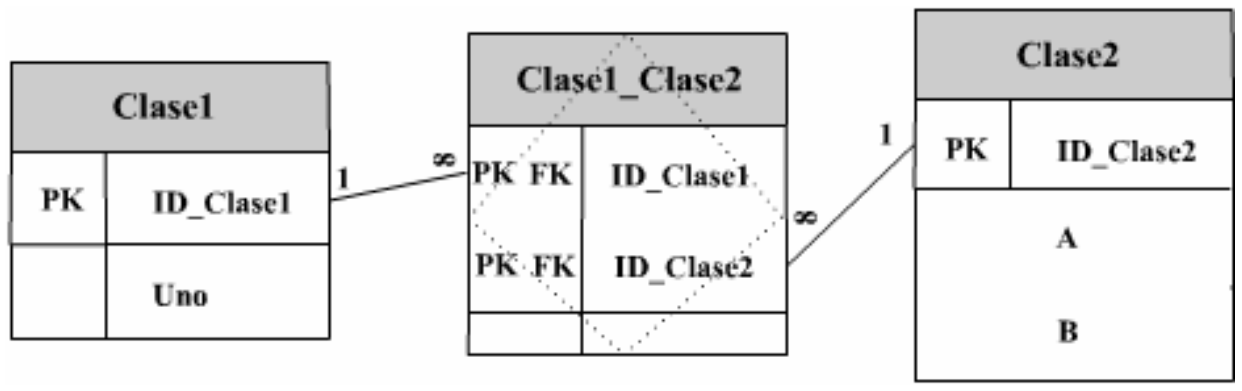

Figura 60. Mapeo correspondiente en tablas de un RDBMS del diagrama de clases de la figura 59

### 2.3.3 Mapeo de la herencia

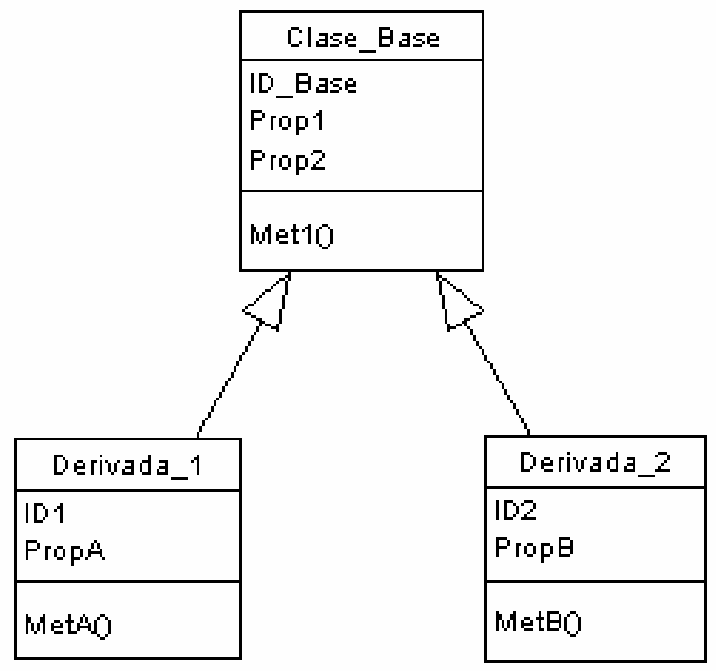

Figura 61. Relación de herencia, de la Clase_Base a las clases derivadas, donde la Clase_Base es abstracta

Para mapear la jerarquía de clases existen tres estrategias:

1. Una tabla por cada clase
2. Una tabla por cada clase concreta (no abstracta *)
3. Una tabla para cada jerarquía de clases

*: Una clase es abstracta cuando no se puede usar para crear un nuevo objeto; es una clase que sólo posee definiciones de propiedades y métodos y no implementa ningún código.

### 2.3.3.1 Una tabla por clase

En este caso se tendría un total de 3 tablas, una para la clase_Base otra para la Derivada_1 y una última para la clase Derivada_2. Cada tabla contendrá los campos que se hayan definido en sus clases correspondientes, pero las tablas correspondientes a las clases derivadas también contendrán el campo clave primaria de la clase base como una clave externa, para así evitar que hayan datos en las tablas de clases derivadas sin sus correspondientes datos en la tabla base.

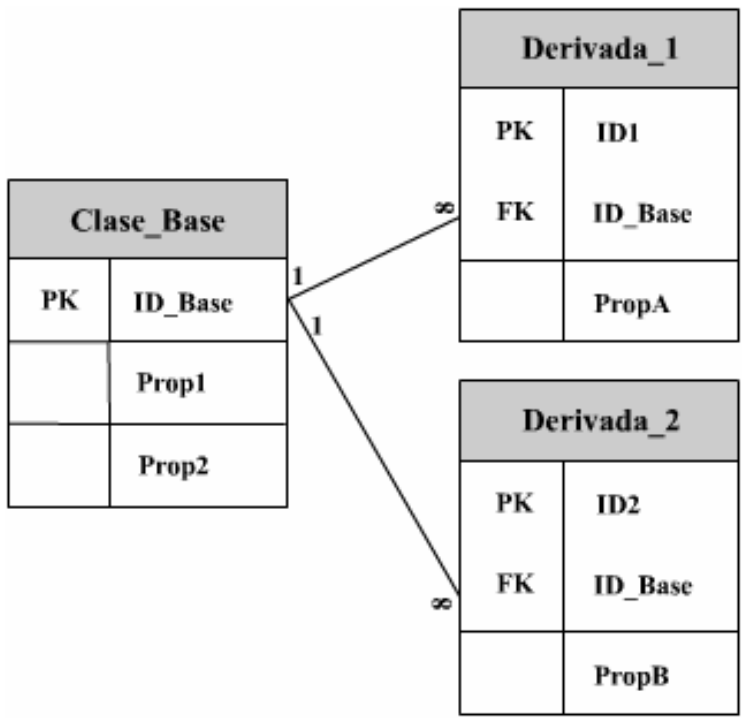

Figura 62. Mapeo correspondiente en tablas de un RDBMS del diagrama de clases de la figura 61

Con este enfoque las propiedades de un objeto están en más de una tabla, generan un listado polimórfico[*2] que se puede lograr mediante consultas con Left Join, y si luego se desea separar los objetos, sólo se deben evaluar los campos correspondendientes a un objeto u otro para observar si poseen o no valores Null.

### 2.3.3.2 Una tabla por cada clase concreta (no abstracta)

Con esta estrategia las clases abstractas no tienen tabla propia. Por lo tanto no habrá una tabla Clase_Base. Esto implicará que las tablas de las clases hijas tendrán más columnas; bajo esta configuración cada una de las tablas correspondientes a las clases derivadas contendrá los campos correspondientes a los atributos de la clase base.

Para obtener un listado polimórfico con este tipo de configuración de tablas se debe usar una consulta con Union, donde se debe tener en consideración que este tipo de consulta sólo acepta igual número de campos en cada Select.

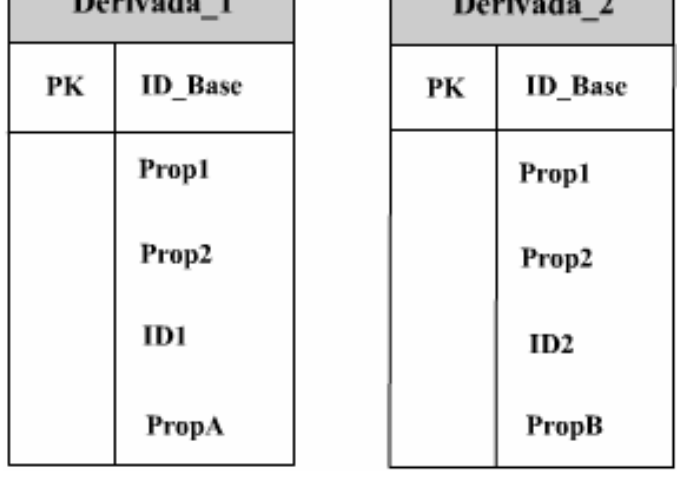

Figura 63. Mapeo2 correspondiente en tablas de un RDBMS del diagrama de clases de la figura 60

*2: Un listado polimórfico es una lista que posee objetos de distintos tipos, especialmente de distintas clases especializadas de una clase base

### 2.3.3.3 Una tabla por cada jerarquía de herencia

Esta estrategia contempla guardar en una sola tabla toda la jerarquía de herencia completa. Una sola tabla con campos correspondientes a los atributos de la clase_base más los campos correspondientes a los atributos de las clases derivadas. Además de eso, se debe agregar una columna de máscara que indique a que tipo de objeto hace referencia cada registro de la tabla. Con esta estrategia no hay que hacer Join (s) ni Unión (s) para realizar un listado polimórfico.

| jerarquía_herencia1 | |
|---|---|
| PK | ID_Base |
| | Prop1 |
| | Prop2 |
| | ID1 |
| | PropA |
| | ID2 |
| | PropB |
| | Tipo_Objeto |

Figura 64. Mapeo3 correspondiente en tablas de un RDBMS del diagrama de clases de la figura 60

**Ventajas y desventajas de un enfoque contra otro**

Cada una de las estrategias para el mapeo de la herencia hacia el RDBMS tiene sus ventajas y desventajas, que se deben tomar en cuenta a la hora de implementar uno u otro tipo de mapeo, además de los inconvenientes a la hora de su normalización.

a) Los últimos tipos presentan la ventaja de tener menos tablas, mientras que el primero presenta la ventaja que es más fácil agregar un nuevo campo correspondiente a un nuevo atributo en cualquiera de las clases.

b) En el primer enfoque es más fácil implementar herencia múltiple con sólo agregar un nuevo nivel más de relación, mientras que en los dos últimos enfoques habría que tener otras consideraciones. En el caso del último enfoque se podría agregar un campo máscara más por cada nivel de herencia múltiple.[60]

c) El primer enfoque se rompe con la clara regla de normalización de los datos referida a que no se deben modelar jerarquías.

d) En el segundo enfoque ocurre que cuando se tiene una tabla que se necesite relacionar a otra que le retorne a un objeto derivado, será necesario algún mecanismo en software que decida cuál camino seguir entre una tabla A o B.

e) En el tercer enfoque va a suceder que existirán registros que tendrán muchos campos vacíos, los correspondientes al objeto que no representa, pero el principal problema en este caso se da con los campos obligatorios.

[60] [GIMENOBLOG] - y - [LHOTKA98]

## 2.4 SISTEMA DE INFORMACION GEOGRÁFICO (SIG)

Un SIG es una herramienta basada en la computadora que integra información de tal forma que ayuda a comprender y encontrar soluciones a problemas. El SIG almacena datos de objetos del mundo real en bases de datos (BD) y realiza representaciones gráficas de estos objetos sobre mapas en pantalla de forma dinámica, de tal manera que cuando un dato es cambiado en la BD, el mapa se actualiza para reflejar los cambios. [61]

Se considera el término SIG como el resultado de la unión de dos tendencias en la evolución computacional. Una de estas tendencias y tal vez las más conocida, es la base de datos alfanumérica, en la cual se manejan grandes volúmenes de información descriptiva y cuantitativa de distintos elementos. El segundo desarrollo importante se refiere a la automatización de elementos espaciales (puntos, líneas o polígonos), de tal forma que el ordenador pueda repetir su forma y tamaño de manera rápida y precisa, comúnmente conocido como cartografía automatizada.[62]

La técnica de producir cartografía y mapas ha evolucionado con la extendida adopción de las computadoras y el desarrollo de la tecnología SIG. Los mapas aparte de ser los documentos impresos con los que estamos familiarizados, ahora también se presentan de forma visualmente interactiva en las computadoras.[63]

El SIG es usado en muchas industrias como un programa utilitario: en negocios comerciales, cumplimiento de la ley, transporte, salud, agricultura, estado, municipalidades, entre otras. Las industrias usan el SIG para cosas como la administración de recursos naturales, estudio de las amenazas naturales, uso y planificación de la tierra, investigaciones demográficas, rutas de vehículos de emergencia, administración de flotilla, valoración ambiental, planificación y mucho más. El número de aplicaciones de los SIG está en aumento.

De forma general, el SIG se usa para cuatro propósitos generales: creación de datos, presentación de datos, análisis y salida. Se pueden presentar objetos de acuerdo a los datos en la BD. Las herramientas de análisis del SIG permiten hacer cosas como: encontrar las viviendas que se encuentran a cierta distancia de una falla sísmica, qué parcelas tienen los tipos de suelos especiales para determinado cultivo, cuáles edificaciones están dentro de una área de inundación. Las opciones de salida incluyen mapas de calidad cartográfica así como reportes, lista y gráficos.[60]

### 2.4.1 Componentes del SIG

Un SIG tiene 5 componentes: personas, datos, hardware, software y procedimientos. Es la combinación de los conocimientos de las personas, datos espaciales y descriptivos, métodos analíticos, software y hardware de computadora, todo organizado para automatizar, administrar y presentar información de forma geográfica.[62]

[61] [CAMPUSESRI]
[62] [BARRIOS99] página 5
[63] [ZEILER99] página 24

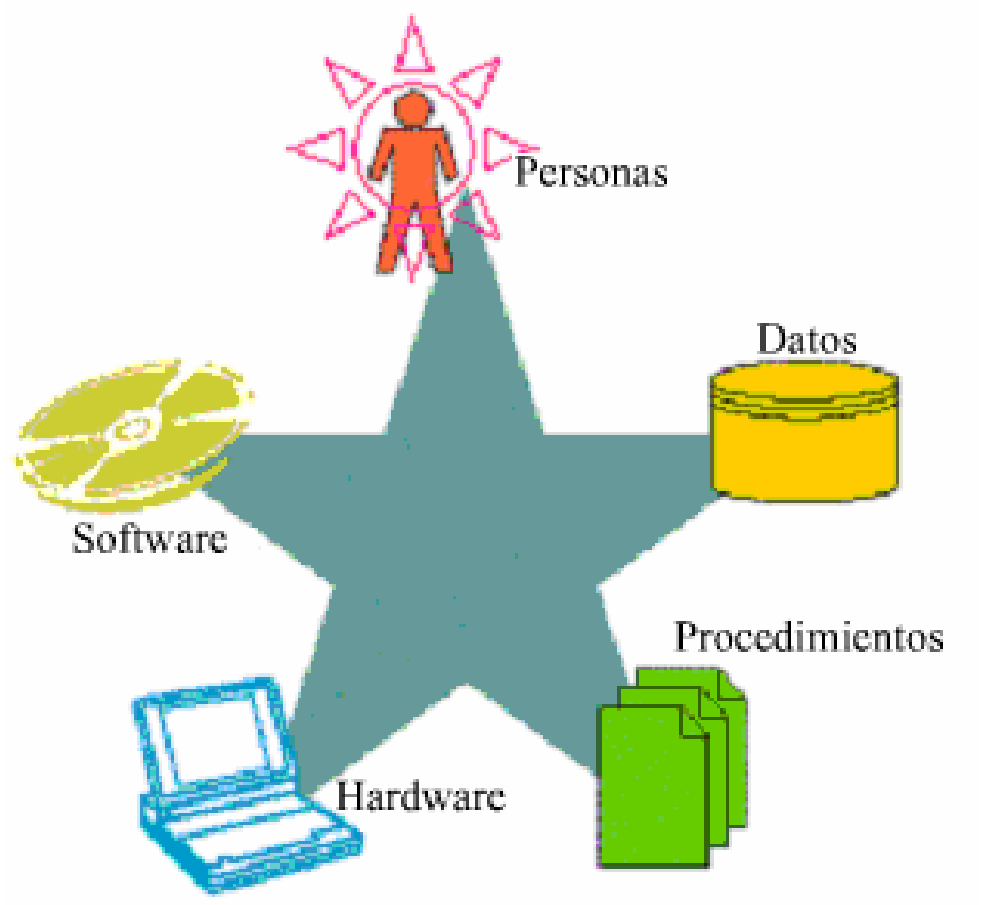

Figura 65. Componentes de un SIS

Como dice su nombre, de hecho un SIG es un sistema y todos sus 5 componentes son requeridos para obtener los resultados deseados.

**Personas**: Son el componente más importante de un SIG. Las personas deben de desarrollar los procedimientos y definir las tareas que el SIG debe realizar.

**Hardware**: Las capacidades del hardware afectan la velocidad de procesamiento, comodidad del uso y tipos de salidas disponibles.

**Software**: Esto no sólo incluye el software de SIG, sino también de BD, dibujo, estadística, simbolización y otros programas.

**Procedimientos**: El análisis en el SIG requiere métodos consistentes y bien definidos para producir resultados correctos y reproducibles.

**Datos**: La disponibilidad y exactitud de los datos afectan los resultados de las consultas y el análisis. Una gran parte de los datos del SIG son de tipo geográfico, pero también puede contener datos alfanuméricos en matrices con formato de tablas de BD (como mdb de Access y dbf de FoxPro) y de hojas de cálculo, que luego se pueden ligar con datos geográficos.

Los datos geográficos están compuestos de tres elementos principales: geometría, atributos y comportamiento. La **geometría** representa las características geográficas asociadas con locaciones del mundo real y son abstraídas en dibujos como puntos, líneas y polígonos (áreas). Los **atributos** describen las características de los objetos geográficos del mundo real. El **comportamiento** significa que los objetos geográficos pueden ser hechos para permitir ciertos tipos de edición, presentación o análisis, dependiendo de circunstancias que el usuario define.

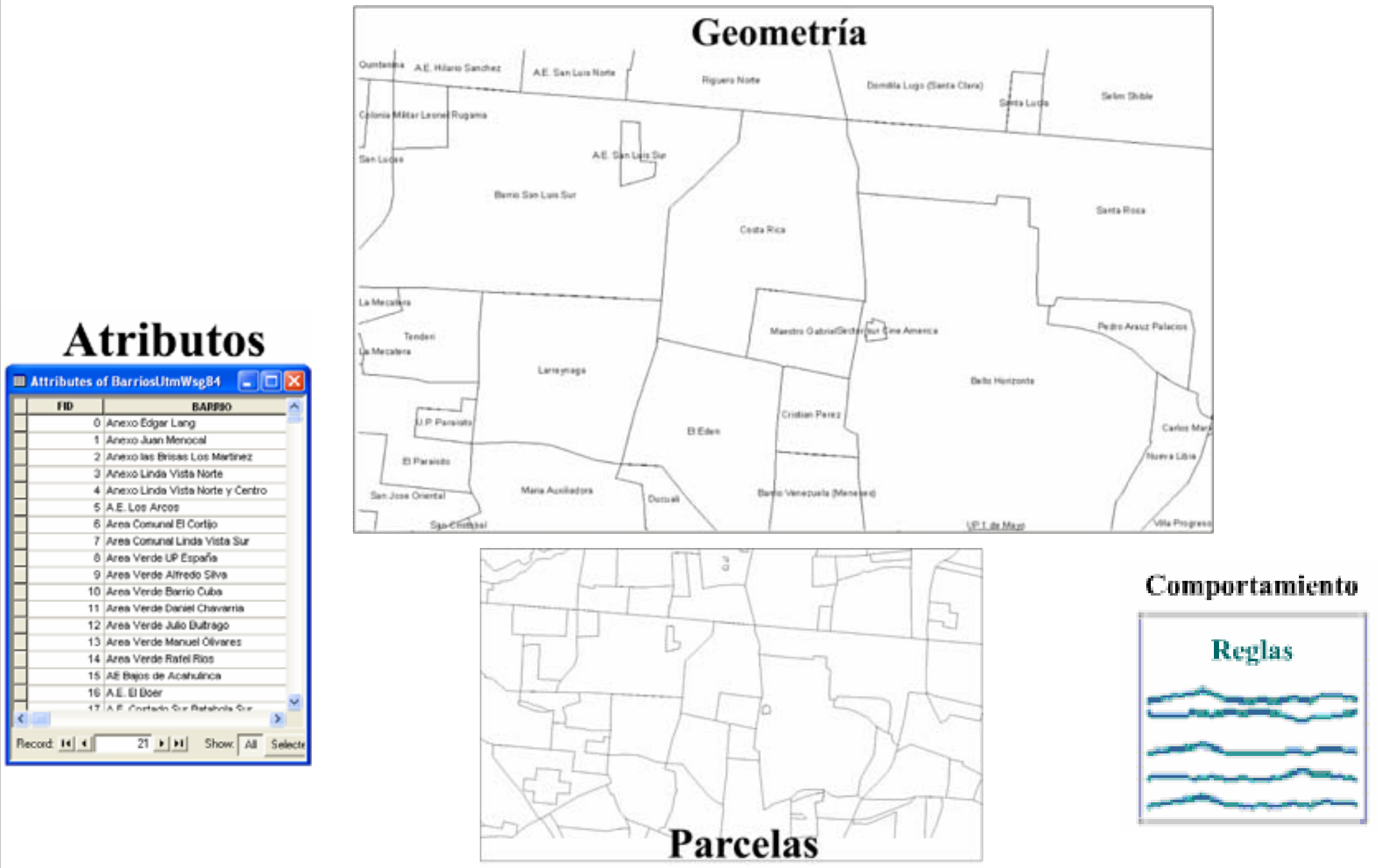

Figura 66. Datos geográficos

### 2.4.2 Organización de los datos geográficos en el SIG

El SIG organiza y almacena la información sobre el mundo como una colección de capas temáticas que pueden ser ligadas geográficamente. Cada capa contiene objetos del mundo real que comparten atributos similares, como calles o ciudades que están localizadas en la misma extensión geográfica. Este concepto simple pero extremadamente poderoso y versátil provee una invaluable característica que ayuda en la solución de problemas del mundo real.

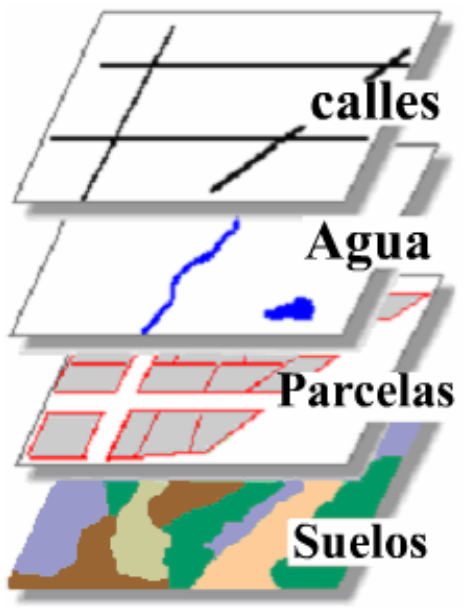

Figura 67. Traslape de capas temáticas

En la figura anterior se muestran cuatro capas temáticas donde se puede observar que cada una contiene objetos del mismo tipo. Todos estos objetos pueden estar ubicados dentro del mismo borde de una ciudad, pero cada una representa un tema distinto.

Sobre un mapa, la posibilidad de saber donde está localizado un espacio relativo a otro (relaciones espaciales), comunica información considerada muy importante.

La topología es un proceso matemático utilizado para determinar relaciones espaciales y propiedades como:

1. Conectividad de líneas
2. Relaciones de líneas
3. Longitud de líneas
4. Adyacencia de áreas
5. Definición de áreas

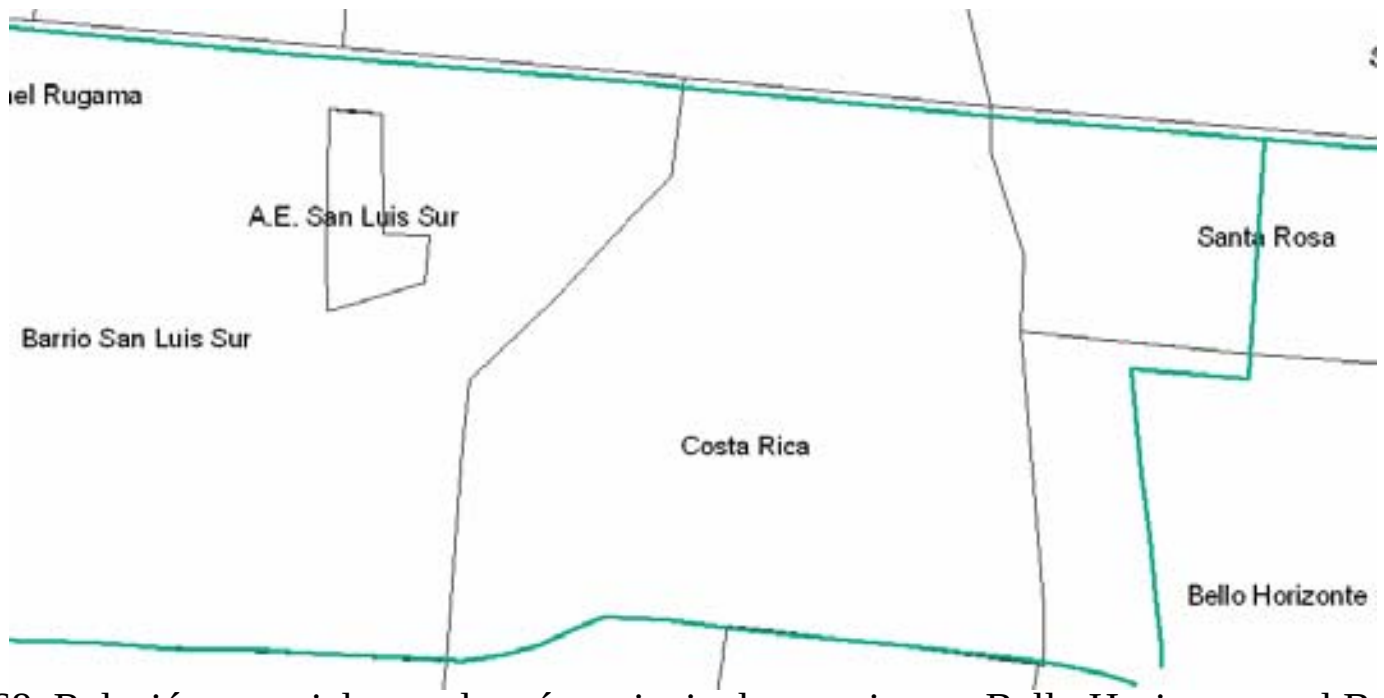

Figura 68. Relación espacial entre las vías principales que juntan Bello Horizonte y el Barrio San Luis

### 2.4.3 Operaciones fundamentales del SIG

Un sistema de información geográfico debe ser capaz de realizar seis operaciones fundamentales:

1. Capturar datos
2. Almacenar datos
3. Consultar datos
4. Analizar datos
6. Presentar los datos
7. Sacar los datos

**Capturar datos**

Los datos que describen elementos geográficos son almacenados en Bases de Datos Geográficas (BDG), las que son un componente costoso y de larga vida; por lo tanto, la captura de datos es una consideración importante.

Un SIG también debe proveer métodos para agregarle coordenadas geográficas y datos tabulares. Mientras más métodos de entradas provea el SIG más versátil se considerará.

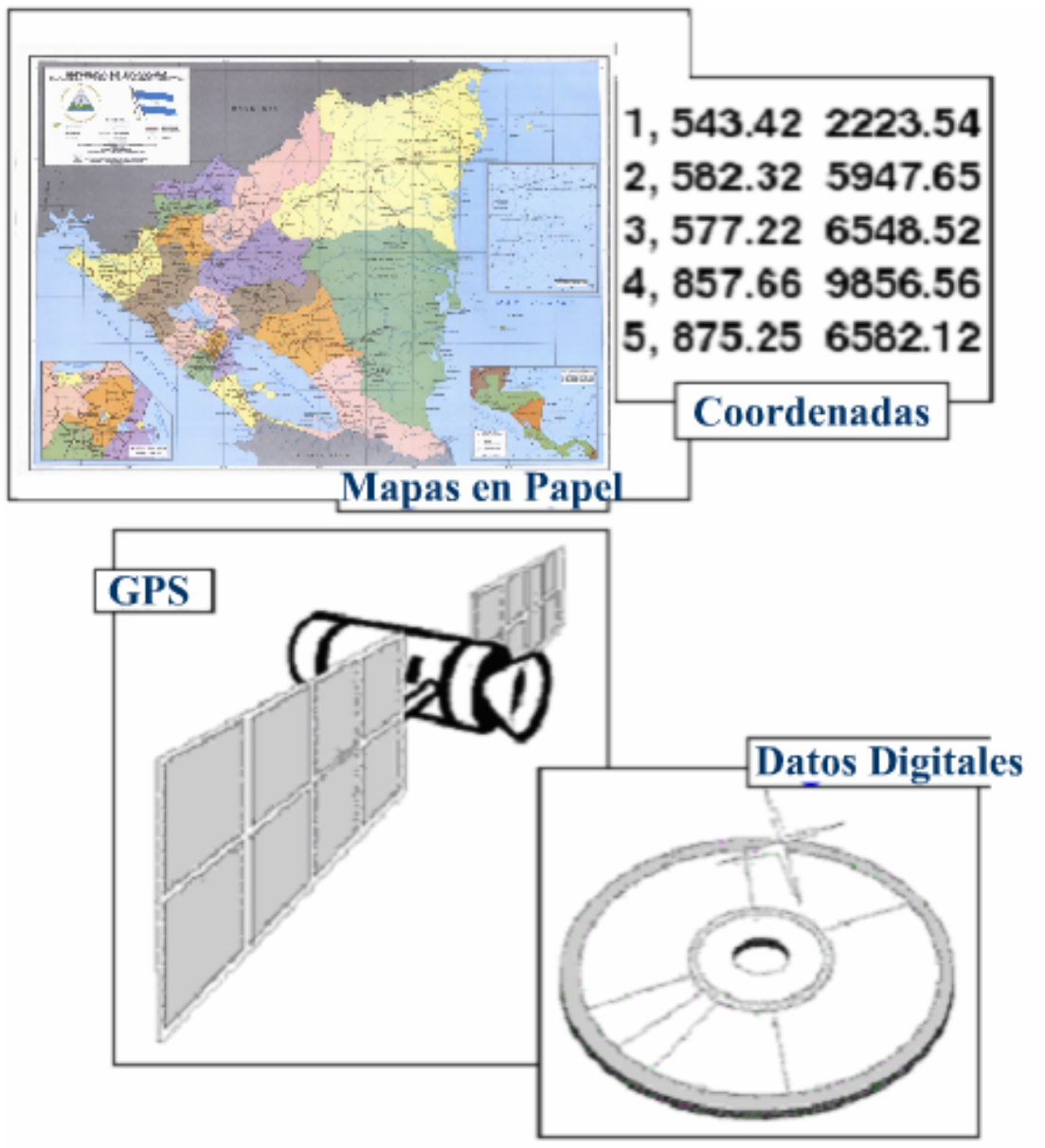

Figura 69. Diferentes fuentes desde las que se pueden traer datos al SIG

## Almacenar datos

Existen dos modelos básicos utilizados para almacenar datos geográficos: vector y raster. Un SIG debería ser capaz de almacenar ambos tipos de datos geográficos. La forma en que aparecen los elementos geográficos en el SIG depende del formato del formato en que se almacenan los datos.

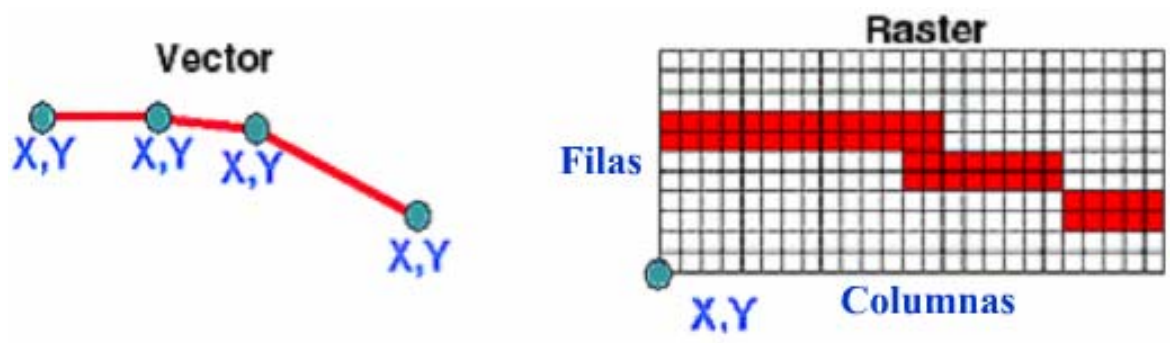

Figura 70. Formatos: vector y raster

Los datos geográficos almacenados en formato de vector son representados similarmente a la forma en que aparecen en los mapas, con puntos, líneas y polígonos. En este formato se utiliza un sistema de coordenadas cartesianas para referenciar las posiciones del mundo real.

En el modelo de datos raster se utiliza una matriz de celdas para almacenar los datos. Se asignan valores a las celdas en las posiciones de la matriz donde existen valores en el mundo real. La aproximación del detalle que se puede observar depende del tamaño de las celdas de la matriz. El formato raster no es apropiado para administración de parcelas, donde debe de conocerse los límites discretos de los objetos, pero si es buen formato para el análisis espacial.

## Consulta a los datos

Un SIG debe proporcionar herramientas para encontrar objetos específicos basados en su posición o atributos. Las consultas que a menudo son creadas como sentencias o expresiones lógicas son utilizadas para seleccionar elementos en el mapa y sus registros o características en la BD.

Una consulta común en SIG es el establecer que objetos existen en determinadas posiciones. En este tipo de consulta el usuario se entera donde están localizados los objetos que le interesan y/o también que características se encuentran asociados a estos. Esta es una de las capacidades del SIG porque los elementos o datos geográficos presentados en un mapa están enlazados a sus atributos (características descriptivas), almacenados en las BD.

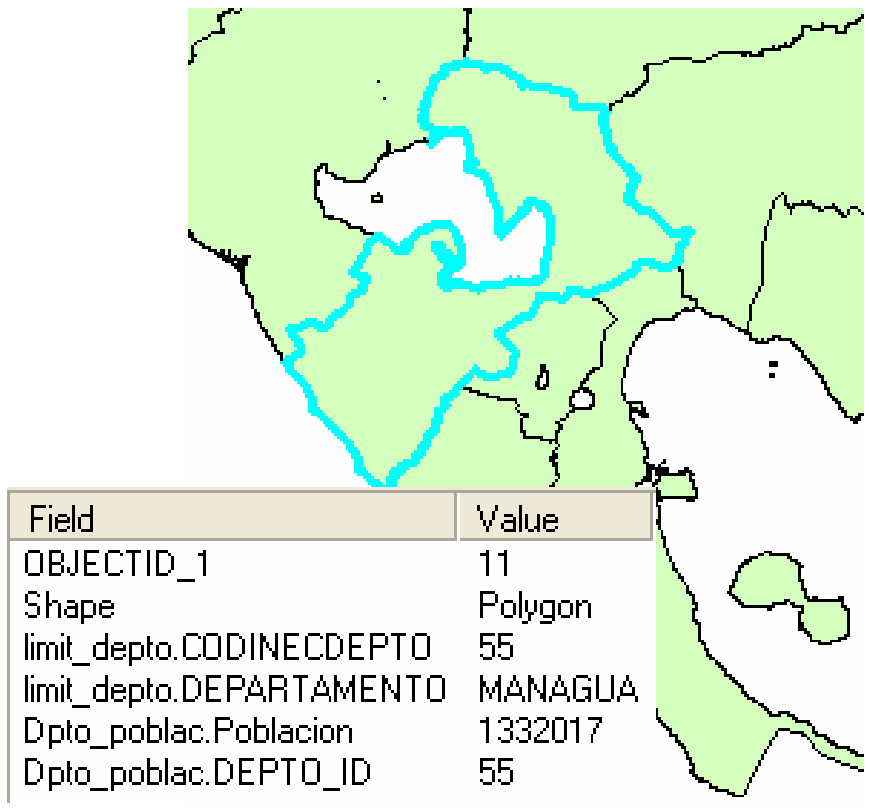

Figura 71. Con un clic sobre un objeto del mapa se puede observar sus atributos asociados en la BD

Otra consulta en SIG es qué objetos geográficos satisfacen ciertas condiciones. En este caso el usuario sabe cuál característica de qué capa le interesa y desea encontrar los objetos geográficos que satisfagan esa característica.[64]

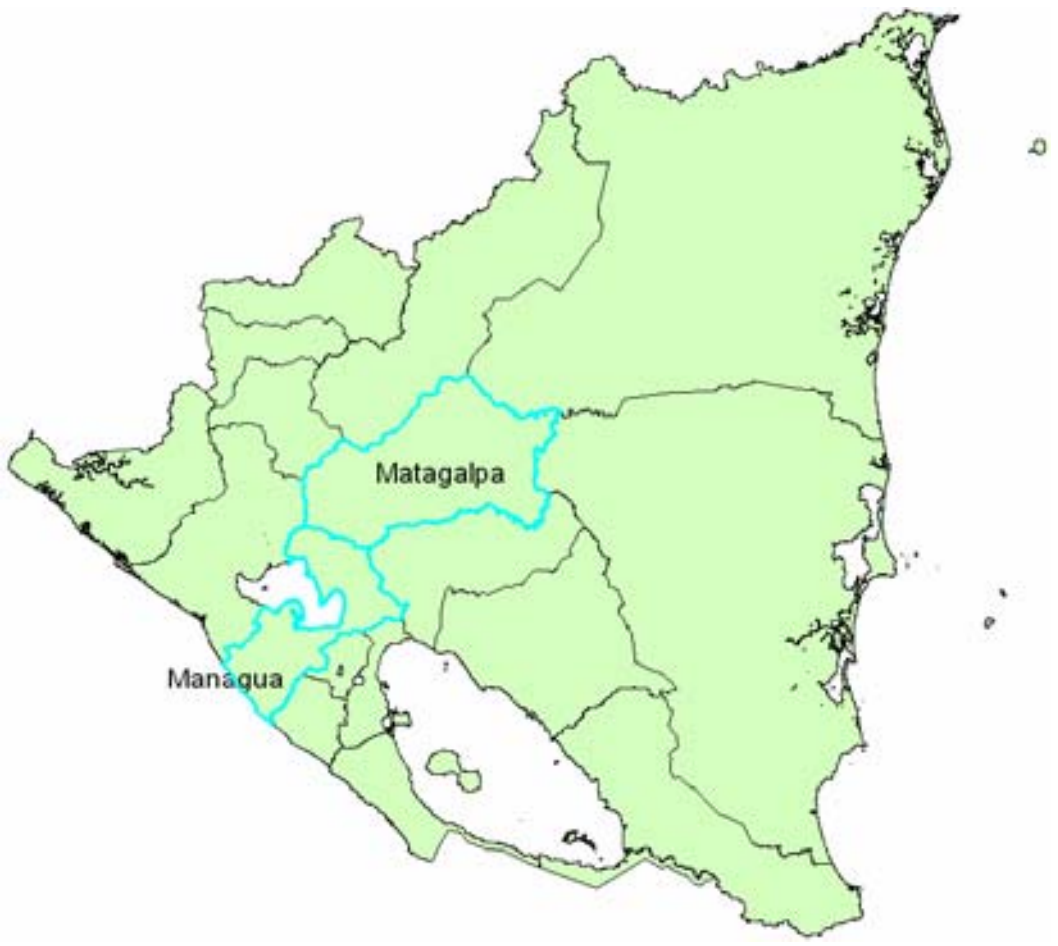

Figura 72. Encontrar todos los departamentos de Nicaragua con más de medio millón de habitantes

[64] [CAMPUSESRI]

### 2.4.4 Programas SIG

En general, un Sistema de Información consiste en la unión de información y herramientas informáticas (programas) para su análisis con unos objetivos concretos. En el caso de los SIG, se asume que la información incluye la posición en el espacio, para lo que se han desarrollado un tipo específico de aplicaciones informáticas dirigidas al manejo de estos sistemas. Estos programas es lo que popularmente (y equivocadamente) se conoce cómo SIG (ArcGIS, IDRISI, ArcInfo, GRASS, etc.), pero que realmente constituyen tan sólo un componente de lo que es realmente un SIG.

Los programas SIG están basados en una serie de capas de información espacial en formato digital que representan diversas variables (formato raster); o bien capas que representan objetos (formato vectorial) a los que corresponden varias entradas en una base de datos enlazada. Esta estructura permite combinar, en un mismo sistema, información con orígenes y formatos muy diversos, incrementando la complejidad del sistema.

Una de las primeras percepciones que se tiene de un SIG son las salidas gráficas a todo color, impresas o en la pantalla de un ordenador. Sin embargo, conviene recordar que hay una diferencia fundamental entre los programas de manejo de gráficos y los SIG. En los primeros, el objeto importante es la imagen que vemos, siendo irrelevante como se codifique; en un SIG la imagen es sólo una salida gráfica sin mayor importancia, lo relevante son los datos que se están representando.

Evidentemente ningún programa de SIG puede ser el mejor de los programas posibles y cubrir todas las expectativas. Por tanto, los programas acaban especializándose en función del tipo de datos, el tipo de aplicaciones y la lógica de trabajo que se supone van a utilizar.

**Agrupados respecto al tipo de datos a utilizar**

1. **SIG Raster**. Incluyen principalmente herramientas para el manejo de variables espaciales (IDRISI, GRASS, ERMapper, SPRING, PCRaster, Ilwis)
2. **SIG Vectorial**. Manejo de objetos (ArcInfo, ArcView, MapInfo, Geomedia, MicroStation)

**Agrupados respecto a la forma de organizar el trabajo**

1. **SIG basados en menús**, orientados normalmente a la gestión tanto en empresas como en administración (ArcView, IDRISI para windows, MapInfo, Geomedia, SPRING)
2. **SIG basados en comandos**, orientados a la investigación (GRASS, ArcInfo, IDRISI para MSDOS, PCRaster)

Finalmente habría que distinguir entre SIG libres (GRASS), comerciales (ArcInfo, IDRISI, Geomedia, ArcVIew, MapInfo) y gratuitos o semigratuitos (SPRING, PCRaster); además entre SIG para Windows o para UNIX.[65]

[65] [MINSALUDCHILE]

### 2.4.5 Generalidades del SIG (ArcGIS) utilizado

ArcGIS es el nombre usado para identificar a la familia insignia de programas y productos SIG de la compañía norteamericana con base en Redlands, California "Instituto de investigación sobre sistemas ambientales" (ESRI) por sus siglas en inglés. ESRI lidera la industria de los SIG desde el año 1988, proveyendo productos de software y soluciones de sistemas de información geográfica a miles de organizaciones alrededor del mundo.[66]

"El software de ESRI ® ArcGIS ® facilita la generación de mapas de alta calidad a través de un amplio rango de capacidades cartográficas y de publicación de mapas, libera a los cartógrafos del duro trabajo de las acciones repetitivas y por lo tanto les permite concentrarse en aplicar las habilidades visuales únicas que tiene el hombre para interpretar y diseñar."[67]

ArcGIS incluye software cliente para ArcGIS, componentes, aplicaciones y software de servidores de datos y aplicaciones. ArcGIS por si solo no es un SIG, más bien es un conjunto de programas para cubrir las necesidades de creación de un SIG. ArcGIS está compuesto de aplicaciones clientes y servidores. Cada software de aplicación puede crear, administrar y servir datos almacenados en más de un formato.

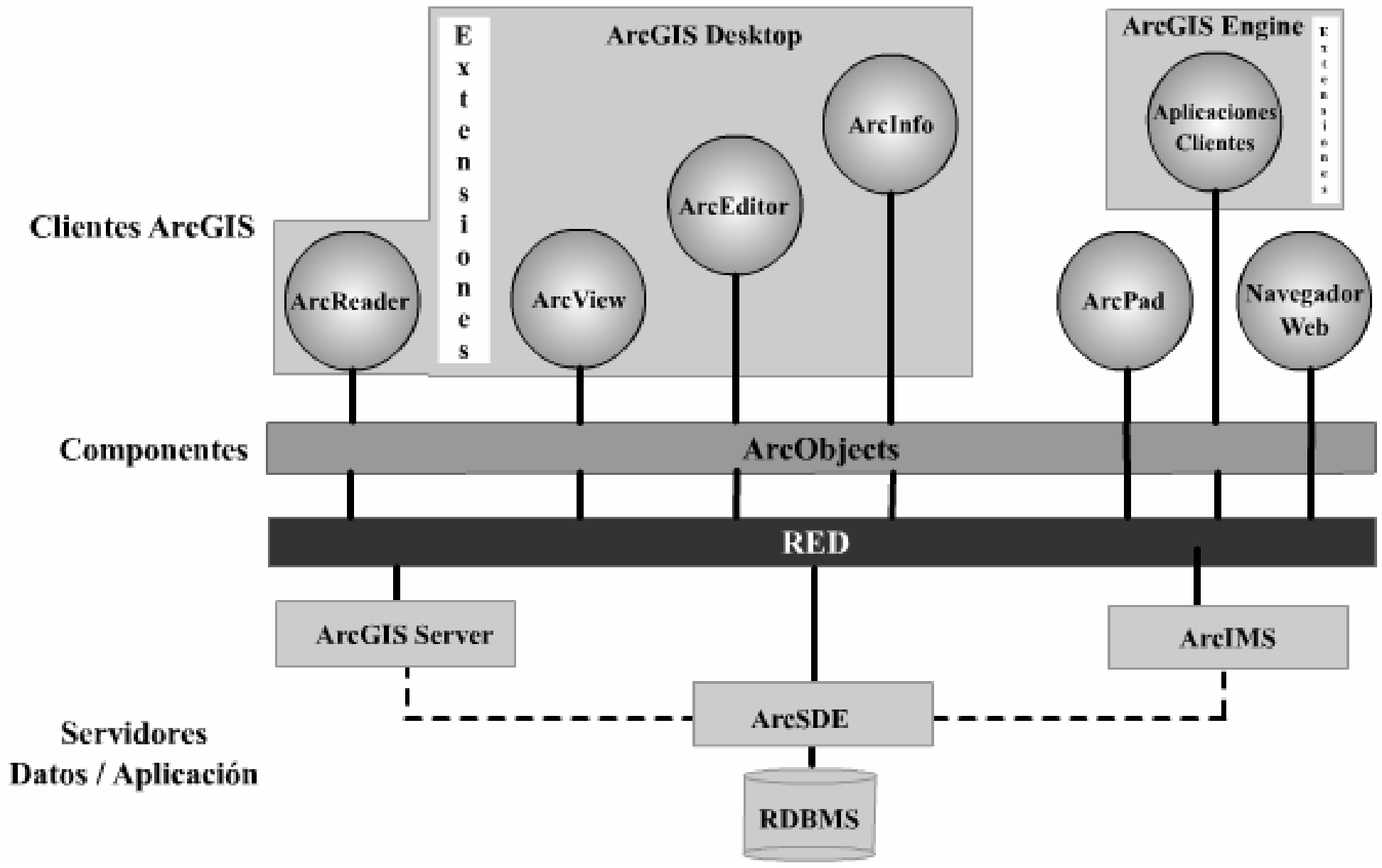

Figura 73. ArcGIS, clientes, componentes y servidores

[66] [POZO04] página 2
[67] [ESRI04 ] página 5

**ArcGIS Desktop**

Integra un conjunto de aplicaciones SIG avanzadas en tres productos: ArcView, ArcEditor y ArcInfo. Estos productos de escritorio poseen el mismo corazón de mapeo, edición y funcionalidades de análisis. El nivel de funcionalidad disponible depende del tipo de licencia que se posea. ArcInfo provee a los usuarios con el mayor nivel de funcionalidad SIG, el que se forma del ArcInfo de escritorio y el ArcInfo de estación de trabajo (ArcInfo WorkStation).

Cada uno de estos tres componentes de ArcGIS Desktop (ArcView, ArcEditor y ArcInfo) consiste de tres aplicaciones separadas que representan los métodos fundamentales que la gente utiliza para interactuar con un GIS. Esos métodos (datos, mapas y herramientas) forman el ambiente de ArcGIS. Los usuarios típicamente tendrán abierto dos o tres de esas aplicaciones al mismo tiempo. Esas aplicaciones son:

> **ArcCatalog:** Utilizado para explorar fuentes de datos geográficos, crear y actualizar metadatos.
> **ArcMap:** Utilizado para desplegar consultas geográficas sobre mapas y para editar y sacar datos.
> **ArcToolbox:** Contiene herramientas poderosas para realizar análisis geográfico y conversión de datos.

**ArcReader:**
Permite a lo usuarios ver mapas publicados en alta resolución (.PMFs) creados en ArcMap.

**ArcGIS Engine:**
Son herramientas de desarrollo de componentes SIG embebidos para crear aplicaciones clientes stand alone usando COM, C++, Java, y .Net.

**ArcPad:**
Es usado con PDAs para la creación y manejo de datos mientras se anda en el campo.

**ArcGIS Server:**
Es una librería compartida de objetos de software GIS utilizada para crear/desarrollar aplicaciones SIG del lado del servidor en ambientes de trabajo empresariales y Web.

**ArcIMS:**
Usado para publicar mapas, datos y meta datos a través protocolos abiertos de Internet.

**ArcSDE:**
Administra y sirve información espacial desde RDBMS externos a los clientes de ArcGIS.[68]

[68] [ESRI05] página 7

### 2.4.6 Personalización y extensión del SIG ArcGIS

Como se pudo observar en la figura x, las aplicaciones clientes ArcGIS descansan sobre la capa de componentes llamada ArcObjects. Estas aplicaciones de escritorio de ArcGIS son construidas a base de estos ArcObjects, por lo que todos los elementos de la interfaz gráfica, como el botón de salvar, la herramienta para dibujar un punto, ejecutan rutinas donde se mandan a llamar a los respectivos ArcObjects que le sirven a la funcionalidad requerida.

Este es un gran ejemplo de aplicaciones en capas, donde todas las rutinas y lógica de las funcionalidades del SIG están implementadas en componentes (llamados ArcObjects) y la interfaz es sólo un cliente de dicha lógica, que implementa la forma en la cual se le presenta al usuario la interacción con los componentes.

Los ArcObjects son la plataforma de desarrollo ArcGIS para la familia de aplicaciones como ArcMap, ArcCatalog, y ArcScene. Estos componentes de software disponen de un completo rango de funcionalidades disponibles en ArcInfo y ArcView para los desarrolladores.

Los ArcObjects son construidos utilizando la tecnología de Microsoft, Component Object Model. Por dicha razón es posible extender los ArcObjects y por ende las funcionalidades del ArcGIS escribiendo componentes COM mediante el uso de cualquier lenguaje de programación que cuente con un compilador COM. Se puede extender cada parte de los ArcObjects de la misma forma en que los desarrolladores de ESRI hacen.[69]

Los usuarios programadores de ArcGIS acceden a los ArcObjects con el propósito de extender, personalizar o mejorar las aplicaciones de ArcGIS. También pueden ser utilizados para programar otras aplicaciones gracias a COM. De esta manera se podrían utilizar los ArcObjects para poner la funcionalidad de un mapa a una aplicación como Microsoft Word o Microsoft Excel; del mismo modo se podrían poner procesos de Word y funcionalidad de hoja de trabajo a ArcMap.

Lo más importante de aprender sobre los ArcObjects son los principios con que trabaja y como leer el modelo de diagrama de objetos, con esto se estará listo para explorar los ArcObjects por uno mismo.[70]

Los ArcObjects están compuestos por más de 2,700 clases. Cada clase representa una parte básica del SIG, y se encuentran empaquetadas en archivos (componentes) Dll.

Los ArcObjects se modelan con 30 diagramas de clases por separado para facilitar su lectura, cada diagrama de clases agrupa clases con que interactúan de forma cercana para cumplir ciertas funciones.

[69] [WCLZ01] páginas 1-2
[70] [BURKE03] páginas 4

| | |
|---|---|
| 3D Analyst Object Model | Application Framework Object Model |
| ArcCatalog Object Model | ArcGIS Object Model |
| ArcMap Editor Object Model | ArcMap Object Model |
| ArcObjects Controls Object Model | ArcPad Object Model |
| ArcScan Object Model | Display Object Model |
| Geocoding Object Model | GeoDatabase Object Model 1 |
| GeoDatabase Object Model 2 | Geometry Object Model |
| IMS Object Model | Labeling and Annotation Object Model |
| Linear Referencing Object Model | Map Layer Object Model |
| Network Object Model | Output Object Model |
| Publisher Extension Object Model | Raster Object Model |
| Spatial Analyst Object Model | Spatial Reference Object Model |
| Styles Object Model | Tin Object Model |
| Tracking Analyst Object Model 1 | Tracking Analyst Object Model 2 |
| StreetMap Europe Object Model | StreetMap USA Object Model |

Tabla 1. Diagramas de clases en ArcGIS 8.3

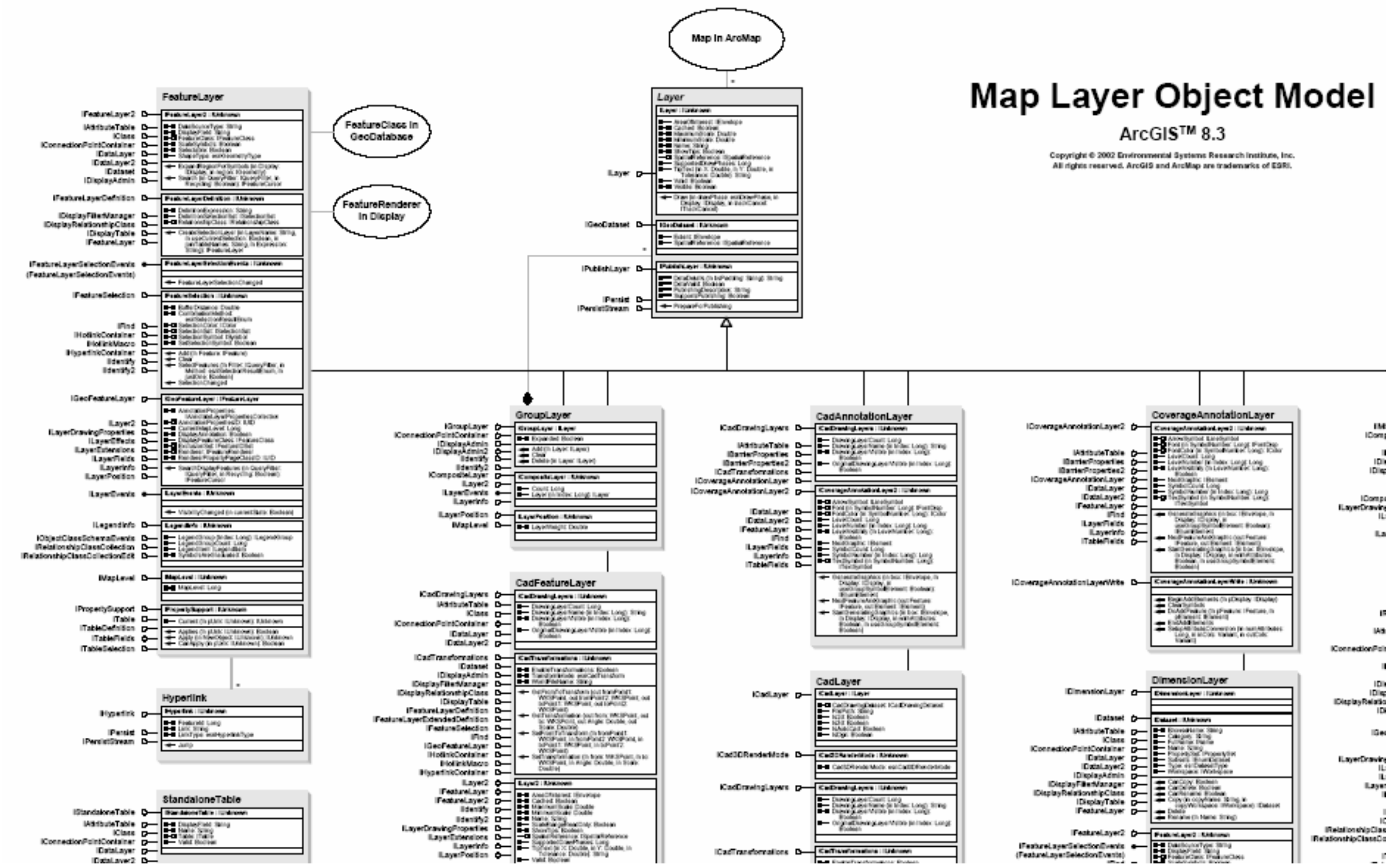

Figura 74. Una cuarta parte del diagrama de clases que modela las capas de un mapa. (Map Layer Object Model)

## 2.5 CÁLCULO DE LA VULNERABILIDAD Y DAÑOS DEBIDO A SISMOS EN VIVIENDAS

La vulnerabilidad sísmica es la susceptibilidad de la vivienda a sufrir daños estructurales en caso de un evento sísmico determinado. De forma general la vulnerabilidad es la fragilidad intrínseca de los elementos en riesgo o grado de exposición de los elementos sociales ante una determinada amenaza. La vulnerabilidad sísmica depende de aspectos como la geometría de la estructura, aspectos constructivos y aspectos estructurales. [71]

El riesgo es la pérdida económica y de vidas humanas que probabilísticamente pueden ser causadas por una determinada amenaza. El riesgo (R) es una función de la amenaza (A) por la vulnerabilidad (V) y se representa con la relación R = f(A*V).

La amenaza es la probabilidad de ocurrencia de un fenómeno potencialmente destructor, dentro de un período de tiempo específico y un área determinada. Debido a que la Amenaza sísmica es una característica de la naturaleza que el hombre todavía no ha podido modificar y mucho menos predecir, la única alternativa disponible para la disminución del riesgo sísmico, consiste en la búsqueda de estrategias adecuadas para medir la Vulnerabilidad de las estructuras y de los sistemas de respuesta de la zona.[72]

Con el objetivo de analizar el daño que puede producir un sismo a las viviendas, se realizan escenarios donde en vez de utilizar la amenaza para los cálculos de los resultados de un valor probabilístico como es el riesgo, se utiliza un valor concreto de sismo y como resultado se obtiene una proporción de daño que genera dicho sismo a cada una de las viviendas debido a la vulnerabilidad que cada vivienda posee. De tal forma, el daño es una función de la vulnerabilidad por el valor concretizado de un sismo.

La predicción del comportamiento de edificios ante eventos sísmicos es importante en la evaluación de pérdidas tanto económicas como de vidas humanas.[73]

### 2.5.1 Aspectos que afectan la vulnerabilidad sísmica de las viviendas

**1. ASPECTOS GEOMÉTRICOS**

### 1a. Irregularidad en planta de la edificación

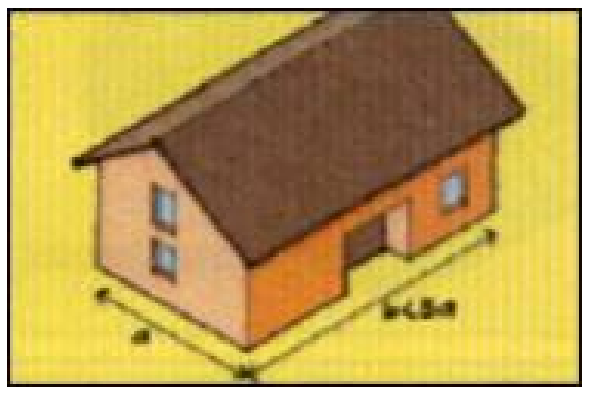

Figura 75. Baja o nula irregularidad en planta

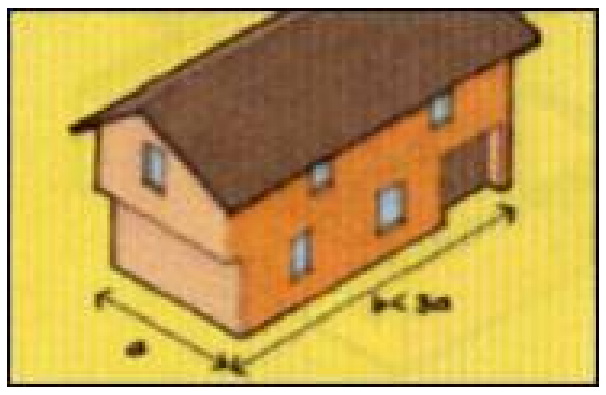

Figura 76. Media irregularidad en planta

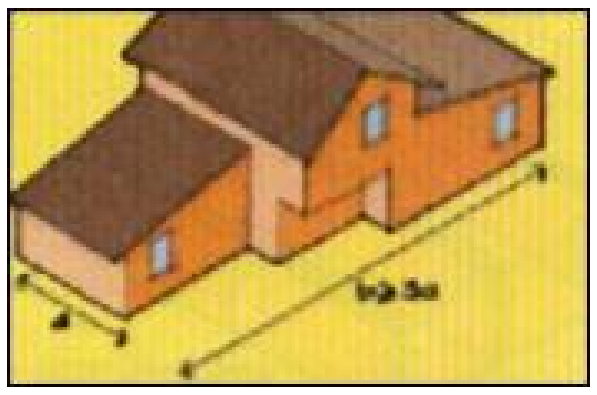

Figura 77. Alta irregularidad en planta

[71] [ CHAVARRIA] capítulo.6.6.2.9
[72] [HERALDO] página 244
[73] [CHAVARRIA] capítulo 6.1

La irregularidad en la planta contribuye proporcionalmente a la vulnerabilidad sísmica de la vivienda. De esta manera, una irregularidad baja contribuirá de forma mínima o nula a la vulnerabilidad, mientras que una irregularidad alta contribuirá de mayor manera.

### 1b. Cantidad de muros en las dos direcciones

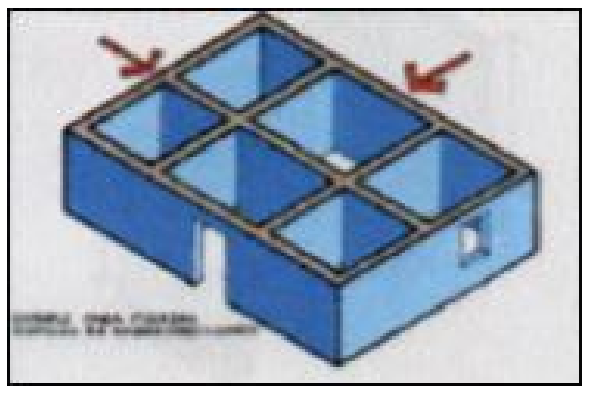

Figura 78. Mayor cantidad de muros en 2 direcciones

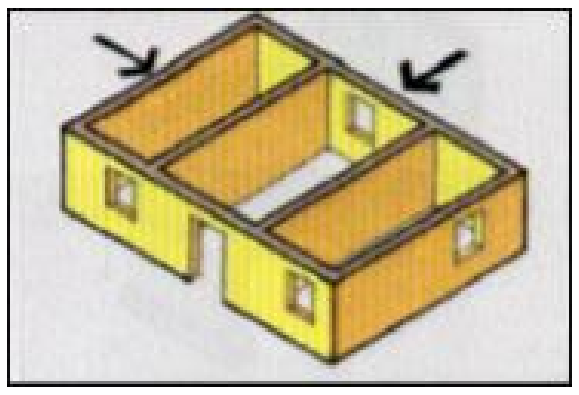

Figura 79. Media cantidad de muros en 2 direcciones

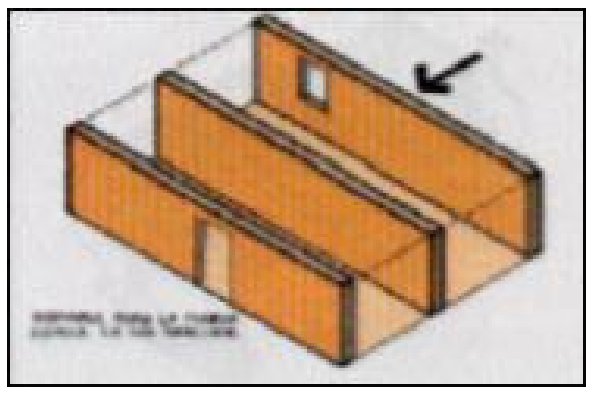

Figura 80. Poca o nula cantidad de muros en 2 direcciones

La cantidad de muros en dos direcciones contribuye de forma inversamente proporcional a la vulnerabilidad sísmica de la vivienda. De este modo, la configuración con mayor cantidad de muros contribuirá de forma mínima o nula a la vulnerabilidad, mientras que la configuración con poca o nula cantidad de muros en dos direcciones contribuirá a la vulnerabilidad de forma mayor.

### 1c. Irregularidad en altura

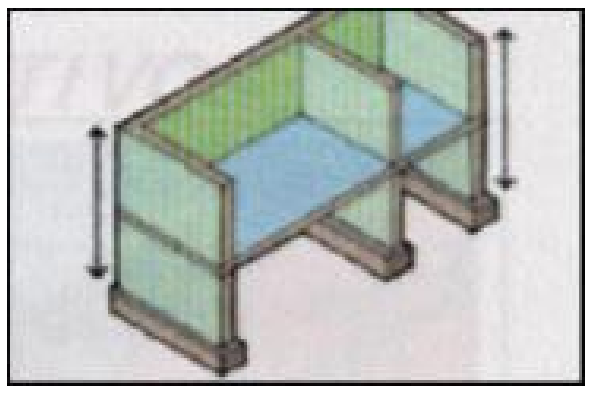

Figura 81. Mayor o total continuidad en los muros desde la cimentación hasta la cubierta

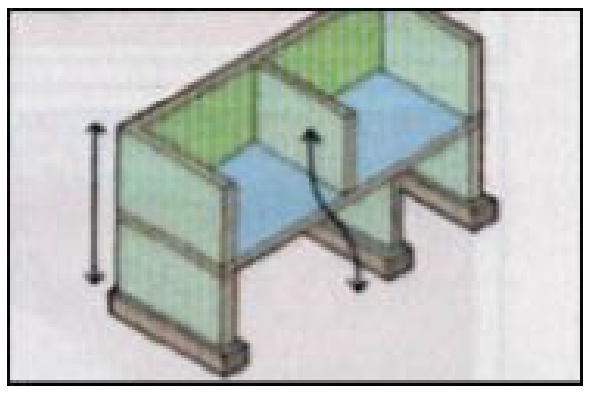

Figura 82. Media continuidad en los muros desde la cimentación hasta la cubierta

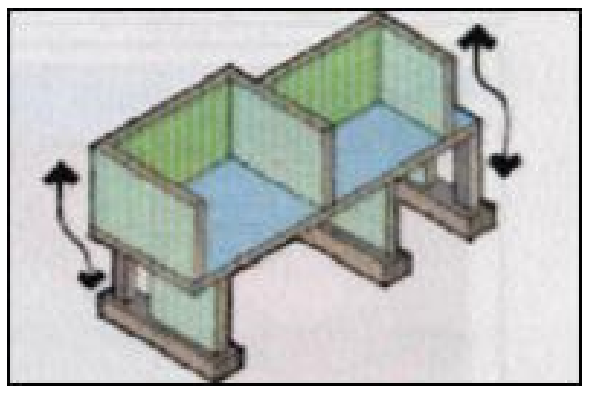

Figura 83. Poca o nula continuidad en los muros desde la cimentación hasta la cubierta

La irregularidad en la altura esta dada por la continuidad o discontinuidad de los muros. La discontinuidad de los muros contribuye de forma directa a la vulnerabilidad sísmica en la vivienda. Así, una vivienda con poca discontinuidad contribuirá de forma mínima, mientras que una vivienda con mayor discontinuidad contribuirá de forma mayor.

## 2. ASPECTOS CONSTRUCTIVOS

### 2a. Calidad de las juntas de pega en mortero

Aquí interviene el espesor y la calidad del mortero (mezcla de concreto), la calidad del proceso de juntar los ladrillos tanto en sus juntas verticales como horizontales.

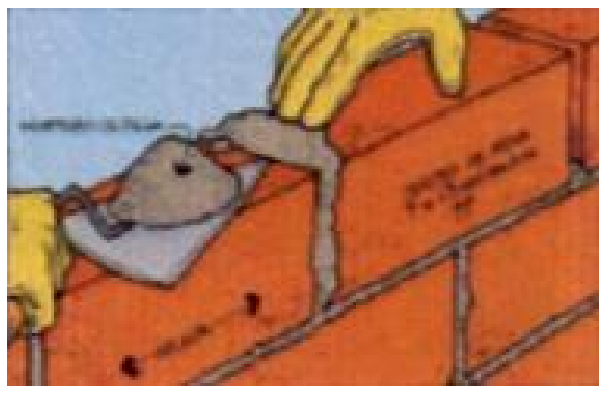

Figura 84. Vulnerabilidad baja: calidad de las juntas de pega en mortero

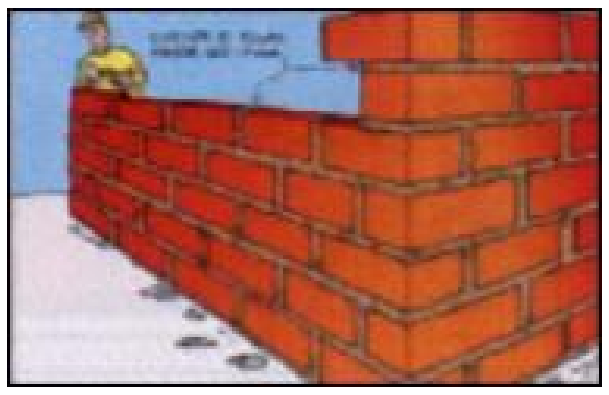

Figura 85. Vulnerabilidad media: calidad de las juntas de pega en mortero

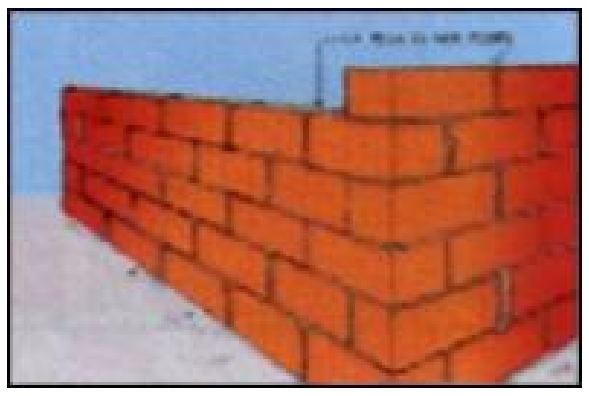

Figura 86. Vulnerabilidad alta: calidad de las juntas de pega en mortero

## 2b. Tipo y disposición de los ladrillos

En esta característica intervienen la buena calidad de las unidades de mampostería (bloques/ladrillos), la uniformidad en su colocación y el hecho de que hayan quedado trabadas unas con otras.

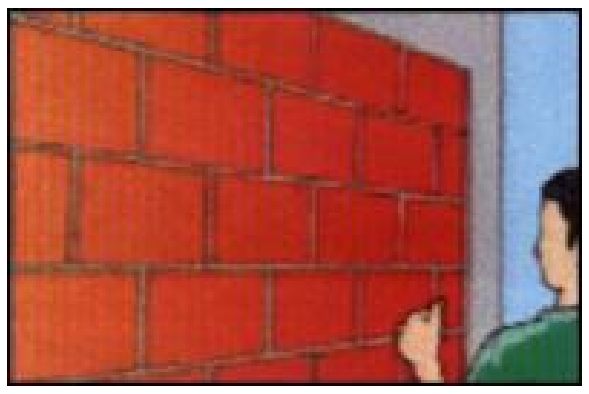

Figura 87. Vulnerabilidad baja: tipo y disposición de las unidades de mampostería

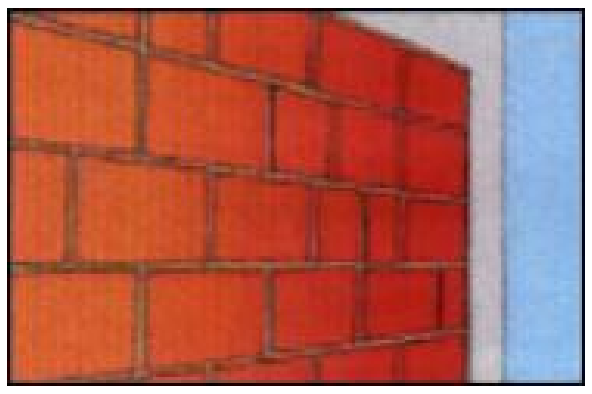

Figura 88. Vulnerabilidad media: tipo y disposición de las unidades de mampostería

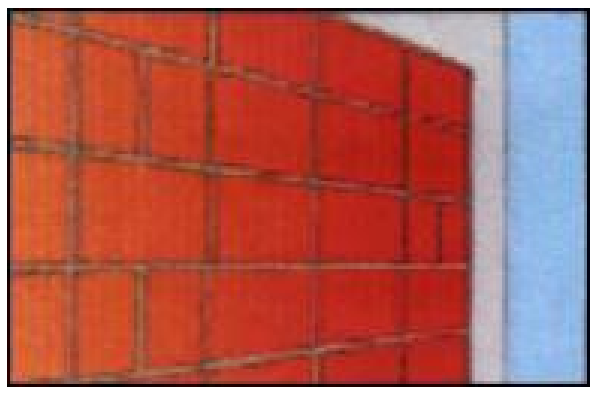

Figura 89. Vulnerabilidad alta: tipo y disposición de las unidades de mampostería

Para cada uno de los siguientes elementos se enumeran una serie de características que se deberían de cumplir para contribuir a una vulnerabilidad sísmica baja o nula en las viviendas. En caso de que no se cumplan todas, se estaría contribuyendo a una vulnerabilidad media y en caso que no se cumplan la mayoría se estaría contribuyendo a una vulnerabilidad alta.

## 2c. Calidad de los materiales

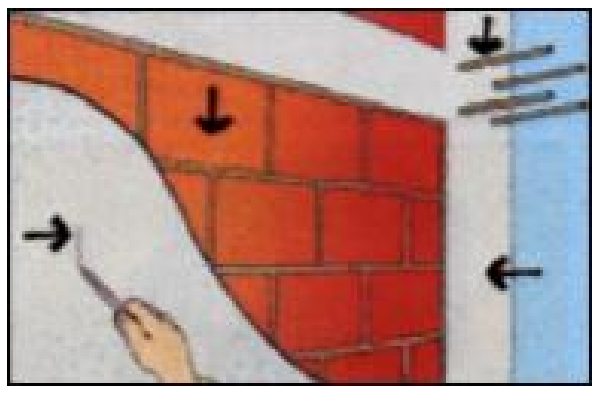

Figura 90. Vulnerabilidad baja: Calidad de los materiales

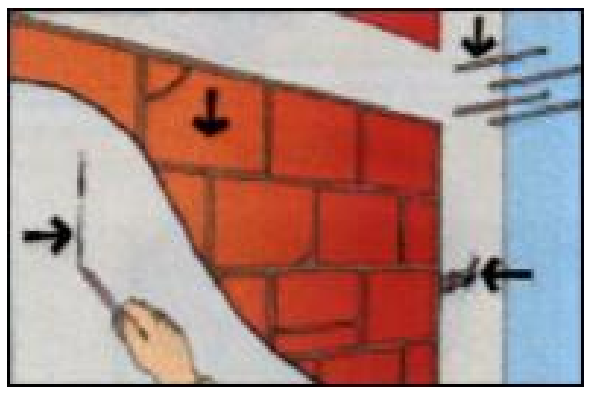

Figura 91. Vulnerabilidad media: Calidad de los materiales

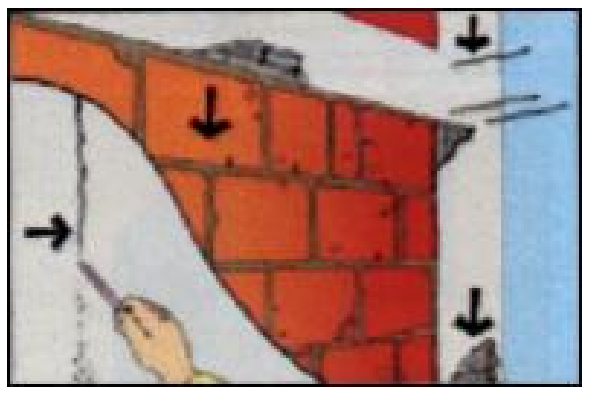

Figura 92. Vulnerabilidad alta: Calidad de los materiales

Requisitos que contribuyen a una vulnerabilidad baja:

- Que el mortero no se deje rayar o desmoronar con un clavo o herramienta metálica.
- El mortero tiene buen aspecto sin hormiguero y el acero no está expuesto.

- La existencia de estribos de por lo menos 3 o 4 barras en sentido longitudinal en el confinamiento del concreto reforzado.
- Que el ladrillo sea de buena calidad, que no esté muy figurado, quebrado, ni despegado y que resista calidad de por lo menos 2 metros de alto sin desintegrarse ni deteriorarse de forma apreciable.

## 3. ASPECTOS ESTRUCTURALES

### 3a. Muros confinados y reforzados

Requisitos que contribuyen a una vulnerabilidad baja:

- Que todos los muros de mampostería de las viviendas estén confinados con vigas y columnas de concreto reforzado alrededor de ellos.
- El espaciamiento máximo entre los espacios de confinamiento o altura de los pisos, es del orden de los 4 metros.
- Todos los elementos de confinamiento deben poseer refuerzo tanto longitudinal como transversal y estar adecuadamente dispuestos.

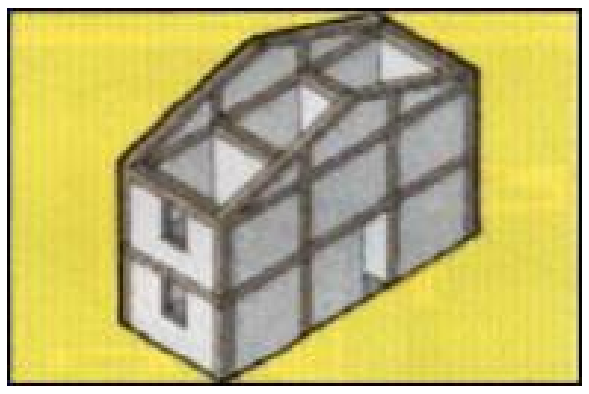

Figura 93. Vulnerabilidad baja: Muros confinados y reforzados

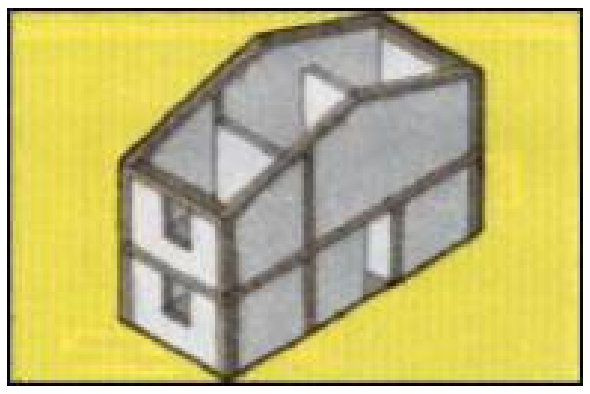

Figura 94. Vulnerabilidad media: Muros confinados y reforzados

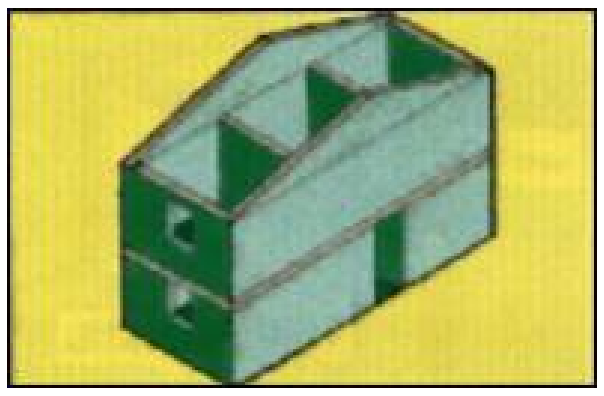

Figura 95. Vulnerabilidad alta: Muros confinados y reforzados

### 3b. Características de las aberturas

Requisitos que contribuyen a una vulnerabilidad baja:

- Las aberturas de los muros deben totalizar menos del 35% del área total del muro.
- La longitud total de las aberturas del muro debe corresponder a menos de la mitad de la longitud total del muro.
- Existe una distancia entre el borde del muro y la abertura adyacente igual a la altura de la misma o 50 cm.

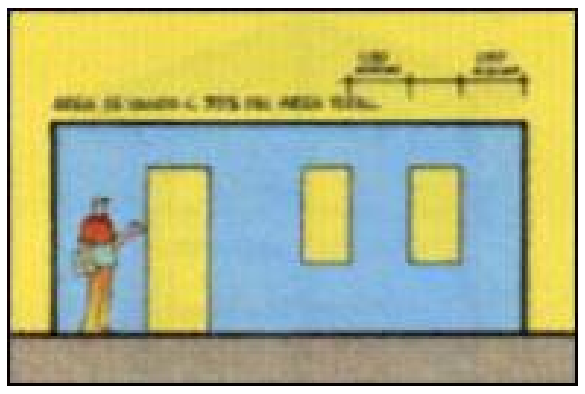

Figura 96. Vulnerabilidad baja: Características de las aberturas

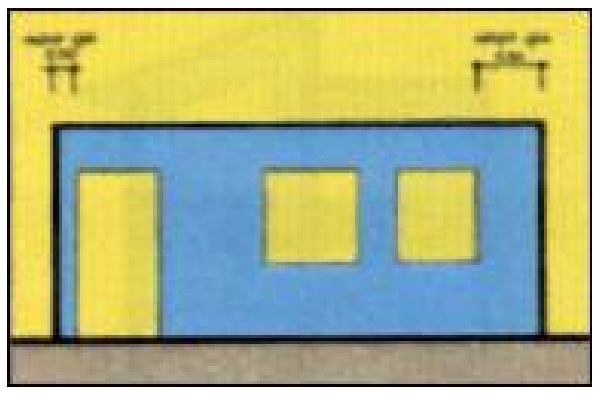

Figura 97. Vulnerabilidad media: Características de las aberturas

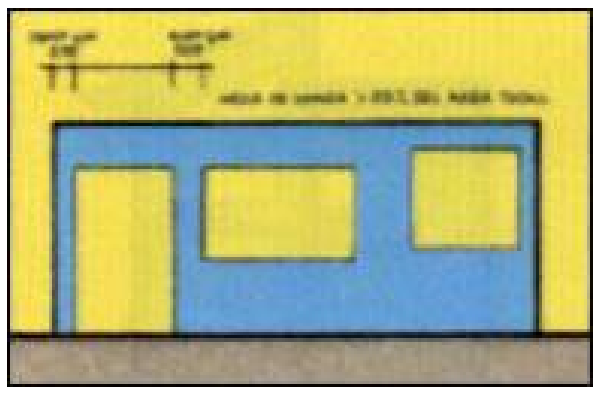

Figura 98. Vulnerabilidad alta: Características de las aberturas

## 3c. Amarre de cubiertas

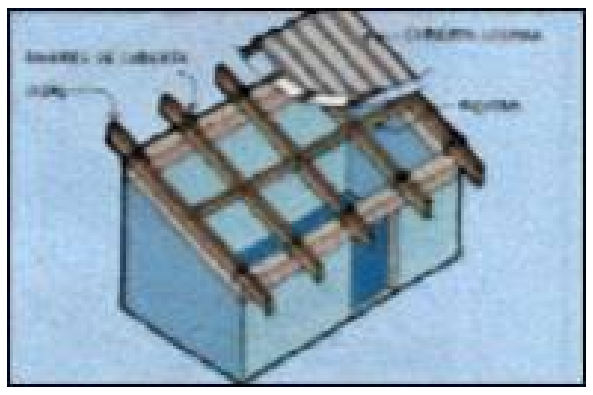

Figura 99. Vulnerabilidad baja: Amarre de cubiertas

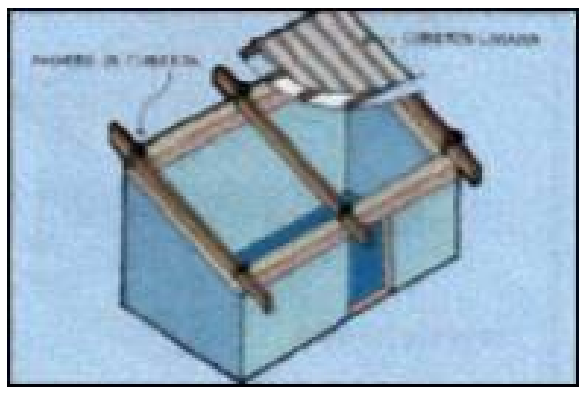

Figura 100. Vulnerabilidad media: Amarre de cubiertas

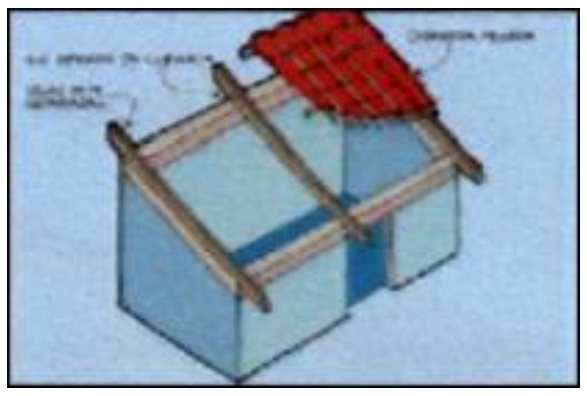

Figura 101. Vulnerabilidad alta: Amarre de cubiertas

Requisitos que contribuyen a una vulnerabilidad baja:

- La existencia de tornillos, alambres o conexiones similares que amarran el techo a los muros.
- Existencia de arriostramiento y distancia no muy grande entre vigas.
- Con cubierta liviana y debidamente amarrada y apoyada a la estructura de la cubierta.

## 4. CIMENTACIÓN

### Vigas de amarre en concreto reforzado

Requisitos que contribuyen a una vulnerabilidad baja:

- La cimentación deberá estar conformada por corridas en concreto reforzado bajo los muros estructurales.
- Las vigas de cimentación que conformen anillos amarrados.

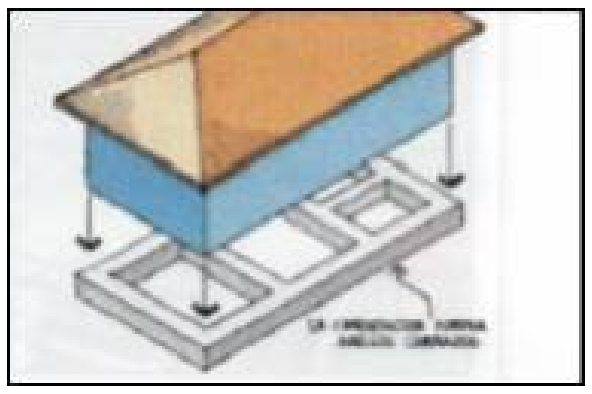

Figura 102. Vulnerabilidad baja: Cimentación

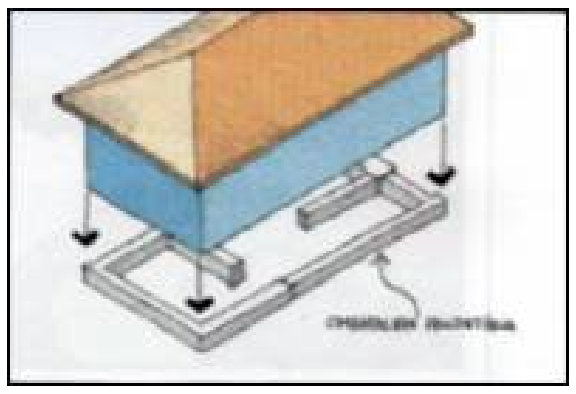

Figura 103. Vulnerabilidad media: Cimentación

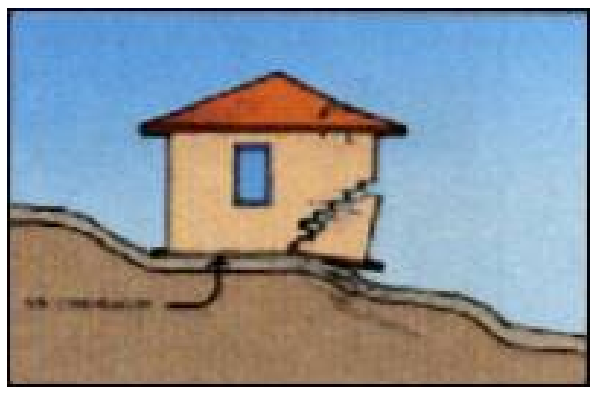

Figura 104. Vulnerabilidad alta: Cimentación

## 5. SUELOS

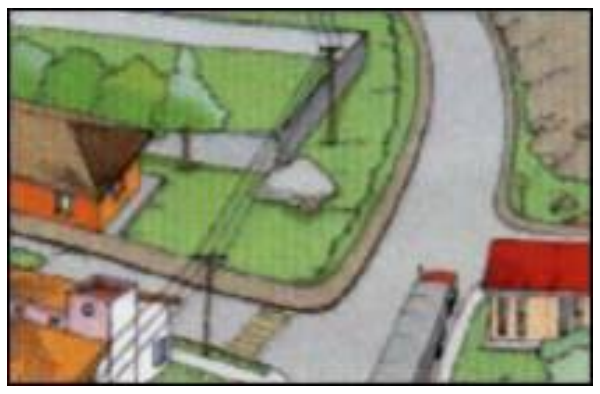
Figura 105. Vulnerabilidad baja: Suelos

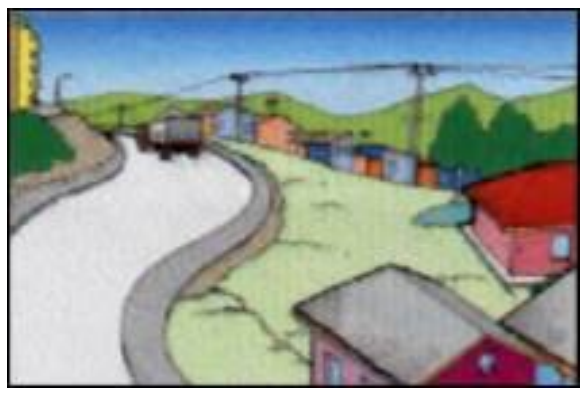
Figura 106. Vulnerabilidad media: Suelos

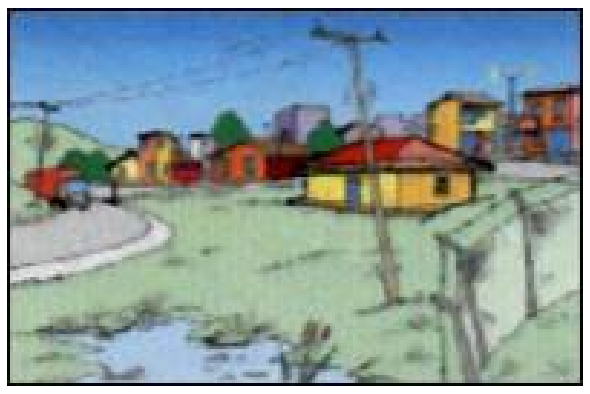
Figura 107. Vulnerabilidad alta: Suelos

Con un suelo duro en la base de la vivienda se contribuye a una vulnerabilidad sísmica baja. Este tipo de suelo se puede identificar cuando alrededor de las edificaciones no existen hundimientos, no se evidencian árboles ni postes inclinados, no se siente vibración cuando pasa un vehículo pesado cerca de la vivienda, o en general las viviendas no presentan agrietamientos o daños generalizados, especialmente agrietamiento en los pisos o hundimientos y desniveles en los mismos.

Un suelo de base de mediana resistencia contribuirá a una vulnerabilidad media. Este tipo de suelo se detecta por la presencia de algunos hundimientos y vibraciones por el paso de vehículos pesados, se pueden presentar algunos daños generalizados en viviendas o manifestaciones de hundimientos pequeños.

Un suelo de base blanda o de arena suelta contribuye a una vulnerabilidad alta. Este tipo de suelo se reconoce por el hundimiento en las zonas vecinas, se siente vibración al pasar vehículos pesados y la vivienda ha presentado asentamientos considerables en tiempo construcción. Así, la mayoría de las viviendas de la zona presentan agrietamientos y hundimientos.

## 5. ENTORNO

### Topografía

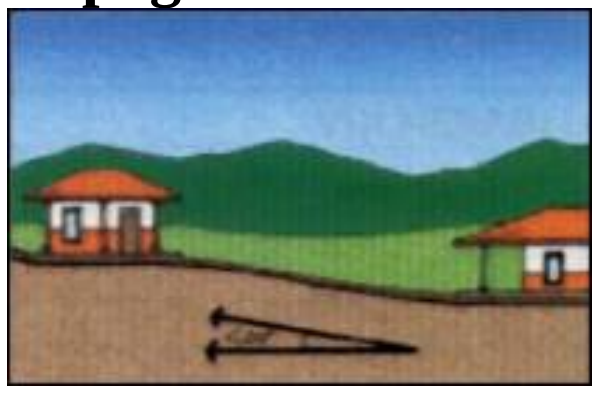
Figura 108. Vulnerabilidad baja: Entorno

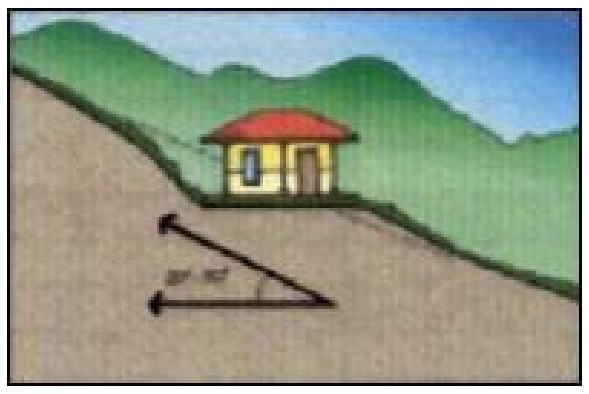
Figura 109. Vulnerabilidad media: Entorno

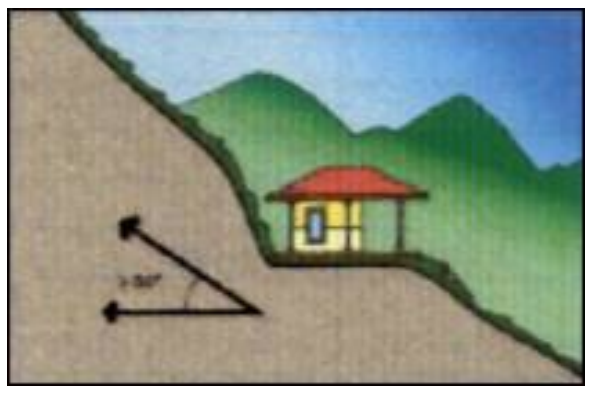
Figura 110. Vulnerabilidad alta: Entorno

El grado de inclinación de la topografía donde se encuentre la vivienda es directamente proporcional en la contribución de la vulnerabilidad sísmica de la misma.[74]

[74] [CHAVARRIA] capítulo.6.6.2.9

### 2.5.2 Metodologías para el cálculo de la vulnerabilidad

Los métodos para el estudio de la vulnerabilidad física de viviendas existentes se dividen en dos grandes grupos, los Métodos "exactos" o Analíticos y los Métodos "aproximados", Cualitativos o subjetivos.

#### 2.5.2.1 Métodos analíticos

La evaluación de la vulnerabilidad de edificios existentes por medio de métodos analíticos está fundamentada en los mismos principios utilizados para el diseño de construcciones sismo resistentes. Es decir, se considera como una evaluación por medio de un método analítico a la arrojada por un modelo previamente calibrado, el cual tiene en cuenta un análisis dinámico inelástico que permite conocer el proceso de plastificación paso a paso y el posterior colapso de la estructura, conocidos los ciclos de histéresis de sus componentes.

Cabe hacer notar que estos métodos no son del todo analíticos, ya que la fase de calibración del modelo requiere de muchos ensayos de laboratorio, los cuales permiten conocer el estado de los materiales y predecir, con un poco más de exactitud, su respuesta ante solicitaciones sísmicas.

Es por esto que la aplicabilidad de estos métodos es discutible por varias razones:

La alta complejidad del modelo que sólo justifica su utilización en casos muy especiales como el de edificaciones esenciales, o para estructuras que después de ser evaluadas con un método cualitativo hayan mostrado tener serias falacias ante una solicitación sísmica.

La necesidad de realizar el análisis utilizando varios tipos de registros de sismos, para cubrir las diferentes posibilidades de acción sobre la estructura. A continuación se enumeran los métodos:

1. Método NSR – 98
2. Método ATC-21
3. Método FEMA-273 [75]

#### 2.5.2.2 Problemas asociados con los métodos analíticos

Los problemas asociados con los métodos analíticos básicamente provienen de las dificultades intrínsecas de la modelación matemática de las estructuras reales.

Se puede afirmar que las propiedades actuales de los materiales y de los elementos estructurales pueden llegar a ser muy diferentes de las asumidas para el análisis y en la mayoría de los casos desconocidas. Es difícil desde un punto de vista matemático tener en cuenta el daño sufrido por la estructura antes del análisis y poder hacer una evaluación precisa de las condiciones iniciales, que permitan una estimación confiable de la respuesta obtenida después del análisis.

[75] [CHAVARRIA] capítulo .6.6.1

La evaluación de estructuras por los métodos analíticos puede llegar a ser una labor muy difícil de implementar a gran escala. Por esta razón, se siguen buscando métodos alternativos que permitan un análisis rápido de la vulnerabilidad sísmica. Sin embargo, la evaluación confiable de la vulnerabilidad de edificios muy particulares nunca podrá ser efectuada por procedimientos distintos a los métodos analíticos y es aquí donde el desarrollo de este tipo de métodos debe proseguir.[76]

#### 2.5.2.3 Métodos cualitativos

Existen distintos métodos aproximados propuestos por diversos autores para la evaluación de la vulnerabilidad de edificaciones, las que comúnmente no cuentan con información detallada acerca de su diseño estructural. Esto impide realizar su análisis mediante los sofisticados métodos modernos que actualmente se utilizan para la evaluación del comportamiento o desempeño y la confiabilidad estructural, de los cuales se hizo mención anteriormente.

Si se tiene en cuenta que en ocasiones es necesario evaluar edificaciones relativamente antiguas, de las cuales no se conservan memorias de su diseño, y que en otras ocasiones es necesario evaluar en forma ágil un amplio número de edificaciones, estas técnicas son realmente útiles, dado que no es posible en la práctica llevar a cabo este tipo de evaluaciones de otra forma. A continuación se enumeran los métodos:

1. Método del ATC – 14
2. Método NAVFAC
3. Métodos Japoneses
4. Método Venezolano
5. Método FEMA-178 = FEMA-310
6. Método del I.S.T.C.
7. Método del Índice de Vulnerabilidad
8. Metodología Propuesta por Hurtado y Cardona[77]

#### 2.5.2.4 Problemas asociados con los métodos cualitativos

Como ya se mencionó, los métodos subjetivos no permiten una evaluación precisa de la vulnerabilidad de las estructuras; sin embargo, llegan a ser la única herramienta disponible en los casos para los que la modelación matemática por medio de los métodos analíticos es muy costosa, compleja o involucra factores cuyo comportamiento es difícil de predecir. En este sentido, la opinión del experto y una base de datos empírico extensa, parece ser la solución a los problemas no resueltos hasta el momento.

[76] [CHAVARRIA] capítulo 6.7
[77] [CHAVARRIA] capítulo 6.6.2

La opinión subjetiva del experto en la mayoría de los casos, resulta ser una base de conocimiento difícil de transmitir y de "calibrar", por ello los métodos que se basan en este tipo de información deben utilizar conceptos simples, generales y fácilmente comprensibles por el usuario.

Lo anterior resulta, generalmente, en metodologías adaptadas a las tipologías y a las prácticas de construcción características de cada país e inclusive de cada región, lo que dificulta su aplicación en otros lugares.[78]

### 2.5.3 Metodología para el cálculo de la vulnerabilidad sísmica a implementar en detalle (Índice de Vulnerabilidad Sísmica)

La metodología **Índice de Vulnerabilidad Sísmica** es una metodología cualitativa creada por los Italianos Benedetti-Petrini en 1982 para realizar cálculos de vulnerabilidad sísmica a gran escala. En el año 1976 ocurrieron numerosos terremotos en diferentes regiones de Italia, lo que llevó a realizar análisis de comportamientos de edificios, gracias a los cuales investigadores de este país identificaron algunos de los parámetros más importantes que controlan el daño en los edificios.

Esos parámetros se han compilado en un formulario de levantamiento, el cual se viene utilizando y mejorando desde el año 1982, con el propósito de determinar de una forma rápida y sencilla la vulnerabilidad sísmica de edificios existentes. La combinación de dichos parámetros, por medio de una escala predefinida en un único valor numérico llamado Índice de Vulnerabilidad, es lo que se conoce hoy en día como el método del Índice de Vulnerabilidad [79]

| PARÁMETRO |
|---|
| 1. Organización del sistema resistente. |
| 2. Calidad del sistema resistente. |
| 3. Resistencia convencional. |
| 4. Posición del edificio y cimentación. |
| 5. Diafragma horizontales. |
| 6. Configuración en planta. |
| 7. Configuración en elevación. |
| 8. Distancia máxima entre los muros. |
| 9. Tipo de cubierta. |
| 10. Elementos no estructurales. |
| 11. Estado de conservación. |

Tabla 2. Parámetros para determinar el índice de vulnerabilidad sísmica

[78] [CHAVARRIA] capítulo .6.7
[79] [CHAVARRIA] capítulo 6.6.2.7

Mediante la ejecución de ciertas adaptaciones, los parámetros para el cálculo de la vulnerabilidad sísmica se relacionan con los aspectos que afectan la vulnerabilidad sísmica de las viviendas descritos en 2.5.1, como se muestra en la siguiente tabla.

| ASPECTOS QUE AFECTAN LA VULNERABILIDAD SÍSMICA | PARÁMETRO CORRESPONDIENTE EN EL ÍNDICE DE VULNERABILIDAD SÍSMICA |
|---|---|
| **Aspectos Geométricos** | |
| -- Irregularidad en la planta de la edificación | 6. Configuración en la planta |
| -- Cantidad de muros en dos direcciones | 8. Distancia máxima entre los muros |
| -- Irregularidad en la altura | 7. Configuración de la elevación |
| **Aspectos constructivos** | |
| -- Calidad de las juntas de pega de mortero | 2. Calidad del sistema resistente |
| -- Tipo y disposición de las unidades de mampostería | 2. Calidad del sistema resistente |
| -- Calidad de las juntas de los materiales | 2. Calidad del sistema resistente |
| **Aspectos estructurales** | |
| -- Muros confinados y reforzados | 1. Organización del sistema resistente |
| -- Detalles de las columnas y vigas de confinamiento | ----- (Difícil de obtener su valor en una vista rápida) |
| -- Vigas de amarre o corona | 9. Tipo de cubierta |
| -- Características de las aberturas | (Implícito en parámetro 3) |
| -- Entrepiso | 5. Diafragmas horizontales |
| -- Amarre de cubiertas | 9. Tipo de cubierta |
| **Cimentación** | ----- (Difícil de obtener como información básica)<br>----- (Involucrar en el parámetro 4) |
| **Suelos** | 3. Resistencia convencional |
| **Entorno** | 4. Posición del edificio y cimentación |

Tabla 3. Relación entre los parámetros del Índice de Vulnerabilidad y los aspectos que afectan la vulnerabilidad sísmica de las viviendas

### 2.5.3.1 Cálculo del Índice de Vulnerabilidad

El Índice de Vulnerabilidad se obtiene mediante la suma ponderada de los valores numéricos que expresan la "calidad sísmica" de cada uno de los parámetros estructurales y no estructurales, que se consideran juegan un papel importante en el comportamiento sísmico de la vivienda. Cada parámetro es afectado por un coeficiente de peso Wi, que varía entre 0.25 y 1.5. Este coeficiente refleja la importancia de cada uno de los parámetros dentro del sistema resistente de la vivienda.

A cada parámetro se le asigna durante la investigación de campo una de las cuatro clases A, B, C o D siguiendo una serie de instrucciones, con el propósito de minimizar las diferencias de apreciación entre los observadores. A cada una le corresponde un valor numérico Ki, que varía entre 0 y 45. De forma que el Índice de Vulnerabilidad se define por la siguiente expresión:

$$VI = \sum_{i=1}^{11} K_i W_i$$

| Parámetros | Clase $K_i$ | | | | Peso $W_i$ |
|---|---|---|---|---|---|
| | A | B | C | D | |
| 1. Organización del sistema resistente. | 0 | 5 | 20 | 45 | 1.00 |
| 2. Calidad del sistema resistente. | 0 | 5 | 25 | 45 | 0.25 |
| 3. Resistencia convencional. | 0 | 5 | 25 | 45 | 1.50 |
| 4. Posición del edificio y cimentación. | 0 | 5 | 25 | 45 | 0.75 |
| 5. Diafragmas horizontales. | 0 | 5 | 15 | 45 | 1.00 |
| 6. Configuración en planta. | 0 | 5 | 25 | 45 | 0.50 |
| 7. Configuración en elevación. | 0 | 5 | 25 | 45 | 1.00 |
| 8. Distancia máxima entre los muros. | 0 | 5 | 25 | 45 | 0.25 |
| 9. Tipo de cubierta. | 0 | 15 | 25 | 45 | 1.00 |
| 10. Elementos no estructurales. | 0 | 0 | 25 | 45 | 0.25 |
| 11. Estado de conservación. | 0 | 5 | 25 | 45 | 1.00 |

Tabla 4. Ejemplo de escala de vulnerabilidad de Benedetti-Petrini

El Índice de Vulnerabilidad define una escala de valores continuos desde 0 hasta 382.5, que es el máximo posible.

Los parámetros 3, 6, 7 y 8 son de naturaleza cuantitativa y requieren ciertas operaciones matemáticas muy sencillas, mientras que los parámetros 1, 2, 4, 5, 9, 10 y 11, son de naturaleza descriptiva.

### 2.5.3.2 Formulario para el levantamiento de la vulnerabilidad

| |
|---|
| No. edificio: ________<br>Dirección: ________________________________________________<br>Fecha: _________ d/m/a No. observador: ____ |
| 1. Organización del sistema resistente: __ |
| 2. Calidad del sistema resistente: __ |
| 3. Resistencia convencional<br>1. Número de pisos N: ____<br>2. Area total cubierta $A_t$: ________.__ $m^2$<br>3. Area resistente sentido x $A_x$: ______.__ $m^2$<br>sentido y $A_y$: ______.__ $m^2$<br>1. Resistencia cortante mampostería $t_k$: ______.__ $Ton/m^2$<br>2. Altura media de los pisos h: __.__ m<br>3. Peso específico mampostería $P_m$: ____.__ $Ton/m^3$<br>4. Peso por unidad de área diafragma $P_s$: __.____ $Ton/m^2$ |
| 4. Posición del edificio y de la cimentación: __ |
| 5. Diafragmas horizontales: __ |
| 6. Configuración en planta $b_1 = a/L$: __.____ $b_2 = b/L$: __.____ |
| 7. Configuración en elevación. Superficie porche %: ____.__<br>T/H: __.____ $\pm$ DM/M %: ______.__ |
| 8. Distancia máxima entre los muros L/S: ____.__ |
| 9. Tipo de cubierta: __ |
| 10. Elementos no estructurales: __ |
| 11. Estado de conservación: __ |

Tabla 5. Formulario para el levantamiento de la vulnerabilidad en las viviendas

La asignación de cada uno de los once parámetros del formulario con cada una de las cuatro clases A, B, C, D, la realizan especialistas en el área como ingenieros civiles y se lleva a cabo con la ayuda de una serie de instrucciones. En éstas se describe de forma muy breve el fundamento teórico de cada uno de los parámetros, con el objetivo de proporcionar al observador de campo un cierto criterio de selección.[80]

#### 2.5.3.3 Cálculo del Índice de Daño

Los valores del Índice de Vulnerabilidad por sí solos no aportan información suficiente para estimar el riesgo sísmico en las zonas de estudio, por lo tanto es conveniente la realización de escenarios de sismos que dejen entrever los daños que estos causarían en las viviendas.

El Índice de Daño se obtiene mediante las funciones de vulnerabilidad, que relaciona el índice de vulnerabilidad normalizado, con el índice de daño condicionado sobre la aceleración horizontal del suelo debido a un sismo propuesto.

El índice de vulnerabilidad normalizado se obtiene mediante la división del índice de vulnerabilidad entre 3.825, con el objetivo de poseer un índice de vulnerabilidad con un rango de valores entre 0 y 100.

Esta correlación ha sido revisada y se ha modificado desde su primera versión publicada en 1989, además se han calibrado datos procedentes de centros donde se han realizado extensivos estudios.

[80] [CHAVARRIA] capítulo .6.6.2.7

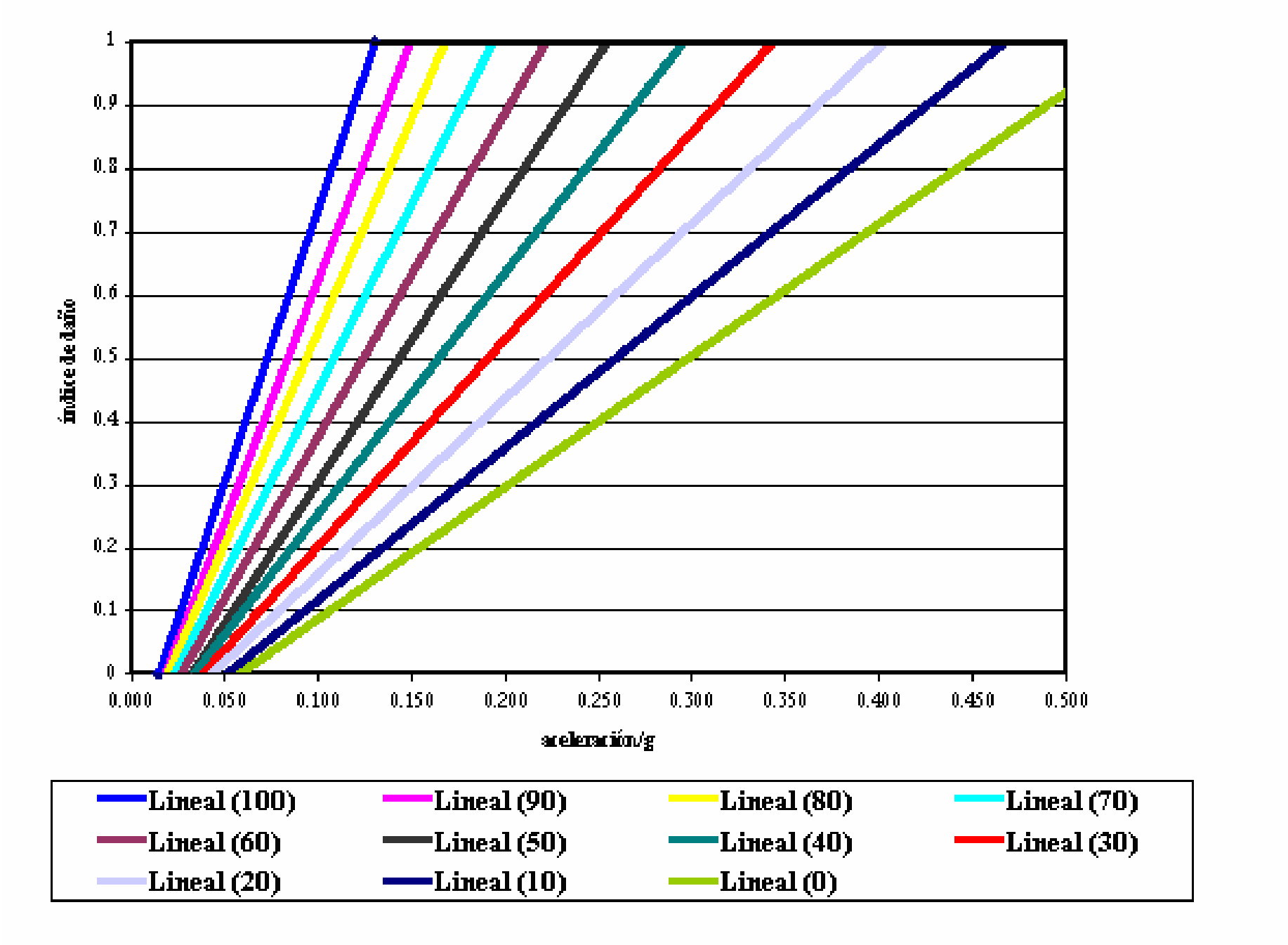

Figura 111. Funciones de vulnerabilidad-daño-aceleración

Estas curvas poseen una forma tri-lineal definida por dos puntos: la aceleración en la cual el daño comienza (d > 0) y la aceleración en la cual la vivienda colapsa completamente (d = 1). Por lo tanto, el daño se expresa en una escala normalizada (0 < d < 1) y representa la proporción dañada con respecto a la condición inicial del edificio.

En la siguiente tabla se describen las rectas mostradas en la figura x:

| Índice de Vulnerabilidad normalizado | Ecuación de la recta |
|---|---|
| 100 | Índice de Daño = 8.6154*(a/g) - 0.1231 |
| 90 | Índice de Daño = 7.6712*(a/g) - 0.1371 |
| 80 | Índice de Daño = 6.7470*(a/g) - 0.1325 |
| 70 | Índice de Daño = 5.8947*(a/g) - 0.1368 |
| 60 | Índice de Daño = 5.1376*(a/g) - 0.1376 |
| 50 | Índice de Daño = 4.5161*(a/g) - 0.1452 |
| 40 | Índice de Daño = 3.8356*(a/g) - 0.1301 |
| 30 | Índice de Daño = 3.2845*(a/g) - 0.1261 |
| 20 | Índice de Daño = 2.7861*(a/g) - 0.1194 |
| 10 | Índice de Daño = 2.4086*(a/g) - 0.1226 |
| 0 | Índice de Daño = 2.0786*(a/g) - 0.1188 |

Tabla 6. Funciones de vulnerabilidad para diferentes índices de vulnerabilidad

## 2.6 Información Catastral en Nicaragua

El catastro de Nicaragua se encuentra dividido en dos, de acuerdo al tipo de información que maneja sobre las notificaciones: el catastro de INETER y el catastro de cada municipalidad.

En el catastro de INETER se lleva información meramente de la cantidad de parcelas y lotificaciones con respecto a la forma y medidas de los mismos, con el objetivo de llevar un control de las lotificaciones a nivel nacional. Mientras que el catastro de las municipalidades lleva información sobre la cantidad de área construida en las lotificaciones y las características de dichas construcciones, con el objetivo de recaudar el impuesto de bienes inmuebles IBI.

La información catastral es importante cuando se desean realizar estudios de vulnerabilidad sísmica, por el hecho de contar de antemano con información sobre las construcciones. Información que fue levantada con otros propósitos, pero que ayuda en los estudios de vulnerabilidad sísmica.

La ayuda que se obtiene de dicha información varía dependiendo de la utilidad que se le desee atribuir en cada proyecto de vulnerabilidad sísmica. Sus alcances van desde el hecho de poderse dar una idea somera del tipo de edificaciones existentes en el área de estudio, hasta una agrupación de las viviendas en tipologías constructivas para realizar estudios sobre una muestra considerable de cada tipologia y luego extrapolar los resultados a las otras viviendas no estudiadas.

Las alcaldías municipales poseen el sistema informático llamado SisCat, desde el cual llevan el control de la información de las parcelas necesarias para la recaudación del impuesto IBI. Dicho sistema posee una estructura de tablas, entre ellas se encuentra una donde se almacena información relevante para la agrupación de viviendas en tipologías constructivas como son el tipo de pared, tipo de techo, año de construcción, estado de la edificación, tipo de uso, entre otros.

# CAPÍTULO 3 DESARROLLO DEL SISTEMA

## 3.1 PLATAFORMA DE DESARROLLO UTILIZADA

Cada una de las herramientas informáticas utilizadas para el desarrollo de esta tesis y el sistema resultante de ella VULNESIS (Vulnerabilidad Sísmica Sistema – o – Sistema Vulnerabilidad Sísmica), tienen una justificación en su selección, no se seleccionaron de forma arbitraria o por preferencia de los tesistas/desarrolladores.

### 3.1.1 Utilización del ArcGIS de ESRI

Los motivos por los cuales se seleccionó el ArcGIS de ESRI son variados, pero la selección de este software SIG es un elemento importante que influye en la selección de las demás herramientas informáticas, como el lenguaje de programación, la tecnología de componentes y la base de datos. A continuación se enumeran en orden de importancia los motivos por los cuales se utiliza el SIG ArcGIS de ESRI.

1. El software SIG para el cual se posee licencias legítimas en las dos instituciones del estado Nicaragüense (INETER y UNI) involucradas en estudios de vulnerabilidad sísmica e interesadas en el desarrollo de la herramienta informática que implemente la metodología propuesta por dichas instituciones.

2. El ArcGIS es el programa que lidera la industria de los SIG, y es el software SIG del que actualmente se posee mayor documentación y personas capacitadas en Nicaragua.

3. Está construido sobre una capa de componentes COM que permite la extensión y la personalización desde cualquier lenguaje de programación que soporte COM, o desde el interior a través del lenguaje VBA (Visual Basic for Aplications), que viene integrado

### 3.1.2 Acerca el lenguaje de programación y la P.O.O. en Visual Basic 6 (VB6)

Se utilizó VB6 como lenguaje de programación por distintos motivos, **el principal** de ellos, es el hecho que el ArcGIS contiene internamente un VBA integrado desde el cual se puede extender directamente el SIG sin necesidad de utilizar otro lenguaje de programación, de esta forma fue posible hacer pruebas del funcionamiento de las bibliotecas AcrObjects del ArcGIS que corrieran dentro del SIG sin necesidad de agregarle un componente COM, y de esta manera poder atrapar los errores justos en el código fuente para depurarlos.

A continuación de enumeran los motivos para la selección del VB6.

1. El ArcGIS posee el lenguaje de programación VBA integrado, mediante el cual se realizaron las primeras versiones de cada una de las rutinas de software implementadas que tenían que ver con ArcObjects, esto nos sirvió para comprobar si era posible implementar todas nuestras ideas sobre la personalización del SIG. Otra gran ventaja que nos proveyó el VBA, fue su facilidad de depuración al programar dentro del ambiente del SIG, este automáticamente nos lleva a la línea donde ocurrió el error, mientras que los componentes compilados desde VB6 únicamente nos pueden mostrar un error en tiempo de ejecución y nuestro manejador de errores sólo nos indicaba el número del error, una pequeña descripción técnica y el módulo en cual aconteció.

2. Era posible utilizar los ArcObjects desde .NET, a través del uso de interoperatividad COM/.NET, sin embargo se detectaron muchos problemas de mal funcionamiento de los ArcObjects consumidos a través de este método, problema que es inherente de la forma en la cual trabaja esta interoperatibilidad. Es por eso, que la mayoría de desarrolladores de ArcObjets continuaron usando lenguajes de programación basados en COM, por no poderse dar el lujo de desarrollar sistemas inestables.

3. La documentación del ArcGIS que se entrega por la compra de la licencia está enfocada en la utilización del Visual Basic 6 como lenguaje de programación para la extensión/personalización del software.

4. Los autores de la tesis poseen bastos conocimientos en el lenguaje de programación y experiencia en los COM (ArcObjects) para la extensión/personalización del ArcGIS.

El sistema se desarrolló siguiendo técnicas de diseño orientadas a objetos, lo que implica una implementación en clases dentro del lenguaje de programación (VB6), basada directamente en el diagrama de clases generado en la etapa de diseño del software.

Todas las versiones de Visual Basic incluyendo VB6 antes de VB.NET han sido fuertemente criticadas de no ser lenguajes de programación orientados a objetos, lo que consideramos una aseveración sólo desde el punto de vista purista, sino fuese así, el lenguaje nos hubiera presentado el inconveniente de no permitirnos implementar el diagrama de clases elaborado en el diseño del sistema, y menos realizar clases empaquetadas y reutilizables dentro de componentes (COM).

La mayoría de los renombrados diseñadores orientados a objetos, como el Dr. David West[81], Rockfor Lhotka[82], entre otros. Concuerdan en que, el lenguaje de desarrollo o la herramienta es algo secundario, según ellos la clave para el éxito es tener un buen análisis, diseñar correctamente e implementar con el lenguaje de programación de su preferencia.

[81] [WEST04] páginas x, xix, 15
[82] [LHOTKA98] página 1

Por lo que según ellos es teóricamente posible crear programas basados en este paradigma (O.O.), usando Fortran o Cobol. Por supuesto, la correcta elección de la herramienta nos ahorrará muchos problemas al momento de implementar nuestro diseño.

Técnicamente hablando, un lenguaje de programación orientado a objetos debe de incluir la herencia. Siguiendo esta definición, VB6 no es orientado a objetos y para algunas personas esto seria el fin de la discusión. Sin embargo, en VB6 se puede simular la herencia mediante la utilización de interfaces, delegación y jerarquías de objetos, que al fin de cuentas, siempre permitirán que el programador orientado a objetos vea al sistema de software como un conjunto de objetos que interactúan entre si, mediante las distintas relaciones que implica un diseño basado en esta metodología. Por lo tanto, aunque VB6 no permita herencia no significa que no se pueda realizar programación orientada a objetos con él.

Se utilizo VB6 stand alone (no VBA) para generar los componentes que empaquetan las clases con la lógica del negocio, la capa de acceso a los datos, y las dos interfaces de usuario implementadas. La primera UI que consume la parte de la lógica del negocio que es independiente del SIG y la segunda UI, que consume la parte de la lógica del negocio que es posible ejecutar dentro del SIG. Esta última UI, accede a los ArcObjects desde fuera del ArcGIS (a diferencia del VBA que trae incorporado la herramienta SIG).

La utilización de VB6 nos amarró directamente a la utilización de componentes COM, DCOM, COM+ o MTS (fuera de la tecnología .NET y de los web services, XML, serializaciones o remoting), lo que implicó la implementación manual de la serialización de los objetos para dejar abierta la comunicación entre componentes remotos con un rendimiento aceptable.

### 3.1.3 Bases de datos

El ArcGIS permite almacenar los datos geográficos en distintos formatos, tales como: Coverage (formato utilizado en las versiones WorkStation para UNIX), shape-file (formato Standard que se empezó a utilizar con ArcView para Windows) y geo-databases (formato introducido a partir del ArcGIS 8.0, donde los datos geográficos se pueden almacenar en RDBMS como MS Access, MS SQL-Server u Oracle).

Debido a que la dirección de geofísica de INETER posee un servidor de base de datos con MS SQL-Server adecuadamente licenciado (en el cual se almacena datos para otras aplicaciones), se utilizo este RDBMS para almacenar la Base de Datos alfanumérica de VULNESIS. Sin embargo, para almacenar la información y datos espaciales para el sistema antes mencionado, se utilizó el formato Personal geo-database de ArcGIS, la cual posee la característica técnica intrínseca de utilizar MS-Acces como fuente para su almacenamiento.

Además estos dos tipos de RDBMS poseen las características de poder ser fácilmente unidas en una sola Base de Datos, una vez que en el futuro se adquiera la licencia ArcSDE, software que permite la utilización de RDBMS empresariales para guardar los datos geográficos. De tal forma que se puedan subir todas la personals geo-databases desde Access hacia SQL-Server.

Cuando se dé este traslado de los datos geográficos de personal geo-database (sobre Access) a geo-databases Empresariales (sobre SQL-Server), también se pudiese realizar la adaptación del VULNESIS para que tanto sus datos geográficos como alfanuméricos utilicen geo-databases Empresariales, lo que no sería un trabajo tan complejo gracias a que sólo se tendría que adaptar la capa de **Centrados-en-Datos objetos de negocios** para que en vez de hablar Access hable SQL-Server.

## 3.2 DESCRIPCIÓN DEL PROCESO DE DESARROLLO DE SOFTWARE UTILIZADO

Se siguió un proceso de desarrollo de software iterativo incremental de dos espirales basados en la ingeniería de software orientada a objetos. Donde en ambas espirales se identifican los riesgos y se procede a superarlos, pero en la primer espiral se iteró sobre las etapas de análisis de requerimientos, análisis como tal, diseño e implementación de prototipos y rutinas de software ligadas a la eliminación de riesgos. Hasta que se consideró que el diseño estaba completado se procedió a iterar sobre la segunda espiral. Donde se abarcan las etapas de implementación del diseño, pruebas y distribución con algunos regresos a la primer espiral, especialmente en la parte de diseño.

Al seguir este enfoque se apostó por el diseño. Algo coherente con la teoría que expresa que un buen diseño orientado a objetos incurrirá en un proceso de programación más sencillo y sin tantos errores ni parches. Gracias a que se anticipan la agregación de particularidades en etapas anteriores a la programación, evitando así la introducción código de último momento que vuelve inestable toda la aplicación (código spaghetti).

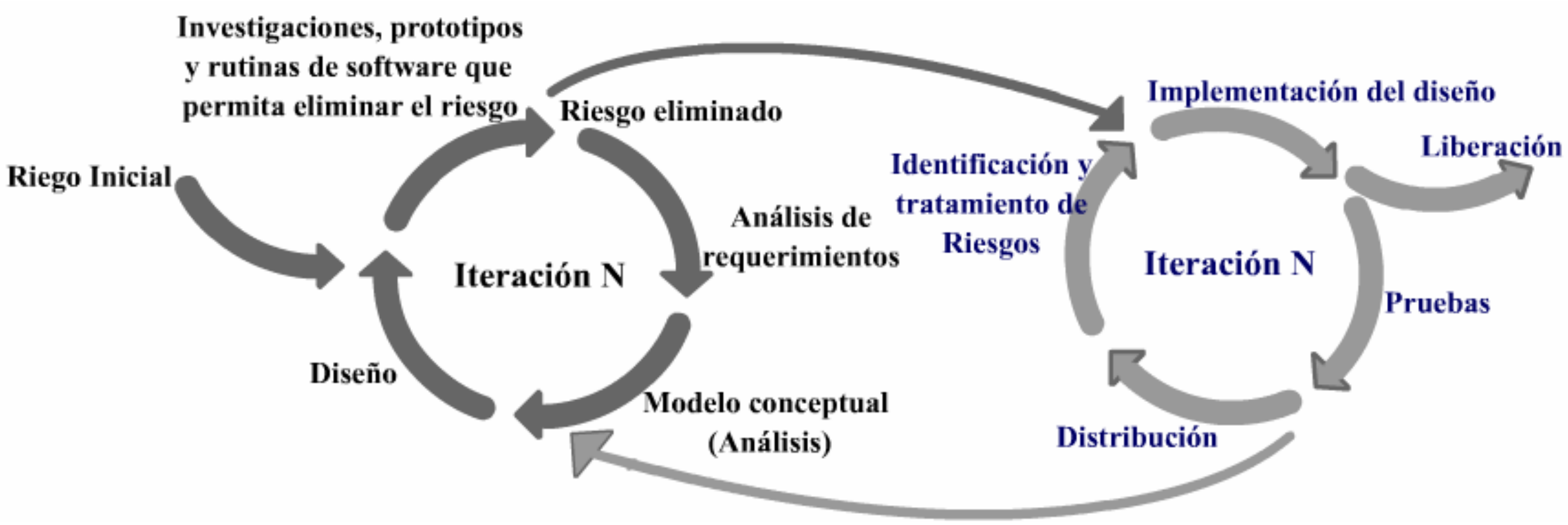

Figura 112. Proceso Iterativo incremental en dos espirales

| | |
|---|---|
| 1. Redacción de los casos de uso de requisitos | 2. Identificación de requerimientos |
| 3. Identificación de casos de usos funcionales | 4. Diagrama de casos de uso |
| 5. Diagramas de actividad para cada caso de uso | 6. Flujos de eventos para casos de uso |
| 7. Identificación de objetos de negocios potenciales | 8. Escenarios para los casos de uso |
| 9. Diagramas de secuencia | 10. Diagramas de colaboración |
| 11. Consolidación de objetos de negocios | 12. Diagrama de clases |
| 13. Clases/Atributos/Métodos/Eventos | 14. División de clases en 2 (separar persistencia) |
| 15. Agregación de clases auxiliares (serialización) | 16. Nuevos diagramas de clases |
| 17. Diagramas de Estado | 17. Diagramas de actividades |
| 19. Diagramas de componentes | 20. Diagramas de despliegue |

Tabla 7. Documentos, diagramas y modelos generados en la primer espiral.

## 3.3 ANÁLISIS DEL SISTEMA

### 3.3.1 Casos de Uso de Requerimientos

Se realizaron numerosas reuniones, entrevistas y conversaciones con los usuarios interesados en el desarrollo del sistema. Auxiliados de los casos de usos de requerimientos se fueron identificando y refinando una a una las necesidades a automatizar. De forma general, estos casos de uso los utilizamos con el objetivo de definir requerimientos, más que para describir funcionalidades.

A continuación se enumeran y describen todos los casos de usos de requerimientos definidos:

**1. Crear nuevo proyecto**
Se mantendrá un inventario de todos los estudios de vulnerabilidad y daños debido a sismos, se podrá acceder a ver cualquier proyecto anterior.

El usuario deberá ser capaz de crear un nuevo proyecto y adjuntarle a este la tabla catastral del área en estudio, para que automáticamente se descubra todo el dominio de tipologías existentes para el proyecto actual; llamadas subtipologías, cuantas edificaciones hay de esas subtipologías y que así mismo dichas edificaciones queden marcadas a cual subtipología pertenece.

**2. Definir tipologías**
El usuario deberá definir tipologías genéricas para agrupar las edificaciones del proyecto actual dentro de estas mediante la inclusión de las subtipologías. Esto se realiza con el fin de poder seleccionar las edificaciones de cada tipología a las que se les realizara el trabajo de campo.

El usuario deberá definir estas tipologías y agregarlas al maestro de tipologías para el proyecto actual. También existirán los maestros de tipos de pared, tipo de techo, estado el edificio y uso del edificio, estos además existirán tanto para todo el sistema como para el proyecto actual.

**3. Seleccionar muestra**
El usuario indicará la cantidad o porcentaje, de edificaciones a realizarle el trabajo de campo por tipologías y el sistema automáticamente las seleccionara al azar.

**4. Imprimir tabla y formatos para trabajo de campo**
El sistema imprimirá una tabla que contendrá información sobre la ubicación de la casa, campos vacíos a ser llenados en el trabajo de campo como las coordenadas (x,y) y nombre de la imagen (foto tomada a la vivienda). Esta tabla también indicará las viviendas a las que habrá de levantárseles los parámetros para el cálculo del IV, para los que también el sistema imprimirá el formato para cada una de las edificaciones debidamente ligada con la tabla mediante indicadores.

**5. Definir pesos de los parámetros para el cálculo del índice de vulnerabilidad**
El usuario deberá ser capaz de indicar el valor de los pesos de cada parámetro para calcular el IV. Si este valor se cambia en etapas posteriores a esta en el sistema, podría afectar algunos cálculos por lo tanto se tendrán que volver a hacer, si este valor se cambia deberá de avisar.

**6. Cargar cartografía**
El usuario podrá cargar elementos cartográficos para presentar los resultados sobre el mapa, elementos como: las parcelas, las manzanas y un polígono con el área del proyecto. Estos elementos el sistema los utilizará automáticamente para presentar los resultados de forma espacial y con distintos niveles de granularidad.

**7. Subir tablas y formatos con la información recopilada en el trabajo de campo**
El usuario basándose en su tabla y formatos llevados al trabajo de campo procederá a subir toda esta información al sistema donde automáticamente se ligarán las viviendas de la tabla catastral con su información espacial. Además se realizarán los cálculos de la vulnerabilidad para las edificaciones que así lo requerían, mientras se van presentando el mapa de vulnerabilidad sísmica por viviendas.

**8. Definir sismos para presentar escenarios de daños**
El usuario definirá escenarios de sismos para observar el índice de daños acorde al sismo y la vulnerabilidad de las viviendas. Cada proyecto constará con un maestro de escenarios de sismos de tal forma que no se repitan sismos dentro del proyecto.

**9. Presentar resultados de la vulnerabilidad sísmica, daños debido a sismos y las tipologías de forma espacial.**
El usuario podrá seleccionar entre diferentes tipos de mapas a presentar: mapas de vulnerabilidad sísmica y mapas de daños por cada sismo que haya definido. Ambos tipos de mapas podrán ser presentados (si acaso tiene toda la base cartográfica) dentro de 3 niveles de granularidad: a nivel de vivienda, nivel de cuadras y nivel de todo el proyecto.

### 3.3.2 Requerimientos identificados

A partir de cada caso de uso de requerimiento se procedió a identificar requerimientos más específicos, estos requerimientos identificados son un mapeo directo desde los casos de uso de requerimientos por lo que a continuación sólo se presenta uno. (La lista completa se encuentra en los anexos)

**1. Crear nuevo proyecto**

1) El sistema mantendrá un inventario de todos los proyectos desarrollados.

2) El inventario de proyectos podrá ser explorado para seleccionar algún proyecto, ya sea concluido como en desarrollo.

3) Se deberá adjuntar a cada proyecto una tabla de las viviendas para dicho proyecto.

4) Se deberá encontrar automáticamente todo el dominio de tipologías (subtipologías) existentes para el proyecto actual a través de la tabla catastral e indicar cuantas edificaciones existen de cada subtipologías además de marcar a que subtipología pertenece cada edificación para el posterior maestro de tipologías.

Mediante estos requerimientos se lograron identificar las funcionalidades del sistema, las que representan mediante los diagramas de casos de uso y sus respectivos documentos de flujo de eventos.

### 3.3.3 Diagrama de Casos de Uso

El diagrama de caso de uso representa una vista desde afuera del sistema y las funcionalidades de cada caso de uso del diagrama (funcionalidades del sistema) es presentado mediante cada documento de flujo de evento asociado a cada caso de uso.

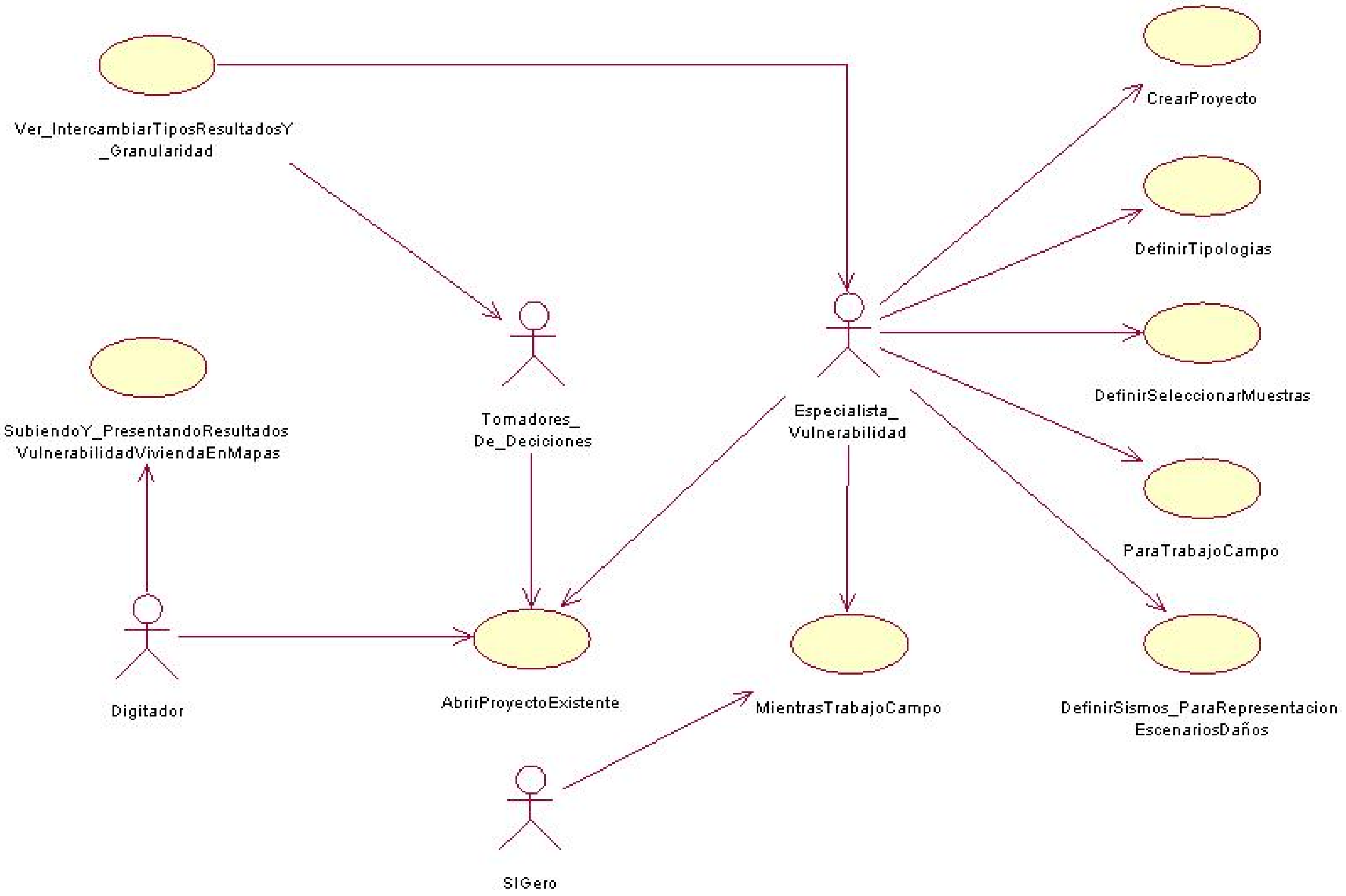

Figura 113. Diagrama de Casos de Uso para *Vulnesis*

### 3.3.4 Flujos de eventos de Casos de Uso (Casos de Uso funcionales)

Representan las funcionalidades del sistema y son más que los casos de usos de requerimientos. La mayoría de los casos de usos de requerimientos van a derivar en casos de usos funcionales detallados (Flujos de eventos de Casos de Uso), estos son usados más para determinar funcionalidades que para identificar requerimientos del sistema y también son útiles para luego identificar los objetos potenciales.

Para ejemplificar los Flujos de eventos de Caso de Uso, se ha incluido el correspondiente al caso de uso 8: SubiendoY_PresentandoResultadosVulnerabilidadViviendasMapas, el resto de los flujos de eventos fueron incluidos en los anexos.

## Flujo de Eventos para el caso de uso SubiendoY_PresentandoResultadosVulnerabilidadViviendasMapas

### .1 Condiciones previas:

Se deberá haber concluido el caso de uso MientrasTrabajoCampo y se deben tener a mano una parte del trabajo recopilado del campo para ser introducida en este caso de uso.

### .2 Flujo principal:

Este caso de uso inicia cuando el usuario está dentro del sistema en la pantalla principal donde se le presenta una matriz con los proyectos existentes a los que tiene acceso y selecciona un proyecto que se encuentra en el estado subiendo información de campo. Se presentará el mapa de índice de vulnerabilidad en su granularidad fina, en la misma ventana en que se presenta el mapa aparecerá una barra de herramientas con distintas opciones, entre las que se observará la opción de **Viviendas**, la cual al ser invocada llama la ejecución del subflujo S-1: **ViviendasProyecto**.

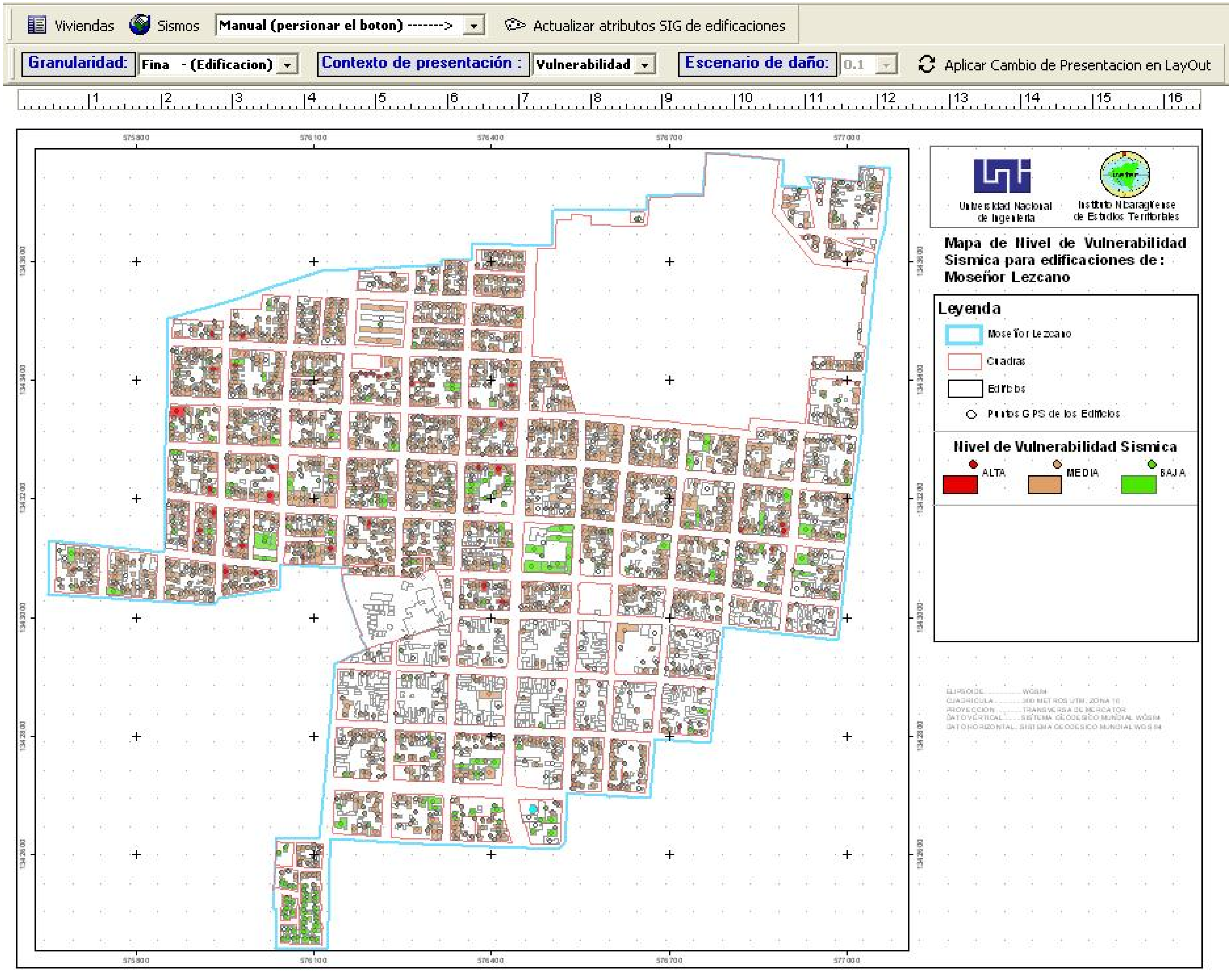

Figura 114. Pantalla principal de VULNESIS en ambiente SIG

**S-1: ViviendasProyecto**
Con este subflujo se despliega una ventana de dialogo donde se podrán seleccionar criterios (ID's de viviendas, Tipos de viviendas "Encuestadas, No Encuestadas", Editadas, Tipologia, Nivel de Vulnerabilidad) para la búsqueda de viviendas que los cumplan dichos criterios pueden ser mezclados para potenciar las búsquedas de forma ortogonal.

Se prestaran las opciones de aceptar filtros la que de ser seleccionada invocaría al subflujo de S-2: SelecciónVivienda con las viviendas que cumplen con los criterios seleccionados por el usuario, también se presentara la opción de mostrar todas la cual al ser invocada también ejecutaría el subflujo S-2: SelecciónVivienda mostrando todas las viviendas sin tomar en cuenta ningún criterio de selección, una tercera opción es cancelar que simplemente cierra la ventana de dialogo de actual.

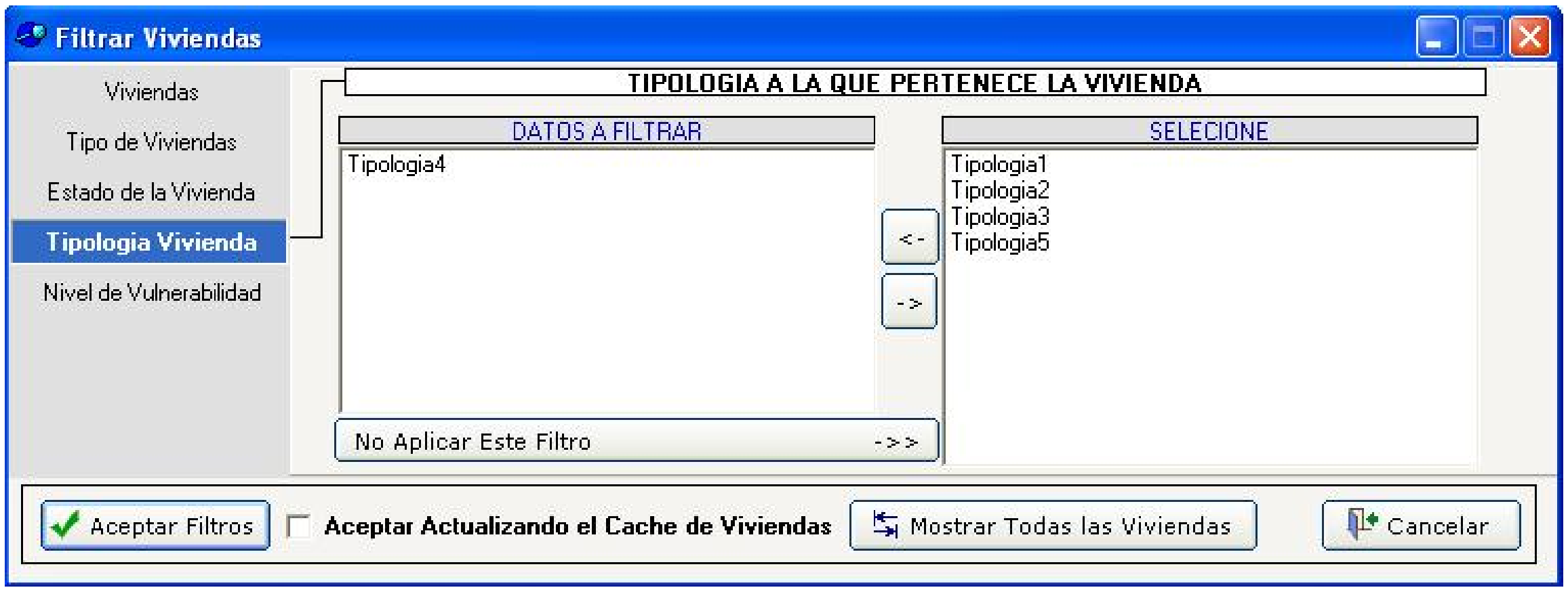

Figura 115. Formulario para la selección de criterios a filtrar

**S-2: <u>SelecciónVivienda</u>**
Al ejecutarse este subflujo se abre un formulario presentándose una matriz con todas las viviendas del proyecto que aplican a los filtros seleccionados por el usuario, se informa sobre la cantidad total de viviendas del proyecto, la cantidad filtrada, los filtros utilizados y se disponen de una lista desplegable desde la cual el usuario podrá seleccionar la escala a la cual se hará el acercamiento grafico de la vivienda seleccionada.

Luego que el usuario seleccione alguna vivienda, puede seleccionar las opciones de editar vivienda seleccionada la que al ser invocada ejecuta el caso de uso **<u>EditarVivienda</u>**, también se presenta las opciones observar vivienda, Filtrar, Salir y refrescar viviendas. Para los que se ejecutan sus subflujos de eventos correspondientes.

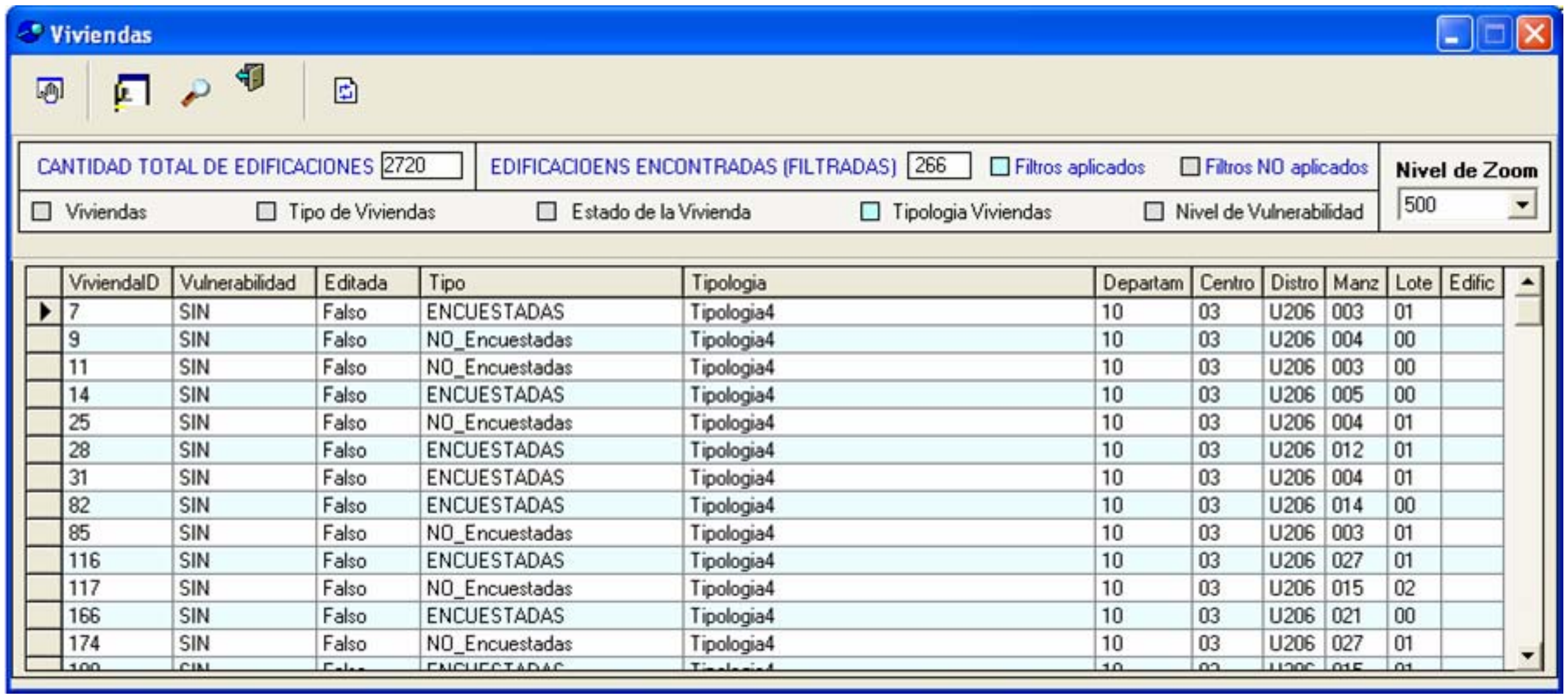

Figura 116. Formulario Viviendas (donde se presentan todas las viviendas que cumplen los criterios de los filtros)

### 3.3.5 Diagramas de Actividad para los Casos de Uso

Los diagramas de actividad se pueden utilizar para especificar distintos elementos del desarrollo de software como: la descripción de un flujo de trabajo, el comportamiento de los objetos de una clase, la lógica de una operación o método y el modelado del flujo de eventos de un caso de uso.

Para el desarrollo de VULNESIS se creó un diagrama de actividad por cada caso de uso, como una forma de complementar el flujo de evento de cada uno de ellos. De esta forma, es posible observar el flujo de eventos a través de un diagrama y así tener otra visión de las funcionalidades del sistema.

Para ejemplificar los diagramas de actividades para los Casos de Uso, se ha incluido el diagrama de actividad correspondiente al caso de uso 8:
SubiendoY_PresentandoResultadosVulnerabilidadViviendasMapas, el resto de los flujos de los diagramas de este tipo fueron incluidos en los anexos.

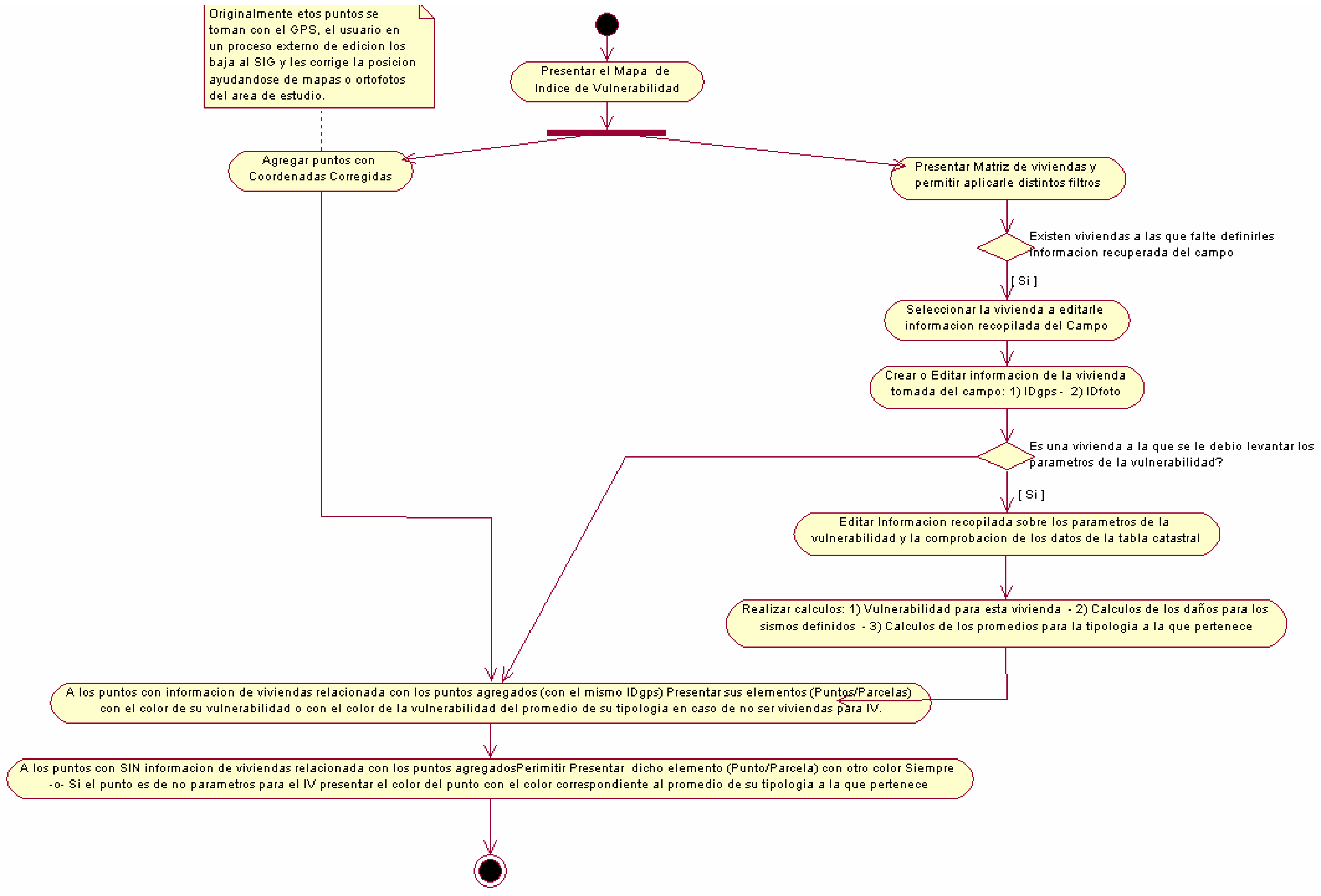

Figura 117. Diagrama de actividad para el Casos de Uso: SubiendoY PresentandoResultadosVulnerabilidadViviendasMapas

### 3.3.6 Objetos de negocios potenciales

A continuación se procedió a identificar los objetos de negocio que soportarían las funcionalidades encontradas.

Estos objetos de negocio se identificaron mediante la recopilación de los nombres que existen en la redacción de los casos de uso, más la utilización de un nivel de juicio para no recoger aquellos elementos que claramente no representaban objetos potenciales, con esto se ubicó a los posibles objetos del mundo real que necesitaban ser modelados en el software para cumplir sus funcionalidades.

No todos los objetos identificados en esta parte se convirtieron en objetos del software, algunos se convirtieron en atributos de otros objetos, mientras que otros se borraron enteramente, sin embargo fue un buen punto de inicio.

A continuación se enumeran las clases candidatas y una descripción somera de cada una de las ellas, identificadas a través del análisis detallado de cada caso de uso.

| | |
|---|---|
| 1. Proyecto | *CreaListaVivienda |
| 2. Proyectos | Colección de objetos 1-Proyecto |
| 3. Tipo | |
| 4. TiposPared | Colección de objetos 3-Tipo |
| 5. TiposTecho | Colección de objetos 3-Tipo |
| 6. TiposUso | Colección de objetos 3-Tipo |
| 7. TiposEstado | Colección de objetos 3-Tipo |
| 8 .TiposProyecto | Compuesto de 4, 5, 6 y 7. *Compara al actual con 9-TiposSistema y de encontrar diferencias en algun tipo se permitirá definir como nuevo o actualizar de uno existente.<br>*Una vez sin diferencias con 9 se permitira salvar y guardado creará un objeto SubTologia. |
| 9. TiposSistema | Compuesto de todos los TiposProyecto |
| 10. ViviendaCatastral | Fila de la tabla catastral |
| 11. ViviendaIndependiente | Vivienda agregada arbitrariamente en el trabajo de campo (por lo general van a ser instalaciones del gobierno a las que por no cobrarsele impuesto no aparacen en la tabla catastral) |
| 12. ViviendaBase | Clase base para 10 y 11 (IV, IDvivienda) |
| 13. ViviendasCatastrales | Colección de objetos tipo 11.<br>*Crea tipo: pared, techo, uso y estado |
| 14. ViviendasIndependientes | Colección de objetos tipo 12. |
| 15. Viviendas | Compuesta de 13 y 14 |
| 16. SubTipolgia | (asignada) |
| 17. SubTipolgias | Colección de 16 |
| 21. ParámetroVulnerabilidad | (Numero, Nombre, Peso, Valores A...D) |
| 22. ParaetrosVulnerabilidad | Colección de 21 |
| 23. ParámetrosVulnerabilidadPlantilla | Colección de 21-Especial, Siempre existentes y propuestos para cada proyecto. |

Tabla 8. Clases candidatas identificadas del análisis del caso de uso “Proyecto”

| | |
|---|---|
| 18. Tipolgía | *AsignarSubTipologia, *DesacignarSubTipologia (CantidadMuestrear, Avg-IV, Tot_IV, Niv_IV, ID) |
| 19. TipolgíasProyecto | *Muestrear |
| 20. TipologíasSistema | |

Tabla 9. Clases candidatas identificadas del análisis del caso de uso “DefinirTipologias”

| | |
|---|---|
| 21. ParámetroVulnerabilidad | (Numero, Nombre, Peso, Valores A...D) |
| 22. ParámetrosVulnerabilidad | Colección de 21 |
| 23. ParámetrosVulnerabilidadPlantilla | Colección de 21-Especial, Siempre existentes y propuestos para cada proyecto. |
| 24. ElementosCartográficos | |
| 25. ElementoCartográfico | 0...3 [Parcelas, Cuadras, Proyecto] |

Tabla 10. Clases candidatas identificadas del análisis del caso de uso “MientrasTrabajoCampo”

26. Sismo
27. Sismos
28. DañoBase
29. DañoTipología
30. DañoVivienda

Tabla 11. Clases candidatas identificadas del análisis del caso de uso “DefinirSismos_ParaRepresentacion EscenariosDaños”

| | |
|---|---|
| 31. MapaVulenerabilidad | (granularidad) |
| 32. Punto | *agregarMas *SiEsNecesarioConvertir *CorregirPropiedades X,Y de Punto |
| 33. Puntos | Colección de 32-Punto |

Tabla 12. Clases candidatas identificadas del análisis del caso de uso “SubiendoY_PresentandoResultadosVulnerabilidadViviendaEnMapas”

| | |
|---|---|
| 34. MapaDaño | (granularidad) |
| 35. MapaTipologia | (granularidad) |

Tabla 13. Clases candidatas identificadas del análisis del caso de uso “IntercambiarTiposResultadosY Granularidad”

### 3.3.7 Diagramas de secuencia

Los diagramas de secuencia fueron creados a partir de los casos de uso. Así como los casos de uso representan el lado del cliente de la aplicación, los diagramas de secuencia están fuertemente enfocados en la perspectiva del análisis de la aplicación para luego ayudar con el diseño.

Estos diagramas se utilizaron para definir la interacción en el tiempo entre distintos objetos encargados de cumplir la funcionalidad requerida por el caso de uso, como mínimo cada uno de los caso de uso que identificamos para VULNESIS contiene un escenario al que está ligado un diagrama de secuencia, también a cada escenario de los casos de uso se liga un diagrama de colaboración, mediante el cual se observa la interacción de los objetos desde otra perspectiva.

Para ejemplificar los diagramas de secuencia, se ha utilizado el correspondiente al escenario: DefinirSismos_ParaRepresentacionEscenariosDaños, el cual se presenta en la figura x, el resto de los diagramas de este tipo fueron incluidos en los anexos.

### 3.3.8 Diagramas de colaboración

Al igual que los diagramas de secuencia los diagramas de colaboración también modelan objetos. En este tipo de diagramas también fueron creado partiendo de los casos de uso y los requerimientos del usuario, el objetivo de este diagrama es el mismo que el de secuencia, la diferencia entre ambos radica en que aquí no importa tanto el orden cronológico de la comunicación entre los objetos, este está enfocado en las relaciones y comunicación de los objetos. La distribución de los objetos en el diagrama permite observar adecuadamente la interacción de un objeto con respecto a los demás. La estructura estática viene dada por los enlaces, la dinámica por el envió de mensaje a través de los enlaces.

Con el desarrollo de los dos tipos de diagramas (secuencia y colaboración) fue posible identificar y comprender mejor los objetos que intervenían en la realización de las funcionalidades requeridas por los casos de uso. Estos diagramas también nos ayudaron a definir el diagrama de clases resultante para el sistema, eliminando así clases innecesarias o combinando clases que en algunos casos se convirtieron en atributos de otras clases. Gracias a estos diagramas fue posible identificar los atributos y métodos de las clases resultantes y aunque en teoría los diagramas de secuencia y actividad son similares en la práctica ambos se complementaron para ayudar a identificar todo lo anteriormente mencionado.

Para ejemplificar los diagramas de colaboración, se ha incluido un diagrama figura x+1 que complementa el diagrama se secuencia del mismo nombre correspondiente al escenario: DefinirSismos_ParaRepresentacionEscenariosDaños. El resto de los diagramas de este tipo fueron incluidos en los anexos.

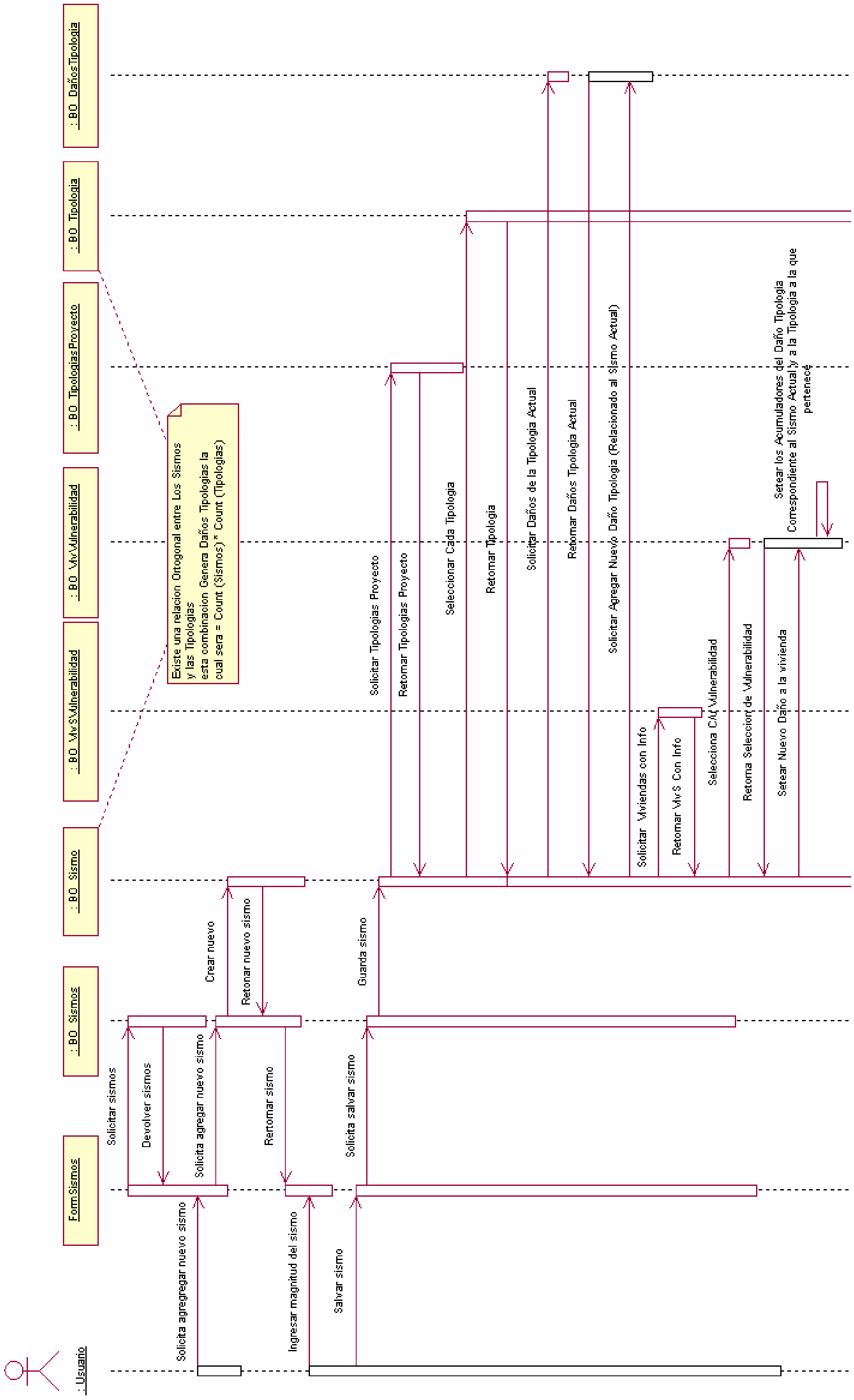

Figura 118. Diagrama de Secuencia para el escenario DefinirSismos ParaRepresentacionEscenariosDaños

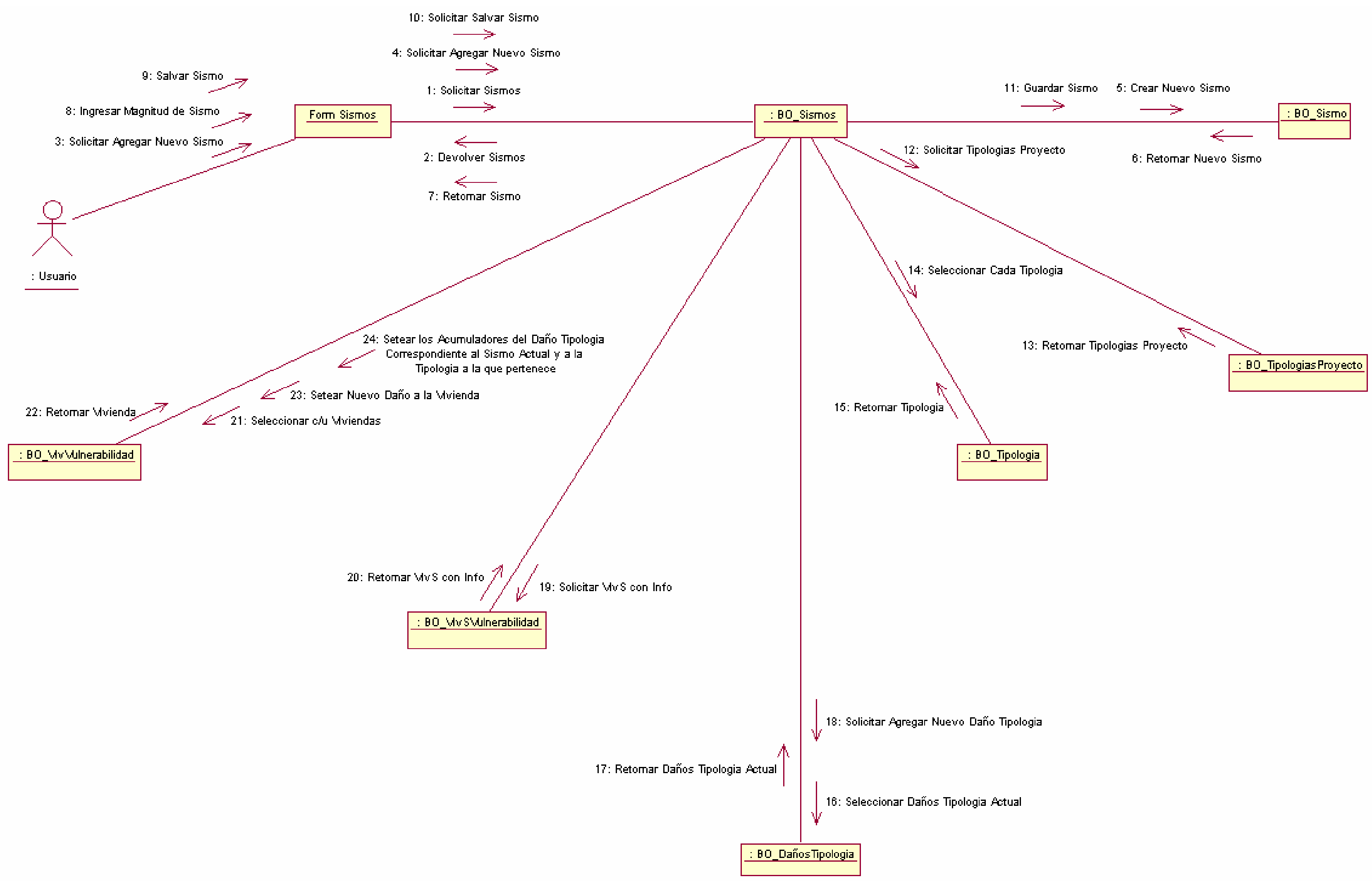

Figura 119. Diagrama de Colaboración para el escenario DefinirSismos_ParaRepresentacionEscenariosDaños

## 3.4 DISEÑO DEL SISTEMA

Una vez que se entendió el sistema lo suficiente nos movimos hacia el diseño, donde se analizó el trabajo interno del sistema. En el paradigma de análisis y diseño orientado a objeto esto significa que se debió de empezar a crear jerarquía de clases que van a formar la columna vertebral del sistema, justamente lo que siguió.

### 3.4.1 Diagrama de Clases de Objetos Potenciales "Con representación simple"

En los diagramas anteriores se han modelado los requerimientos del usuario y que funcionalidades se iban a necesitar para llenar esos requerimientos, ahí se modelo el análisis o el ¿Qué? Ahora con el diagrama de clases se empieza a modelar el diseño o el ¿Cómo?

Gracias a un análisis orientado a objetos se creo esta primer versión de clases del sistema, las cuales representan completas entidades del mundo real que relacionadas implementan todas las funcionalidades, reglas y requerimientos de la aplicación. También representan la forma en como el desarrollador de la UI (Interfaz de usuario) ve el sistema.

Las clases representadas en estos diagramas de VULNESIS son más que simples repositorios de funciones globales con parámetros que se pasan a través de sus atributos, y más que entidades de tablas de una base de datos relacional.

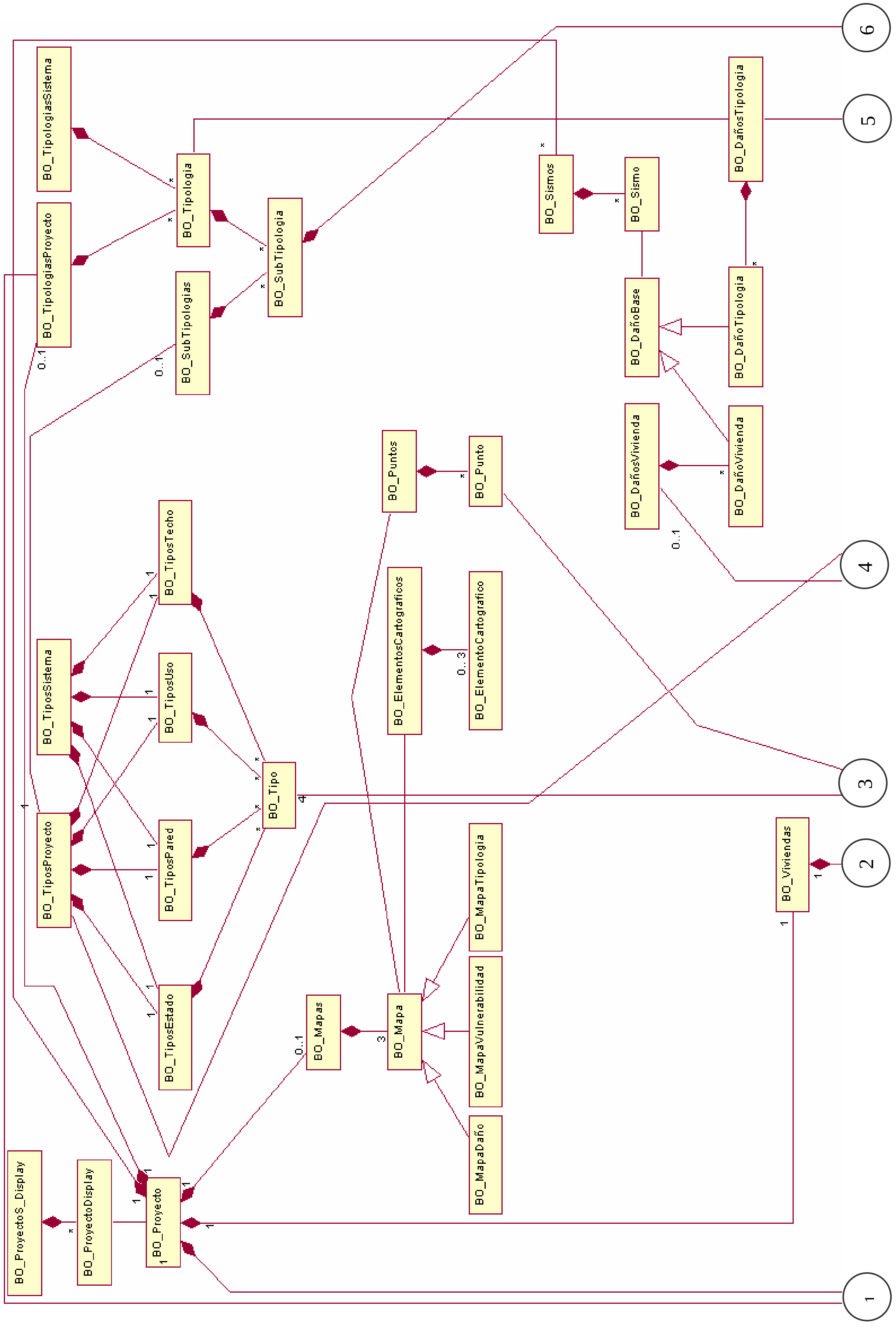

Figura 120. Diagrama de  Clases con representación simple (Objetos de Negocio)    Parte superior del diagrama (1/2)

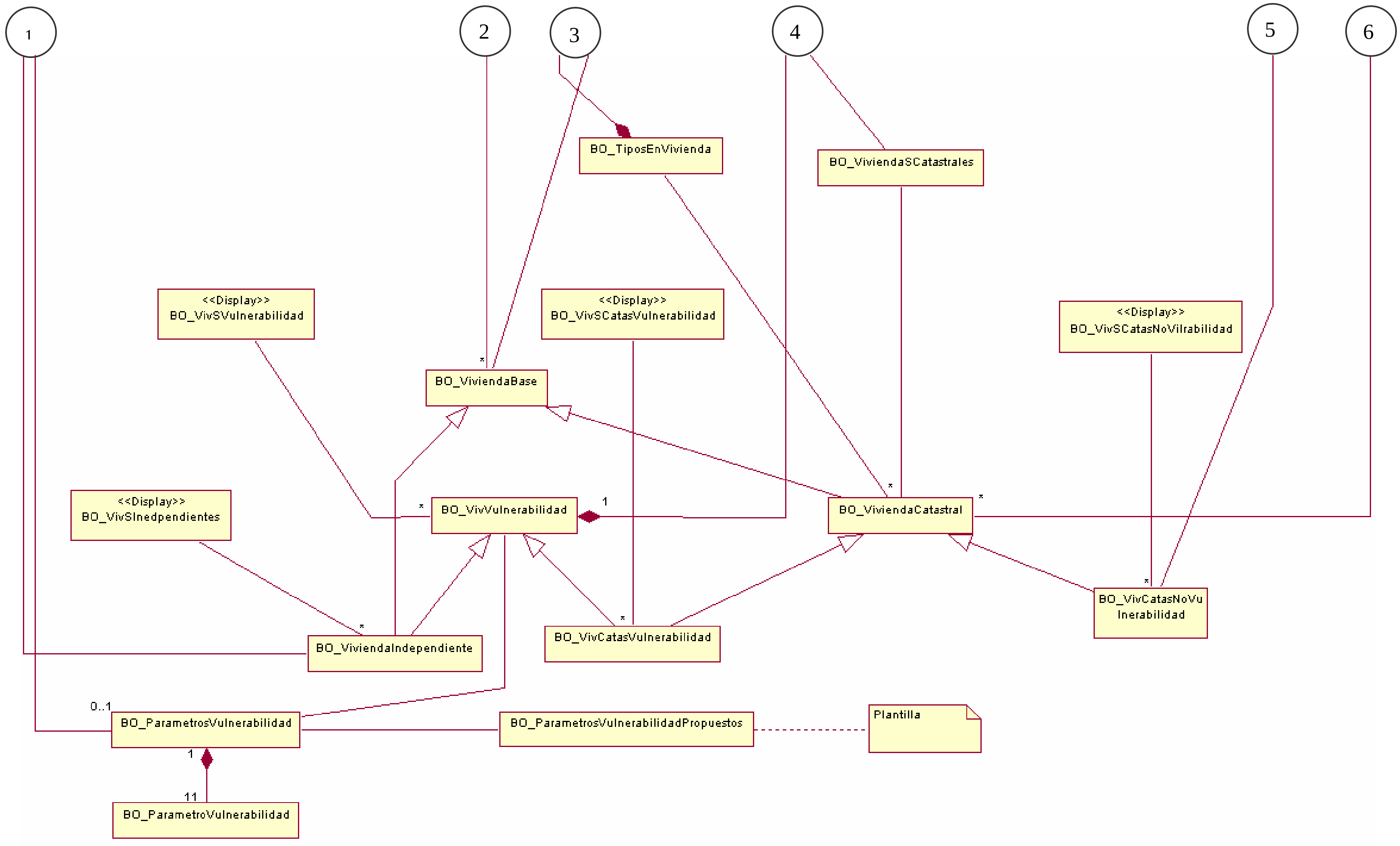

Figura 121. Diagrama de Clases con representación simple (Objetos de Negocio) Parte inferior del diagrama (2/2)

### 3.4.2 Diagrama de Clases Final (Clases implementadas) "Con representación simple"

De forma natural y como lo contempla el proceso de desarrollo de software orientado a objetos e iterativo incremental, esta primera versión de las clases (3.4.1) sufrió modificaciones a medida que se iba avanzando en la elaboración del software. A continuación se presenta el diagrama de clases final, el cual lo acomodamos en sub-diagramas donde se agrupan las clases que tienen más comunicación entre si, para facilitar su proceso de lectura, aunque también se generó un diagrama mayor donde están integradas todas las clases.

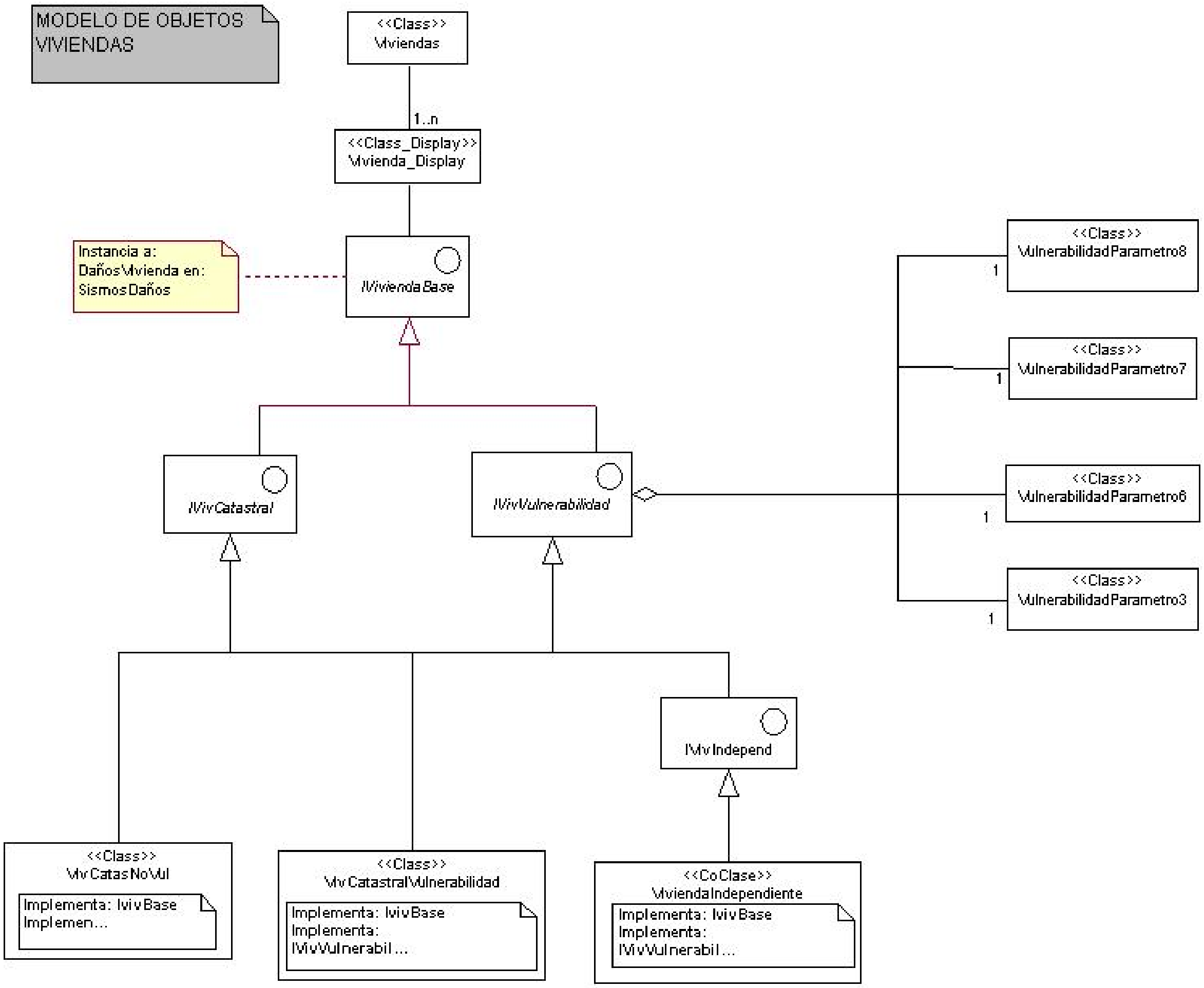

Figura 122. Diagrama de lógica del negocio, sub-diagrama de clases VIVIENDAS

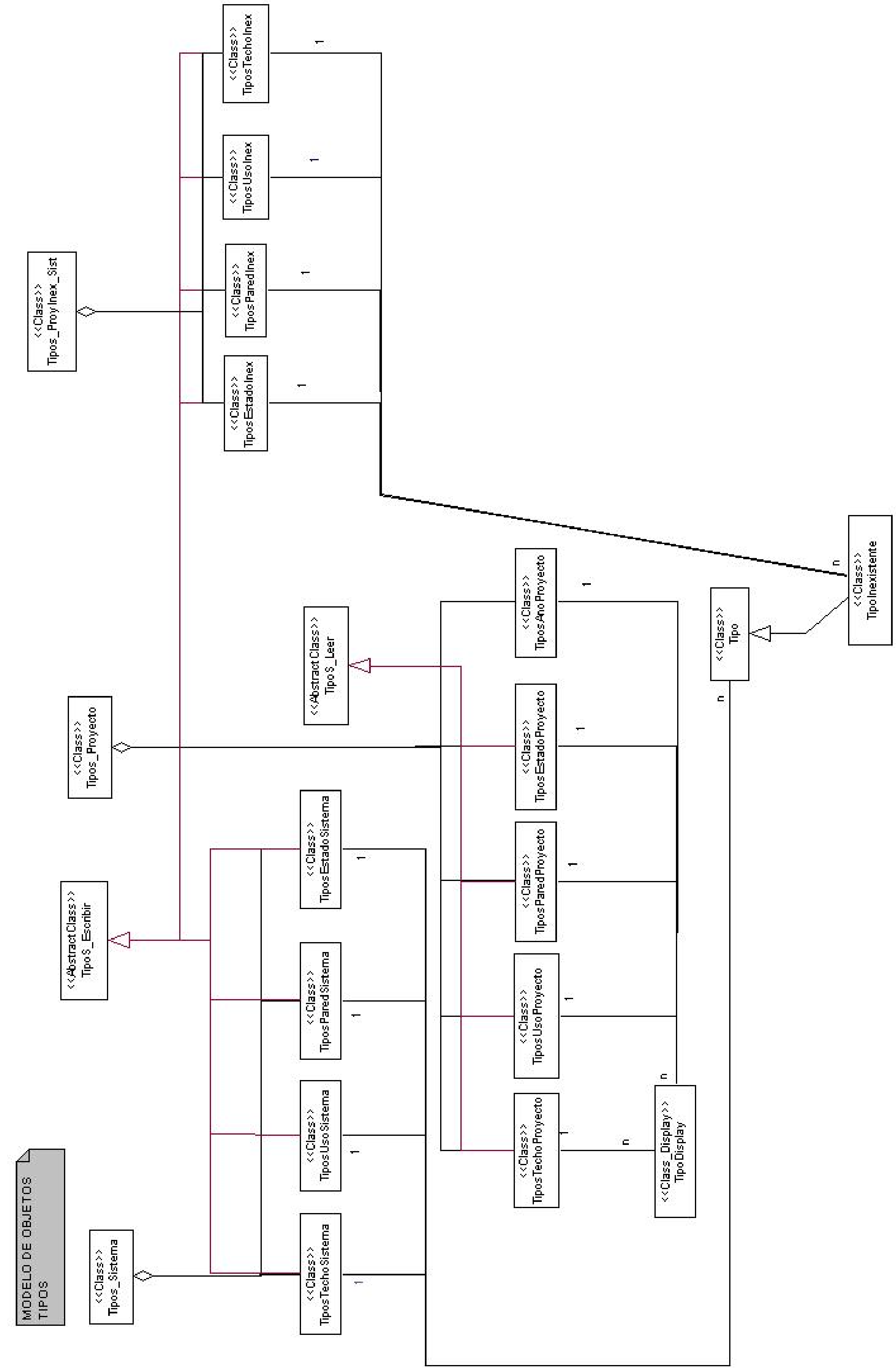

Figura 123. Diagrama de lógica del negocio, sub-diagrama de clases TIPOS

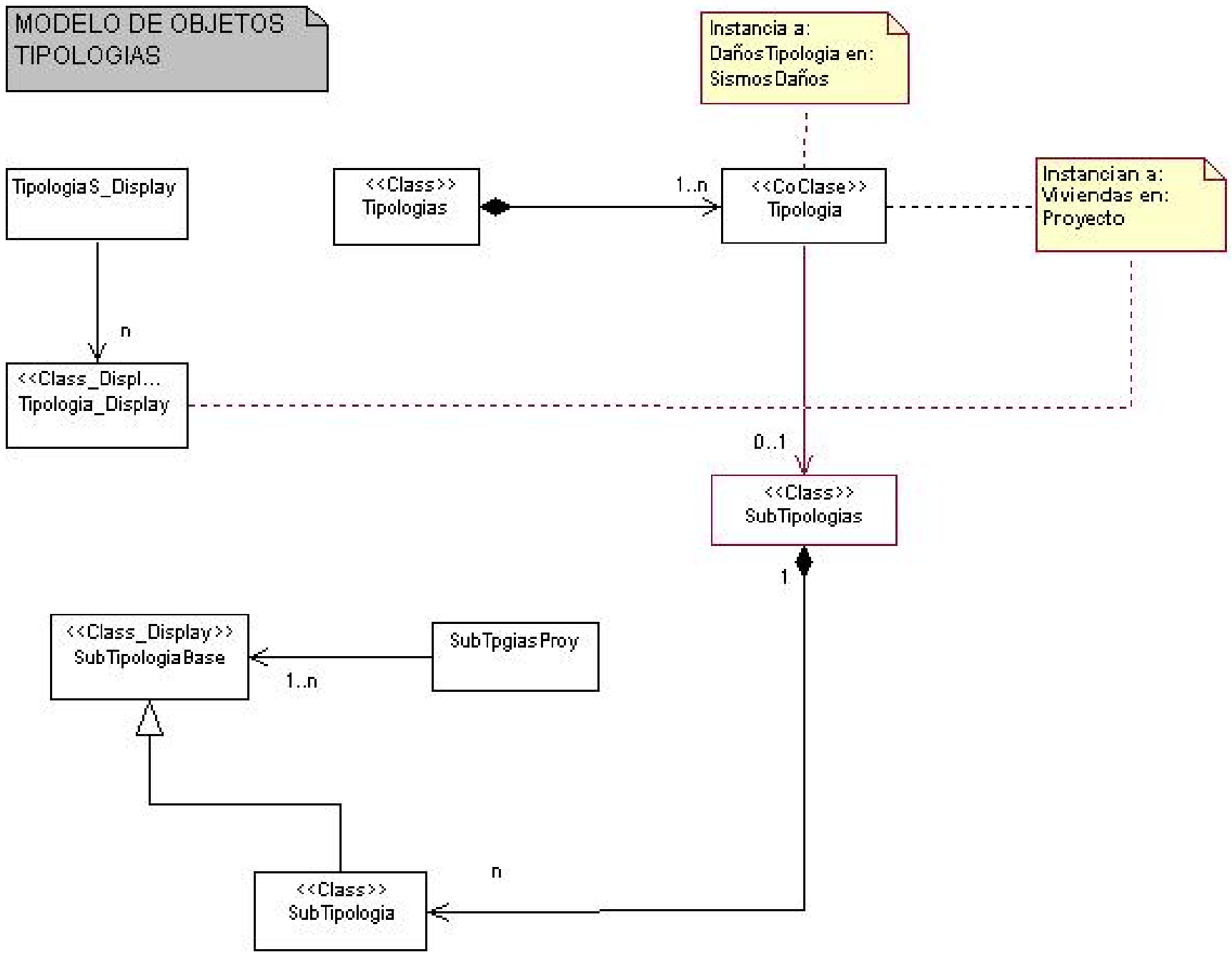

Figura 124. Diagrama de lógica del negocio, sub-diagrama de clases TIPOLOGÍAS

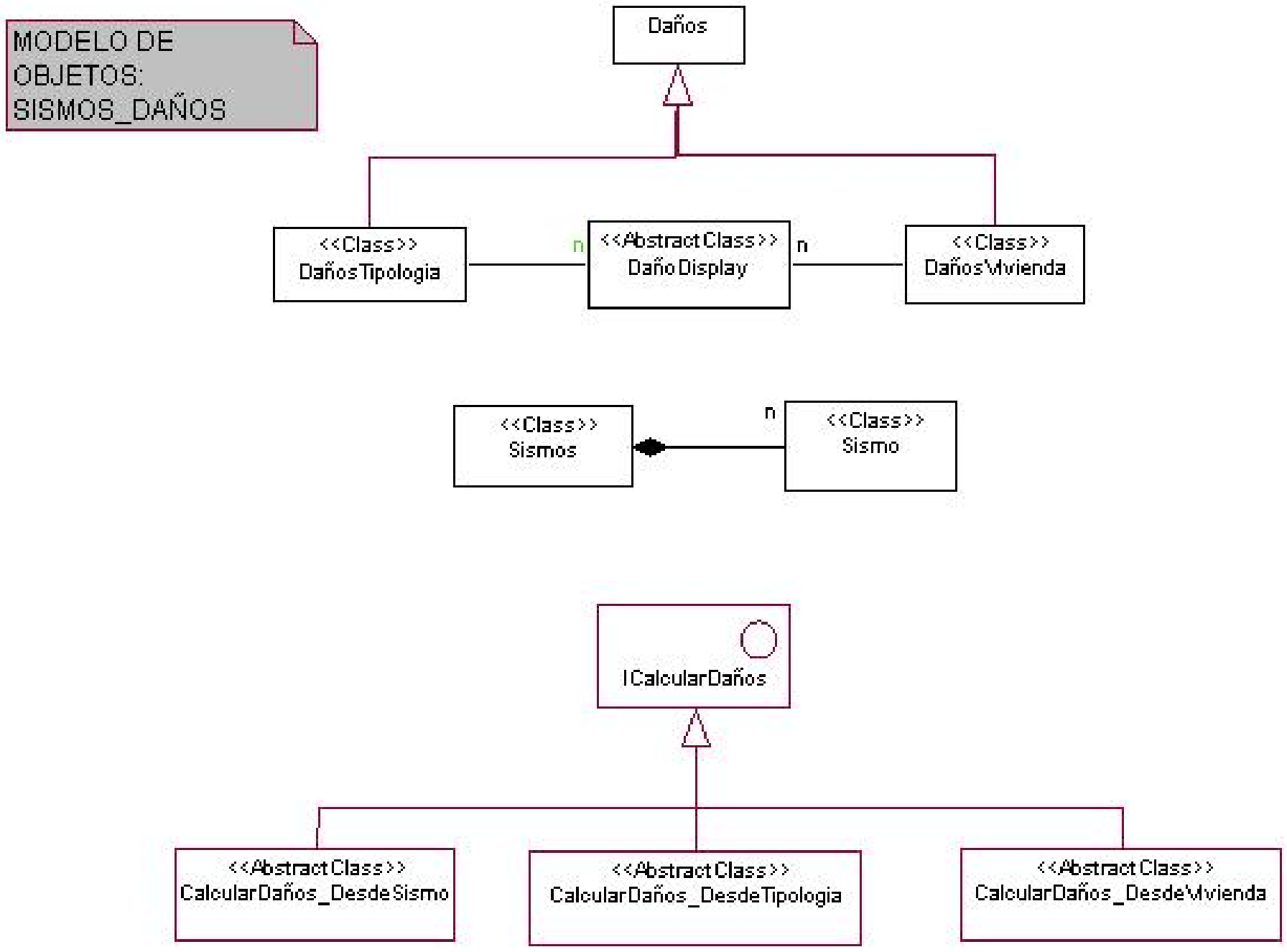

Figura 125. Diagrama de lógica del negocio, sub-diagrama de clases SISMOS_DAÑOS

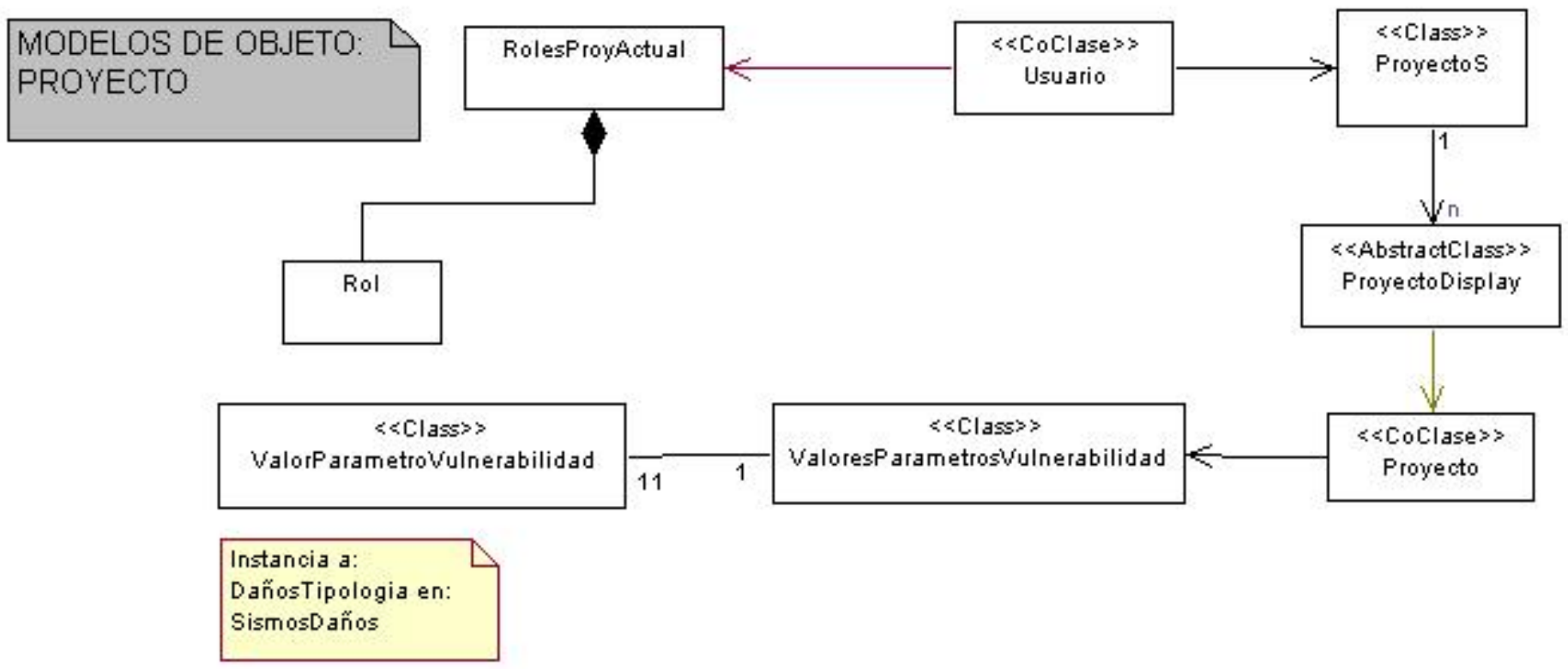

Figura 126. Diagrama de lógica del negocio, sub-diagrama de clases PROYECTO

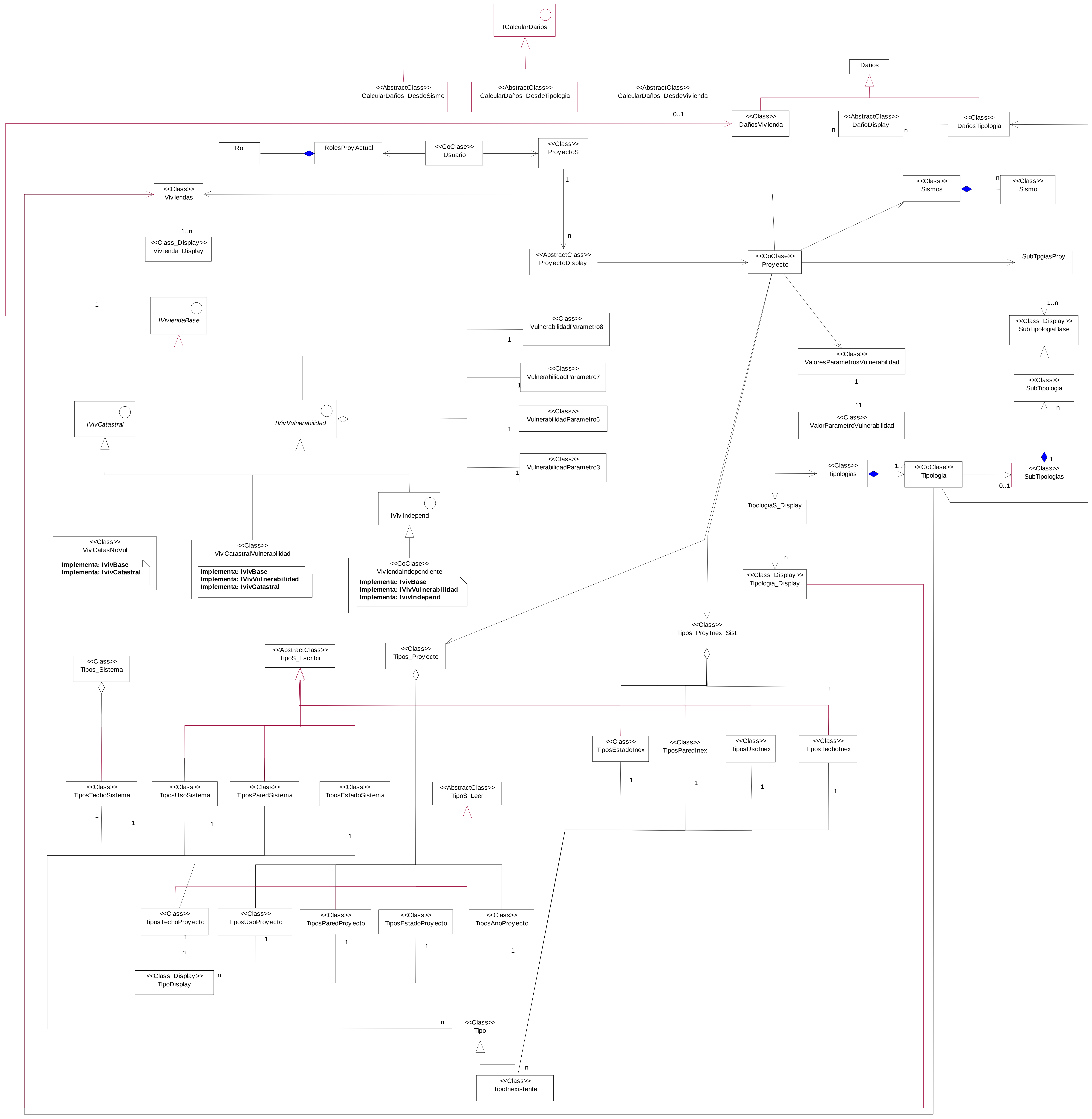

Figura 123. Vista completa del diagrama de clases de lógica del negocio (Con representación simple)

### 3.4.3 Diagrama de clases de la lógica del negocio – “Con representación típica”

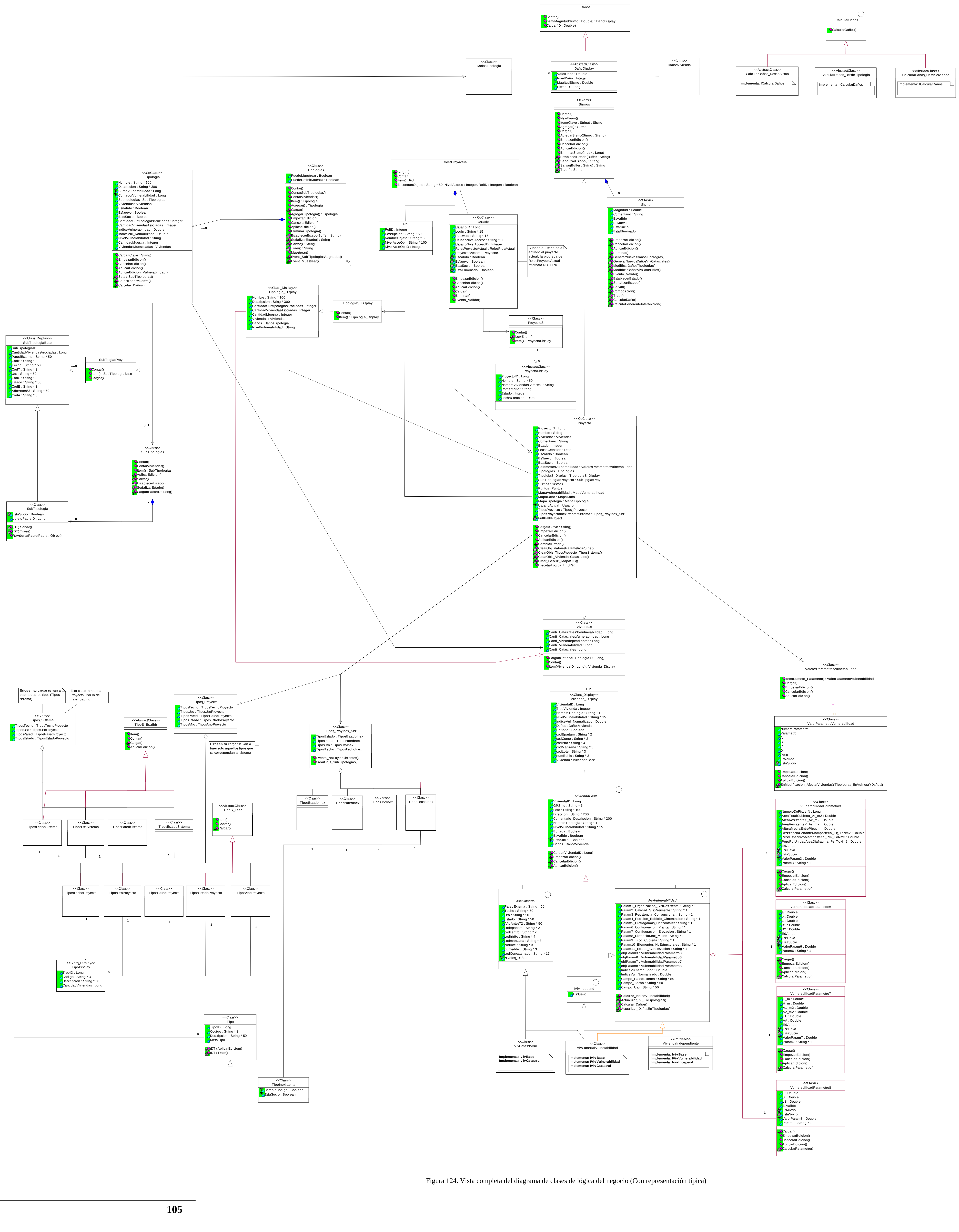

Figura 124. Vista completa del diagrama de clases de lógica del negocio (Con representación típica)

### 3.4.4 Diagramas de Estado

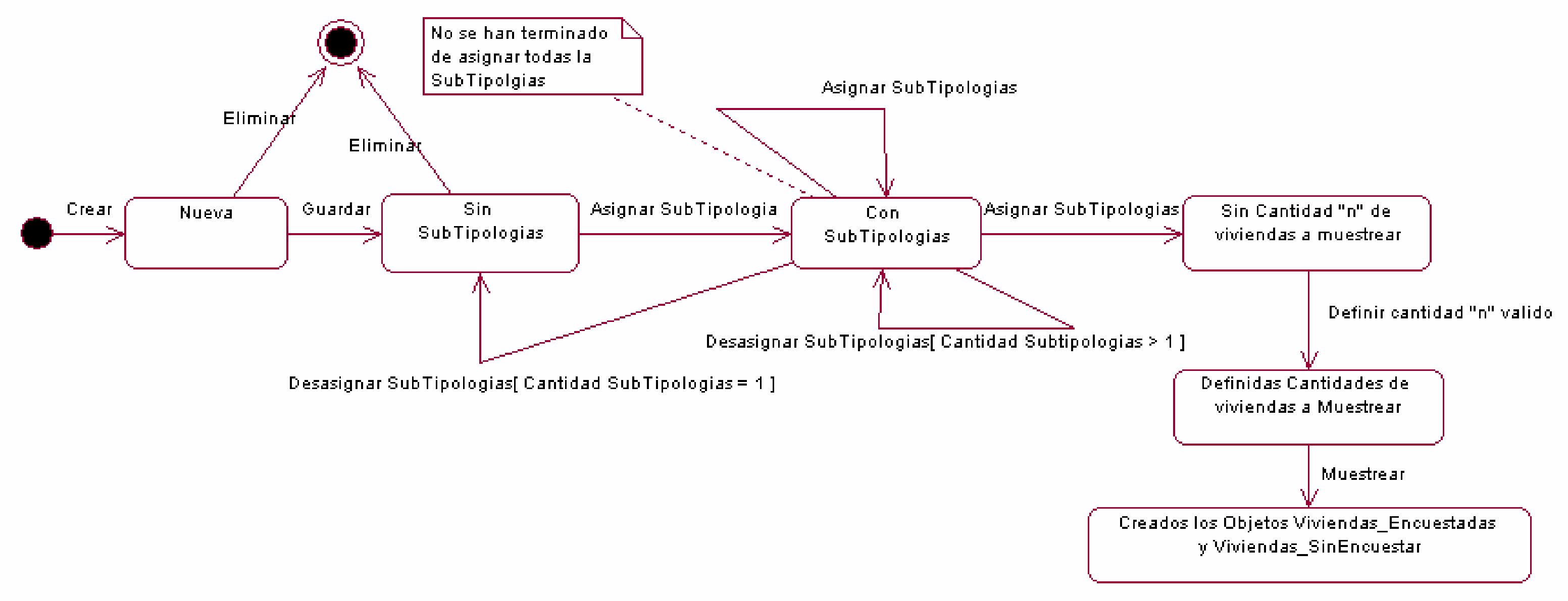

Figura 129. Diagrama de estado para la clase Tipología.

### 3.4.5 Diagramas Actividad

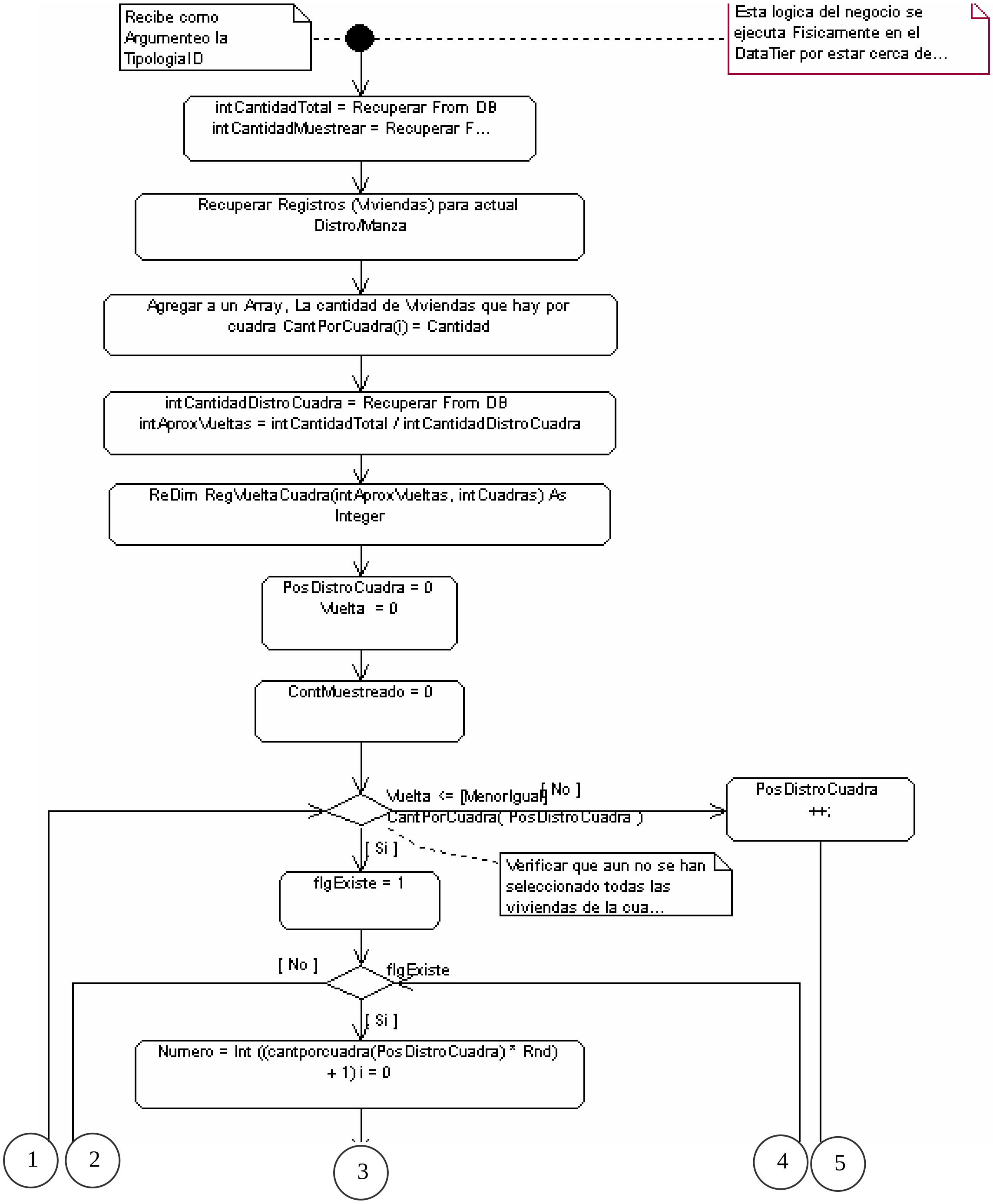

Figura 130. Diagrama de actividad para el método muestrear de la clase Tipología. _Parte superior del diagrama (1/2)

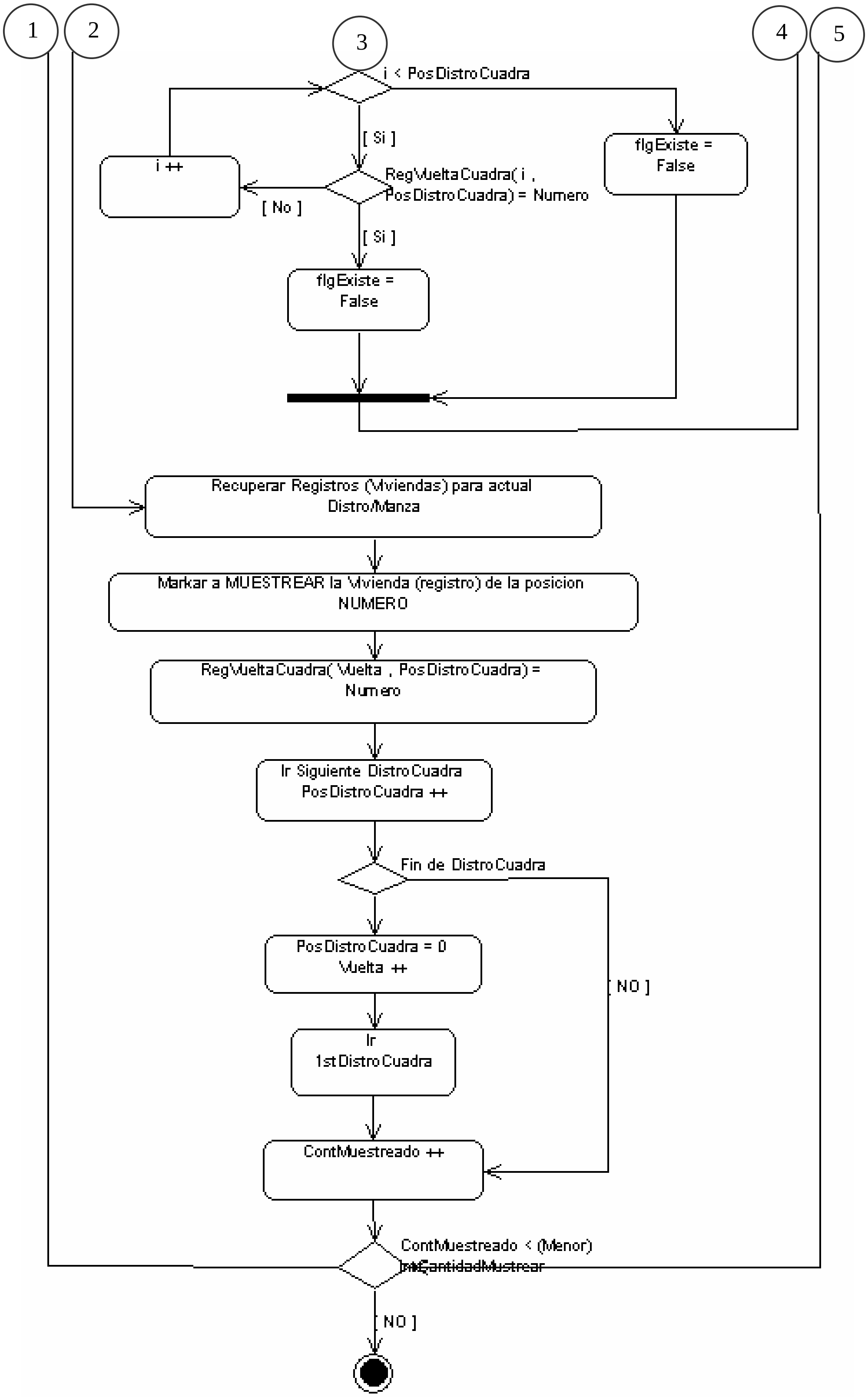

Figura 131. Diagrama de actividad para el método muestrear de la clase Tipología. _Parte inferior del diagrama (2/2)

### 3.4.6 Diagrama de Componentes

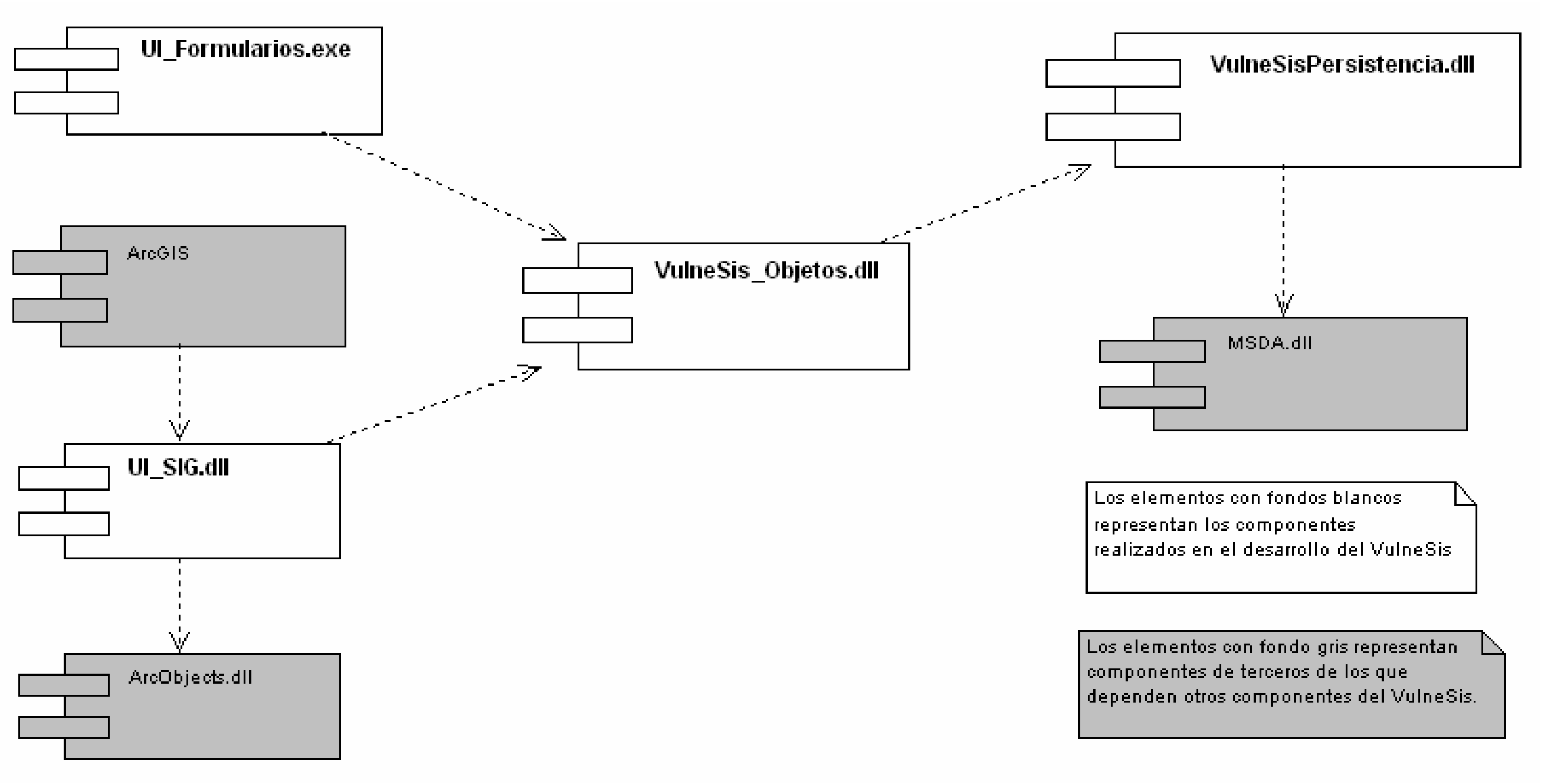

Figura 132. Diagrama de componentes de *Vulnesis*

## 3.4.7 Diagrama de Despliegue

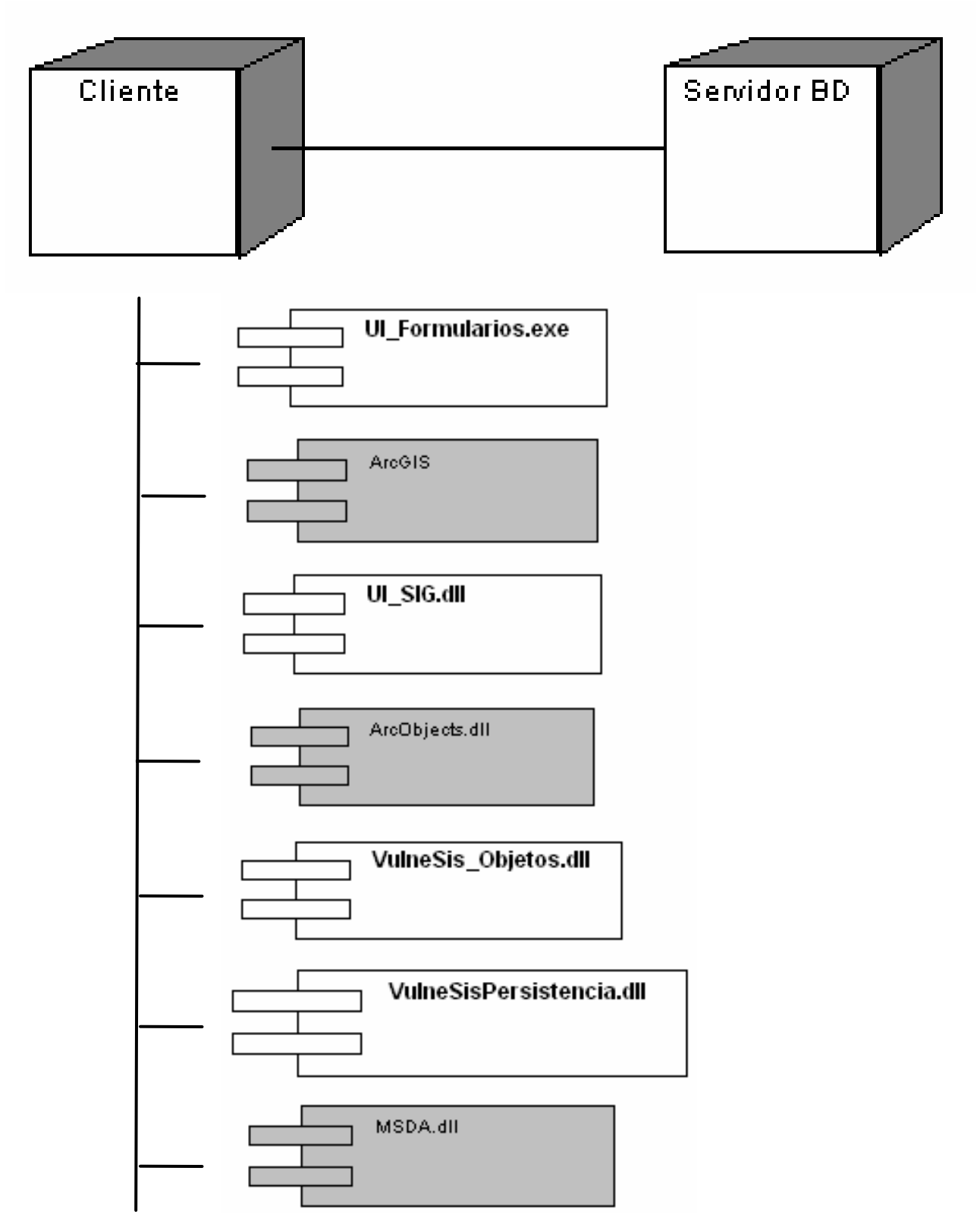

Figura 133. Diagrama de despliegue usado para *Vulnesis* con una arquitectura física en 2 capas y cliente gordo

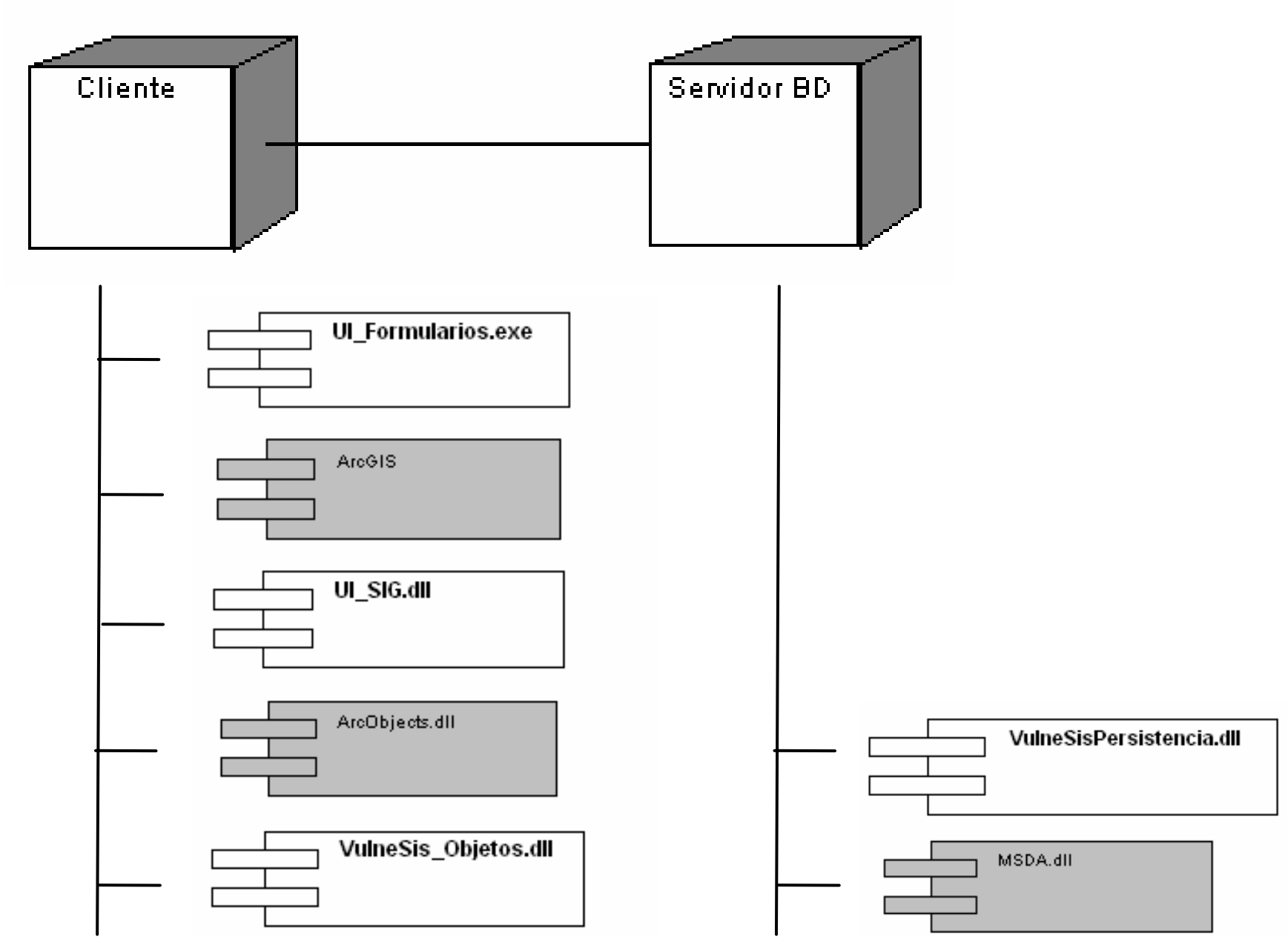

Figura 134. Diagrama de despliegue para *Vulnesis* con una arquitectura física en 2 capas, delegando al servidor de BD el trabajo de la capa lógica de persistencia (Esta delegación responde bien cuando el Servidor de BD tiene suficiente poder como para hacer sus labores de BD y correr una capa lógica de una aplicación)

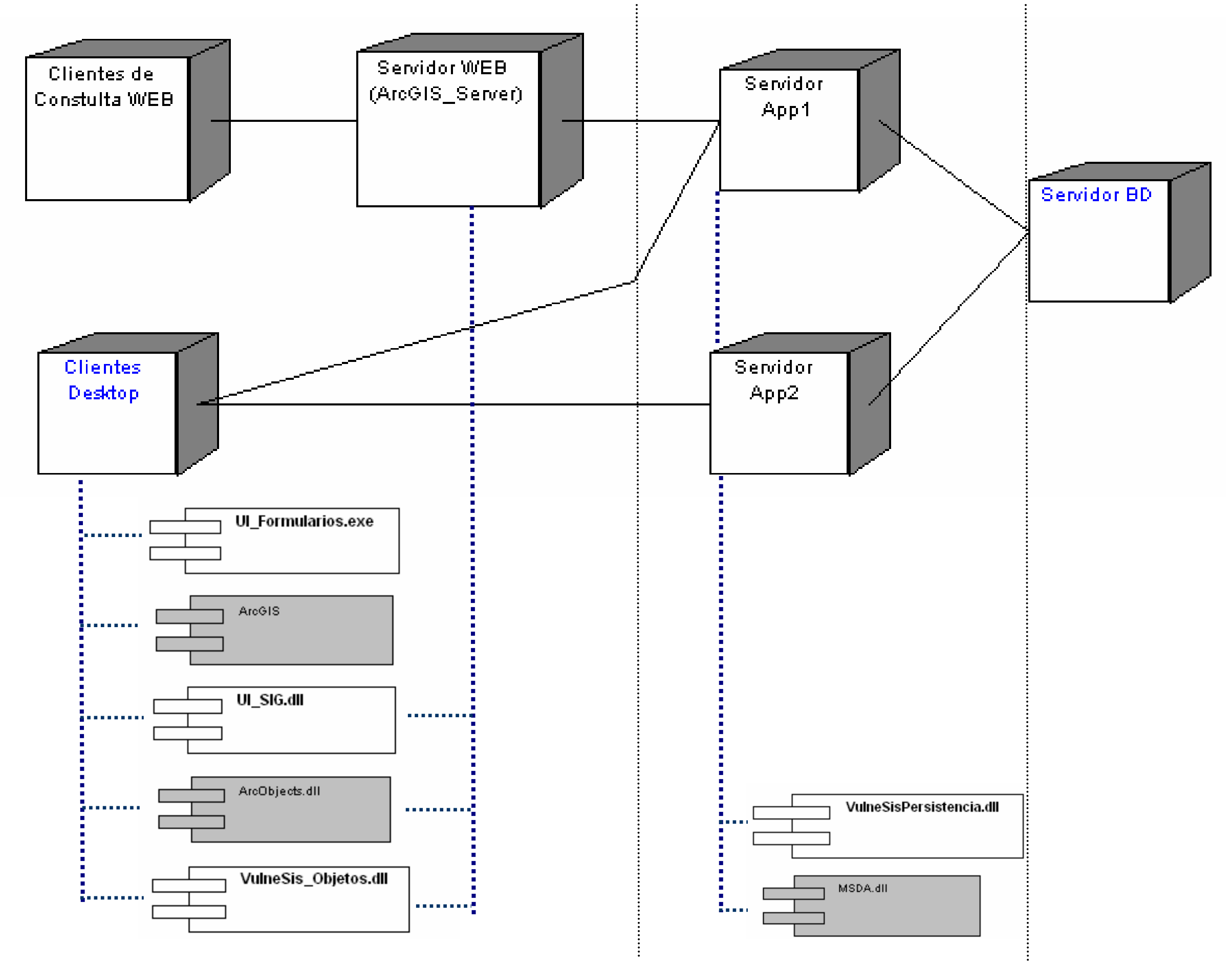

Figura 135. Diagrama de despliegue para *Vulnesis*
con un arquitectura física en 3 capas, donde a través de un servidor Web y ArcGIS Server se permiten clientes Web de *Vulnesis*
este seria el caso para un ambiente con muchos usuarios por lo que la intervención de Servidores Apps permitirían balancear la carga del Servidor de BD con el que se pudieran conectar a través de un BackBone

# CAPÍTULO 4 COMPROBACIÓN DE LA HIPÓTESIS

## 4.1 Planteamiento de la hipótesis:

Una herramienta informática que implemente la nueva metodología propuesta por funcionarios de INETER y la UNI para el estudio a gran escala de la vulnerabilidad y daños debido a sismos en las edificaciones, reducirá los pasos y el tiempo necesario para la realización de los cálculos y procedimientos que permitirán presentar los resultados de dicho estudio para su posterior análisis.

## 4.2 Resumen de variables

| Nombre de la variable | Tipo | Indicador | Escala de medición |
|---|---|---|---|
| X1: Cantidad de pasos con herramienta Geoinformática. | Indepen-diente | Enumeración de cada uno de los pasos a utilizar en estudios de vulnerabilidad y daños debido a sismos cuando se utilice la herramienta informática de esta tesis. | Cualitativa Nominal |
| X2: Cantidad de pasos resultantes sin utilizar la herramienta Geoinformática. | Indepen-diente | Enumeración de cada uno de los pasos utilizados en estudios de vulnerabilidad y daños debido a sismos cuando no se utilizó la herramienta Geoinformática generada en esta tesis. | Cualitativa Nominal |
| X3: Tiempo a utilizar en el caso de usar la herramienta Geoinformática. | Depen-diente de X1 | Cada valor de X1, va a tener asociado un peso de tiempo, y la sumatoria del producto de ese peso multiplicado por la cantidad de veces que se repita cada paso, dio como resultado el valor de esta variable. | Cuantitati-va Continua |
| X4: Tiempo invertido cuando no se hace uso de la herramienta Geoinformática. | Depen-diente de X2 | Cada valor de X2, va a tener asociado un peso de tiempo, y la sumatoria del producto de ese peso multiplicado por la cantidad de veces que se repita cada paso, dio como resultado el valor de esta variable. | Cuantitati-va Continua |

Tabla 14. Cuadro resumen de variables de la tesis

## 4.3 Medición de variables

### X1: Cantidad de pasos con herramienta Geoinformática.

| Nombre del paso | Descripción |
|---|---|
| Pa1: | Importar tabla catastral del área de estudio al sistema. |
| Pa2: | Editar información general del nuevo proyecto a crear. |
| Pa3: | Editar información de los tipos que el sistema no reconoce en su maestro. |
| Pa4: | Crear tipología y asignar subtipologias (viviendas). |
| Pa5: | Definir las cantidades a muestrear. |
| Pa6: | Imprimir reporte con el formato adecuado de las viviendas a encuestar parámetros de vulnerabilidad en el campo. |
| Pa7: | Imprimir reporte con el formato adecuado de todas las viviendas, viviendas a las que se les recuperará punto GPS y se le tomará una foto. |
| Pa8: | Agregar elementos Cartográficos.<br>- En ArcCatalog, ubicarse sobre la capa destino y llamar a la función cargar.<br>- Ubicar y seleccionar las capas de fuentes.<br>- Realizar la correspondencia de los campos ID de viviendas de la capa fuente con la capa destino. |
| Pa9: | - Levantamiento en el campo de la coordenada, foto de las edificaciones y los parámetros de vulnerabilidad a ser valorados. |
| Pa10: | - Subiendo la información recopilada del campo y presentando automáticamente los resultados. |
| Pa11: | - Definir sismos para escenarios de daños. |
| Pa12: | - Presentación dinámica de los escenarios de daños y rangos de vulnerabilidad sísmica y en distintos niveles de granularidad (parcela, manzana y proyecto) |

Tabla 15. Cuadro donde se enumeran cada uno de los pasos de la variable X1.

Del análisis presentado en la tabla anterior se concluye que el valor de la variable X1 es 12 pasos. X1= 12

## X2: Cantidad de pasos resultantes sin utilizar la herramienta Geoinformática.

| Nombre del paso | Descripción |
|---|---|
| Pb1: | Importar tabla catastral del área de estudio desde FoxPro hacia Access. |
| Pb2: | Imprimir reporte que muestra los distintos tipos (pared, techo, topografía, estado) de las viviendas. |
| Pb3: | Análisis de los tipos de la tabla catastral para el proyecto usado. |
| Pb4: | Adaptar los algoritmos que crean las tipologías, asignándoles niveles de vulnerabilidad a las viviendas contenidas en ellas (en función a los tipos particulares del proyecto). |
| Pb5: | - Levantamiento en el campo de la coordenada, foto de las edificaciones y los parámetros de vulnerabilidad a ser valorados. |
| Pb6: | - Alimentación del programa que calcula la vulnerabilidad sísmica y daños, con los datos de campo e introducción manual de los resultados en una tabla en excel, junto con la coordenada de las viviendas inspeccionadas y el código de las fotos de dicha vivienda. |
| Pb7: | - Importar la tabla anterior a Access y definirles a sus campos el formato adecuado. |
| Pb8: | - Subir la tabla de Access anterior al ArcGIS donde se representaran las edificaciones como puntos (utilizando los campos de las coordenadas x,y) con su tabla de atributos asociada. |
| Pb9: | - Convertir el índice de vulnerabilidad, en un valor numérico a 3 rangos (baja, media y alta). |
| Pb10: | - Calcular el valor promedio de vulnerabilidad por manzanas mediante la superposición en ArcMap de la capa viviendas (Puntos) y la capa manzanas (polígonos). |
| Pb11: | - Preparar mapa donde se presentan las viviendas con sus niveles de vulnerabilidad. |
| Pb12: | - Preparar mapa donde se presentan las manzanas con sus niveles de vulnerabilidad. |
| Pb13: | - Preparar mapa donde se presentan el área del proyecto con su nivel de vulnerabilidad. |
| Pb14: | - Convertir el índice de daño de cada escenario, en un valor numérico de 5 rangos(menor, moderado, severo, total y colapso) |
| Pb15: | - Preparar un mapa por cada escenario de daño donde se presentan las viviendas con sus respectivos niveles de daños. |
| Pb16: | - Preparar un mapa por cada escenario de daño donde se presentan las manzanas con sus respectivos niveles de daños. |
| Pb17: | - Preparar un mapa por cada escenario de daño donde se presente el área del proyecto con su respectivo nivel de daño. |
| Pb18: | - Adaptación del mapa de daños a la columna de la tabla de las propiedades del nivel de escenario correspondiente. |

Tabla 16. Cuadro donde se enumeran cada uno de los pasos de la variable X2.

Del análisis representado en la tabla anterior se concluye que el valor de la variable x2 es 18 pasos. X2= 18

Condiciones deseables, para asegurar la aproximación de los valores de las variables de tiempo x3 y x4:

- ✓ Usuario con nivel de operador de computadoras
- ✓ Edad entre 18 años y 65años.
- ✓ Computadora con al menos las siguientes especificaciones:
  - Procesador: 1 Ghz
  - RAM: 512 MB
  - Espacio libre en disco: 1 Gb

**X3: Tiempo a utilizar en el caso de usar la herramienta Geoinformática.**

| Nombre del paso | Dependien-te de la cantidad de elementos | Cantidad de elementos para prueba inicial | Tiempo con la prueba inicial | Cantidad de elementos para la prueba final | Tiempos para la prueba final | Valor de tiempo estimado en segundos |
|---|---|---|---|---|---|---|
| Pa1: | No | 100 (registros) | 5 segs. | 2720 (registros) | 50 segs. | 27.5 (Proceso) |
| Pa2: | No | 1 (proyecto) | 37.73 segs. | 1 (proyecto) | 37.73 segs. | 37.73 (Proceso) |
| Pa3: | Si | 4 (tipos) | 47.11 segs. | 12(tipos) | 229.73 segs. | 15.46 (por tipo) |
| Pa4: | Si | 2 tipologías<br>28 subTipologías | 224 segs. | 4 tipologías<br>28 subTipologías | 340 segs. | 98.5 (por tipología) |
| Pa5: | Si | 2 tipologías | 14.32 segs. | 4 tipologías | 30.11 segs. | 7.34 (por tipología) |
| Pa6: | No | Total de viviendas | 0 segs. | Total de viviendas | 0 segs. | 0 |
| Pa7: | No | Total de viviendas | 0 segs. | Total de viviendas | 0 segs. | 0 |
| Pa8: | Si | 1 capa | 71 segs. | 3capas | 81 segs. | 71 base y 5 mas por cada extra. |
| Pa9: | No | Se identificó una equivalencia de este paso con el paso Pb5 por lo que la presencia o no ausencia de un sistema no genera variación en el tiempo que toma realizar este paso. | | | | 0 |
| Pa10: | Si | 1 vivienda | 196 segs. | 1 vivienda | 196 segs. | 196 (por vivienda) |
| Pa11: | Si | 1 sismo | 29.1 segs. | 1 sismo | 29.1 segs. | 29.1 (por sismo) |
| Pa12: | No | Vulnerabilidad<br>1 escenario daño<br>por vivienda<br>por cuadra<br>por proyecto | 36.70 segs. | Vulnerabilidad<br>4 escenario daño<br>por vivienda<br>por cuadra<br>por proyecto | 147 segs. | 23.87 (por escenario) |

Tabla 17. Cuadro donde se enumeran el tiempo estimado que tomó cada uno de los pasos de la variable X3.

El valor de tiempo de la variable X3, va a estar en función de la cantidad de elementos de algunos pasos. Esto es debido a que en diferentes proyectos, la cantidad de elementos conformantes de los pasos marcados con “Si” en la columna “Dependiente de la cantidad de elementos” varía.

Todos los pasos que tienen “No” en la columna “Dependiente de la cantidad de elementos”, poseerán en la formula para X3, un coeficiente constante de 1.

Del análisis representado en la tabla anterior se concluye que la variable X3 esta en función de: tipo, tipolgia, Capa, Vivienda, Sismo.

$x3 = y = f(ti, ta, ca, vi, si)$

$donde:$
$ti = tipo$
$ta = tipo\log ia$
$ca = capa$
$vi = vivienda$
$si = sismo$

$$y = 27.5 + 37.73 + 0 + 0 + 0 + 23 + (16.46 * ti) + [(98.5 + 7.34) * ta] + [71 + 5(ca - 1)] + (196 * vi) + (29.1 * si)$$

**Valor resultante de la variable X3:**

$$x3 = y = 158.1 + (16.46 * ti) + (105.84 * ta) + (5 * ca) + (196 * vi) + (29.1 * si)$$

A continuación se muestra la tabulación de y, con valores aproximados utilizados en proyectos anteriores, según ha sido indicado por los asesores y pruebas en los sistemas utilizados:

| ***No.*** | ***ti*** | ***ta*** | ***ca*** | ***si*** | ***vi*** | ***Valor en Días Lab.*** | ***Observación*** |
|---|---|---|---|---|---|---|---|
| 1 | 30 | 12 | 10 | 4 | 550 | **3.81** (109,585 s.) | Proyecto Quezalguaque en León |
| 2 | 30 | 12 | 10 | 4 | 8,000 | **54.51** (1,570,085 s.) | Proyecto Juigalpa |
| 3 | 20 | 10 | 8 | 4 | 2,720 | **18.57** (534,819 s.) | Proyecto de prueba usado por los tesistas (barrio Monseñor Lezcano) |
| 4 | 1 | 1 | 1 | 1 | 1 | **0.017** (570 s.) | Valores de prueba, para indicar en cuanto tiempo se obtiene resultados con la cantidad mínima de datos. |
| 5 | 30 | 15 | 10 | 4 | 200,000 | **1361** | Valores de prueba, para indicar el resultado cuando vi, tienda a un numero muy elevado (infinito) |
| 6 | 30 | 15 | 10 | 4 | 800 | **5.52** | Valores de prueba para aumentar la cantidad tabulada. |
| 7 | 30 | 15 | 10 | 4 | 1,600 | **10.97** | Valore para aumentar la cantidad tabulada. |
| 8 | 30 | 15 | 10 | 4 | 2,400 | **16.41** | Valore para aumentar la cantidad tabulada. |
| 9 | 30 | 15 | 10 | 4 | 3,200 | **21.86** | Valore para aumentar la cantidad tabulada. |
| 10 | 30 | 15 | 10 | 4 | 4,000 | **27.3** | Valore para aumentar la cantidad tabulada. |
| 11 | 30 | 15 | 10 | 4 | 4,800 | **32.75** | Valore para aumentar la cantidad tabulada. |
| 12 | 30 | 15 | 10 | 4 | 5,600 | **38.19** | Valore para aumentar la cantidad tabulada. |
| 13 | 30 | 15 | 10 | 4 | 6,400 | **43.63** | Valore para aumentar la cantidad tabulada. |
| 14 | 30 | 15 | 10 | 4 | 7,200 | **49.08** | Valore para aumentar la cantidad tabulada. |

Tabla 18. Con valores aproximados utilizados en proyectos anteriores.

Con esta cantidad de elementos tabulados y tomando en cuenta que la mayor parte de los proyectos los valores para ti, ta, ca y si se mantienen similares, fue posible considerarlos como constantes y de esa forma se procedió a crear una grafica que representa la variable x3 en función de su variable más critica como es vi (el número de viviendas), y así observar su comportamiento de forma grafica.

El eje de las x o valores de dominio representa a las viviendas y el eje y la cantidad de horas laborales.

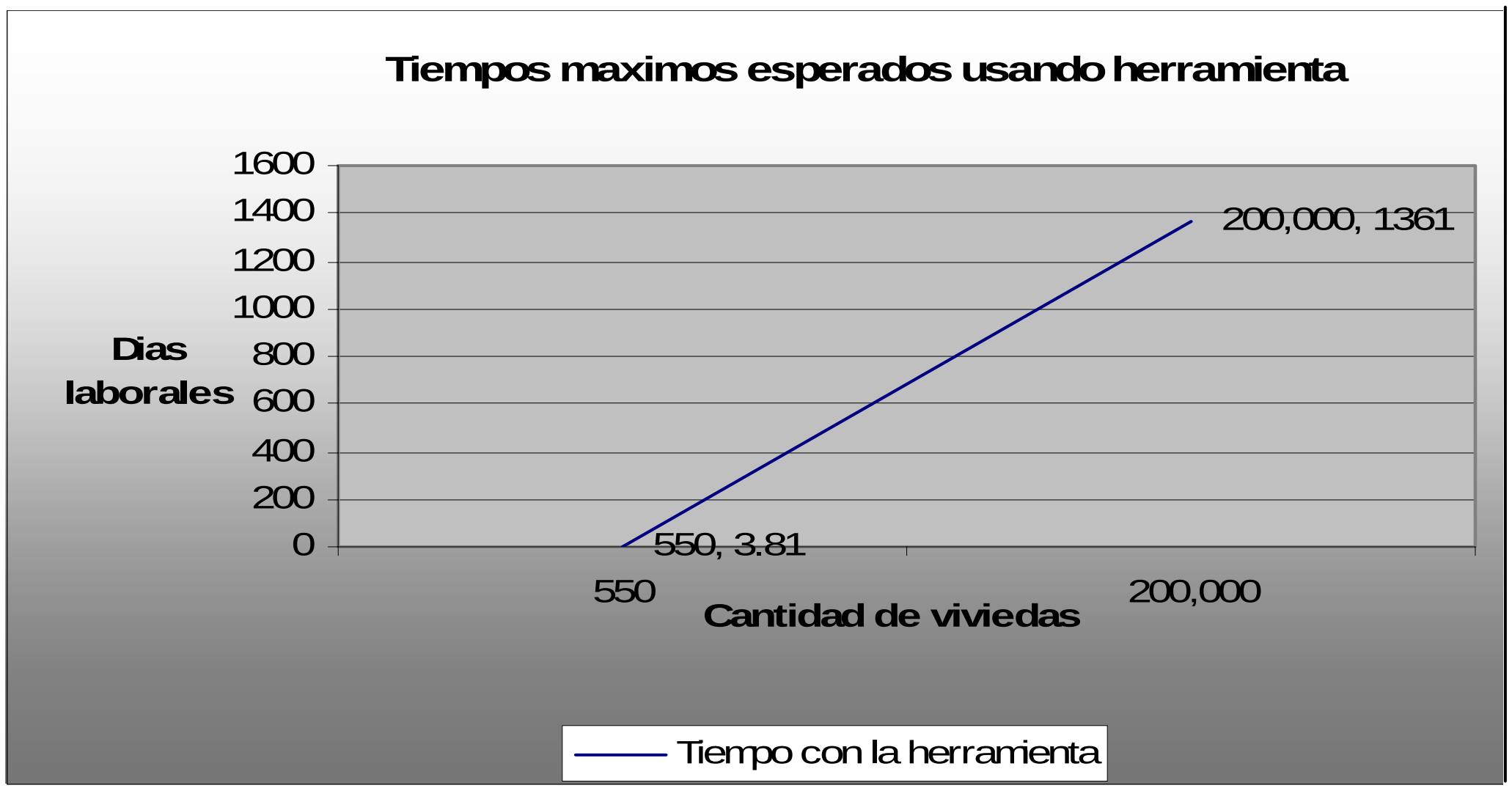

Figura 136. Gráfica de la función que representa a la variable X3.
Con domino de 0 a 200,000

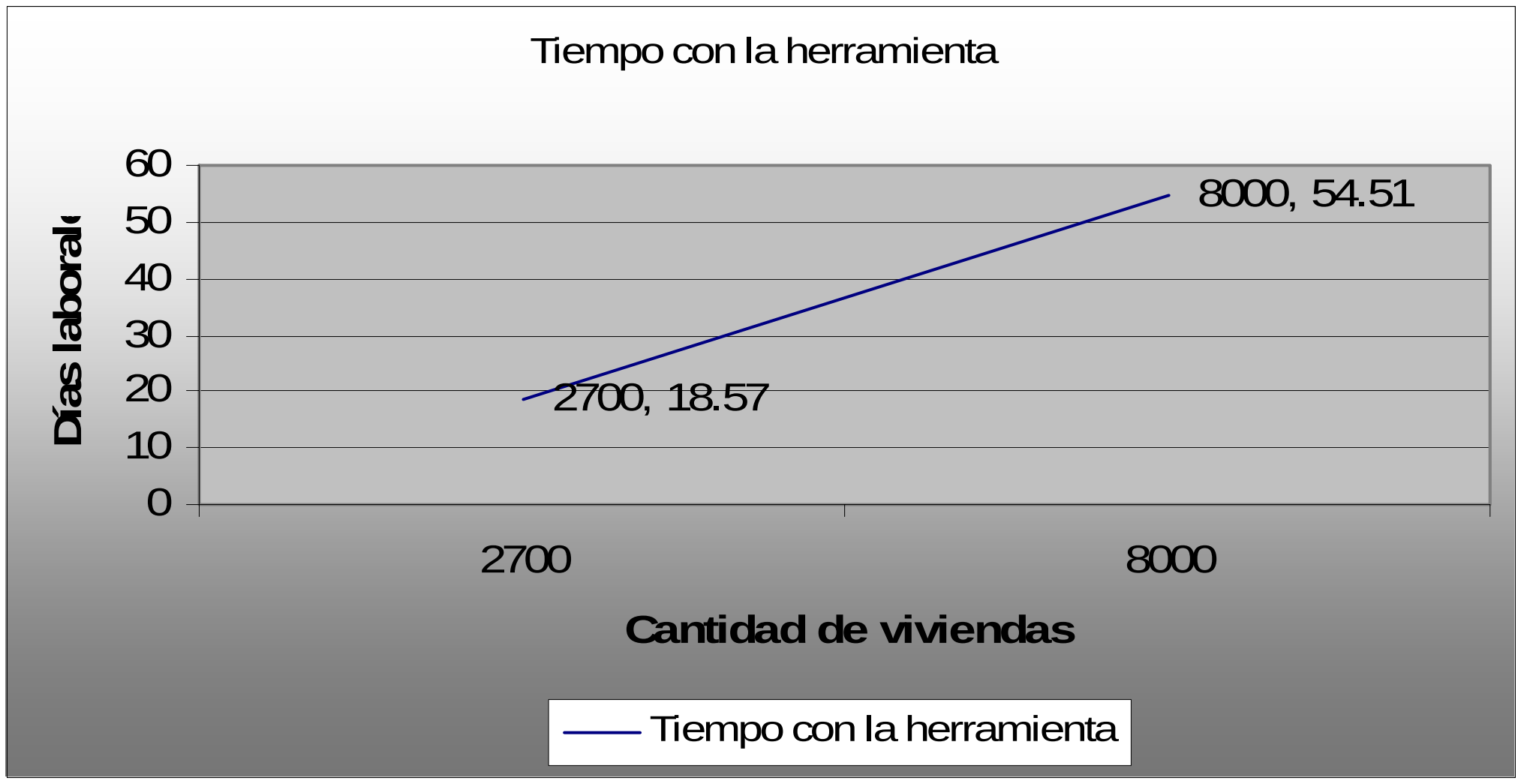

Figura 137. Gráfica de la función que representa a la variable X3.
Con domino de 0 a 8,000

En la gráfica 136, se presenta la función para el valor mínimo esperado y para un valor muy grande, para tratar de entender el comportamiento de la función cuando su valor de dominio tiende al infinito.

La gráfica 137, se realizó con el objetivo de hacer un acercamiento a la función dentro de valores de dominio más comunes que pueden suceder de un proyecto a otro.

## X4: Tiempo invertido cuando no se hace uso de la herramienta Geoinformática.

| Nombre del paso | Dependien-te de la cantidad de elementos | Cantidad de elementos para prueba inicial | Tiempo con la prueba inicial | Cantidad de elementos para la prueba final | Tiempos para la prueba final | Valor de tiempo estimado en segundos |
|---|---|---|---|---|---|---|
| Pb1: | No | 1,000 (registros) | 19.89 segs. | 30,508 (registros) | 27.6 segs. | 22.5 (proceso) |
| Pb2: | No | Total de viviendas | 0 segs. | Total de viviendas | 0 segs. | 0 |
| Pb3: | No | Tabla catastral | 21600 segs. | Tabla catastral | 21600 segs. | 21600 |
| Pb4: | No | Tabla catastral | 7200 segs. La suma de este paso y el anterior da (1 día laboral) | Tabla catastral | 7200 segs. La suma de este paso y el anterior da (1 día laboral) | 7200 |
| Pb5: | No | Se identificó una equivalencia de este paso con el paso Pa9 por lo que la presencia o ausencia de un sistema no genera variación en el tiempo que toma realizar este paso. | | | | 0 |
| Pb6: | Si | 1 Vivienda | 451.48 segs. | 1 Vivienda | 451.48 | 451.48 |
| Pb7: | No | 1 Hoja Excel | 144.20 segs. | 1 Hoja Excel | 144.20 segs | 144.20 |
| Pb8: | No | 1 Tabla de puntos | 93.2 segs. | 1 Tabla de puntos | 93.2 | 93.2 |
| Pb9: | No | 1 Capa puntos | 157.8 segs. | 1 Capa puntos | 157.8 segs. | 157.8 |
| Pb10: | No | 1 Capa puntos 1 Capa manzanas | 180.17 segs. | 1 Capa puntos 1 Capa manzanas | 180.17 segs. | 180.17 |
| Pb11: | No | 1 Mapa | 28800 segs. (1 día laboral) | 1 Mapa | 28800 segs. (1 día laboral) | 28800 |
| Pb12: | No | 1 Mapa | 21600 segs. | 1 Mapa | 21600 segs. | 21600 |
| Pb13: | No | 1 Mapa | 7200 segs. | 1 Mapa | 7200 segs. | 7200 |
| Pb14: | Si | 1 Sismo | 93.2 segs. | 1 Sismo | 93.2 segs. | 93.2 |
| Pb15: | No | 1 Mapa | 28800 segs. (1 día laboral) | 1 Mapa | 28800 segs. (1 día laboral) | 28800 |
| Pb16: | No | 1 Mapa | 21600 segs. | 1 Mapa | 21600 segs. | 21600 |
| Pb17: | No | 1 Mapa | 7200 segs. | 1 Mapa | 7200 segs. | 7200 |
| Pb18: | Si | 1 Sismo | 147.19 segs. | 1 Sismo | 147.19 segs. | 147.19 |

Tabla 19. Cuadro donde se enumeran cada uno de los pasos de la variable X4.

El valor de tiempo de la variable x4, va a estar en función de la cantidad de elementos de algunos pasos. Esto es debido a que en diferentes proyectos, la cantidad de elementos confortantes de los pasos marcados con “Si” en la columna “Dependiente de la cantidad de elementos” varía.

Todos los pasos que tienen “No” en la columna “Dependiente de la cantidad de elementos”, poseerán en la formula para x4, un coeficiente constante de 1.

Del análisis representado en la tabla anterior se concluye que la variable x4 esta en función de: Vivienda, Sismo.

$$x4 = g = f(vi, si)$$

$$donde:$$
$$vi = vivienda$$
$$si = sismo$$

$$g = 22.5 + 21600 + 7200 + 144.20 + 93.2 + 157.8 + 180.17 + 28800 + 21600 + 7200 + 28800 + 21600 + 7200 + (451.48 * vi) + (93.2 * si) + [147.19 * (si - 1)]$$

$$g = 144450.68 + (451.48 * vi) + (240.39 * si)$$

A continuación se muestra la tabulación de y, con valores aproximados utilizados en proyectos anteriores, según ha sido indicado por los asesores y pruebas en los sistemas utilizados:

| ***No.*** | ***si*** | ***vi*** | ***Valor en Días Lab.*** | ***Observación*** |
|---|---|---|---|---|
| 1 | 4 | 550 | **13.67** (393,726 s.) | Proyecto Quezalguaque en León |
| 2 | 4 | 8,000 | **130.46** (3,557,252 s.) | Proyecto Juigalpa |
| 3 | 4 | 2,720 | **47.68** (1,373,437 s.) | Proyecto de prueba usado por los tesistas (barrio Monseñor Lezcano) |
| 4 | 1 | 1 | **5.03** (145,142 s.) | Valores de prueba, para indicar en cuanto tiempo se obtiene resultados con la cantidad mínima de datos. |
| 5 | 4 | 200,000 | **3140** | Valores de prueba, para indicar el resultado cuando vi, tienda a un numero muy elevado (infinito) |
| 6 | 4 | 800 | **17.59** | Valores de prueba para aumentar la cantidad tabulada. |
| 7 | 4 | 1,600 | **30.13** | Valore para aumentar la cantidad tabulada. |
| 8 | 4 | 2,400 | **42.63** | Valore para aumentar la cantidad tabulada. |
| 9 | 4 | 3,200 | **55.21** | Valore para aumentar la cantidad tabulada. |
| 10 | 4 | 4,000 | **67.75** | Valore para aumentar la cantidad tabulada. |
| 11 | 4 | 4,800 | **80.29** | Valore para aumentar la cantidad tabulada. |
| 12 | 4 | 5,600 | **92.83** | Valore para aumentar la cantidad tabulada. |
| 13 | 4 | 6,400 | **105.37** | Valore para aumentar la cantidad tabulada. |
| 14 | 4 | 7,200 | **117.91** | Valore para aumentar la cantidad tabulada. |

Tabla 20. Con valores aproximados utilizados en proyectos anteriores.

Con esta cantidad de elementos tabulados y tomando en cuenta que la mayor parte de los proyectos el valor si (sismo) se mantiene similar, fue posible considerarlo como constante y de esa forma se procedió a crear una gráfica que representa la variable x4 en función de su variable más critica como es vi (el numero de viviendas), y de así observar su comportamiento de forma grafica.

El eje de las x o valores de dominio representa a las viviendas y el eje y la cantidad de horas laborales.

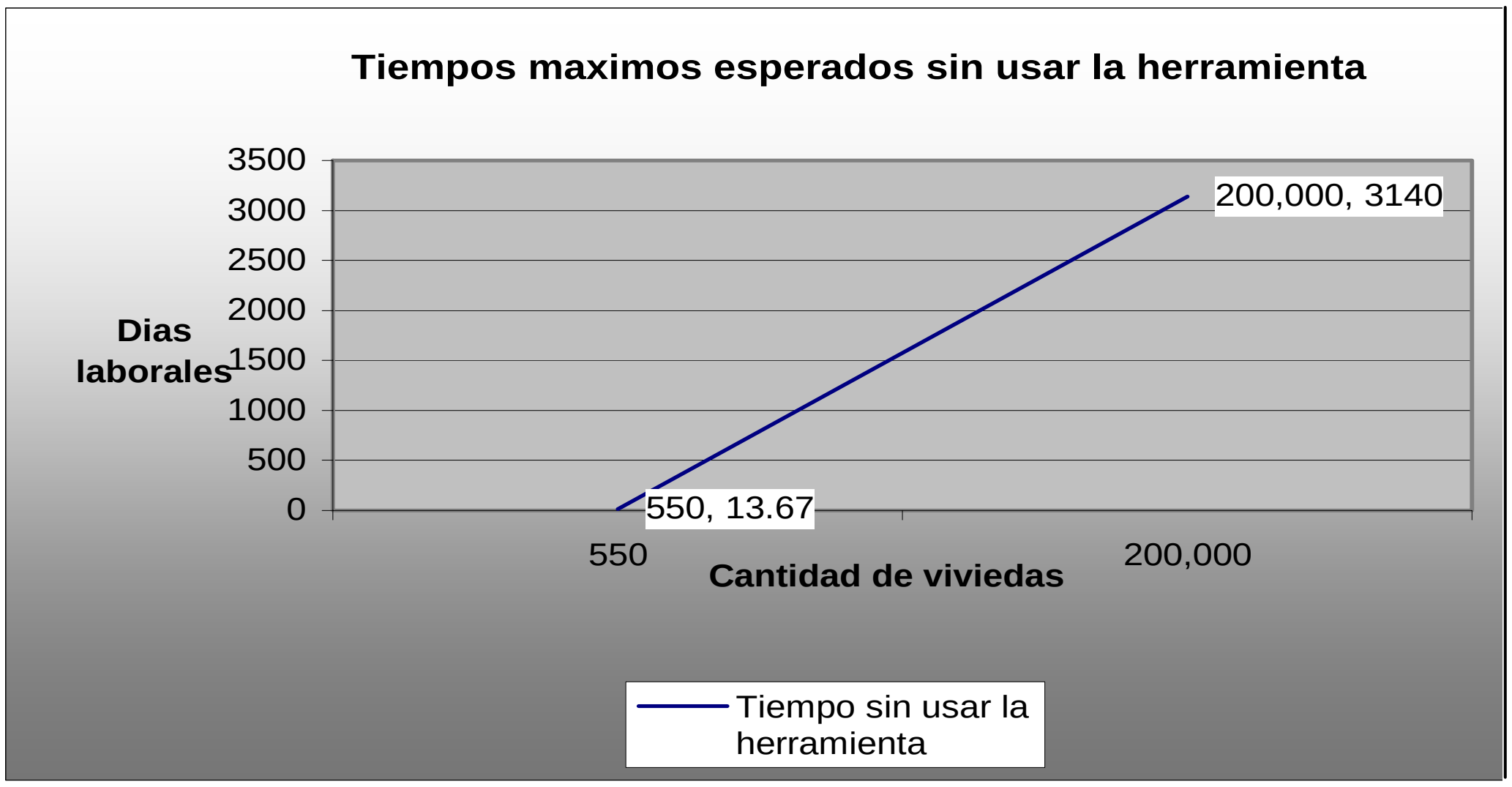

Figura 138. Gráfica de la función que representa a la variable X4.
Con domino de 0 a 200.000

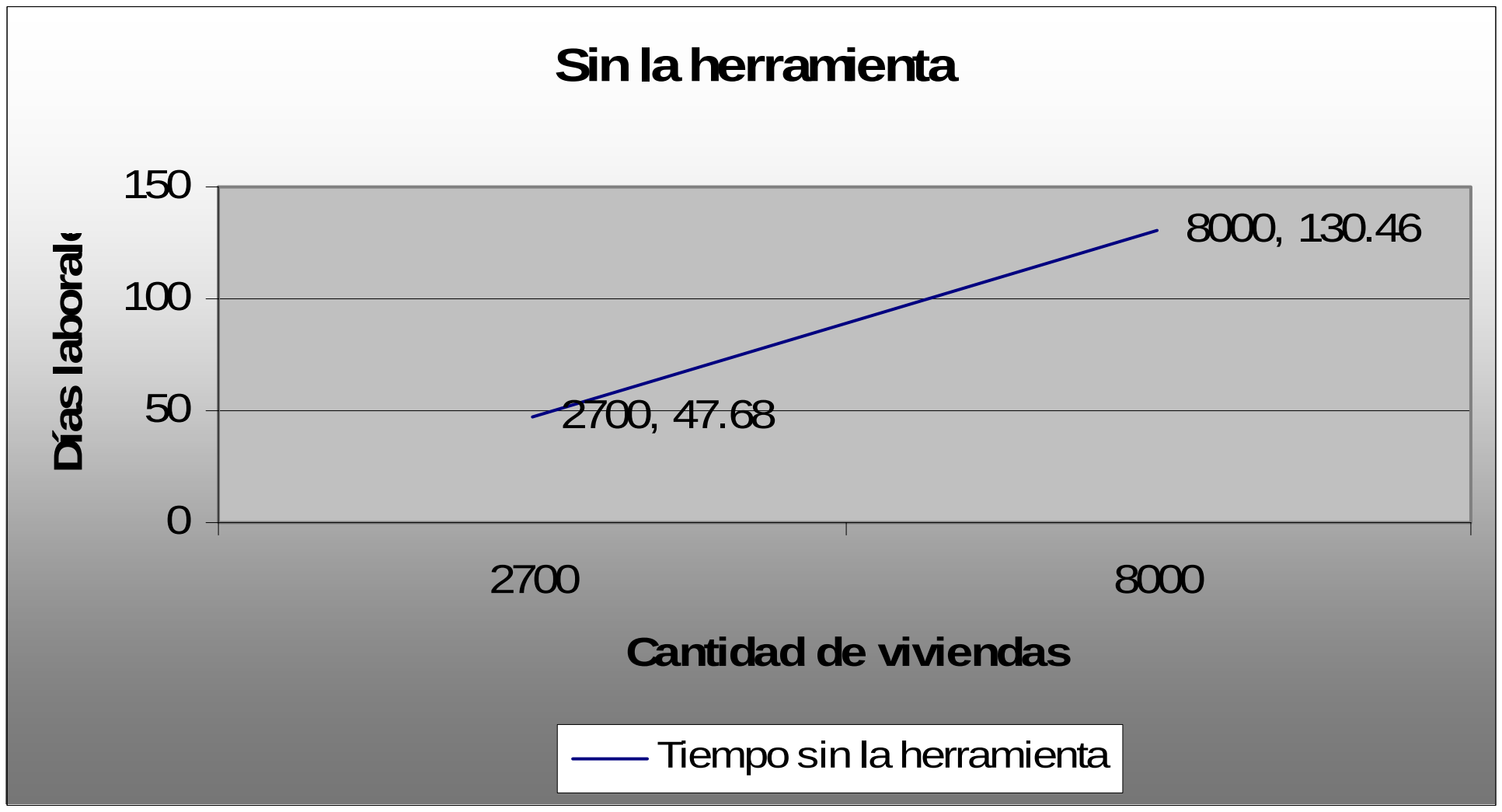

Figura 139. Gráfica de la función que representa a la variable X4.
Con domino de 0 a 8.000

En la gráfica 138, se presenta la función para el valor mínimo esperado y para un valor muy grande, para tratar de entender el comportamiento de la función cuando su valor de dominio tiende al infinito.

La gráfica 139, se realizó con el objetivo de hacer un acercamiento a la función dentro de valores de dominio más comunes que pueden suceder de un proyecto a otro.

## 4.4 Comprobación de la hipótesis:

Fue necesario realizar la comprobación de la hipótesis a través de dos vías una por los pasos y otro por el tiempo, esto debido a que en el planteamiento de la hipotes se habla de la reducción de los pasos y del tiempo.

A continuación se presenta la comprobación de la primera condicionante de la hipótesis (en función a la cantidad de pasos), en los que intervienen la variable x1 y x2:

Variable **x1** (Cantidad de pasos a utilizar con herramienta Geoinformática)

Variable **x2** (Cantidad de pasos resultantes sin utilizar la herramienta Geoinformática – Forma tradicional).

$$x1 = 12\,paso$$
$$x2 = 18\,pasos$$

$$\frac{x1}{x2} = \frac{12\,pasos}{18\,pasos} = \frac{2}{3} = 0.66...$$

Se encontró verdadera la primera condicionante de la hipótesis debido a que la razón entre x1 y x2 es menor que 1. Además que la cantidad de pasos antes del uso del sistema es un 50% mayor. A continuación se presenta la comprobación de la segunda condicionante de la hipótesis (en función al tiempo), en la que intervienen la variable x3 y x4:

Variable **x3** (Tiempo a utilizar en el caso de usar la herramienta Geoinformática)

Variable **x4** (Tiempo invertido cuando no se hace uso de la herramienta Geoinformática – Forma tradicional).

Debido a que estas dos variables están en función a ciertos parámetros, siendo el más importante la cantidad de viviendas, una forma de analizar sus comportamientos es a través de superposición de sus gráficas en un mismo plano.

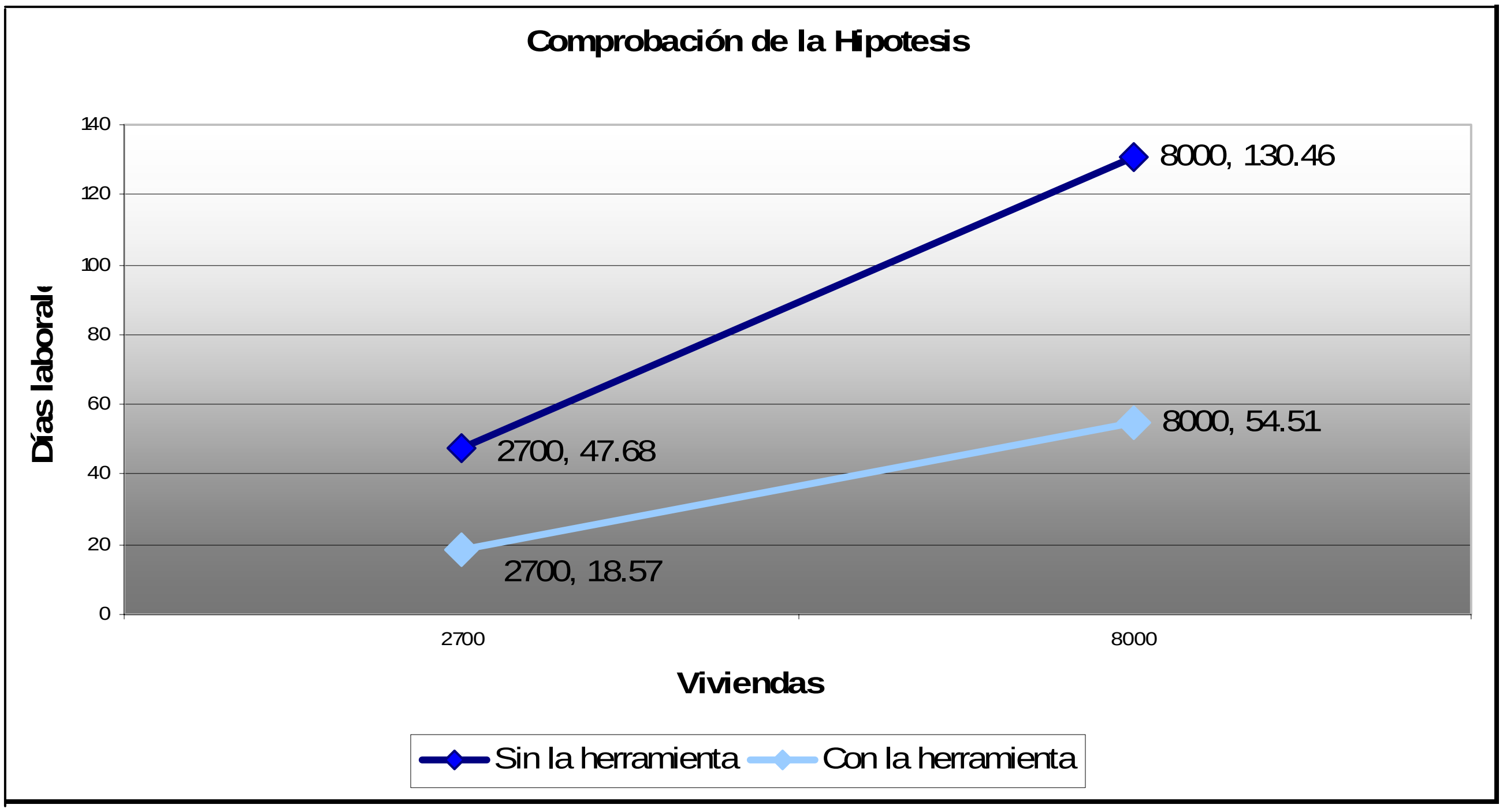

Figura 140. Gráfica de las funciones que representan a las variables X3 y X4 superpuestas.

Estas gráficas se obtienen del trabajo realizado en sección 4.

Datos obtenidos en base a los datos tabulares, para el caso de 550 viviendas

$$x3 = 3.81\,jornadas$$
$$x4 = 13.67\,jornadas$$

$$\frac{x4}{x3} = \frac{13.67}{3.81} = 3.58$$

Datos obtenidos en base a los datos tabulares, para el caso de 8000 viviendas

$$x3 = 130.46\,jornadas$$
$$x4 = 54.52\,jornadas$$

$$\frac{x4}{x3} = \frac{130.46}{54.52} = 2.39$$

Del análisis de la magnitud del tiempo para los casos anteriores, se concluye que el valor de x3 es menor que el valor de x4. Tendiendo en la mayoría de los casos a un 239%

# CAPÍTULO 5 CONCLUSIONES Y RECOMENDACIONES

## 5.1 CONCLUSIONES

Durante el desarrollo de esta tesis se concluyó lo siguiente:

1) Es posible disponer de un sistema Geoinformático que implemente la nueva metodología para el estudio a gran escala de vulnerabilidad y daños en las edificaciones debido a sismos. Además es fácilmente adaptable a metodologías basadas en la vulnerabilidad de las viviendas tomando en consideración factores como riesgo, cuantificación de pérdidas materiales, cuantificación de muertes y planeación de rutas de emergencia en caso de siniestros.

2) La herramienta desarrollada permite visualizar resultados aproximados desde la introducción de los primeros datos, además de contar con un almacén de estudios de vulnerabilidad sísmica donde se puede revisar trabajos realizados con anterioridad

3) En casos de sismos ocurridos, se podrá acceder al sistema y seleccionar del banco de estudios anteriores a los que estén dentro del área del sismo acontecido y someterlos a escenarios de daños del movimiento telúrico en cuestión, esto permitirá darse una idea inmediata de la situación en el área afectada.

4) El sistema desarrollado redujo el tiempo y el número de pasos que se requerían para realizar este tipo de estudios. Obteniendo un tiempo total de 1/3 con respecto al valor registrado para estos estudios sin la utilización de este herramienta.

5) Se reducen errores humanos inherentes al traslado manual de los datos al calcular o procesar datos de una herramienta informática a otra. Este traslado de los datos es uno de los principales inconvenientes de la forma tradicional de realizar este tipo de estudios.

6) Las principales metodologías de Ing. de software orientado a objetos se clasifican por el ciclo de vida que sigue el proyecto en:

1-Lineales o Cascada
2-Evolutivos (Iterativo e Incremental)

Y por la rigidez de planificación o documentación se agrupan en:
1-Metodologías ágiles.
2-Metodologías pesadas.

7) El diseño y la implementación del sistema usando Ingeniería de software orientada a objetos y la Arquitectura lógica en capas, dio como resultado un software de fácil mantenimiento y sencilla corrección de errores.

8) La arquitectura de capas lógicas tiene como objetivo principal el definir que tipos de servicios proveerá el software, sin importar la cantidad de capas físicas (equipos) en las que se implante.

9) El ArcGIS permite extender su funcionalidad por medio del acceso a sus clases y objetos internos llamados ArcObjects.

## 5.2 RECOMENDACIONES

1) Usar el sistema para proyectos futuros de estudios a gran escala de vulnerabilidad y daños en las edificaciones debido a sismos para afinar tanto la nueva metodología como la herramienta creada para auxiliarla.

2) Aprovechar la reducción del tiempo, y por ende de costos, para expandir el área donde se han realizado los estudios aplicando la nueva metodología, proporcionando de esta forma una mejor visión del estado de las edificaciones de Nicaragua ante posibles sismos.

3) Utilizar *VULNESIS* como base para futuras herramientas auxiliares, tanto para los estudios de vulnerabilidad y daños debido a sismos, como para otras disciplinas donde los sistemas de información geográfica jueguen un papel preponderante.

4) Tener en consideración que el sistema no realiza validación alguna sobre el margen de error que se produce, por los valores que se introducen al momento de realizar el muestreo de las viviendas a encuestar.

5) Salir del ámbito de Managua, e ir hacia las municipalidades donde se tenga una tabla catastral y demás condiciones necesarias para la implementación de la herramienta.

6) Promover la enseñanza del diseño y programación orientada a objetos y la Arquitectura lógica escalable basada en componentes, teniendo en cuenta el estudio como un posible material didáctico más.

7) Ilustrar la filosofía de componentes de software y el funcionamiento interno del ArgGIS, con ejemplos teórico-prácticos que bien podrían ser extraídos de este trabajo. Además de utilizar la tesis como auxilio a futuras monografías que deseen profundizar más en estos tópicos.

# Bibliografía

# ANEXOS

## A. MODELOS, DIAGRAMAS Y DOCUMENTOS DEL DESARROLLO DEL SOFTWARE

### A.1 Flujos de eventos de los Casos de uso

A continuación se presentan todos los casos de uso utilizados para el desarrollo del VULNESIS.

## Flujo de Eventos para el caso de uso <u>CrearProyecto</u>

**.1 Condiciones previas:**
Para crear un nuevo proyecto el usuario deberá tener la tabla catastral con la información de las viviendas para el área del proyecto.

**.2 Flujo principal:**
Este caso de uso inicia cuando el usuario está dentro del sistema en la pantalla principal y selecciona el menú de crear un nuevo proyecto. El sistema presentará una pantalla con los campos necesarios para definir la información del nuevo proyecto y una opción que permitirá (IMPORTAR) seleccionar la tabla catastral importada al sistema siempre que ya haya definido la información general del proyecto. Cuando el usuario seleccione dicha opción el sistema procederá a encontrar, enumerar y contar los distintos tipos de pared, techo, uso y estado existentes en las viviendas de la tabla catastral. Si el sistema encuentra que todos los tipos existentes se encuentra en el maestro de tipos del sistema procede a ejecutar el S-1: PresentarCrearMaestroTiposProyecto, sino procederá a realizar el S-2: DefinirTiposNoExistentes. Luego de ejecutarse el S-1 satisfactoriamente el sistema procederá a realizar el S-3: PresentarCrearMaestroSubTipologias.

**.3 SubFlujos:**

**S-1: PresentarCrearMaestroTiposProyecto:**
El sistema procederá a encontrar, enumerar y contar los distintos tipos de pared, techo, uso y estado existentes en las viviendas de la tabla catastral. Luego presentará dichos tipos en 4 matrices correspondiéndose cada matriz con cada tipo.

**S-2: DefinirTiposNoExistentes:**
El sistema presentará los tipos no existentes en 4 matrices distintas, una matriz por tipo. El sistema tendrá la opción de editar el tipo en caso de ser nuevo (no existía en ningún proyecto anterior), o la opción de corregirlo en caso de tratarse de un tipo que esté utilizando una notación distinta para el nuevo proyecto. Luego que se hayan definido todos los tipos el sistema procederá a realizar S-1.

**S-3: PresentarCrearMaestroSubTipologias:**
El sistema procederá a encontrar, enumerar y contar todas las combinaciones de: pared, techo, estado, uso año construcción después o antes del 72. Combinaciones para las viviendas existentes.

# Flujo de Eventos para el caso de uso <u>DefinirTipologias</u>

### .1 Condiciones previas:
Se deberá haber concluido exitosamente el caso de uso CrearProyecto.

### .2 Flujo principal:
Este caso de uso inicia cuando el usuario está dentro del sistema en la pantalla principal donde se le presenta una matriz con los proyectos existentes que tiene acceso, y selecciona un proyecto que se encuentra en el estado para definir tipologías. Se desplegará un formulario que contendrá tres matrices ordenadas colocadas de la siguiente forma: En la parte superior estará la matriz de subtipoligías no asignadas a ninguna tipología (la primera vez que se mande a llamar a este caso de uso para un proyecto esta matriz contendrá a todas las subtipologías encontradas en el caso de uso CrearProyecto en el S-3). En la zona intermedia estará la matriz con las tipologías definidas (creadas por uno o sacadas del maestro de tipologías) para el proyecto. Y en la inferior estarán las subtipologías pertenecientes a la tipología seleccionada en la matriz de la zona media. El sistema presentará las opciones de: Agregar tipologías del maestro de tipologías del sistema (se ejecuta el S-1: Agregar TipologiasDesdeMaestroTipologiasSistema), crear una nueva tipología (se ejecuta el S-2: CrearNuevaTipologia), eliminar una tipología del maestro de tipologías del proyecto actual (se ejecuta el S-3:EliminarTipologiaMaestroTipologiasProyectoActual), desasignar subtipologías de la tipología actual seleccionada (se trasladan a la matriz de subtipologías no asignadas y se rompe el enlace lógico entre la tipología), Asignar subtipologías a una tipología (Se ejecuta el subflujo S-4: AsignarSubtipologias)

### .3 SubFlujos:

**S-1: AgregarTipologiasDesdeMaestroTipologiasSistema**
El sistema abrirá una ventana donde se presentaran todas las distintas tipologías definidas en proyectos anteriores (Maestro de tipologías del sistema) el usuario tendrá la opción de seleccionar una de estas tipologías y agregarlas al maestro de tipologías del proyecto actual.

**S-2: CrearNuevaTipologia**
El sistema desplegará una ventana donde el usuario escribirá el nombre y la descripción de la nueva tipología, que también por ser nueva se agregará al maestro de tipologías del sistema.

**S-3: EliminarTipologiaMaestroTipologiasProyectoActual**
El usuario seleccionará de la matriz tipologías definidas (Matriz de en medio) la tipología a borrar y presionará un botón que procederá con el borrado de esta de la matriz, de las tipologías del proyecto y si es necesario de las maestro de tipologías del sistema.

**S-4: AsignarSubtipologias**
El usuario seleccionará elementos desde la matriz de subtipologías no asignadas, y los asignará a la tipología seleccionada actualmente.

# Flujo de Eventos para el caso de uso SeleccionarMuestra

**.1 Condiciones previas:**
Se deberá haber concluido el caso de uso Definir tipología.

**.2 Flujo principal:**
Este caso de uso inicia cuando el usuario está dentro del sistema en la pantalla principal donde se le presenta una matriz con los proyectos existentes a los que tiene acceso, y selecciona un proyecto que se encuentra en el estado muestreo. El sistema mostrara una ventana donde el usuario elegirá una de las siguientes opciones: 1-Cantidad total de viviendas a muestrear por todo el proyecto (S-1: CantidadTotal), 2-Cantidad de viviendas a muestrear por cada tipología (S-2: CantidadPorTipologia). Luego de haberse ejecutado satisfactoriamente S-1 o S-2, el programa le presentará al usuario una matriz con las viviendas seleccionadas.

**.3 SubFlujos:**

**S-1: CantidadTotal**
Se mostrará una ventana de dialogo donde el usuario podrá ingresar el número entero que representará la cantidad total de las viviendas, o un porcentaje del total, el cual será mayor que cero y menor o igual que cien.

**S-2: CantidadPorTipologia**
Se mostrará una ventana de dialogo donde el usuario podrá ingresar el número entero al lado de cada tipología, el cual representará la cantidad de viviendas a muestrear. O bien, un porcentaje el cual será mayor que cero y menor o igual que cien.

# Flujo de Eventos para el caso de uso ParaTrabajoCampo

**.1 Condiciones previas:**
Se deberá haber concluido exitosamente el caso de uso SeleccionarMuestras

**.2 Flujo principal:**
Este caso de uso inicia cuando el usuario está dentro del sistema en la pantalla principal donde se le presenta una matriz con los proyectos existentes a los que tiene acceso, y selecciona un proyecto que se encuentra en el estado ParaTrabajoCampo. Se desplegará un formulario que contendrá tres opciones de impresión de reportes: 1-Imprimir reporte de Matriz de todas las viviendas de la tabla catastral, con información para identificarlas y los espacios en blanco para la información a ser recuperada en el campo, 2-Imprimir un reporte con el formato para el levantamiento de los parámetros del índice de vulnerabilidad sísmica por cada vivienda seleccionada, este reporte también presentará la información de la Tabla catastral a ser comprobada.

## Flujo de Eventos para el caso de uso <u>MientrasTrabajoCampo</u>

**.1 Condiciones previas:**
Se deberá haber concluido exitosamente el caso de uso SeleccionarMuestra.

**.2 Flujo principal:**
Este caso de uso inicia cuando el usuario está dentro del sistema en la pantalla principal donde se le presenta una matriz con los proyectos existentes a los que tiene acceso, y selecciona un proyecto que se encuentra en el estado MientrasTrabajoCampo. Se desplegará un formulario que contendrá las siguientes opciones: Definir pesos para cada uno de los parámetros del índice de vulnerabilidad (S-1: DefinirPesosIndiceVulnerabilidad)

**S-1: DefinirPesosIndiceVulnerabilidad**
Se le presentará al usuario una ventana en la cual a través de una matriz (grid) definirá el peso cada uno de los once parámetros. Una vez hecho esto, antes de abandonar esta etapa se le preguntará si esta seguro.

## Flujo de Eventos para el caso de uso DefinirSismos_ParaRepresentacion EscenariosDaños

**.1 Condiciones previas:**
Para realizar este caso de uso al menos se deberá haber realizado el flujo de eventos para el caso de uso 2: CrearProyecto.

**.2 Flujo principal:**
Este caso de uso inicia cuando el usuario está dentro del sistema en la pantalla principal
donde se le presenta una matriz con los proyectos existentes a los que tiene acceso, y selecciona un proyecto que se encuentra en un estado donde es posible definir sismos. Se presentará el menú de sismos para la representación de escenario de daños, cuando el usuario selecciona dicha opción le aparecerá una pantalla con la matriz de los sismos anteriormente definidos para el proyecto y la opción de agregar uno nuevo. Dicha opción desplegará un cuadro de dialogo pidiéndole que digite el valor para el sismo, el cual no debe ser un valor ya existente en el proyecto S-1.
El sistema automáticamente calculará los daños para cada vivienda (vulnerabilidad o independientes) para la cual ya se ha agregado información recopilada del trabajo de campo.

**.3 SubFlujos:**

**S-1:** Al introducirse un sismo ya existente el sistema enviará un mensaje avisando de dicha situación con la opción de ingresar nuevamente otro valor para sismo o de finalizar la actividad.

# Flujo de Eventos para el caso de uso SubiendoY_PresentandoResultadosVulnerabilidadViviendasMapas

**.1 Condiciones previas:**
Se deberá haber concluido el caso de uso MientrasTrabajoCampo y se deben tener a mano una parte del trabajo recopilado del campo para ser introducida en este caso de uso.

**.2 Flujo principal:**
Este caso de uso inicia cuando el usuario está dentro del sistema en la pantalla principal donde se le presenta una matriz con los proyectos existentes a los que tiene acceso, y selecciona un proyecto que se encuentra en el estado subiendo información de campo. Se presentará el mapa de índice de vulnerabilidad en su granularidad fina, en la misma ventana en que se presenta el mapa aparecerá una barra de herramientas con distintas opciones, entre las que se observará la opción de **Viviendas**, la cual al ser invocada llama la ejecución del subflujo S-1: **ViviendasProyecto**.

**S-1: ViviendasProyecto**
Con este subflujo se despliega una ventana de dialogo donde se podrán seleccionar criterios (ID's de viviendas, Tipos de viviendas "Encuestadas, No Encuestadas", Editadas, Tipología, Nivel de Vulnerabilidad) para la búsqueda de viviendas que los cumplan dichos criterios pueden ser mezclados para potenciar las búsquedas de forma ortogonal.

Se prestarán las opciones de aceptar filtros la que de ser seleccionada invocaría al subflujo de S-2: SelecciónVivienda con las viviendas que cumplen con los criterios seleccionados por el usuario, también se presentará la opción de mostrar todas la cual al ser invocada también ejecutaría el subflujo S-2: SelecciónVivienda mostrando todas las viviendas sin tomar en cuenta ningún criterio de selección, una tercera opción es cancelar que simplemente cierra la ventana de dialogo de actual.

**S-2: SelecciónVivienda**
Al ejecutarse este subflujo se abre un formulario presentándose una matriz con todas las viviendas del proyecto que aplican a los filtros seleccionados por el usuario, se informa sobre la cantidad total de viviendas del proyecto, la cantidad filtrada, los filtros utilizados y se disponen de una lista desplegable desde la cual el usuario podrá seleccionar la escala a la cual se hará el acercamiento grafico de la vivienda seleccionada.

Luego que el usuario seleccione alguna vivienda, puede elegir las opciones de editar vivienda seleccionada la que al ser invocada ejecuta el subflujo S-3: EditarVivienda, también se presenta las opciones observar vivienda, Filtrar, Salir y refrescar viviendas. Para los que se ejecutan sus subflujos de eventos correspondientes.

**S-3: EditarViviendaInformacionCampo**
Este subflujo desplegará un formulario que para todas las viviendas presentara los campos IDgps e Idfoto pero en caso de tratarse de una viviendas a la que se le levantó información de vulnerabilidad se presentaran también otros campos correspondientes a la forma para recuperar valores de los parámetros de vulnerabilidad sísmica en el campo y campos para la comprobación de los valores de la tabla catastral.

# 9 Flujo de Eventos para el caso de uso IntercambiarTiposResultadosY Granularidad

## 9.1 Condiciones previas:

Se deberá haber empezado a trabajar en el caso de uso SubiendoY_PresentandoResultados VulnerabilidadViviendaEnMapas.

## 9.2 Flujo principal:

Este caso de uso inicia cuando el usuario está dentro del sistema en la pantalla principal donde se le presenta una matriz con los proyectos existentes a los que tiene acceso y selecciona un proyecto que se encuentra en el estado de subiendo y presentado resultados. Automáticamente entrará al sistema donde se presenta el mapa de vulnerabilidad en su granularidad más fina (listo para ver los resultados actuales e introducir nuevos valores). Existirá una barra de herramientas que permitirá 1.cambiar la granularidad en que se presenta el mapa de (Pasar a Media, Gruesa o volver a Fina), 2. cambiar el contexto de presentación de resultados entre escenarios de daños y vulnerabilidad, o hacer una mezcla de ambos.

## A.2 Diagramas de actividad para los Casos de uso

A continuación se presentan todos los diagramas de actividades resultantes del análisis para *VULNESIS*.

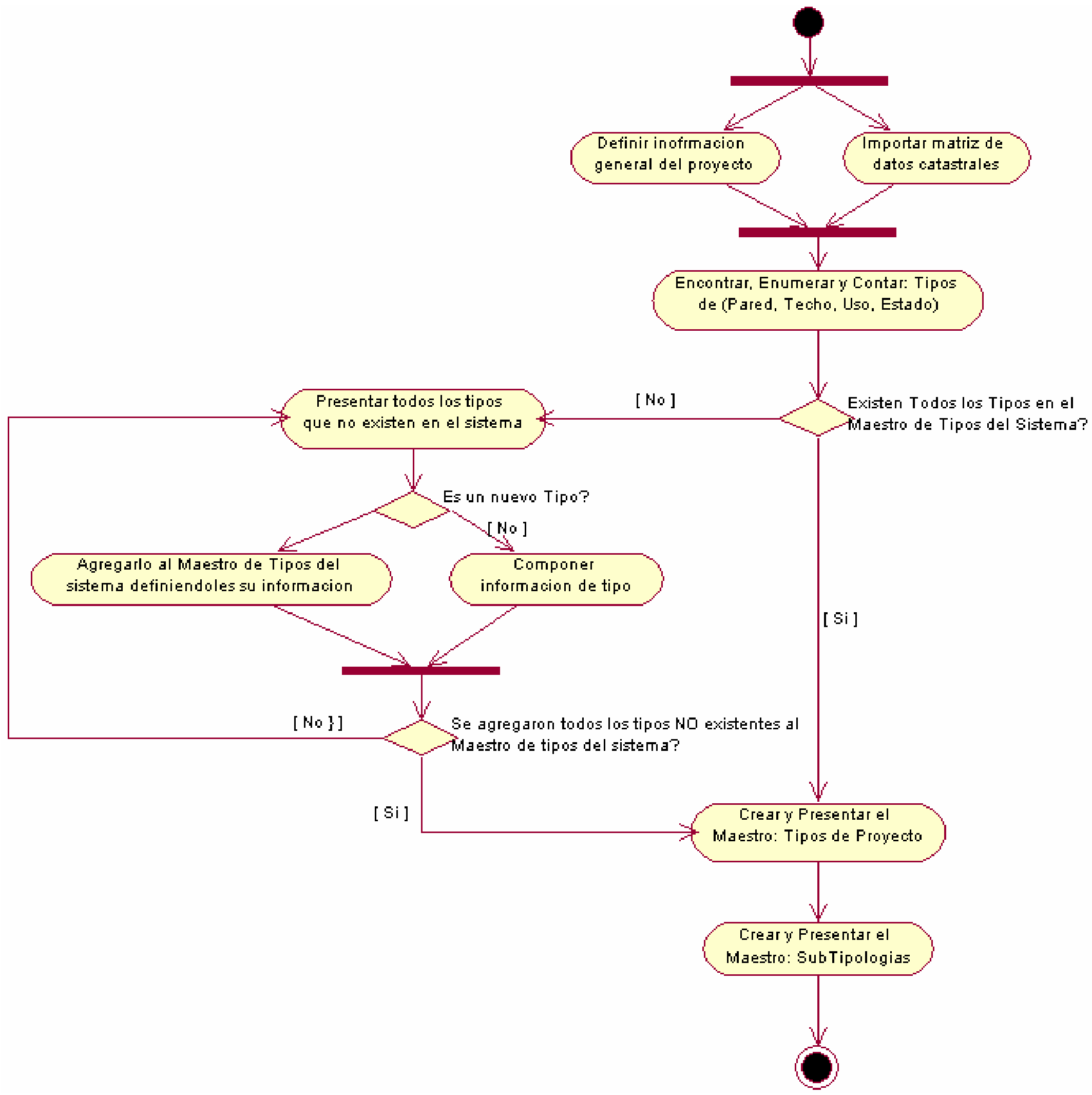

Figura 141. Diagrama de actividad para el Casos de Uso: CrearProyecto

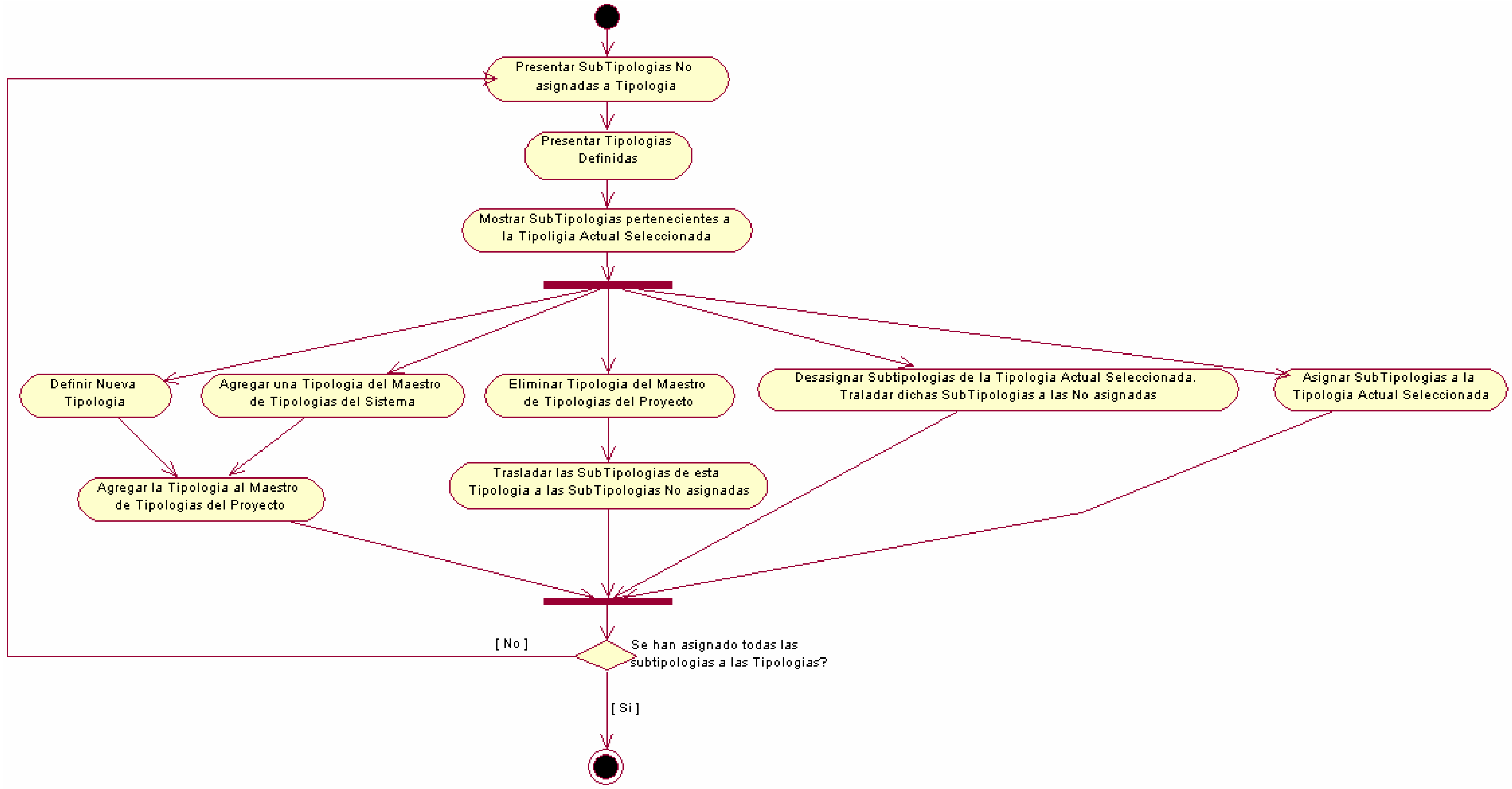

Figura 142. Diagrama de actividad para el Casos de Uso: Definir Tipologías

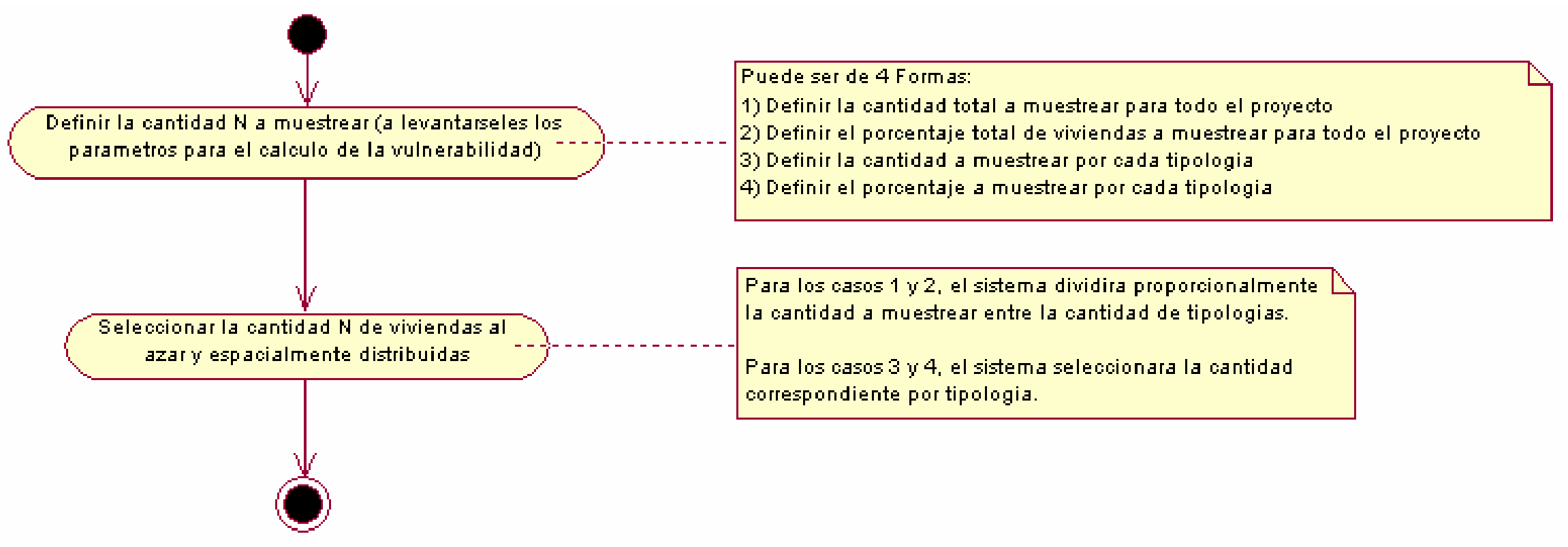

Figura 143. Diagrama de actividad para el Caso de Uso: SeleccionarMuestras

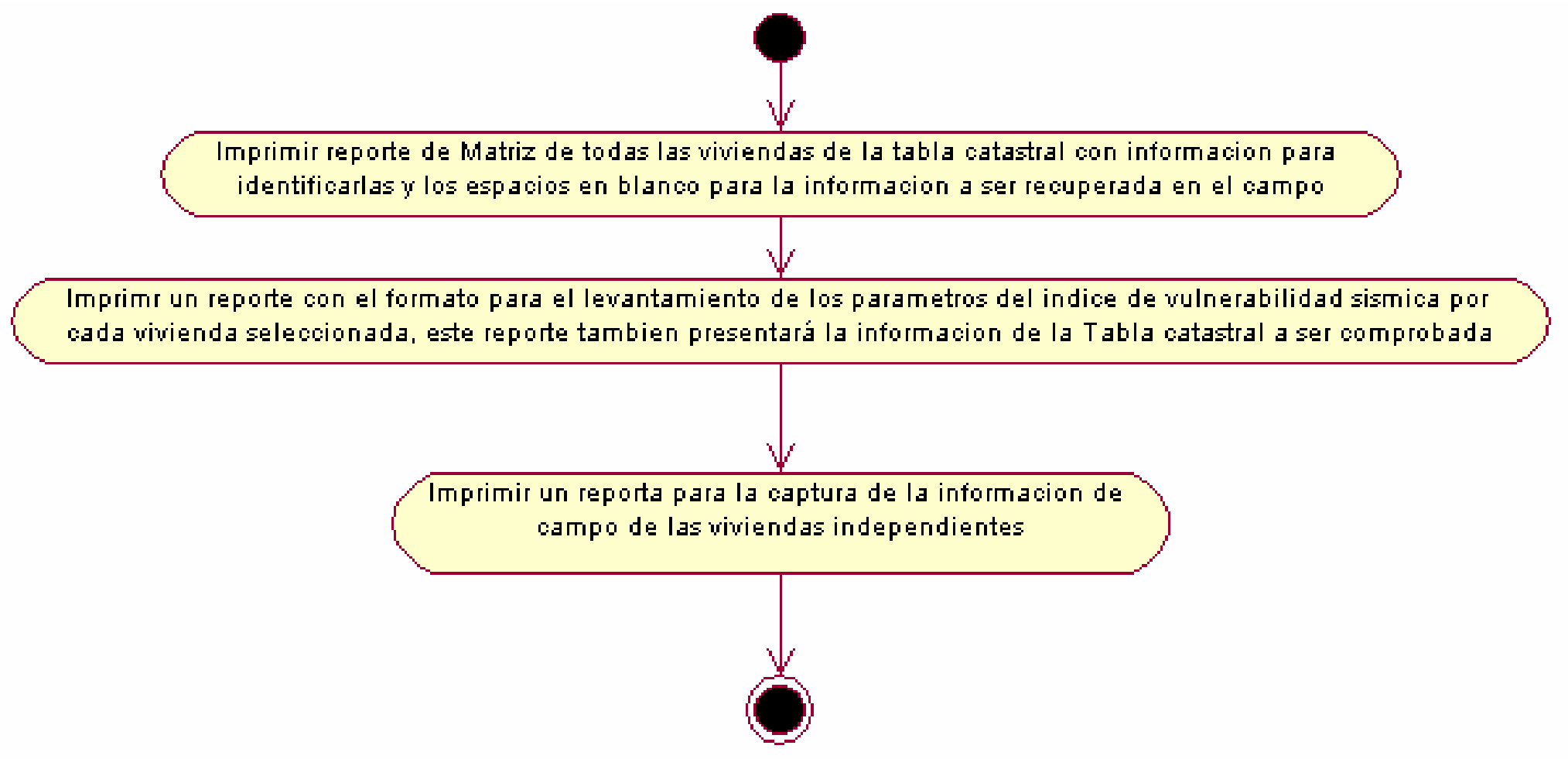

Figura 144. Diagrama de actividad para el Caso de Uso: TrabajoCampoY_Cartografia

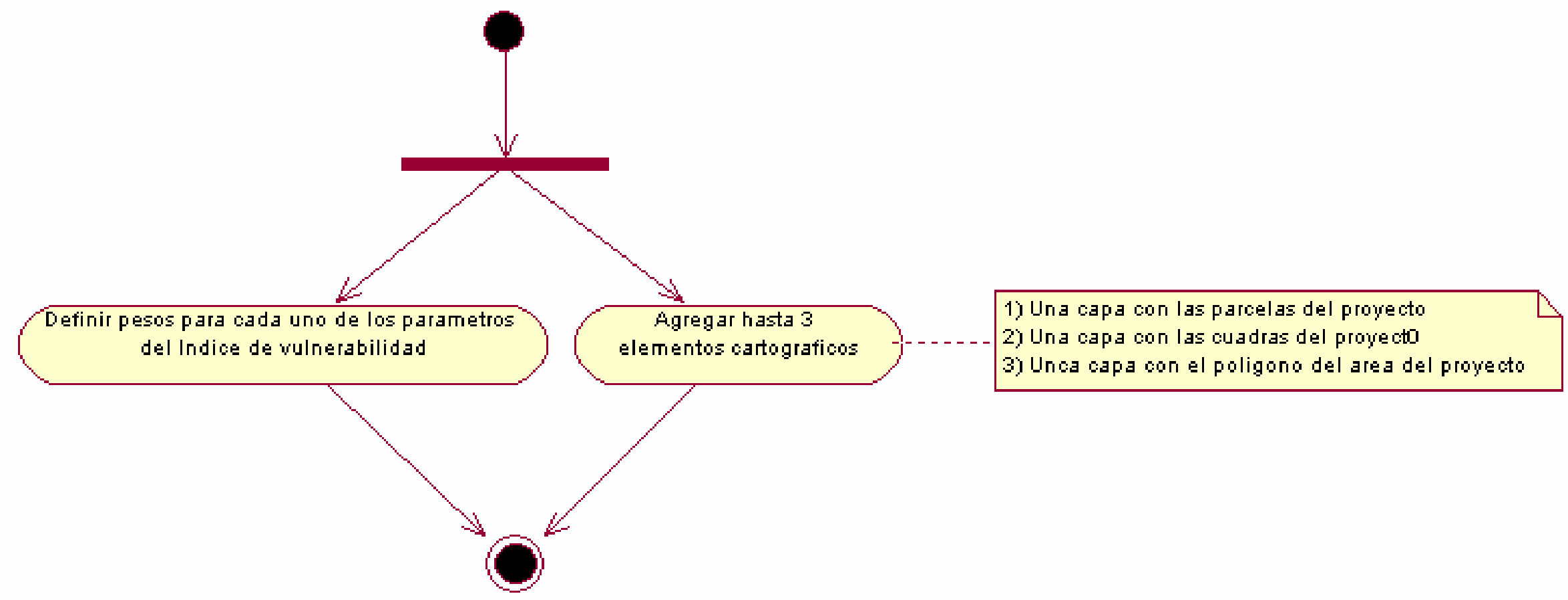

Figura 145. Diagrama de actividad para el Caso de Uso: MientrasTrabajoCampoY_Cartografia

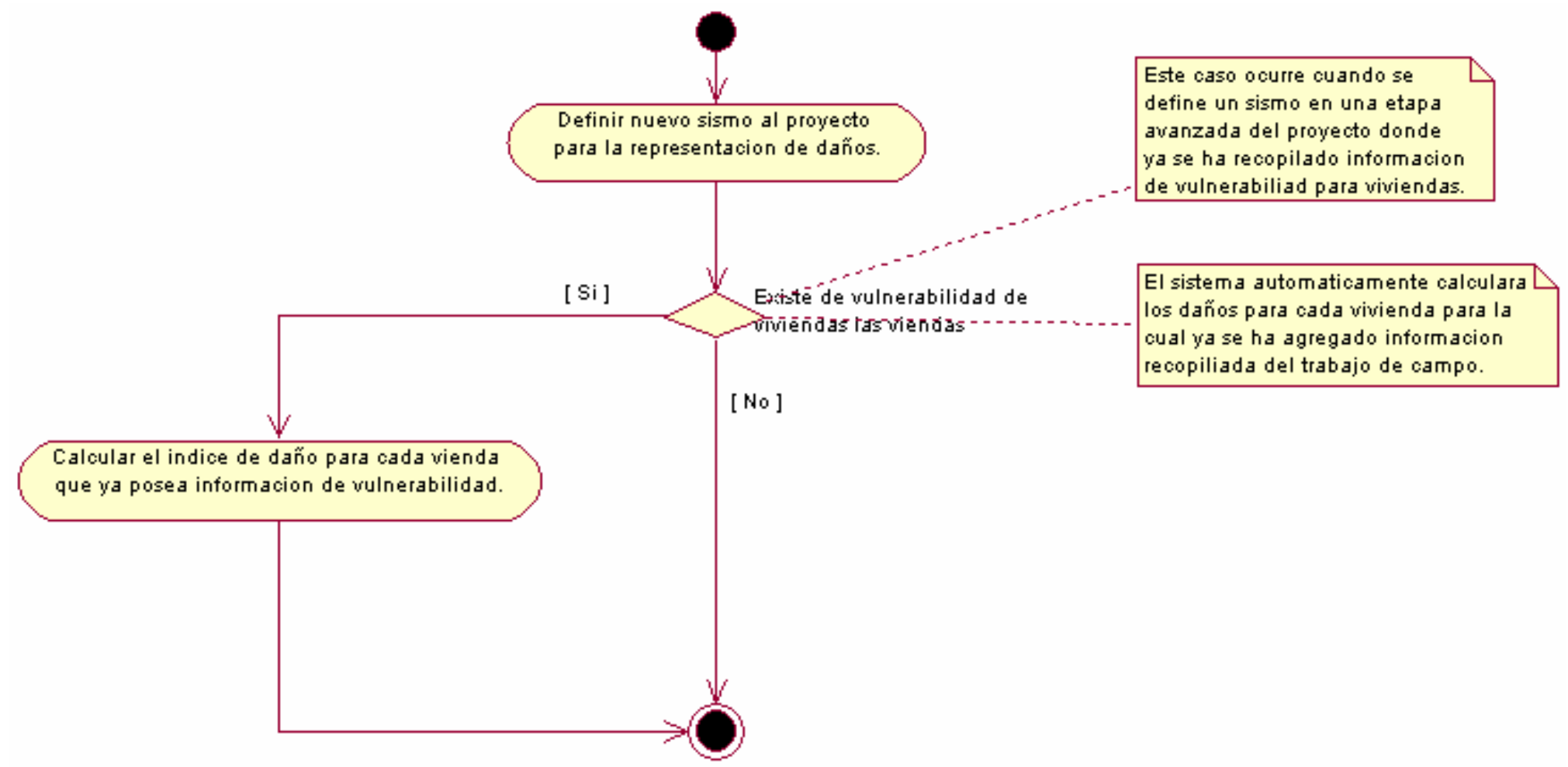

Figura 146. Diagrama de actividad para el Caso de Uso: DefinirSismos_ParaRepresentacionEscenariosDaños

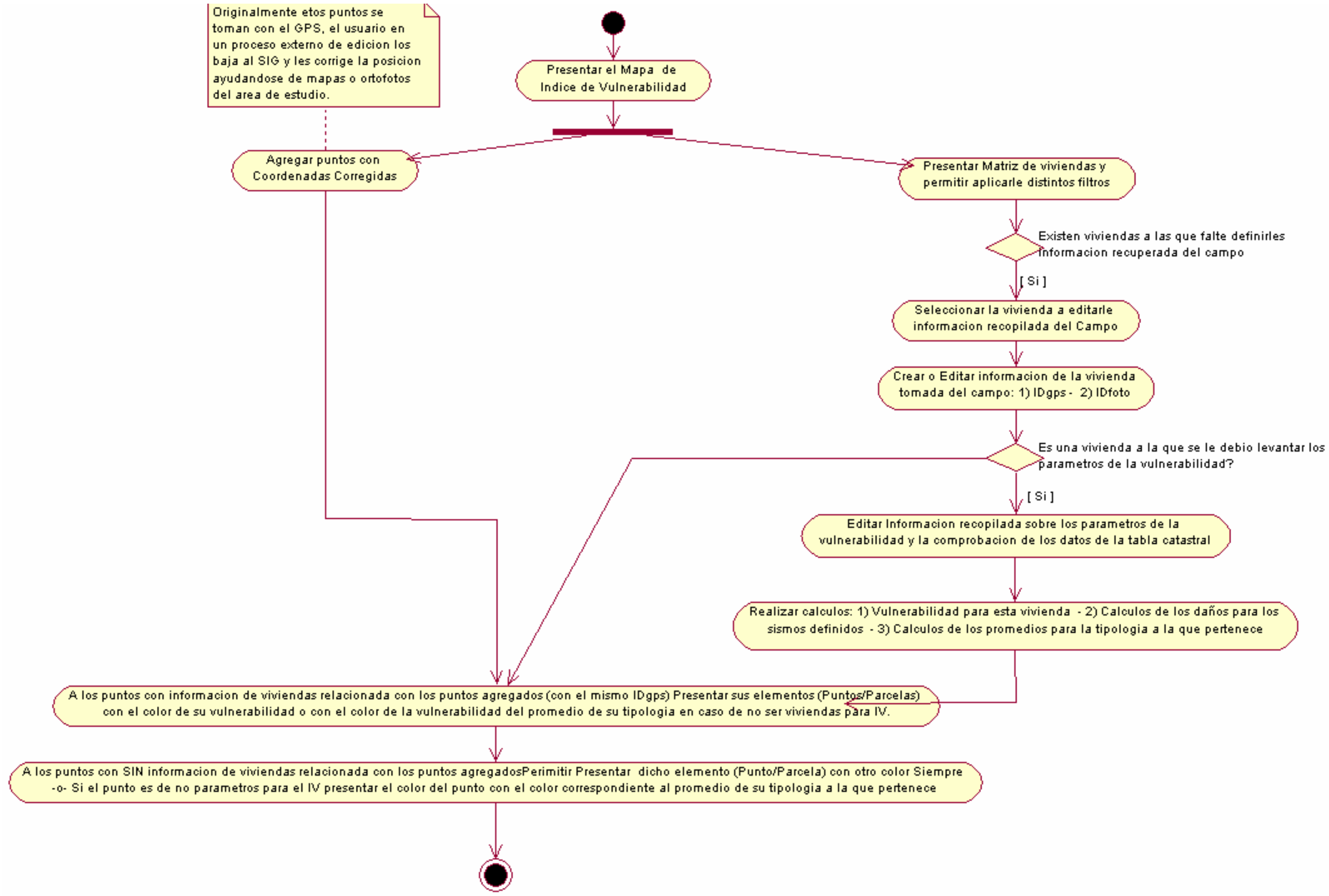

Figura 147. Diagrama (A) de actividad para el Caso de Uso: SubiendoY_PresentandoResultadosVulnerabilidadViviendaEnMapas

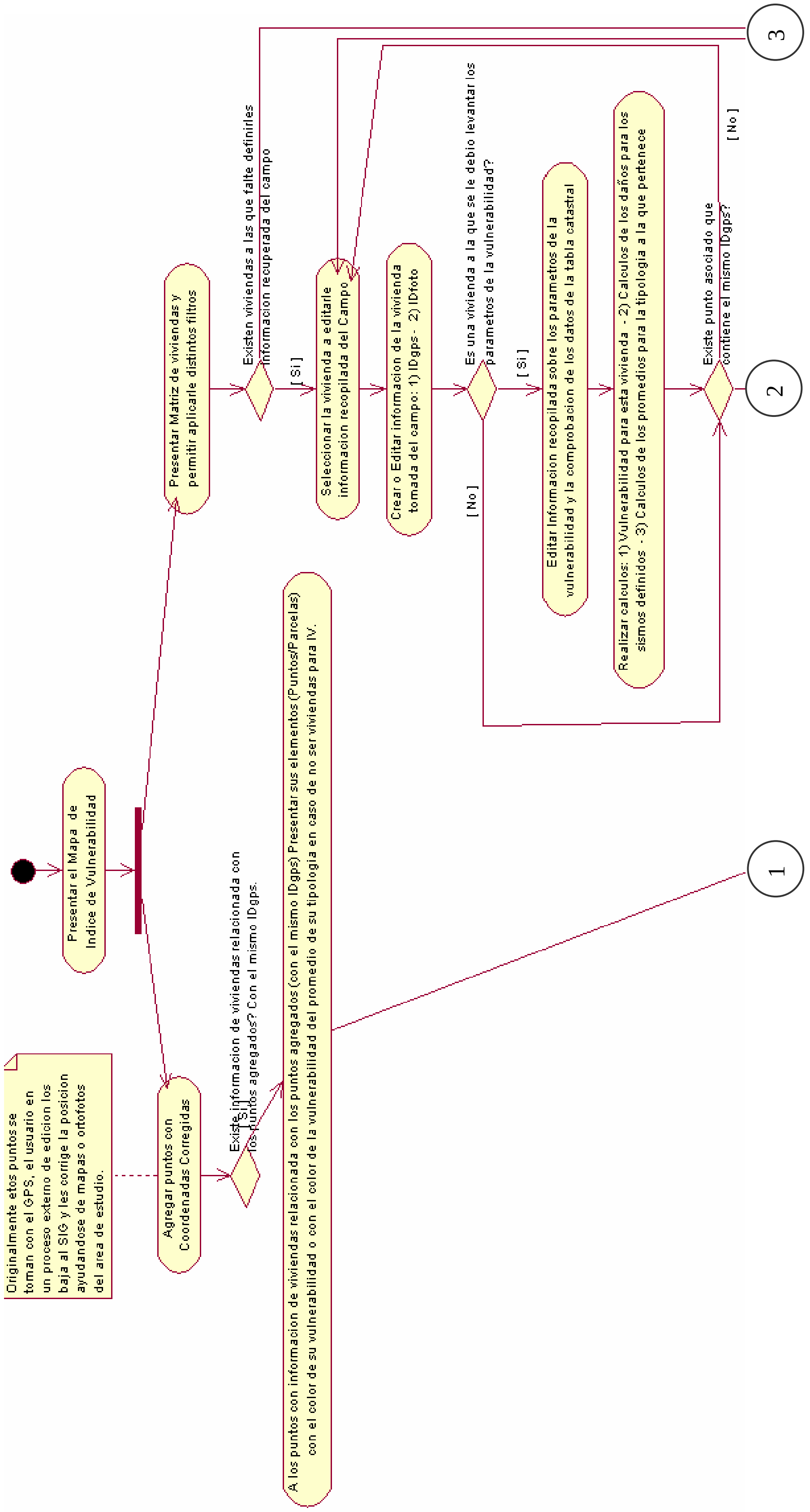

Figura 148. Diagrama (B) de actividad para el Caso de Uso: SubiendoY PresentandoResultadosVulnerabilidadViviendaEnMapas Parte superior del diagrama (1/2)

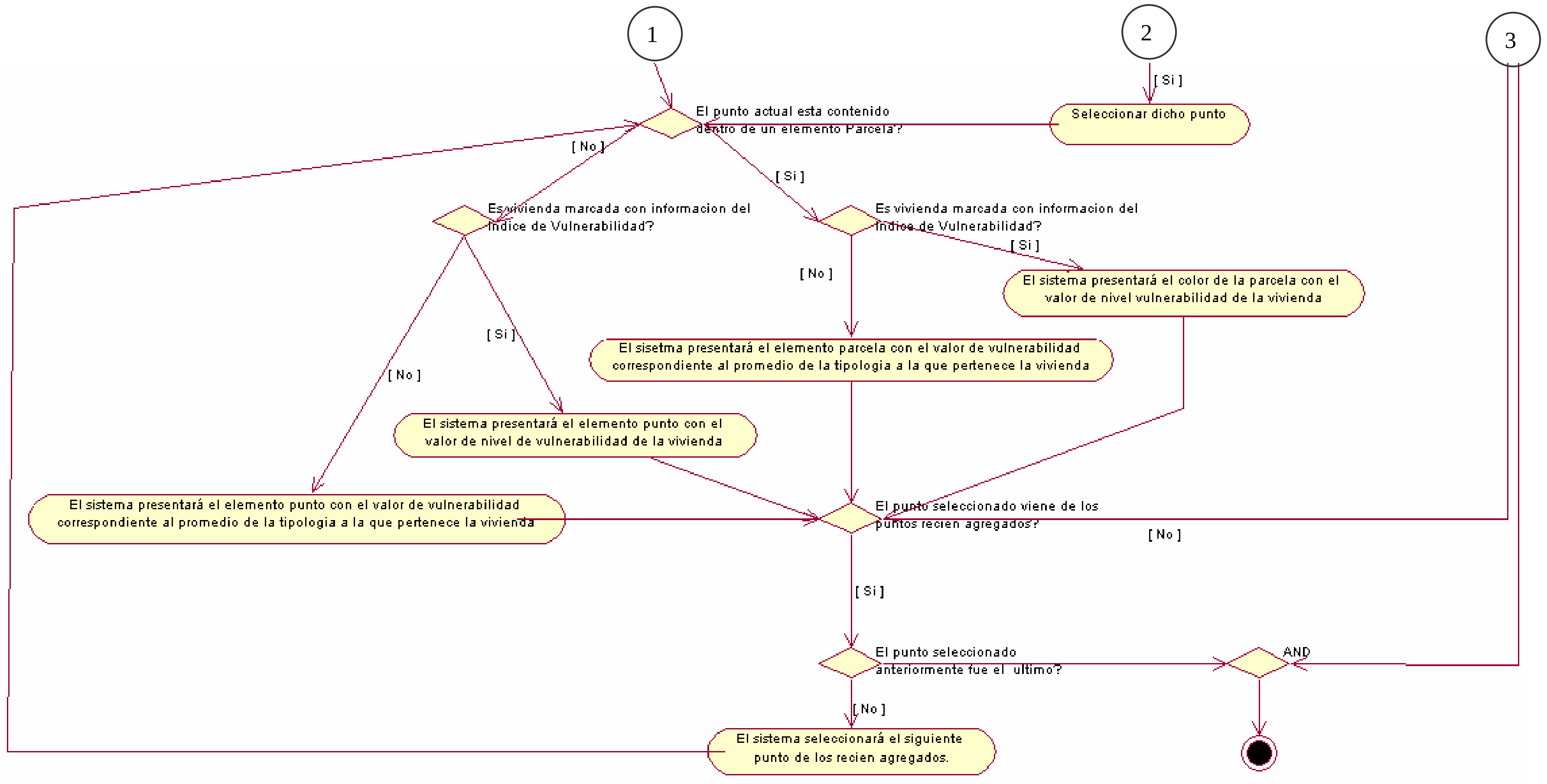

Figura 149 Diagrama (B) de actividad para el Caso de Uso: SubiendoY PresentandoResultadosVulnerabilidadViviendaEnMapas Parte superior del diagrama (2/2)

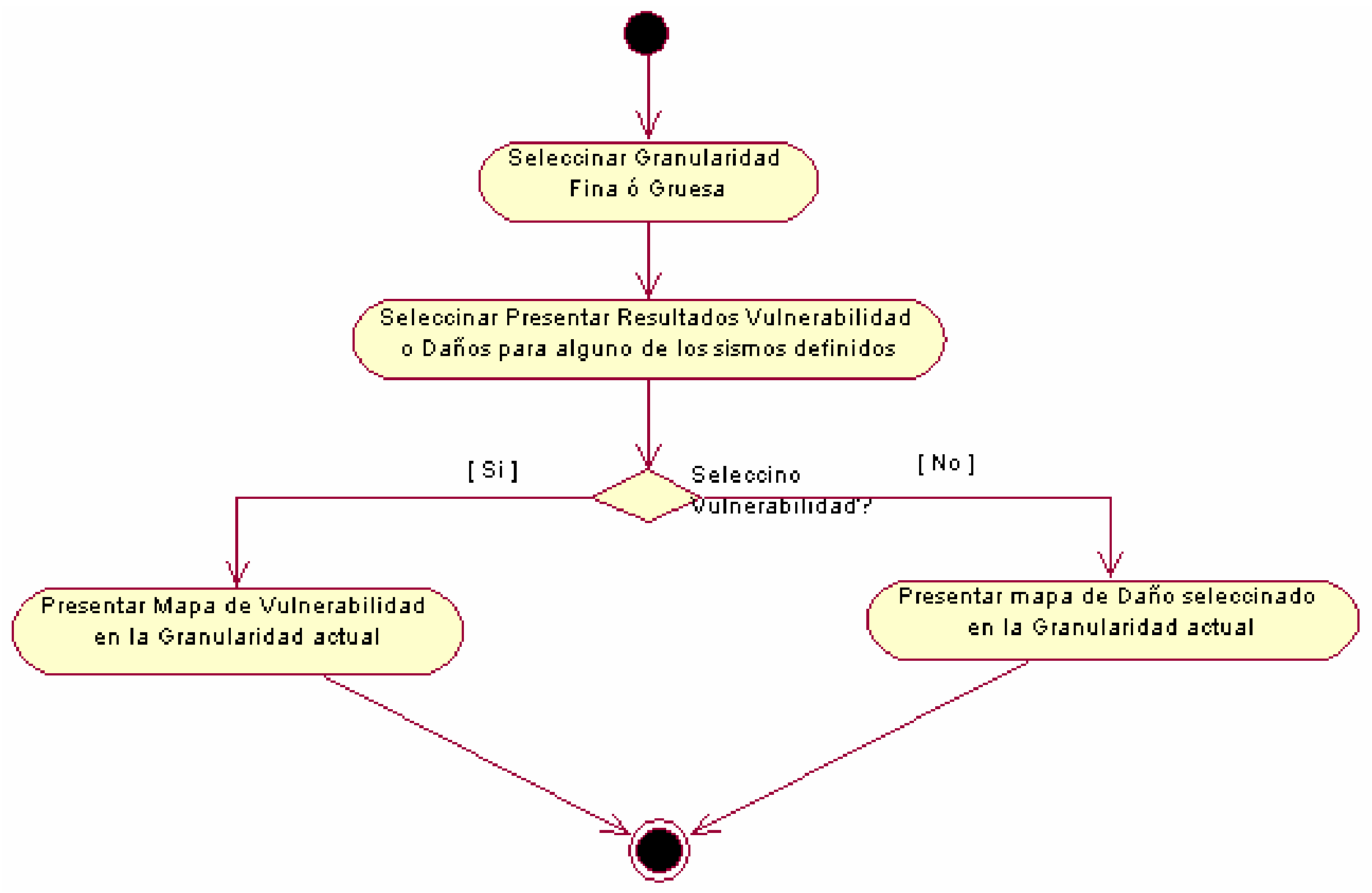

Figura 150. Diagrama de actividad para el Caso de Uso: IntercambiarTiposResultadosY_Granularidad

## A.3 Realización de Casos de uso

Los casos de uso presentan una vista del sistema desde afuera, y la funcionalidad de los casos de uso es presentado mediante los flujos de eventos. La realización de los casos de uso es presentada mediante escenarios que indican como el caso de uso se realiza mediante la interacción de una sociedad de objetos.

Los escenarios son instancias de los casos de uso, son una vía a través del flujo de eventos. Estos escenarios son desarrollados para identificar la interacción entre los objetos y las clases que se deben utilizar para cumplir las funcionalidades requeridas del sistema.

Los casos de uso fueron capturados en documentos de texto (flujo de eventos) y los escenarios son capturados en diagramas de interacción (Diagramas de secuencia y diagramas de colaboración).

A continuación se presentan los diagramas que describen la relación entre los casos de uso y los escenarios (Realización de casos de uso).

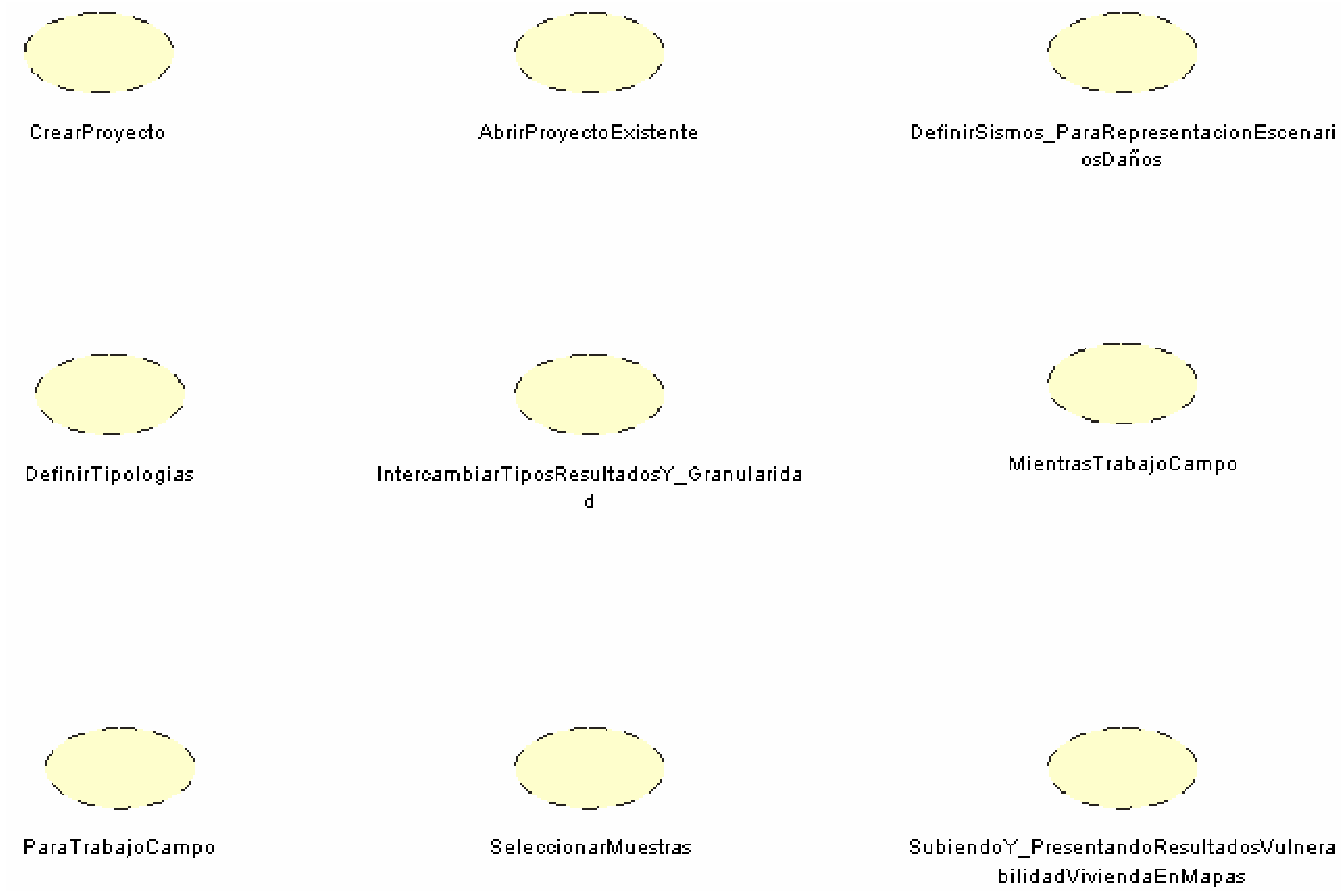

Figura 151. Diagrama de Casos de Uso de Realización

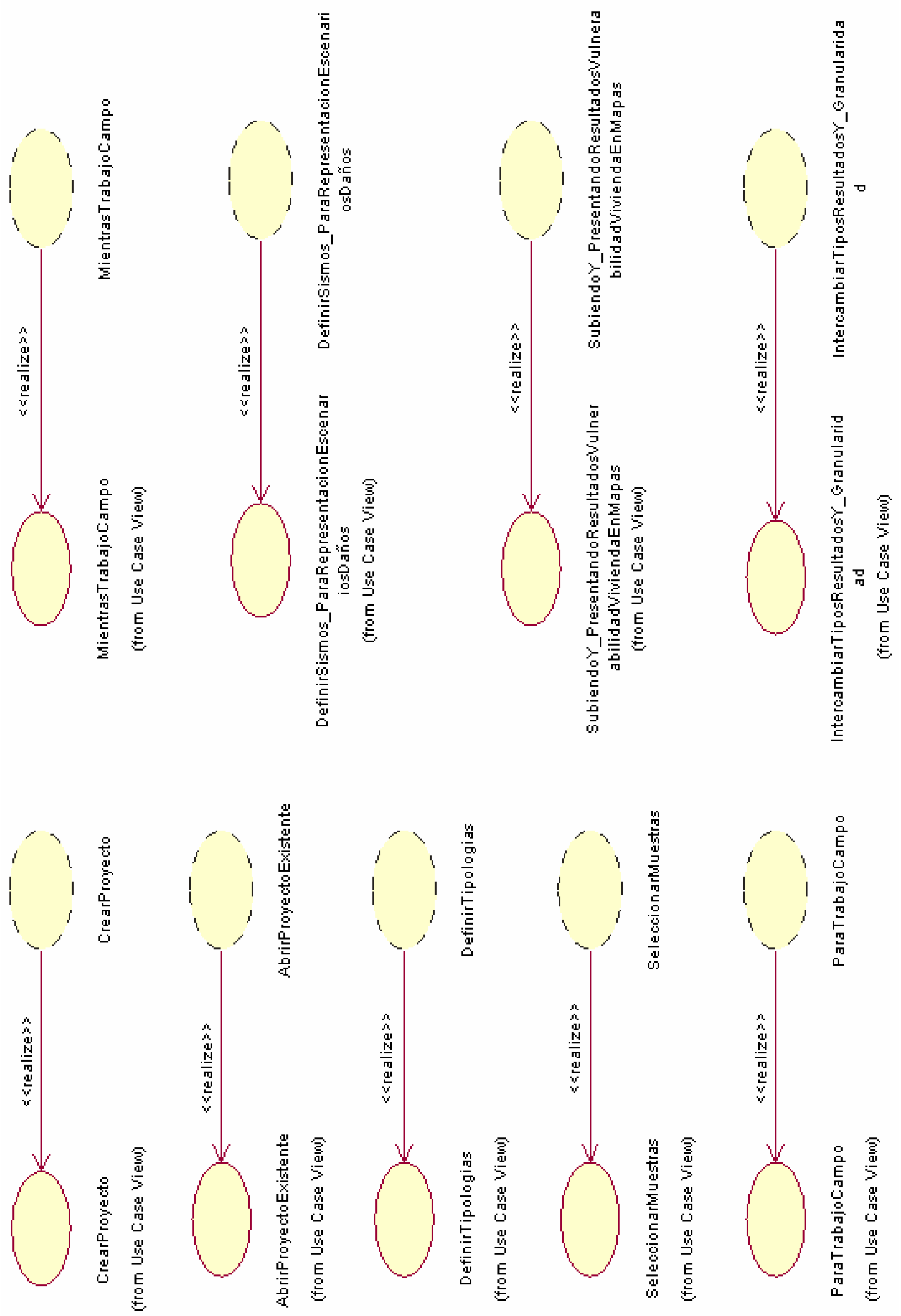

Figura 152. Diagrama de Transición de Casos de Uso hacia realización

## A.4 Diagramas de Secuencia

A continuación se presentan todos los diagramas de secuencia resultantes del análisis para *VULNESIS*.

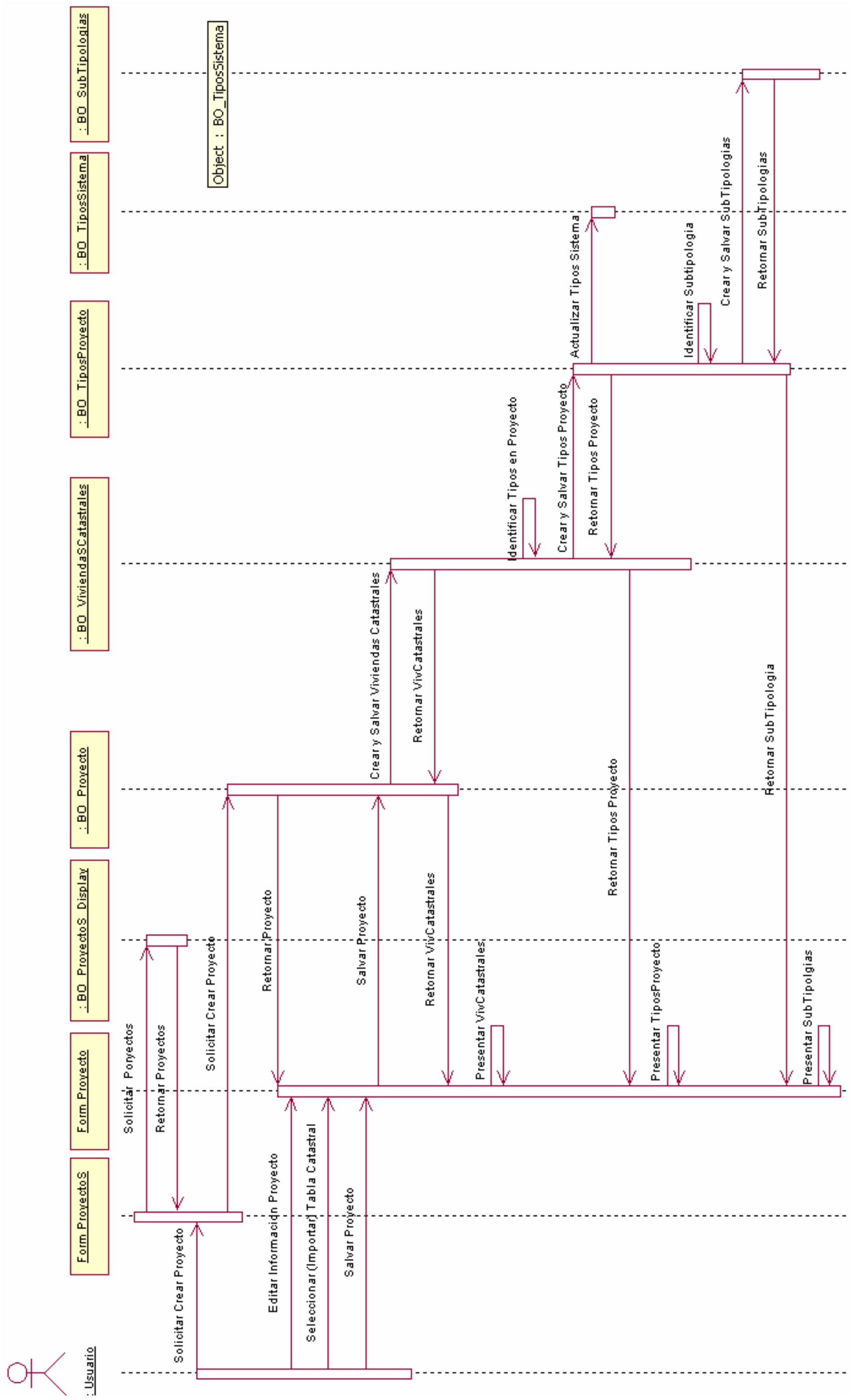

Figura 153. Diagrama de Secuencia CrearProyecto

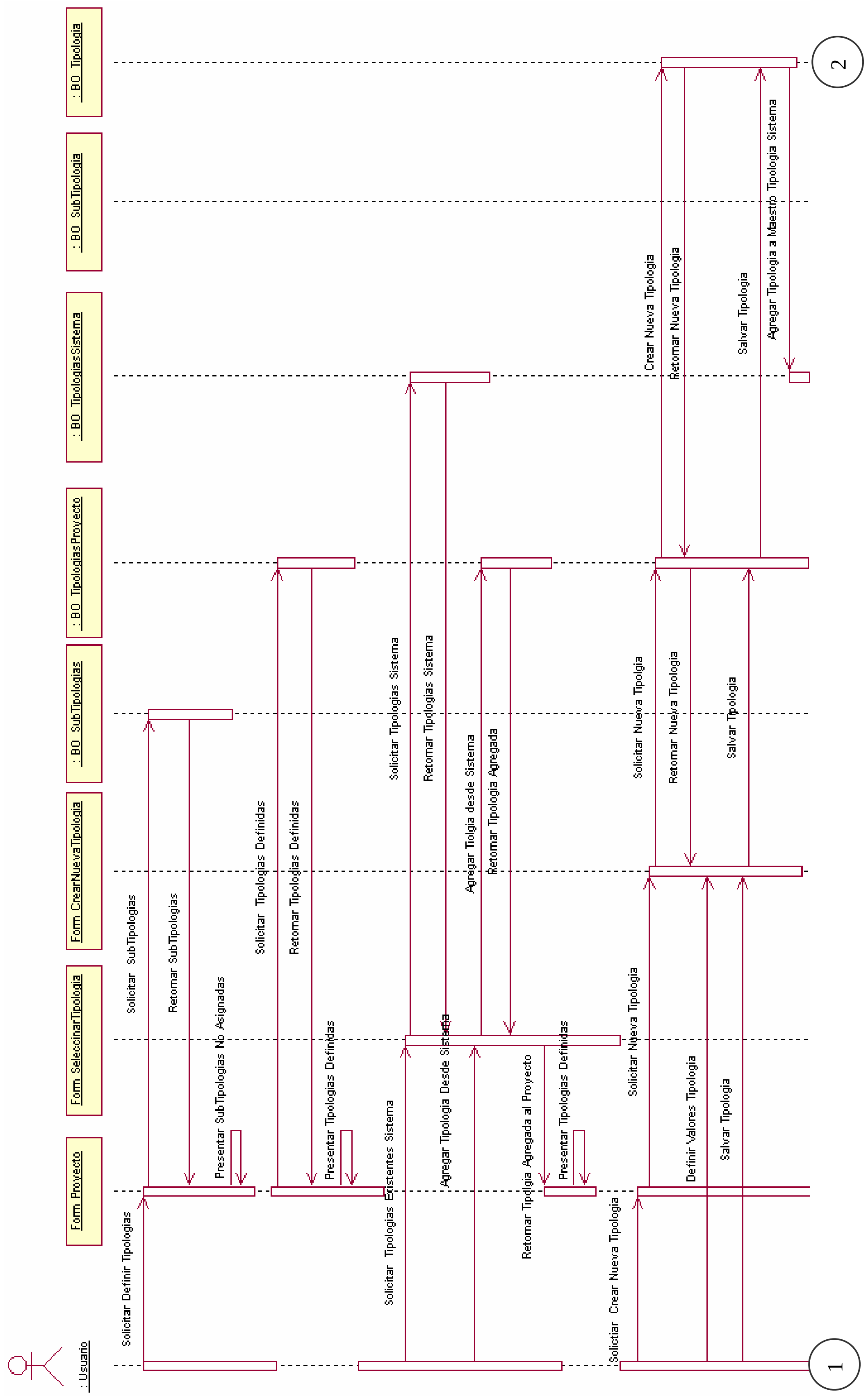

Figura 154. Diagrama de Secuencia DefinirTipologías Parte superior del diagrama (1/2)

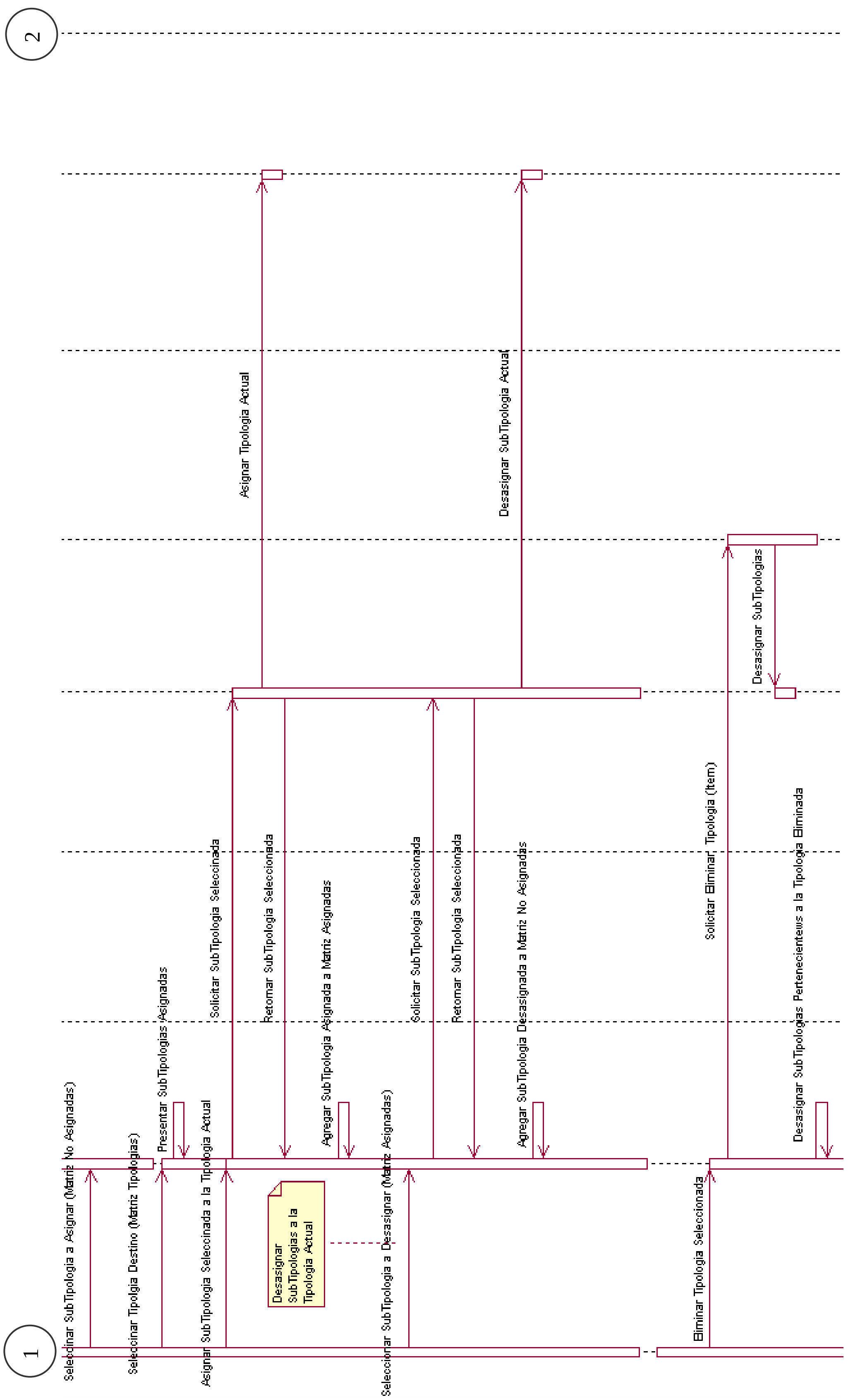

Figura 155. Diagrama de Secuencia DefinirTipologías        Parte inferior del diagrama (2/2)

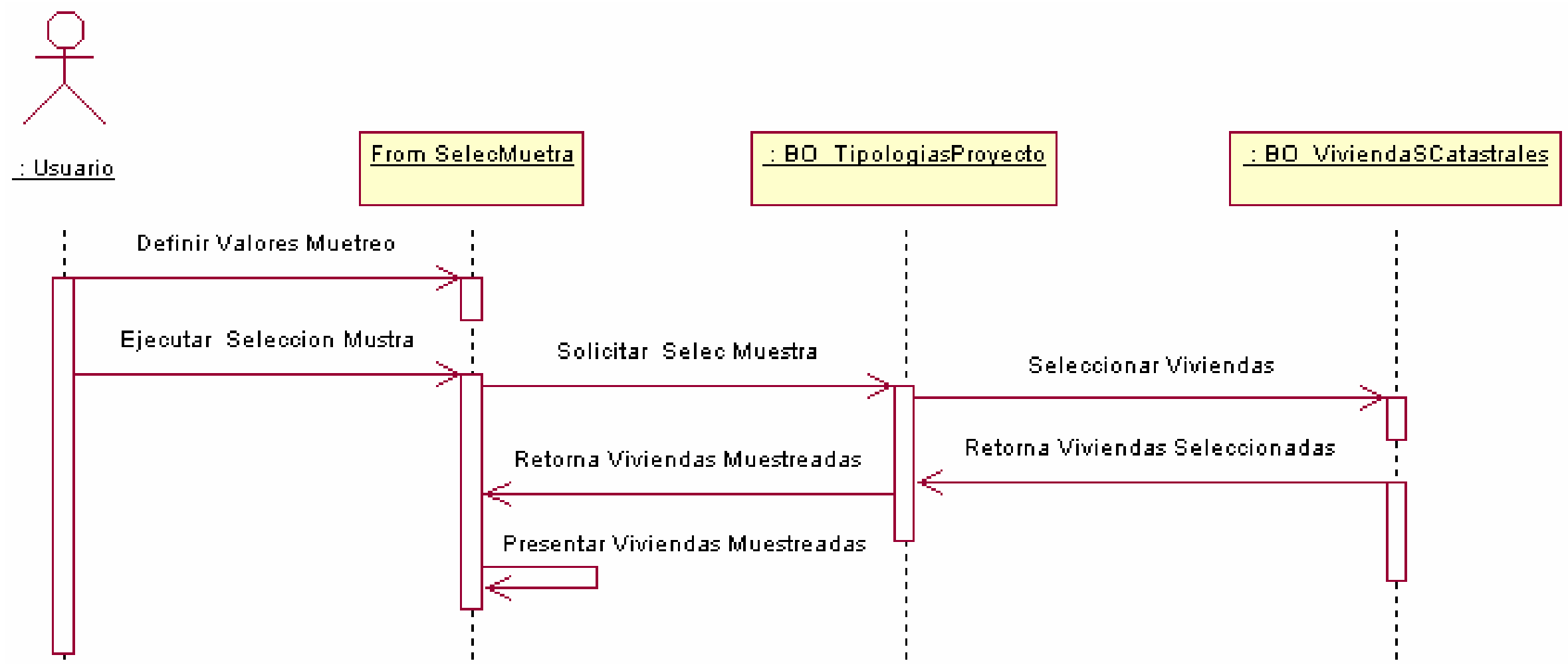

Figura 156. Diagrama de Secuencia SeleccionarMuestra

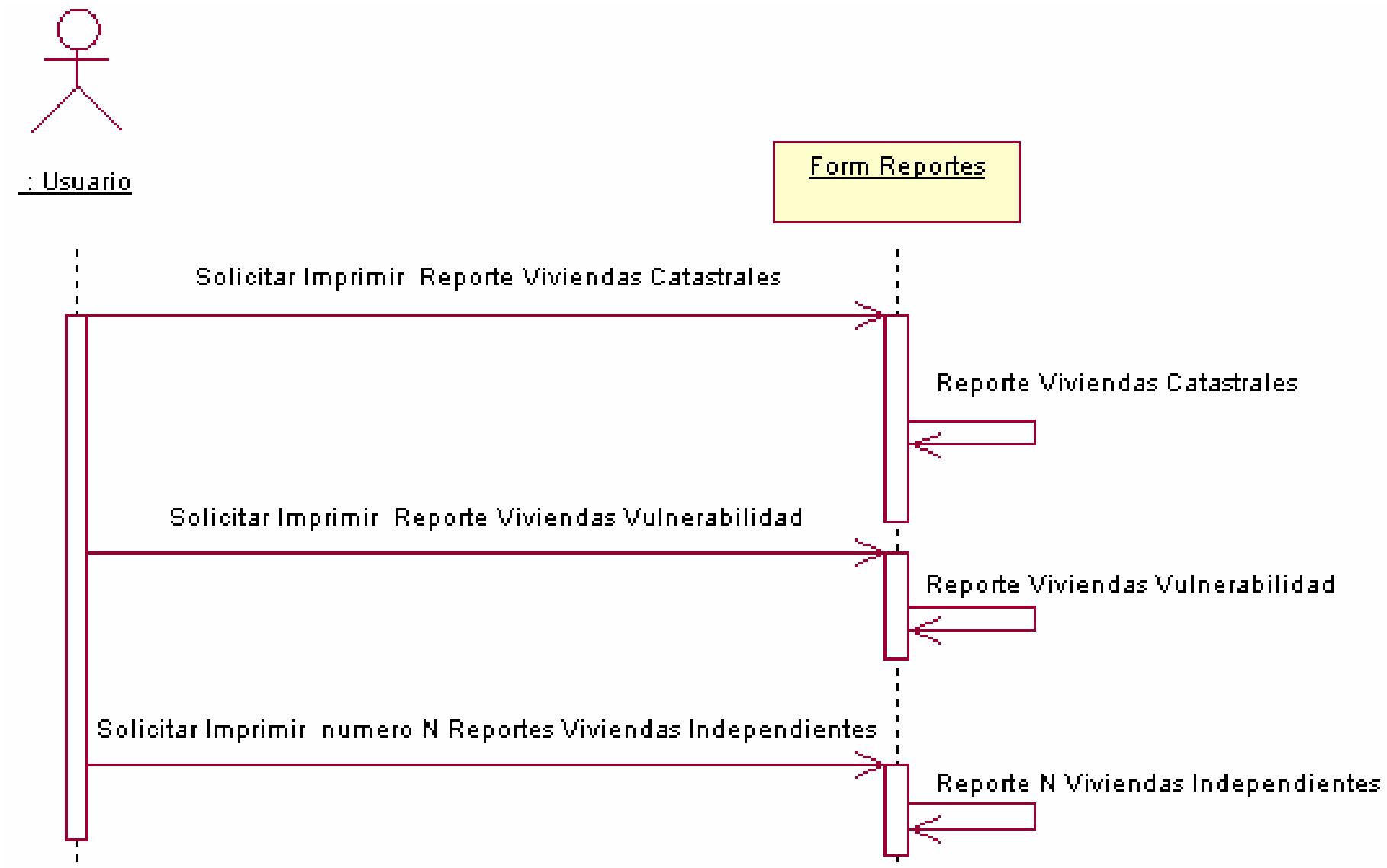

Figura 157. Diagrama de Secuencia ParaTrabajoCampo

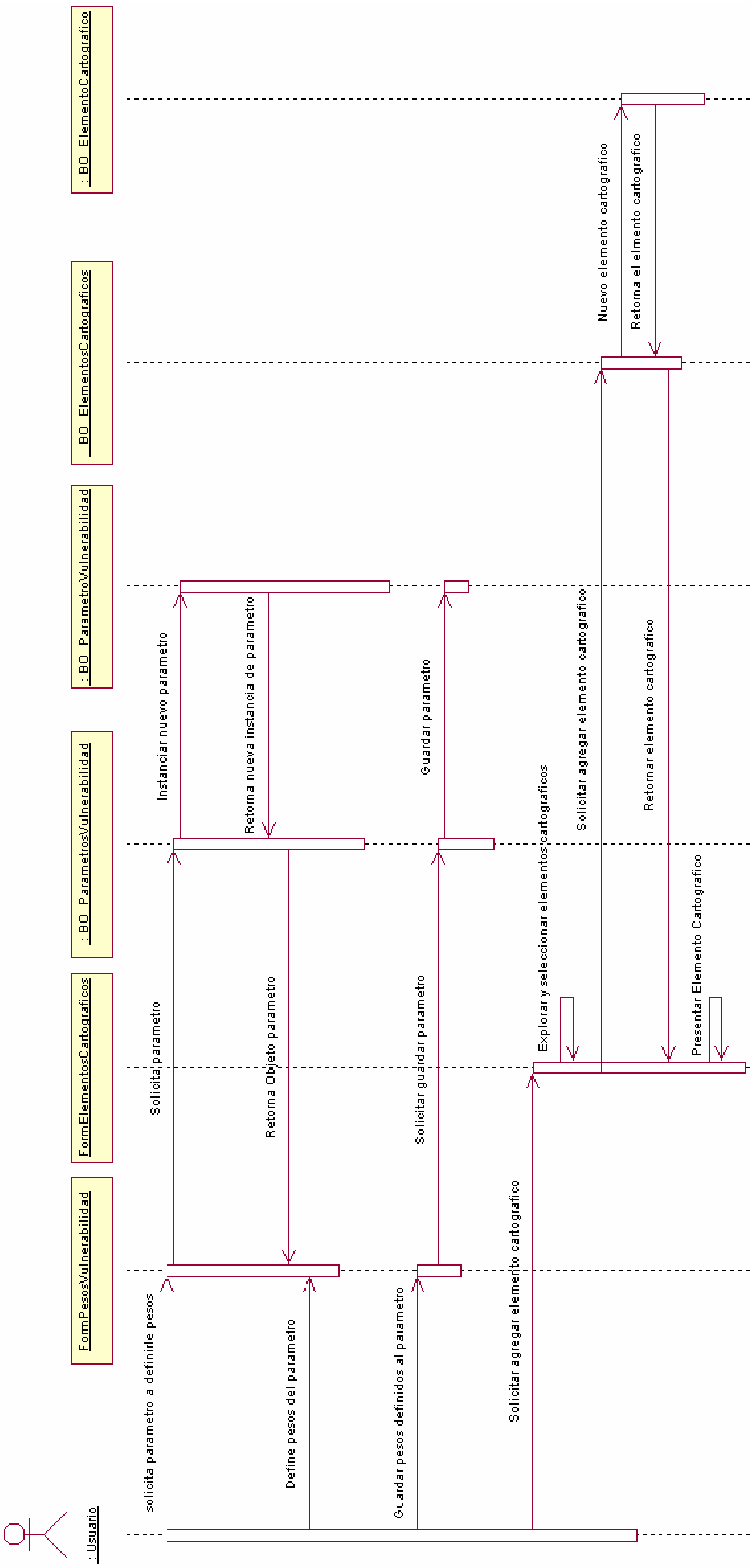

Figura 158. Diagrama de Secuencia MientrasTrabajoCampo

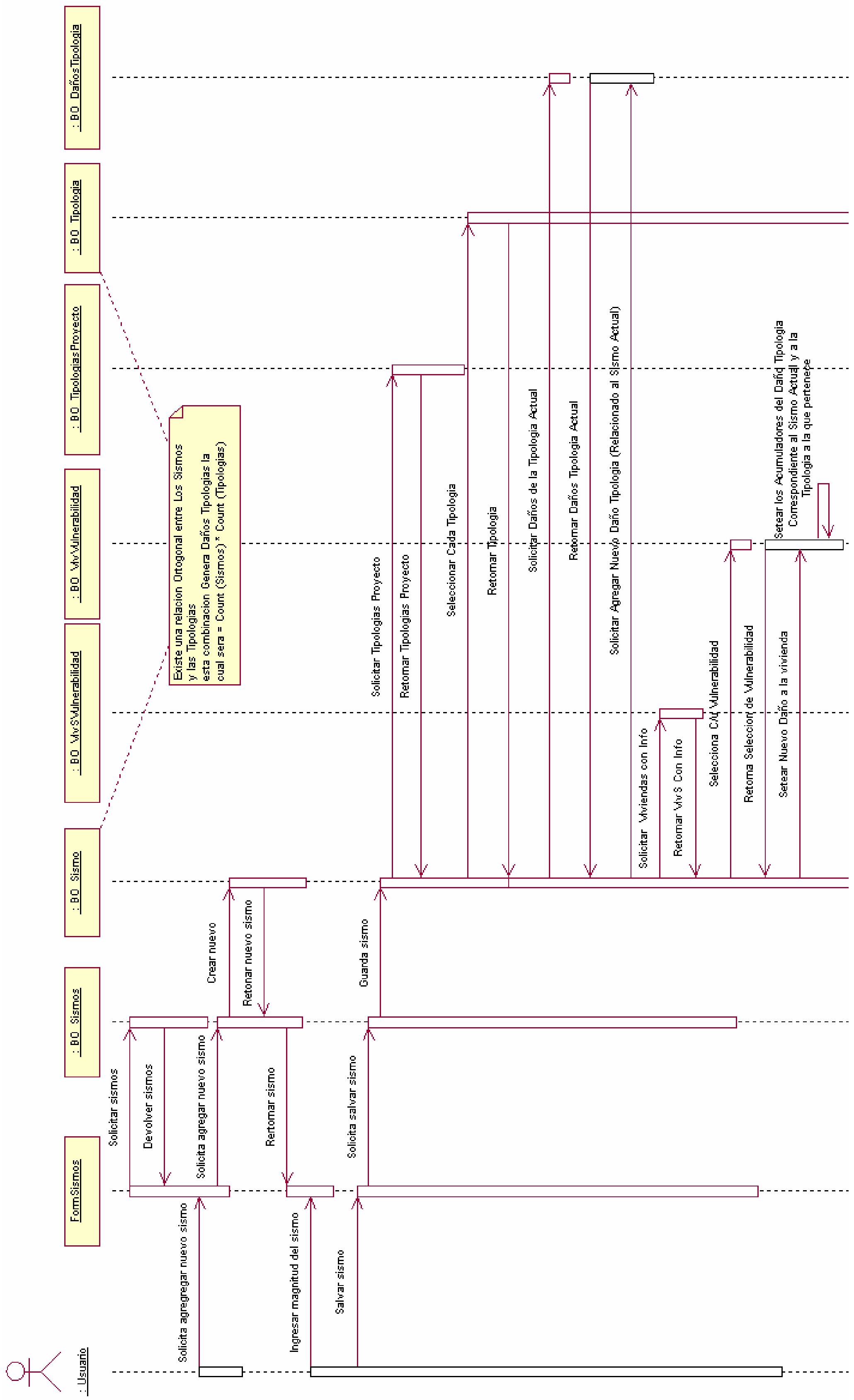

Figura 159. Diagrama de Secuencia MientrasTrabajoCampo

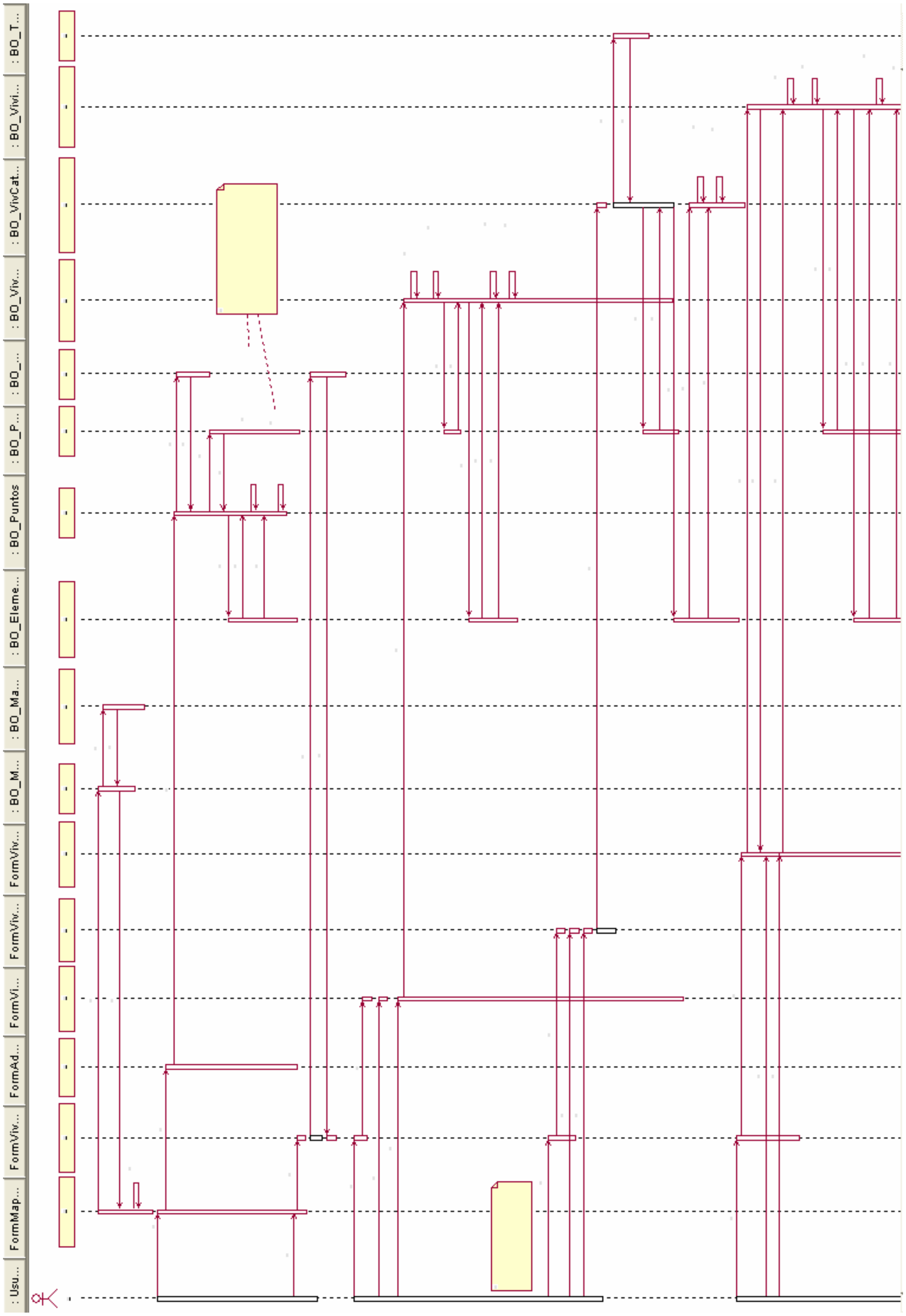

Figura 160. Diagrama de Secuencia SubiendoY PresentandoResultadosVulnerabilidadEnMapas (Vista TOTAL)

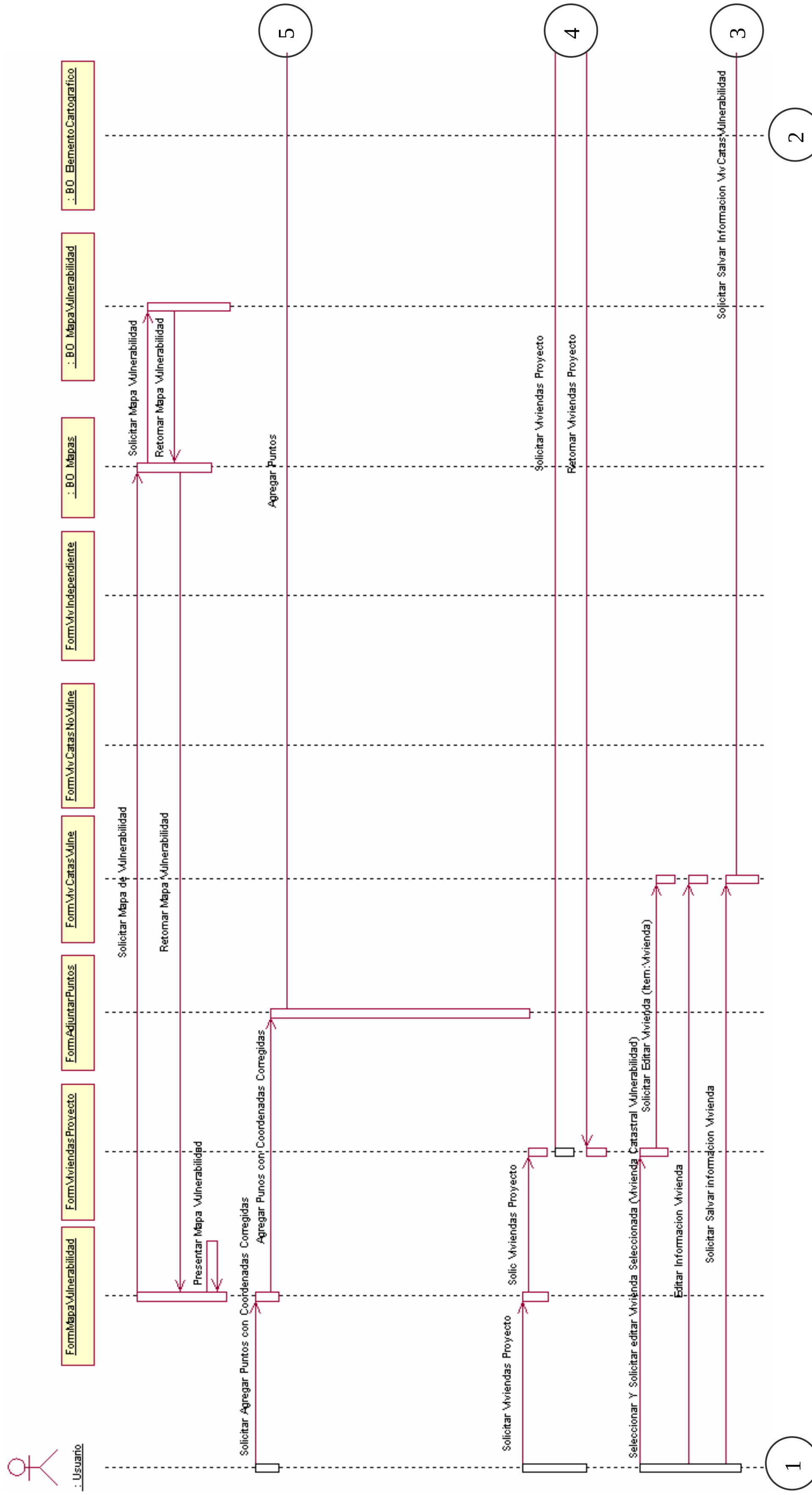

Figura 161. Diagrama de Secuencia SubiendoY PresentandoResultadosVulnerabilidadEnMapas

Parte superior izquierda del diagrama

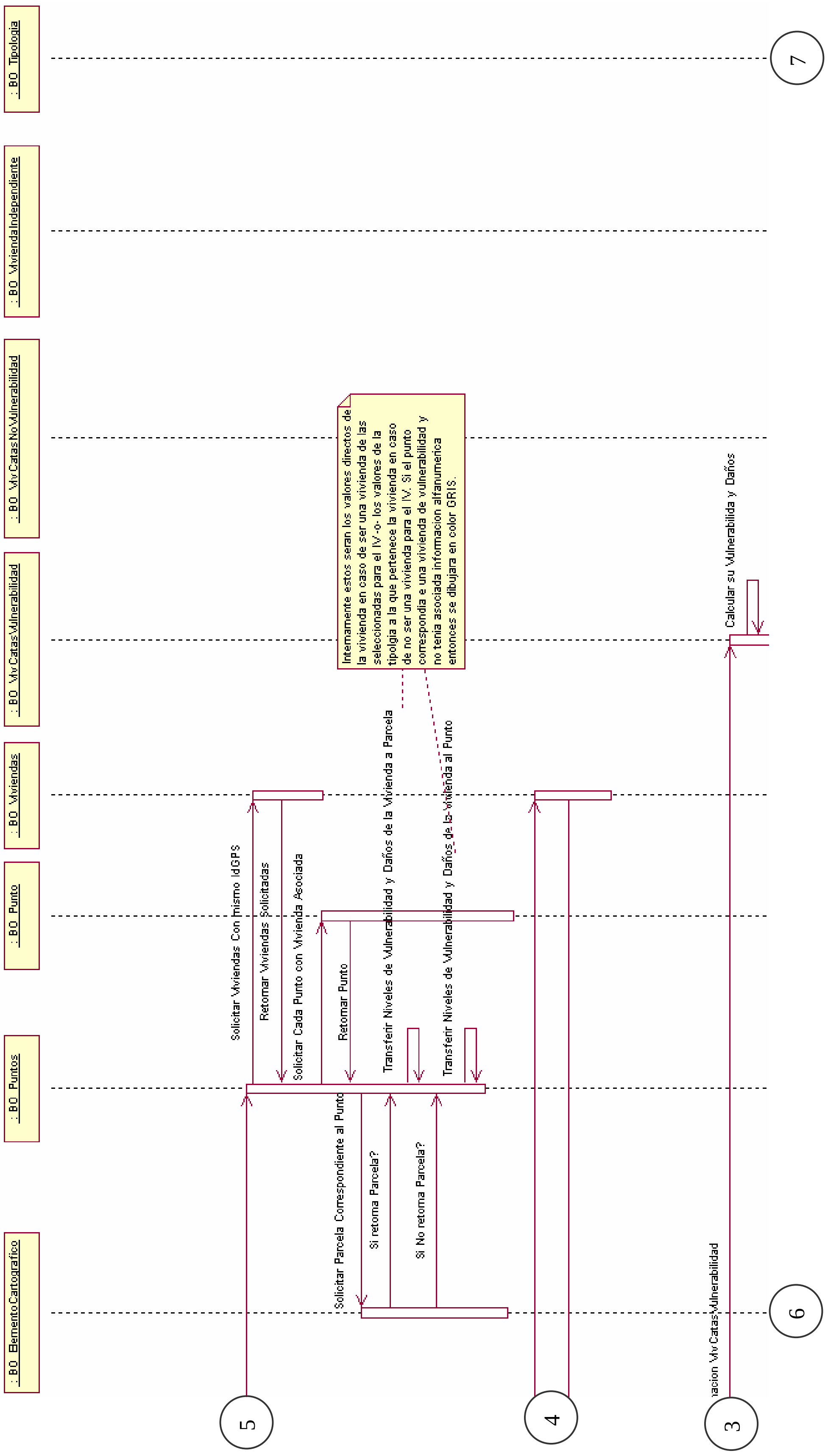

Figura 162. Diagrama de Secuencia SubiendoY PresentandoResultadosVulnerabilidadEnMapas

Parte superior derecha del diagrama (2/4)

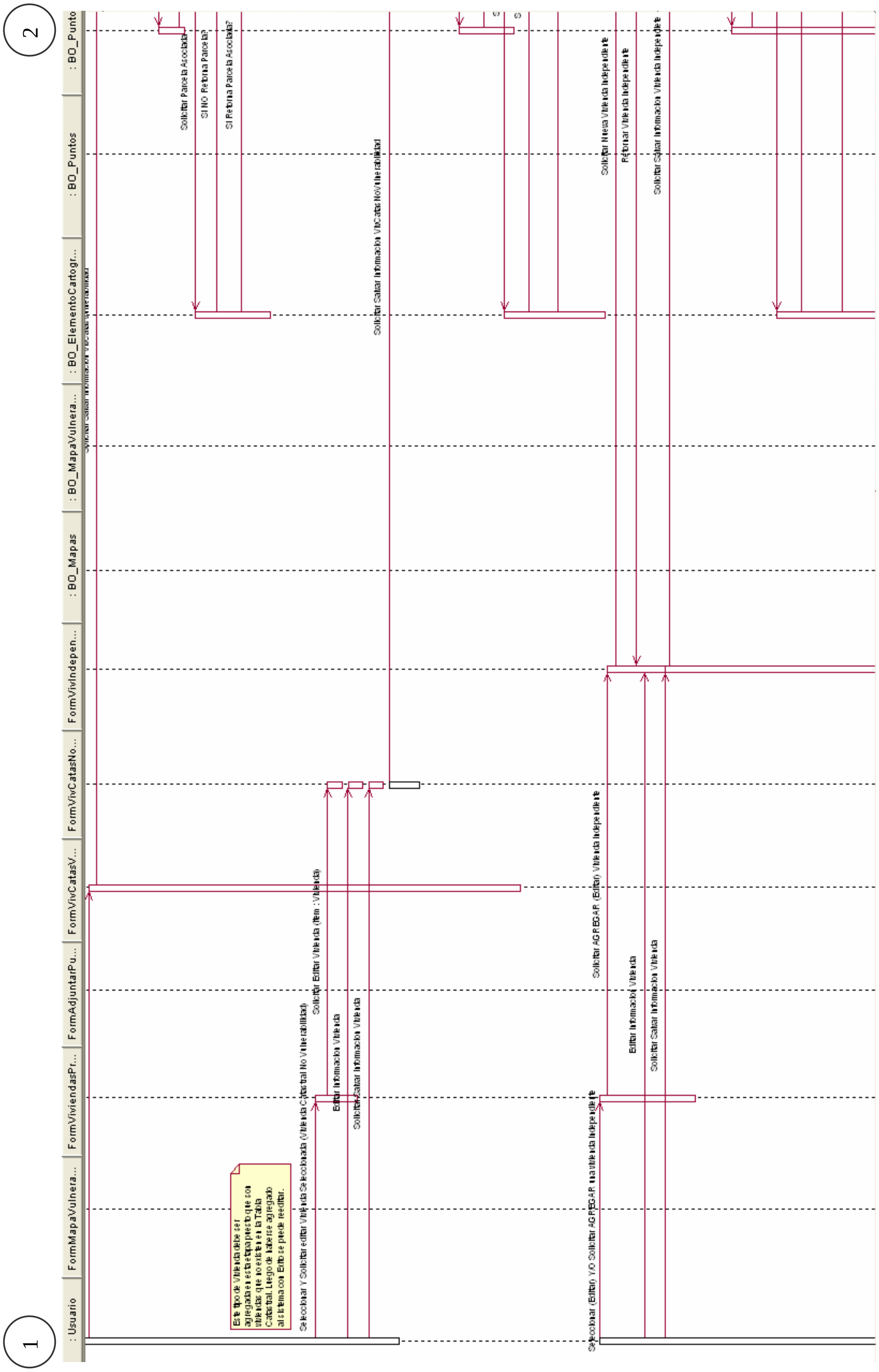

Figura 163. Diagrama de Secuencia SubiendoY PresentandoResultadosVulnerabilidadEnMapas

Parte inferior izquierda del diagrama

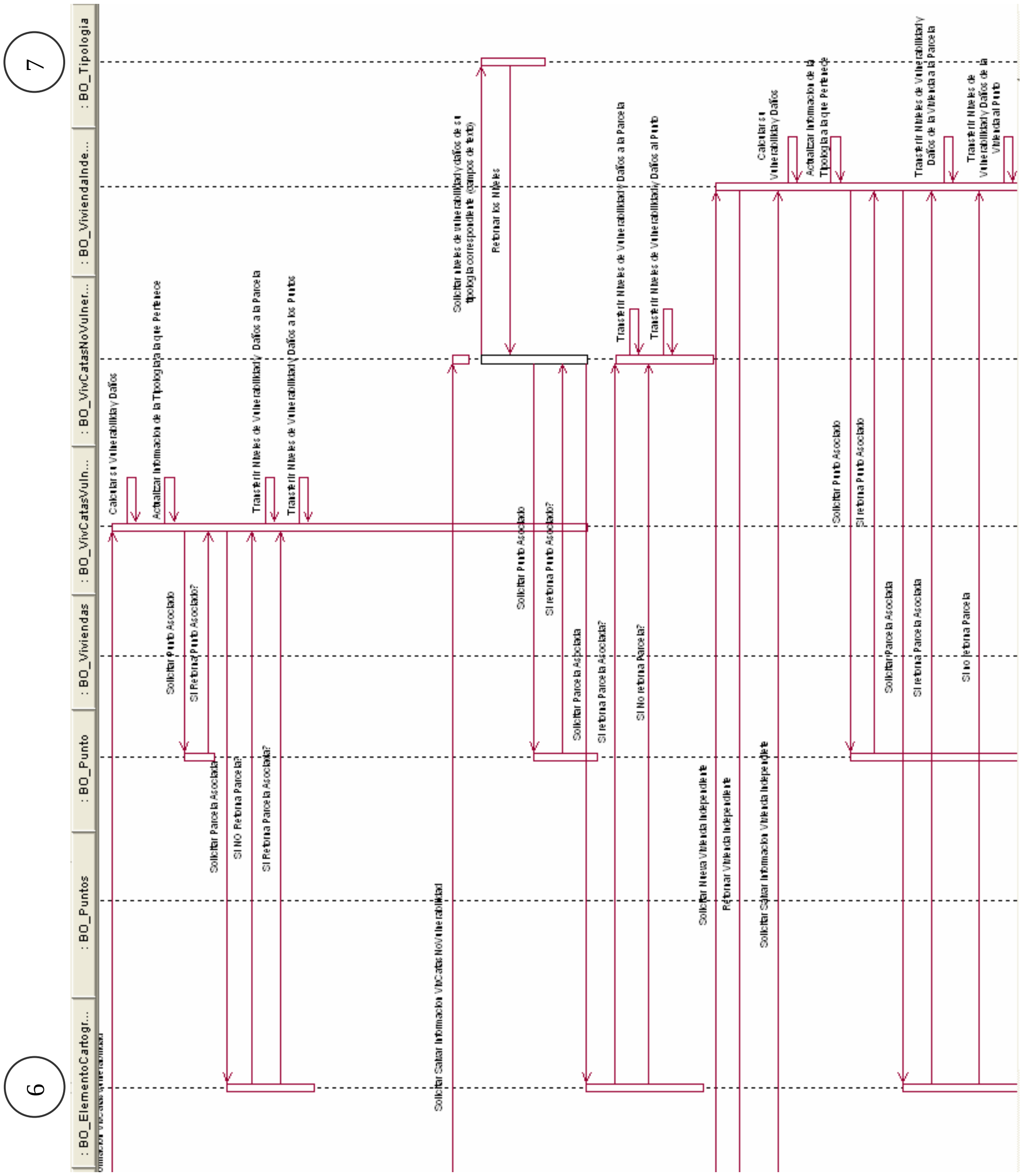

Figura 164. Diagrama de Secuencia SubiendoY PresentandoResultadosVulnerabilidadEnMapas Parte inferior izquierda del diagrama

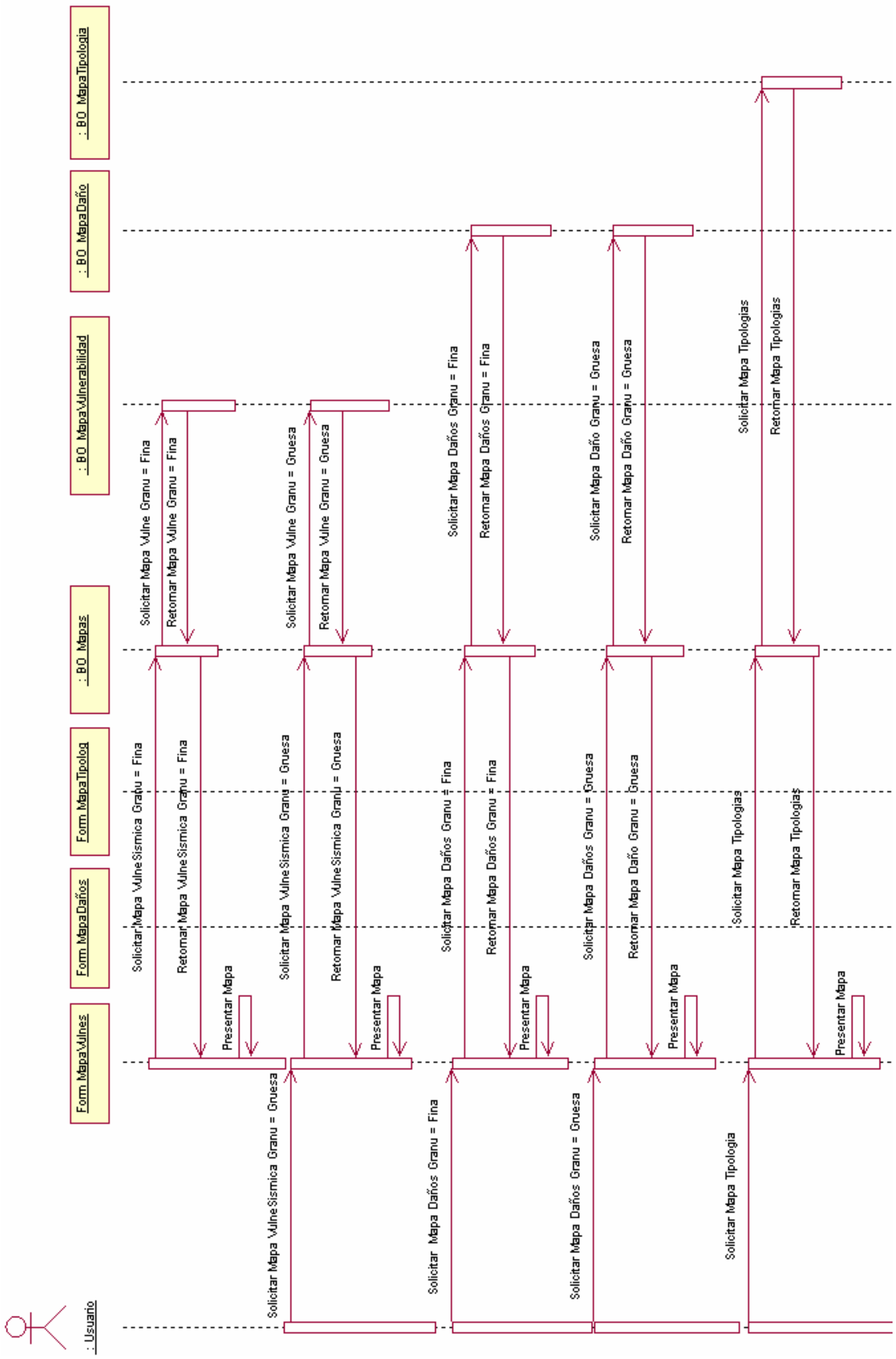

Figura 165. Diagrama de Secuencia IntercambiarTiposResultadosY Granularidad

## A.5 Diagramas de Colaboración

A continuación se presentan todos los diagramas de colaboración resultantes del análisis para *VULNESIS*.

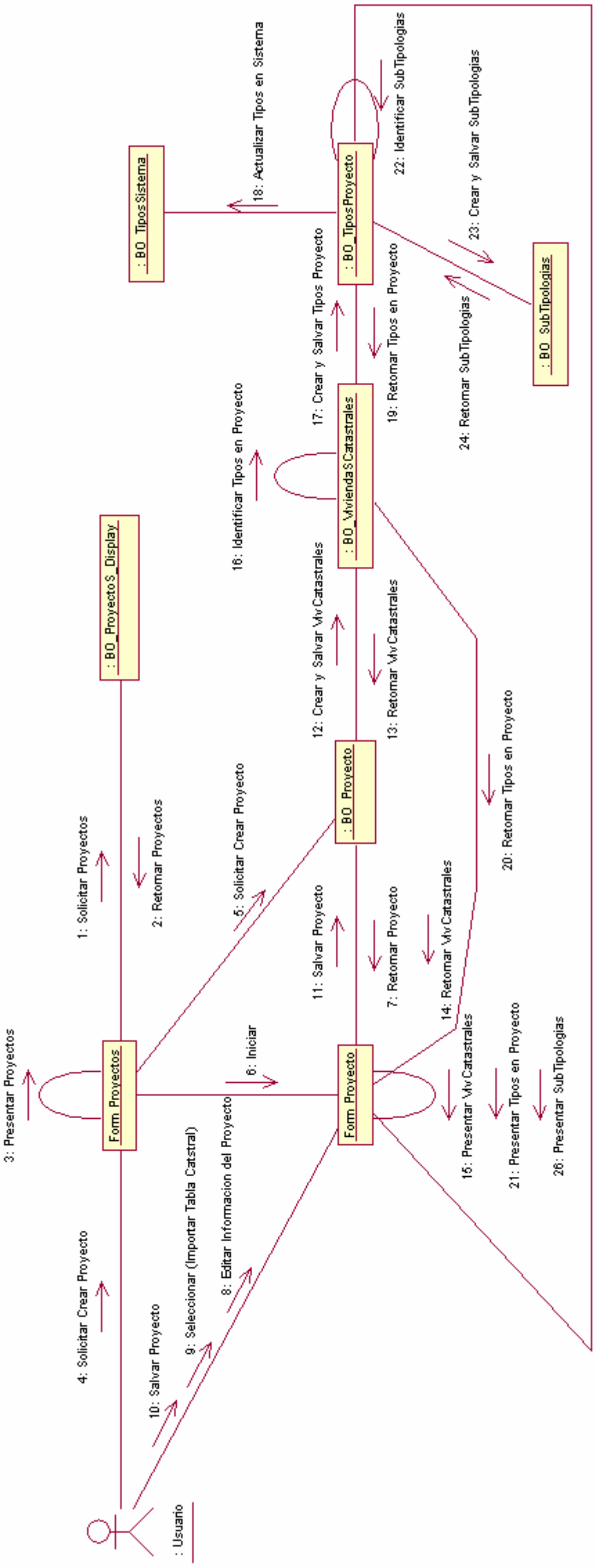

Figura 166. Diagrama de Colaboración CrearProyecto

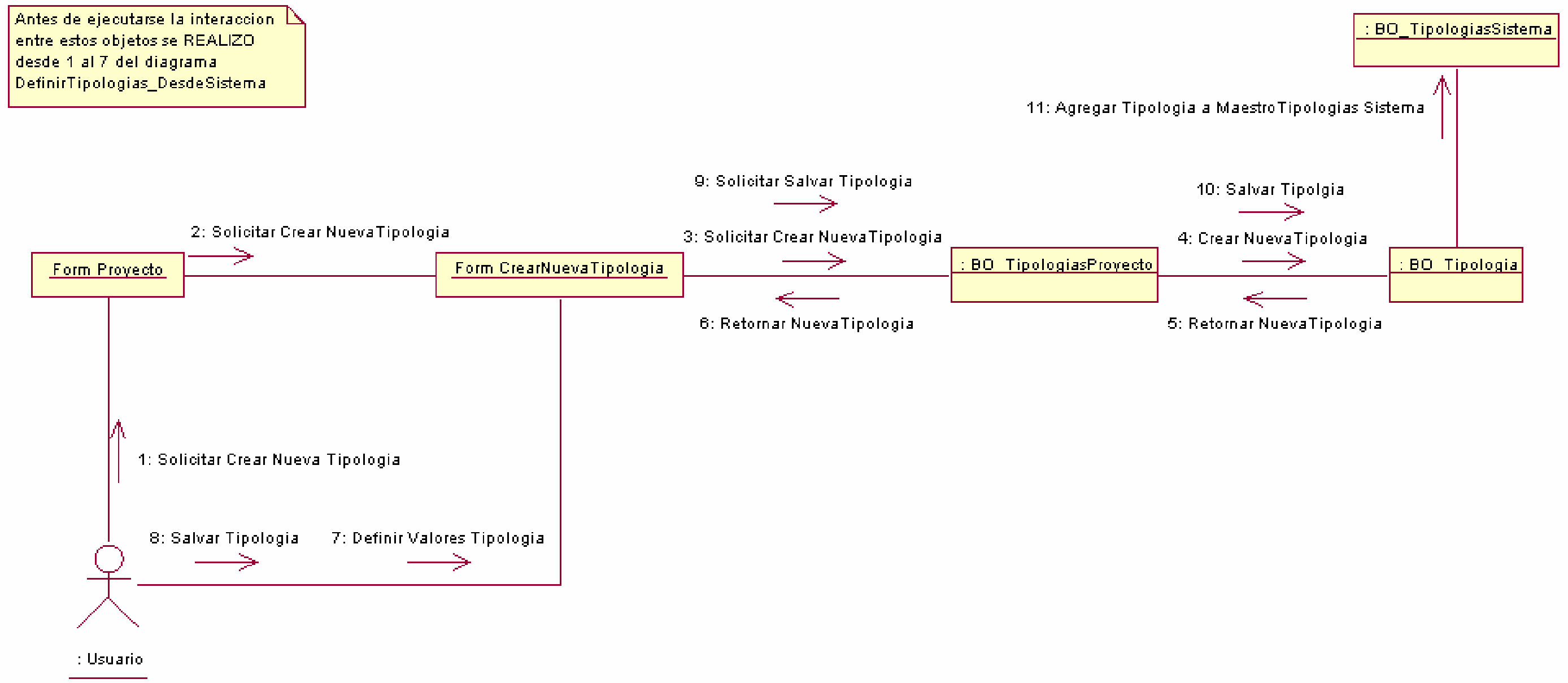

Figura 167. Diagrama de Colaboración Para el Caso de Uso: DefinirTipologias, en el escenario: AgregarNuevea

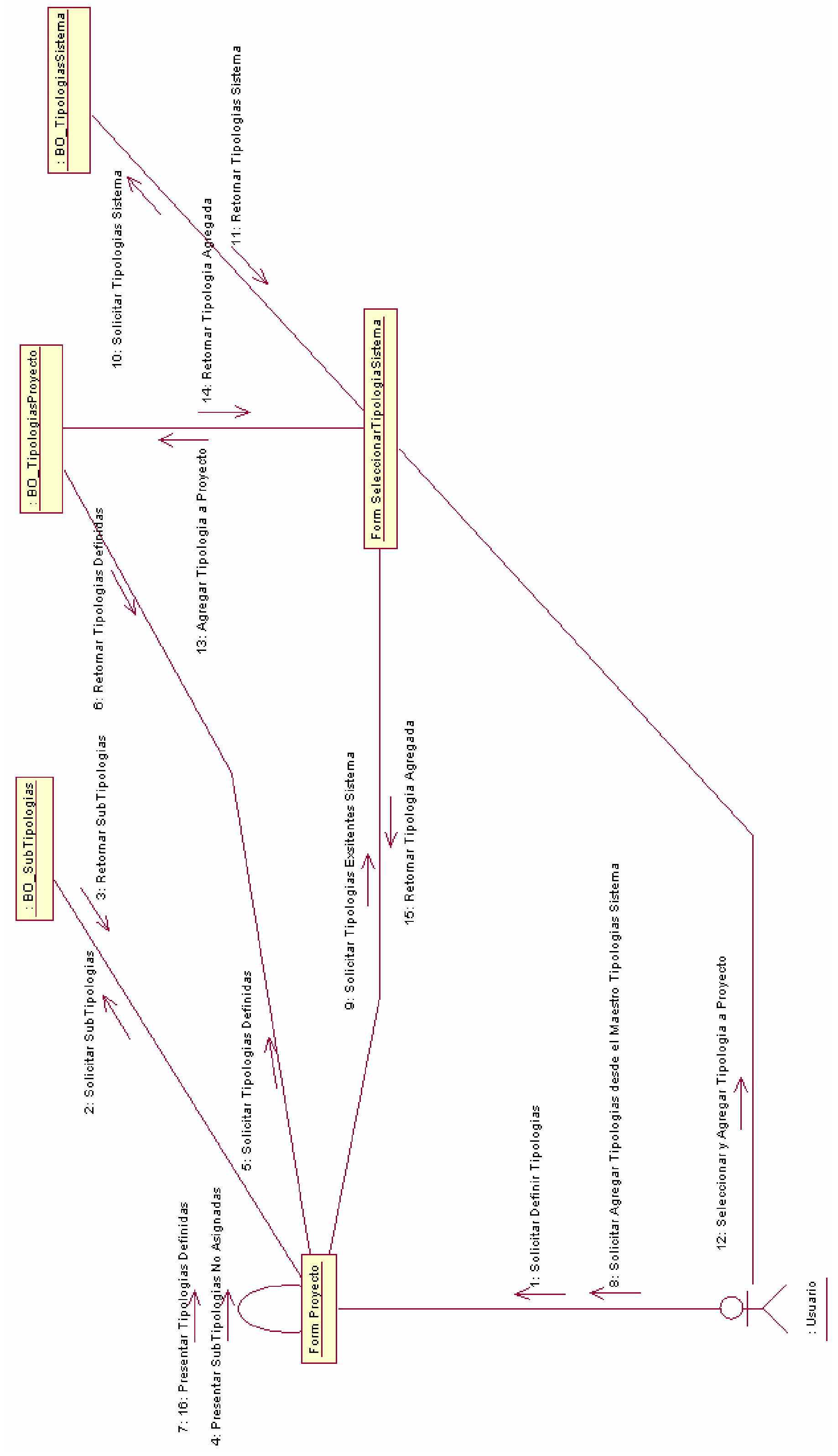

Figura 168. Diagrama de Colaboración Para el Caso de Uso: DefinirTipologías, en el escenario: AgregarTipologías_DesdeExistentesEnSistema

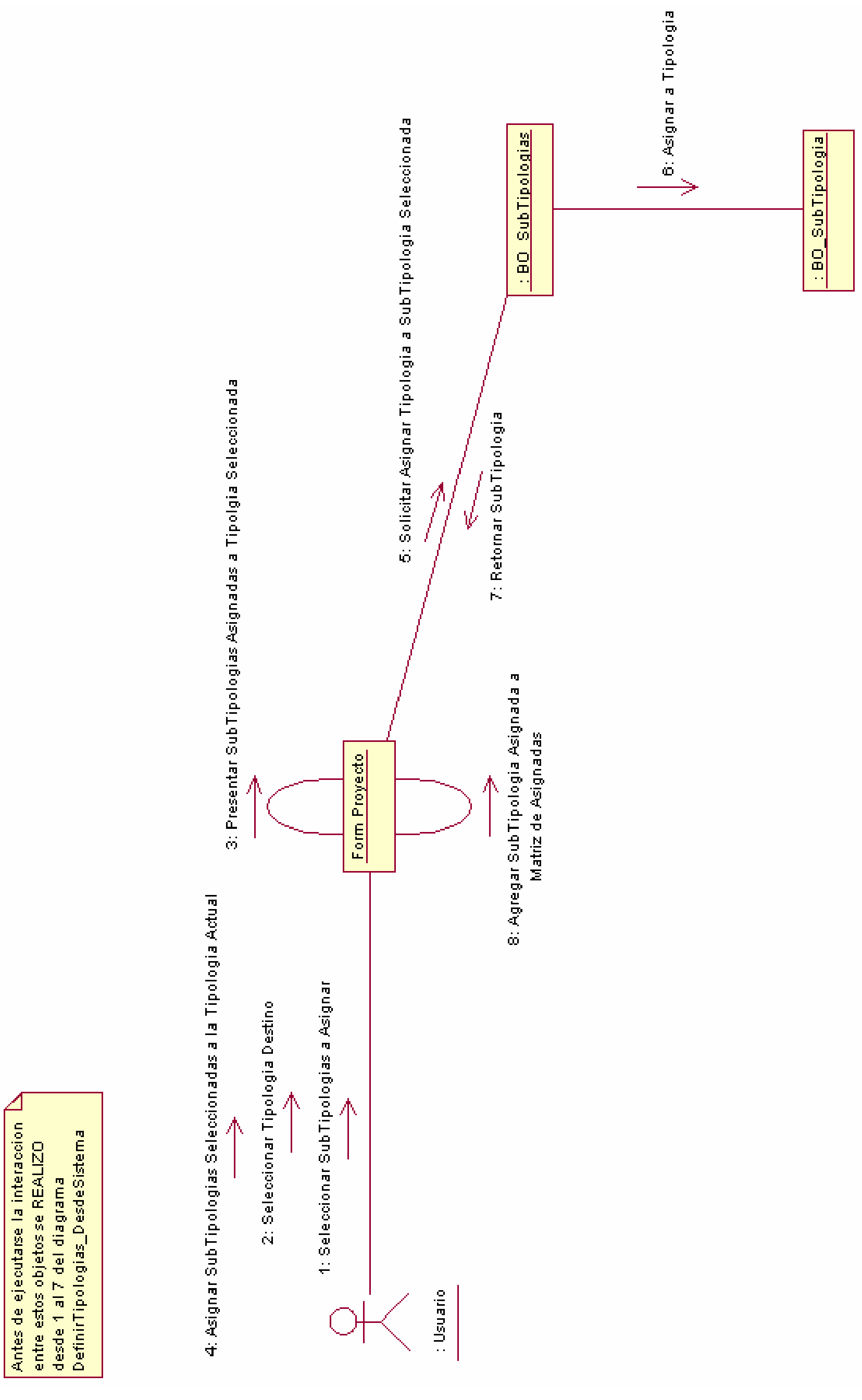

Figura 169. Diagrama de Colaboración Para el Caso de Uso: DefinirTipologías, en el escenario: AsignarSubTipologia_aTipologia

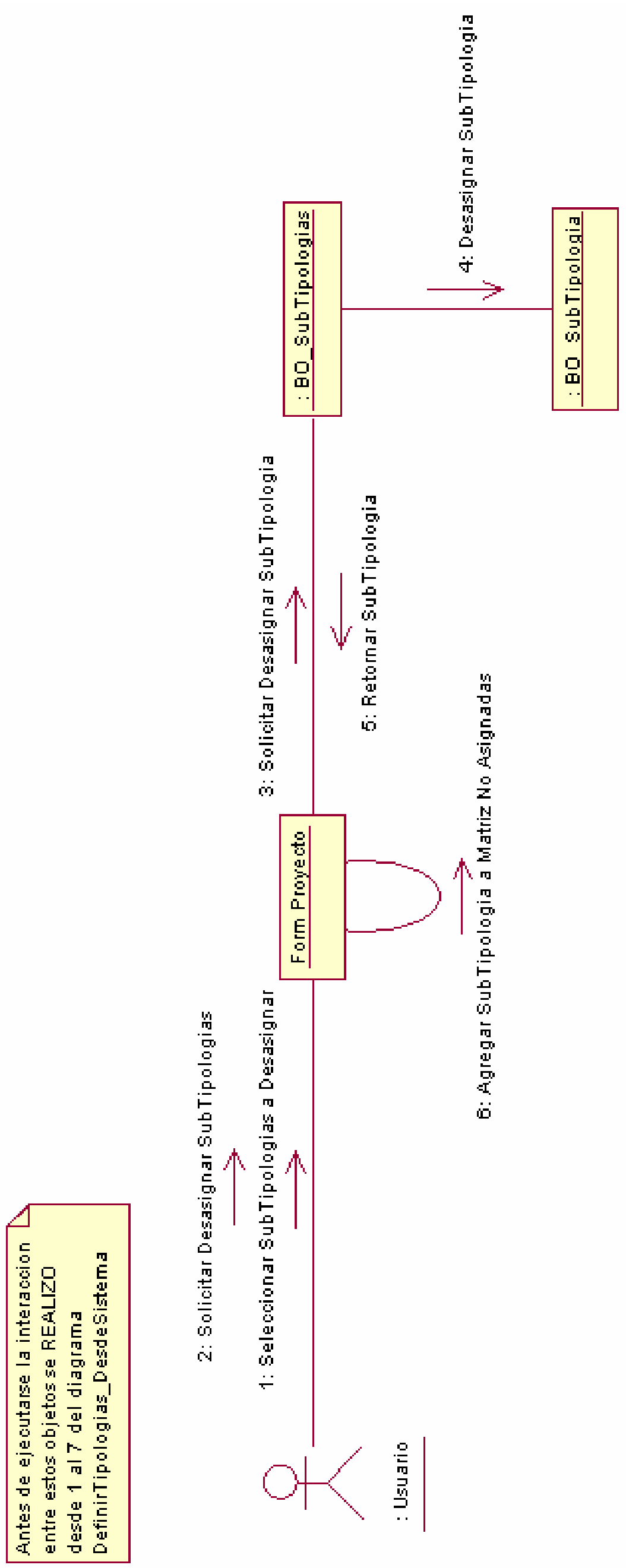

Figura 170. Diagrama de Colaboración Para el Caso de Uso: DefinirTipologias, en el escenario: DesasignarSubTipologias_deTipologia

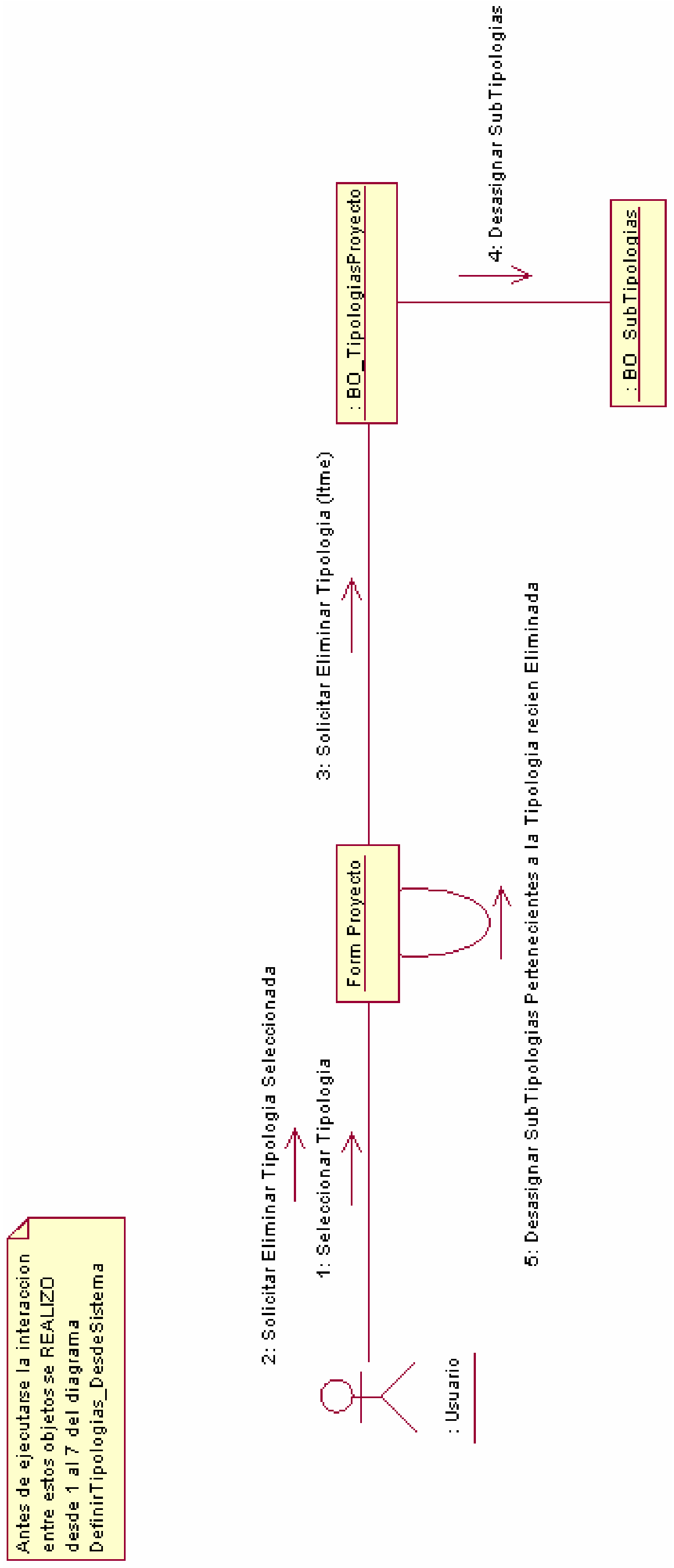

Figura 171. Diagrama de Colaboración Para el Caso de Uso: DefinirTipologias, en el escenario: EliminarTipologiaSeleccionada

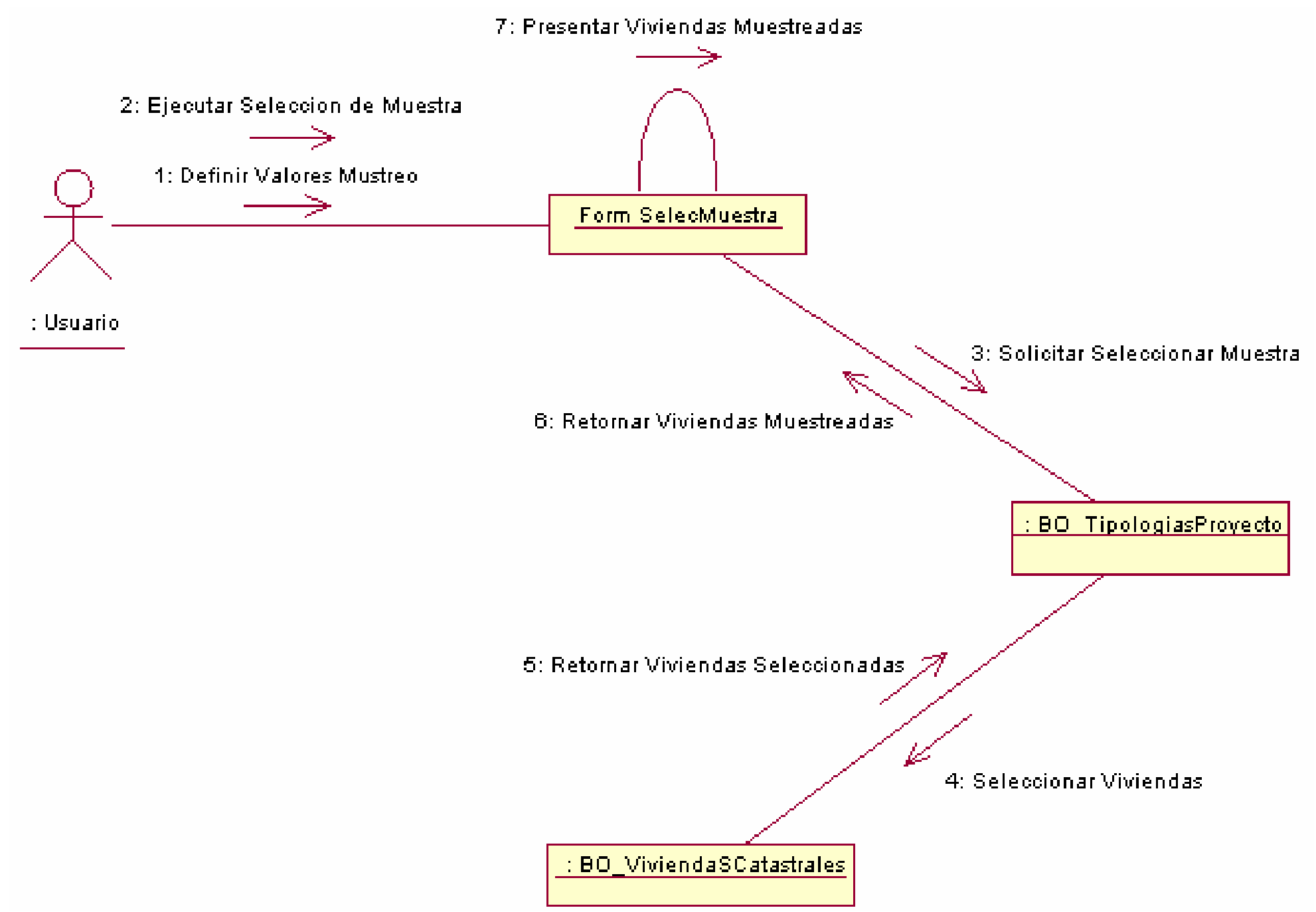

Figura 172. Diagrama de actividad para el Caso de Uso: SeleccionarMuestras

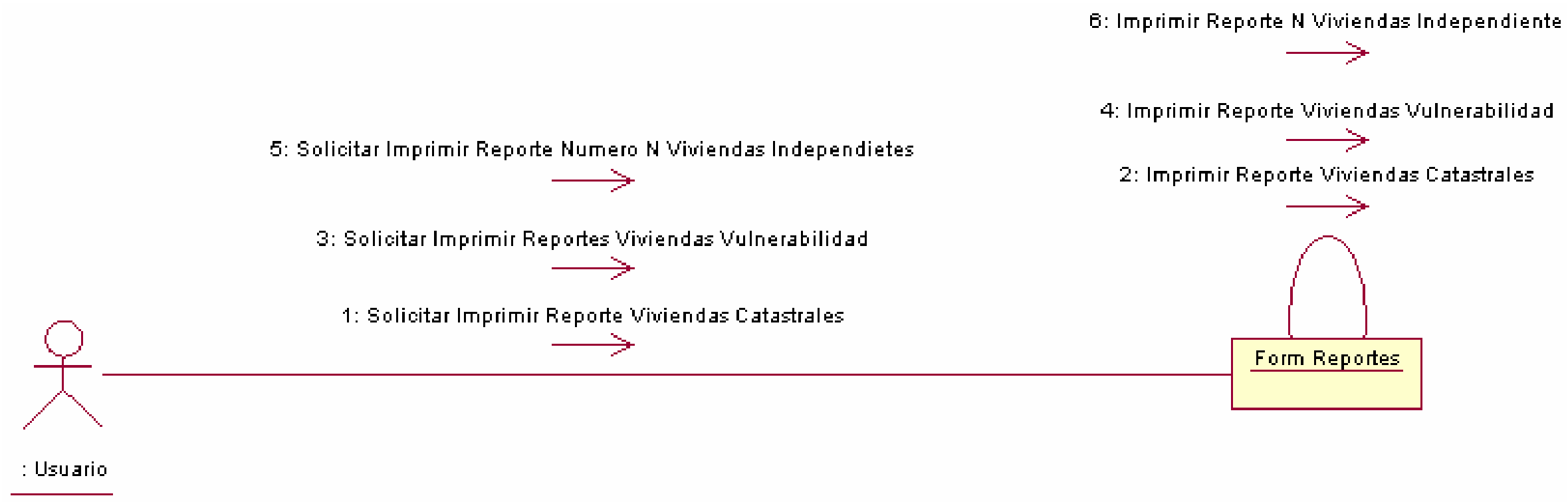

Figura 173. Diagrama de actividad para el Caso de Uso: ParaTrabajoCampo

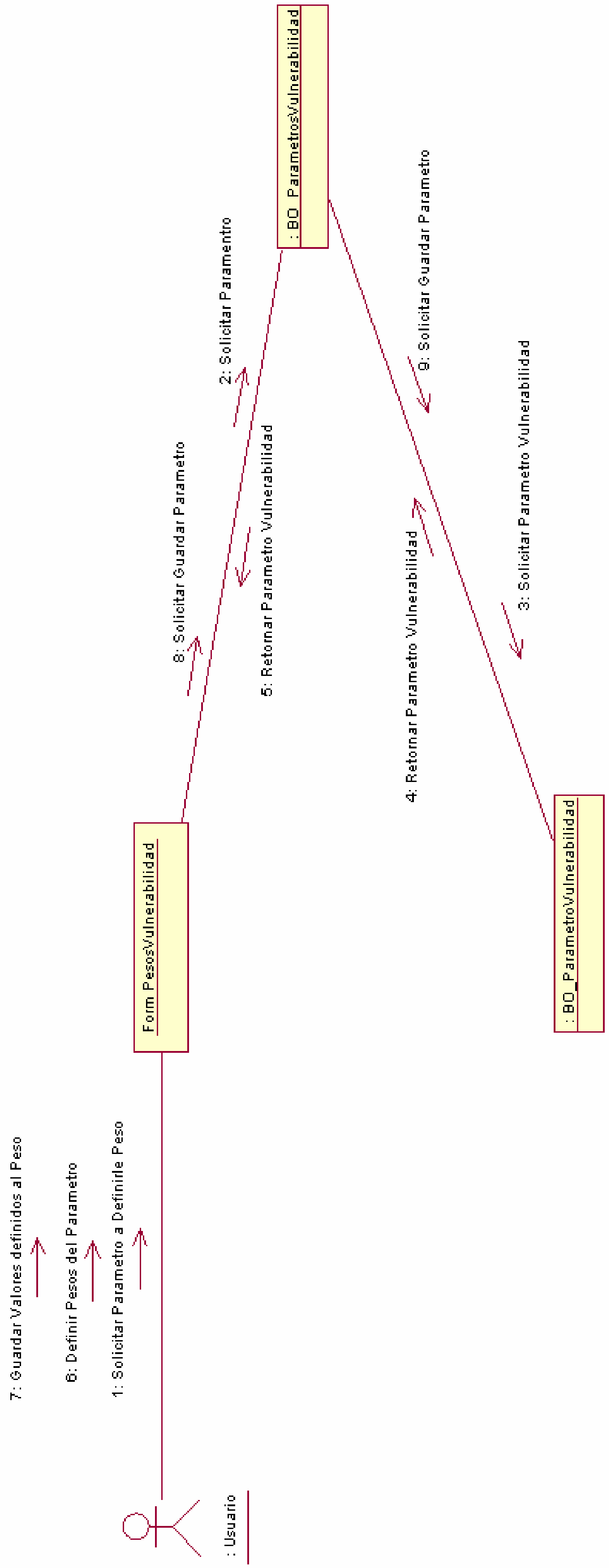

Figura 174. Diagrama de Colaboración Para el Caso de Uso: MientrasTrabajoCampo, en el escenario: ParametrosVulnerabilidad

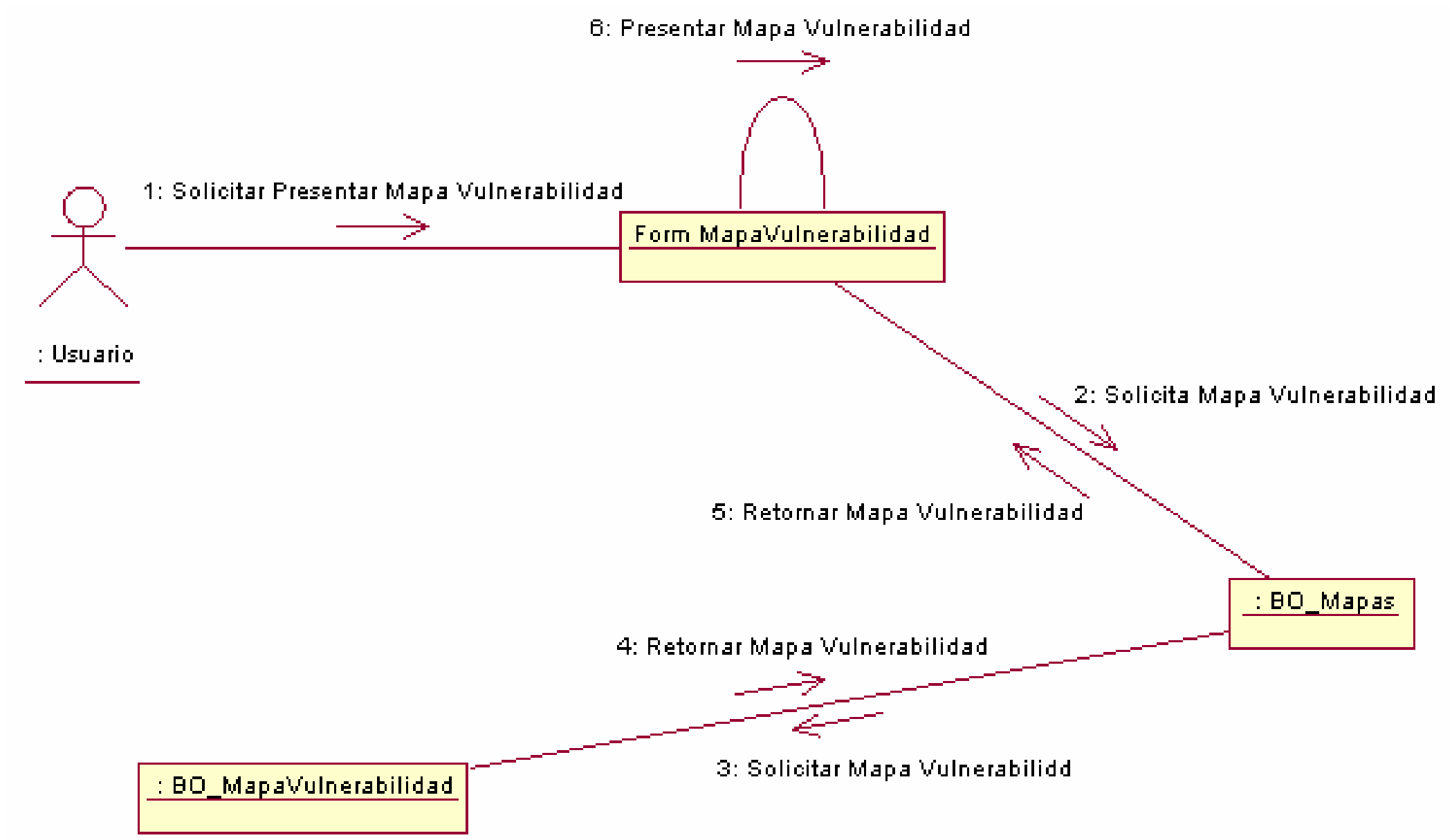

Figura 175. Diagrama de actividad para el Caso de Uso: SubiendoY_PresentandoResultadosVulnerabilidadViviendasEnMapas, En el escenario: INICIO

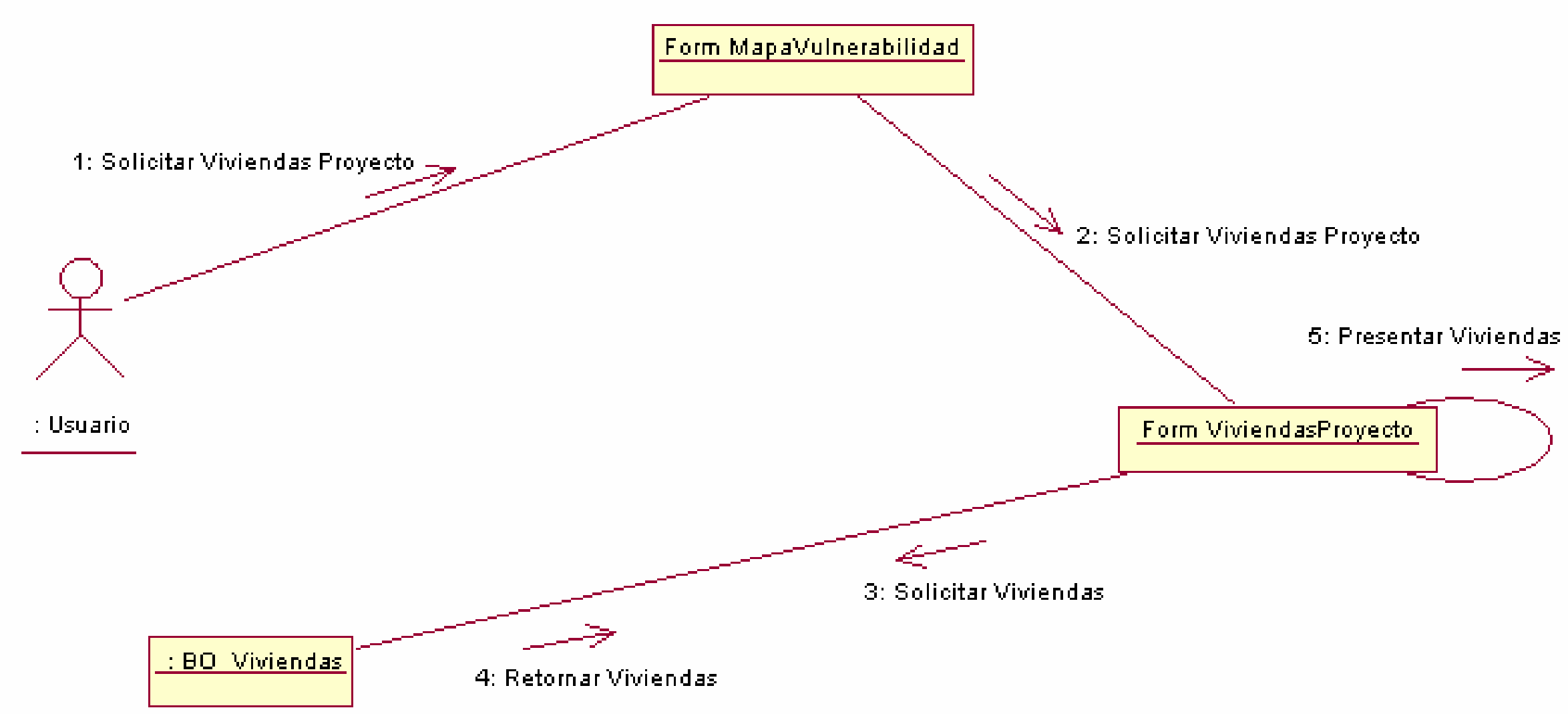

Figura 176. Diagrama de actividad para el Caso de Uso: SubiendoY_PresentandoResultadosVulnerabilidadViviendasEnMapas, En el escenario: PresentarViviendas

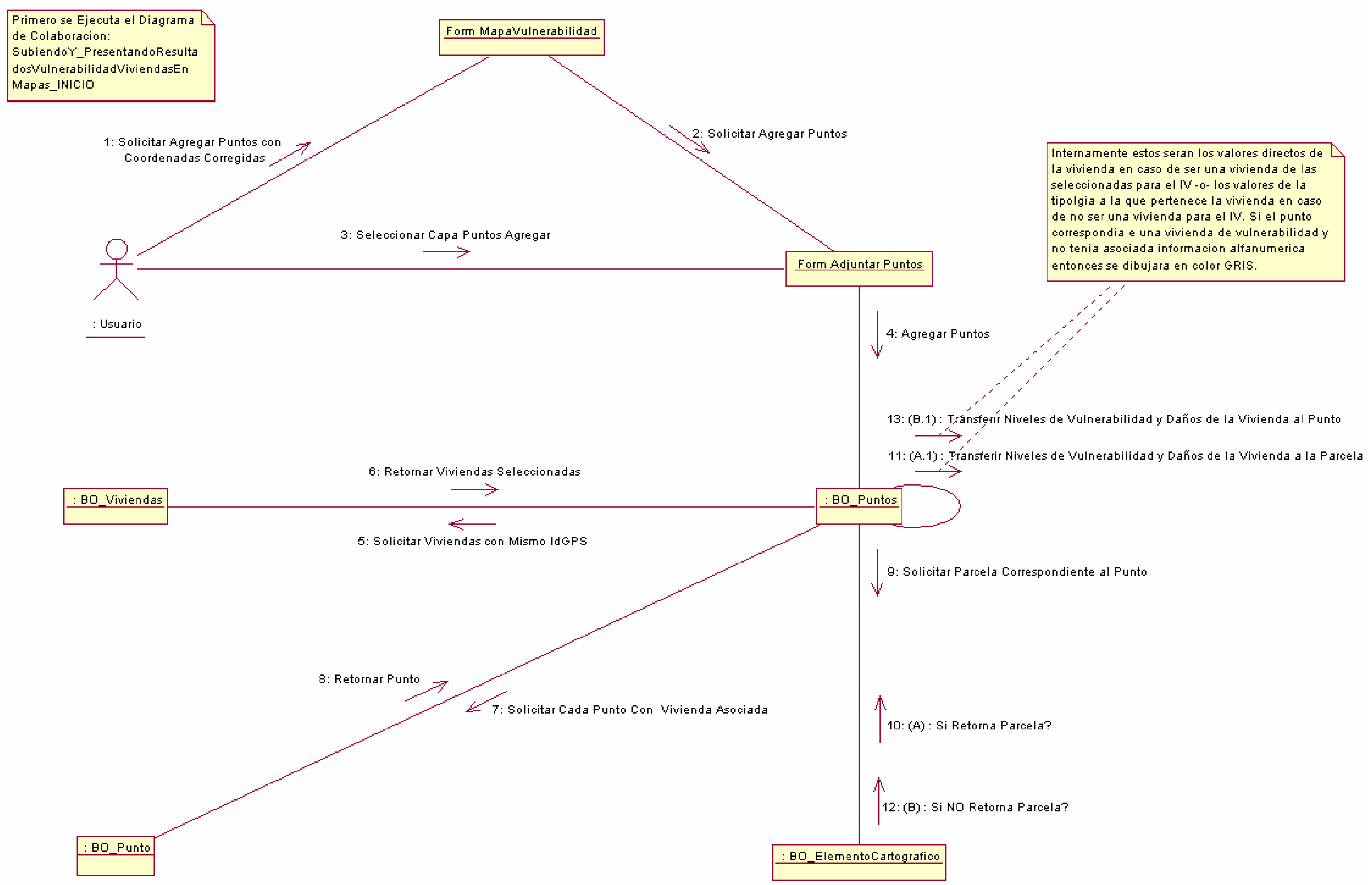

Figura 177. Diagrama de actividad para el Caso de Uso: SubiendoY_PresentandoResultadosVulnerabilidadViviendasEnMapas, En el escenario: AgregarPuntos

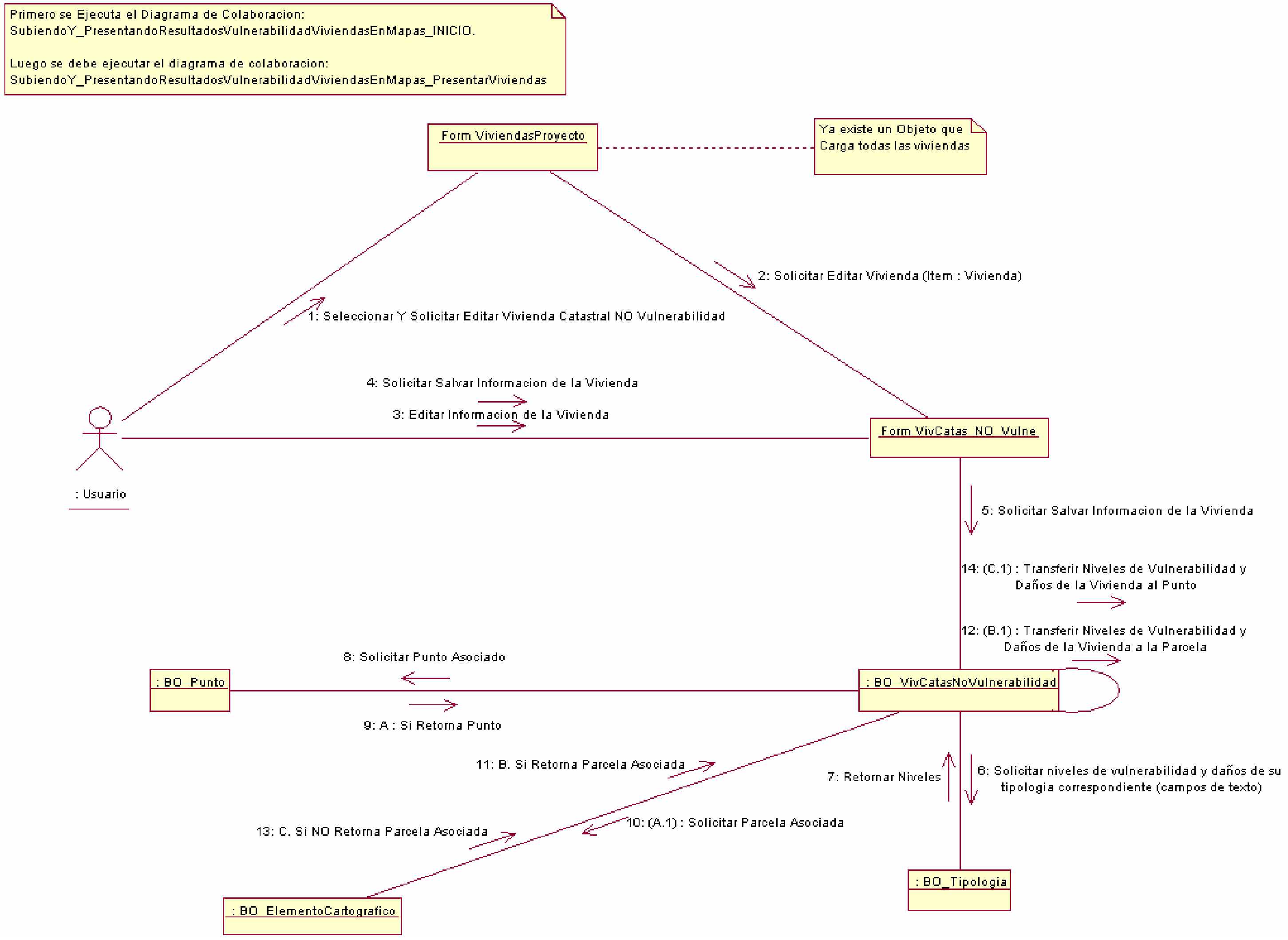

Figura 178. Diagrama de actividad para el Caso de Uso: SubiendoY_PresentandoResultadosVulnerabilidadViviendasEnMapas, En el escenario: EditarViviendaCatastralNoVulnerabilidad

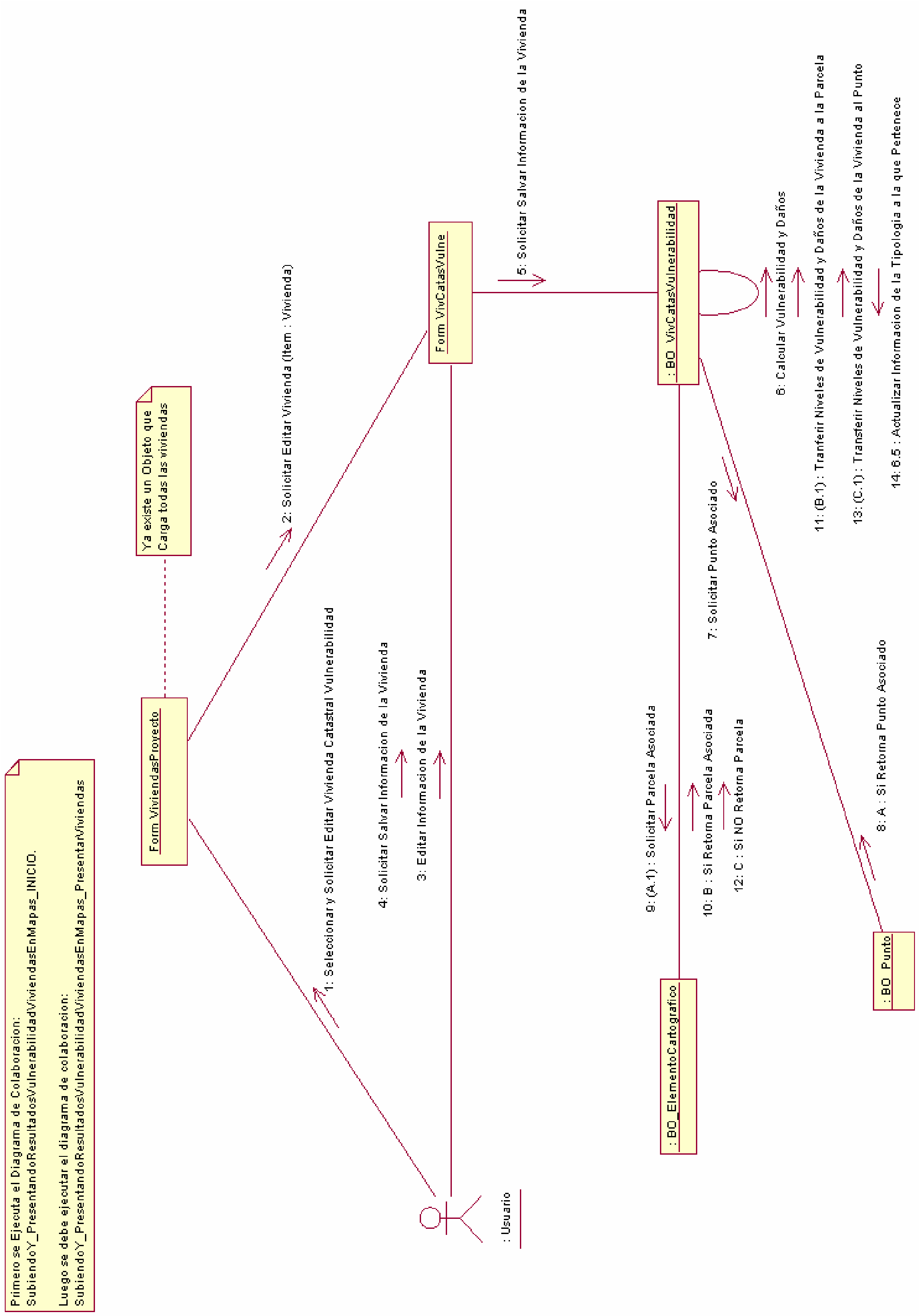

Figura 179. Diagrama de actividad para el Caso de Uso: SubiendoY_PresentandoResultadosVulnerabilidadViviendasEnMapas, En el escenario: EditarViviendaCatastralVulnerabilidad

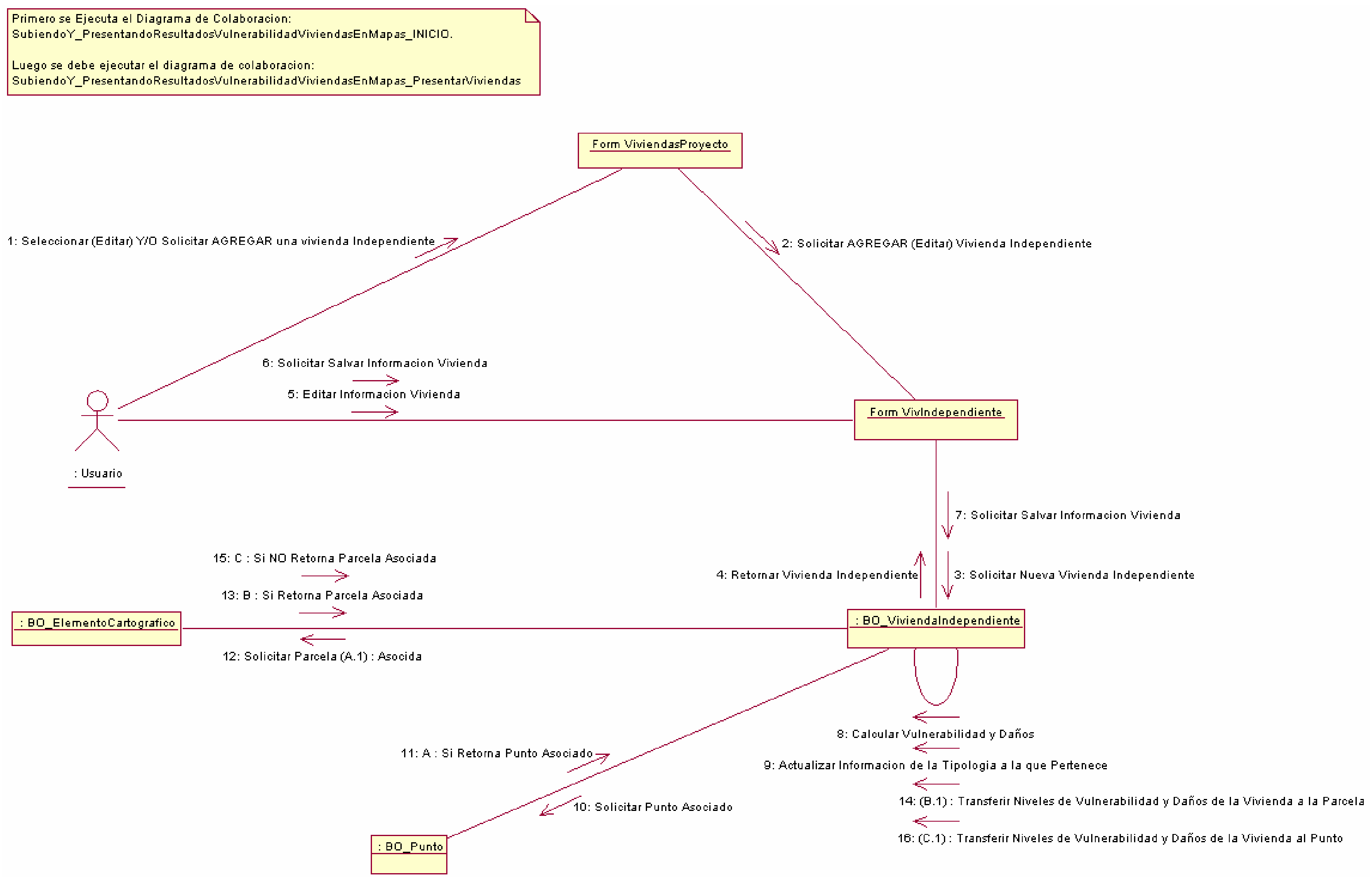

Figura 180. Diagrama de actividad para el Caso de Uso: SubiendoY_PresentandoResultadosVulnerabilidadViviendasEnMapas, En el escenario: EditarViviendaIndependiente

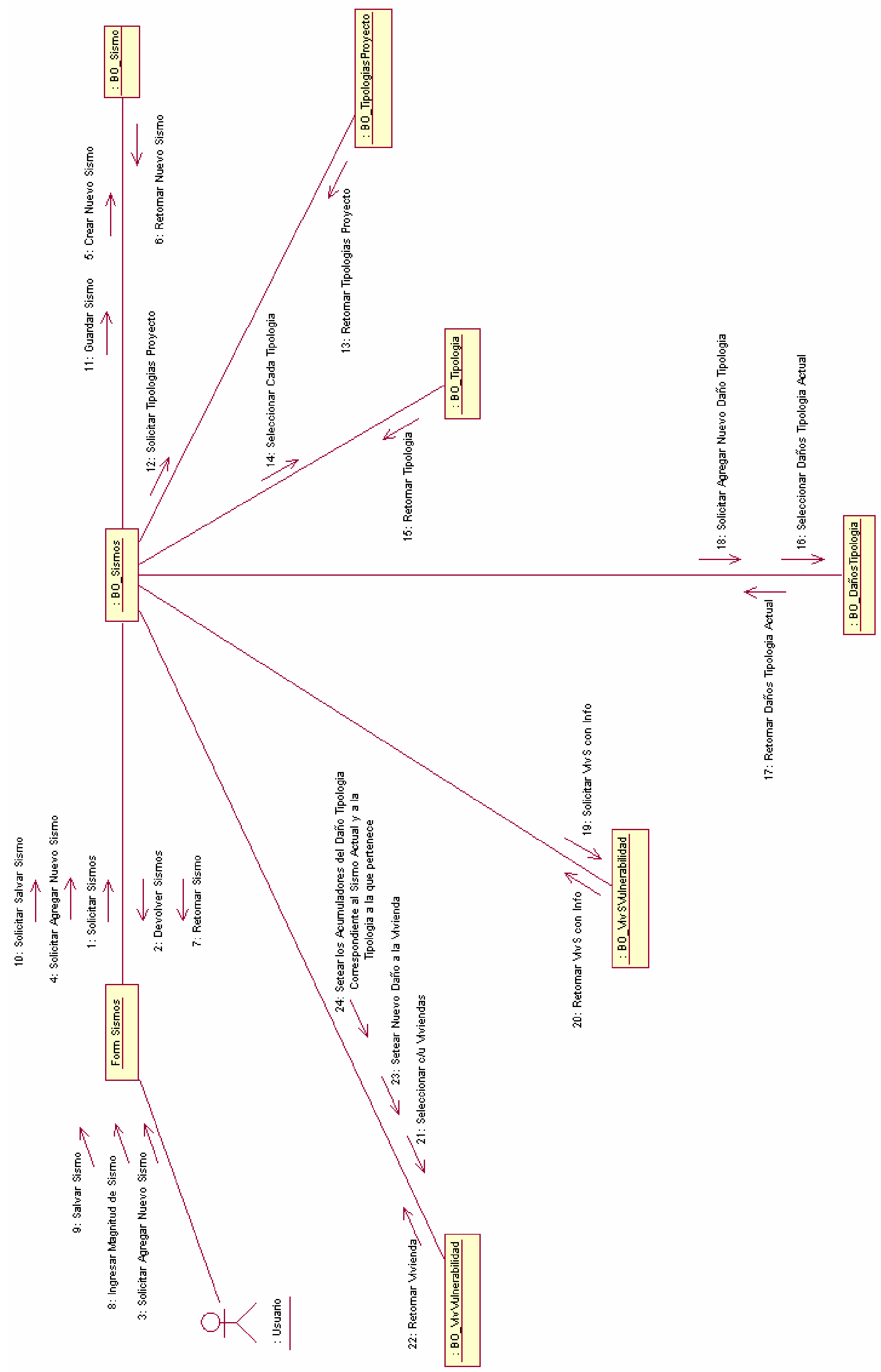

Figura 181. Diagrama de actividad para el Caso de Uso: DefinirSismos_ParaRepresentacionEscenariosDaños

# A.6 Diagramas de Clases

## A.6.1 Objetos Potenciales
## "Con representación simple de objetos de negocio"

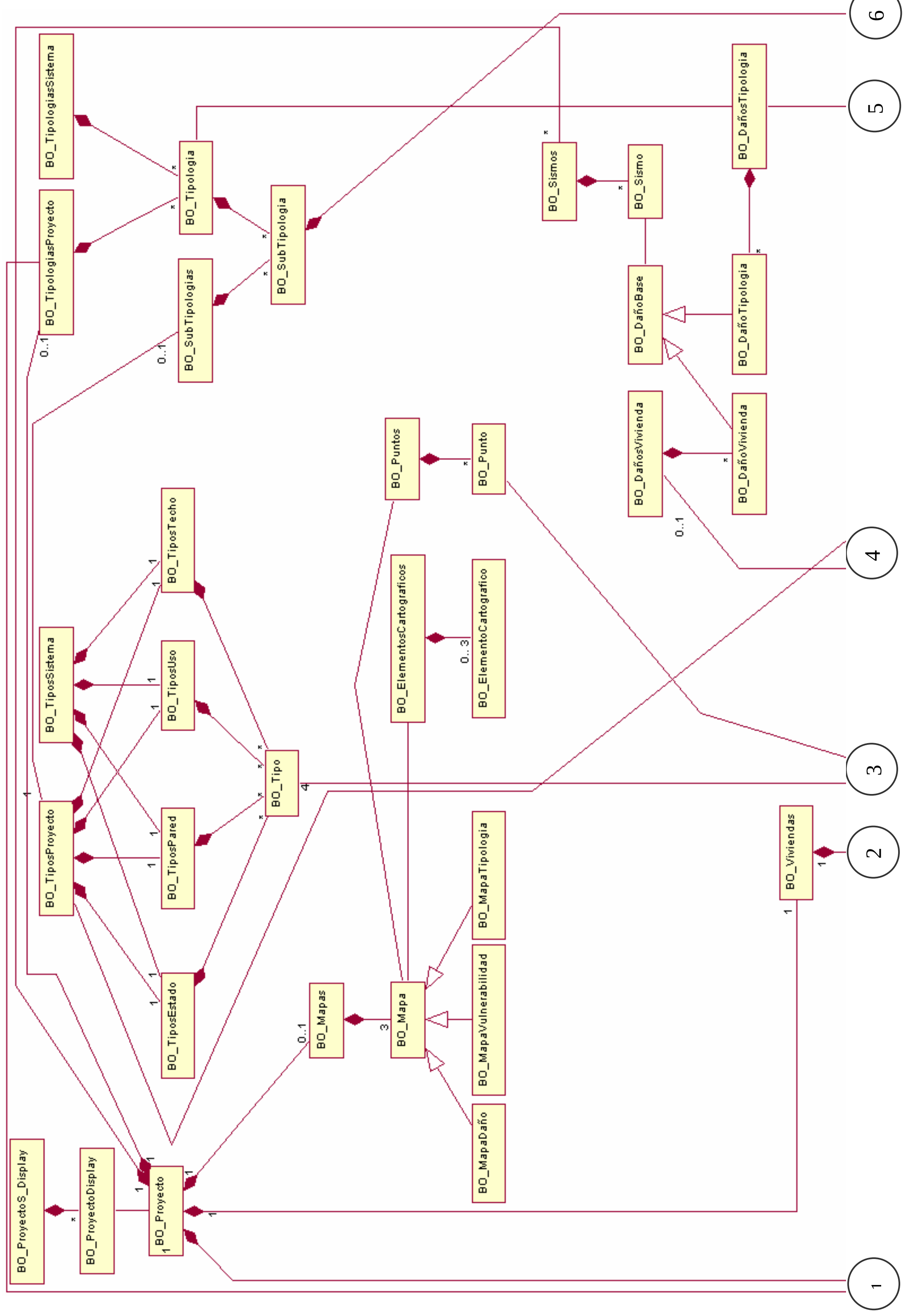

Figura 182. Diagrama de Clases con representación simple (Objetos de Negocio) Parte superior del diagrama (1/2)

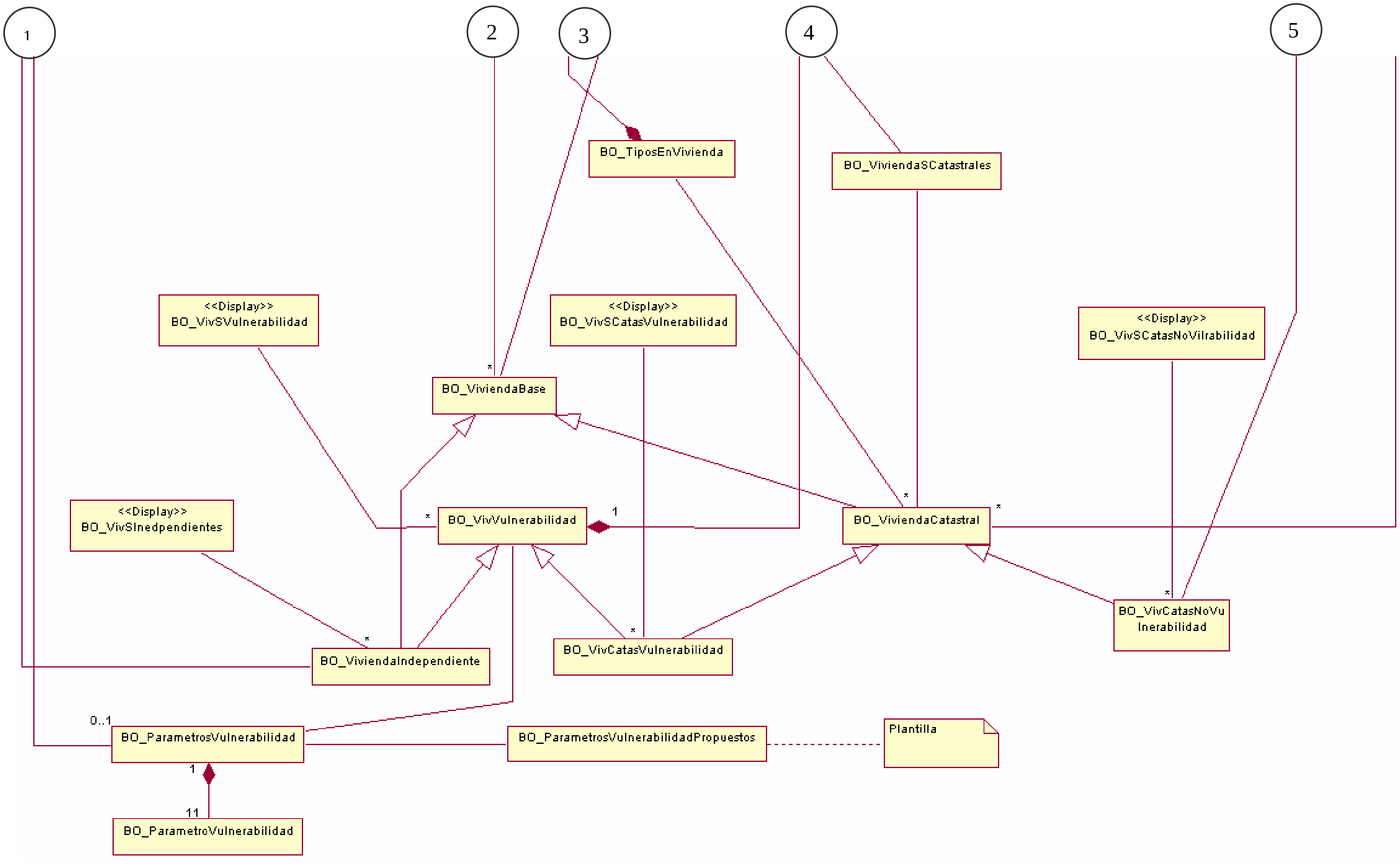

Figura 183. Diagrama de Clases con representación simple (Objetos de Negocio) Parte inferior del diagrama (2/2)

### A.6.2 Clases Final (Clases implementadas) “Con representación simple”

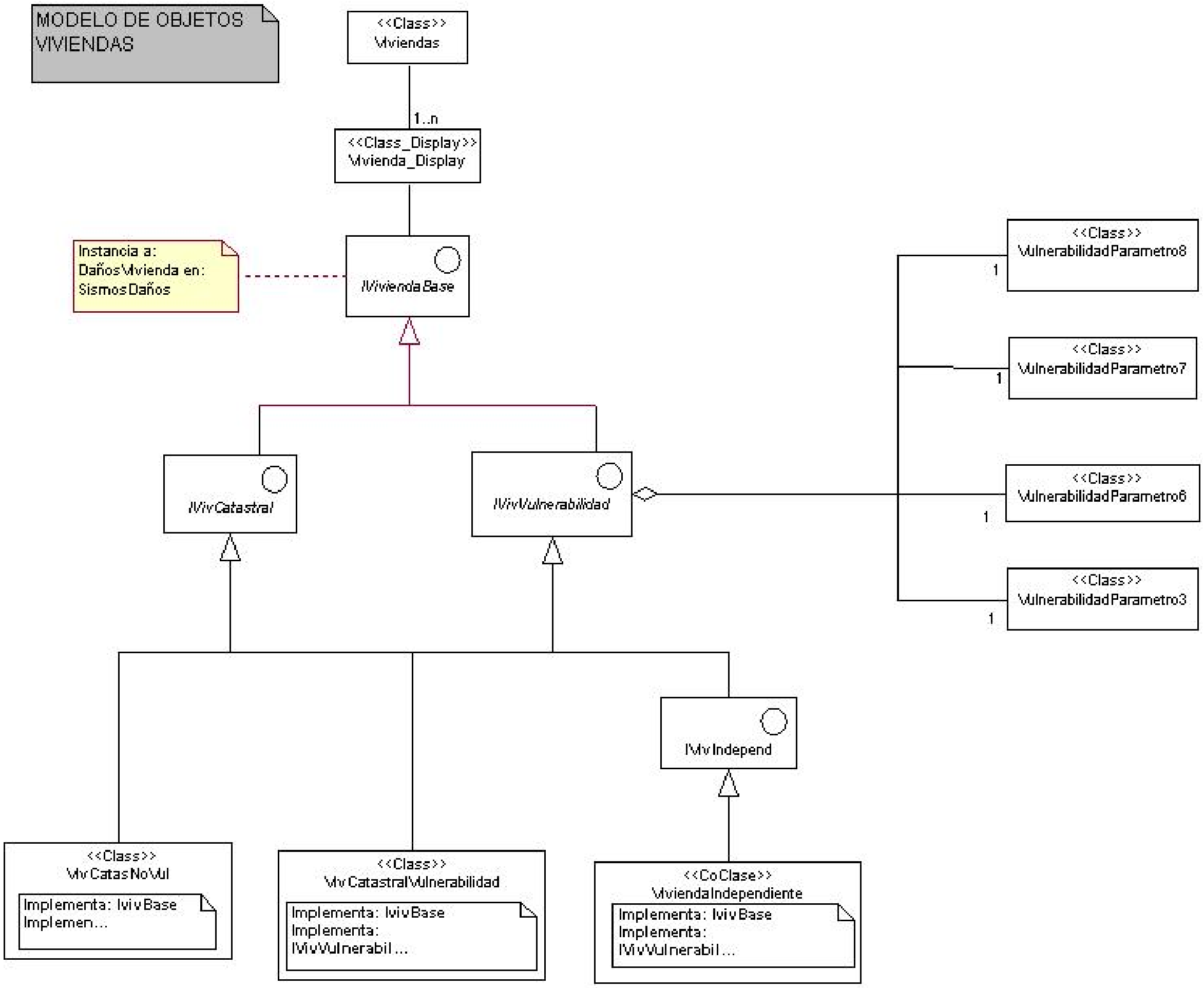

Figura 184. Diagrama de lógica del negocio, sub-diagrama de clases VIVIENDAS

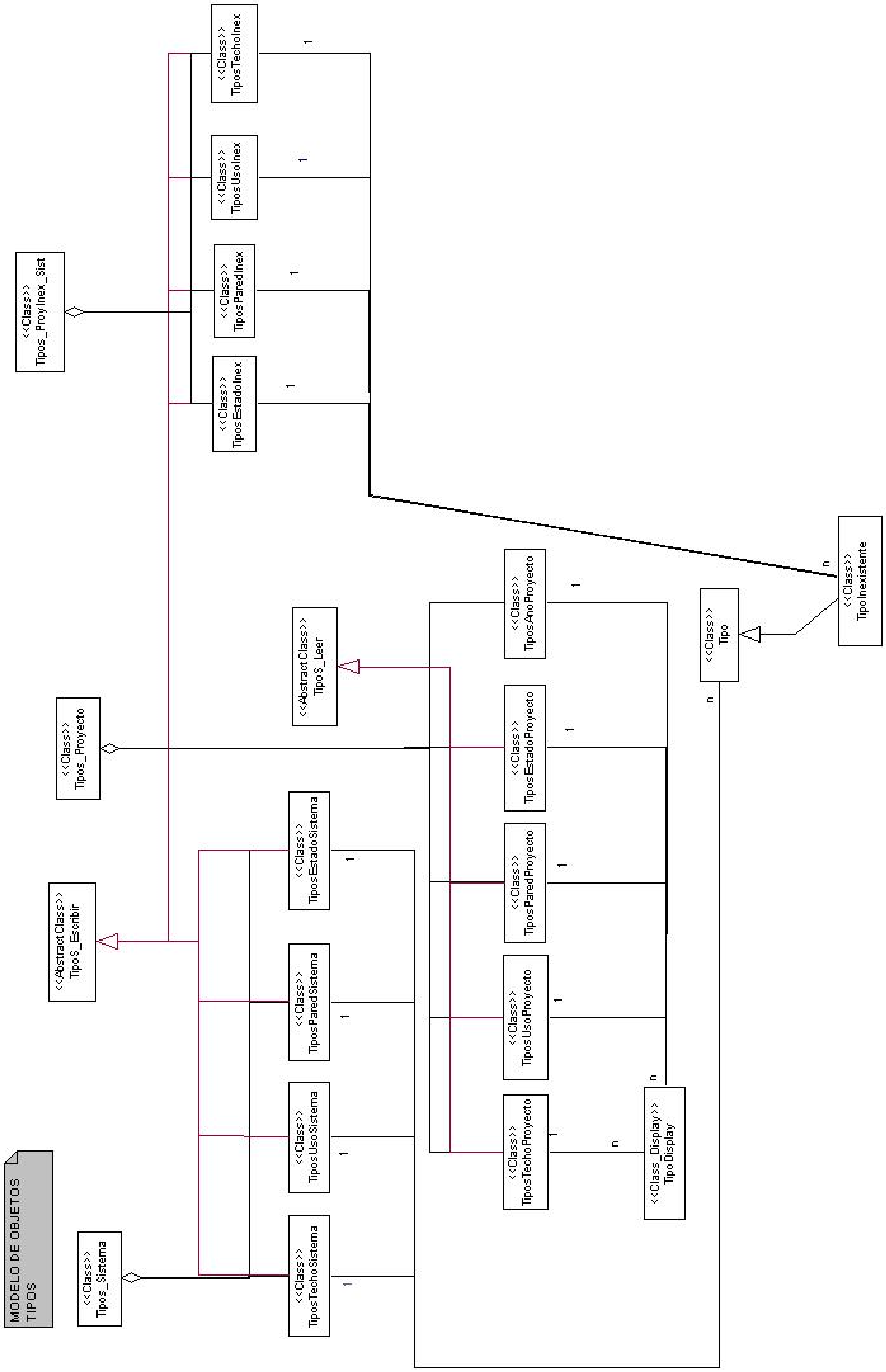

Figura 185. Diagrama de lógica del negocio, sub-diagrama de clases TIPOS

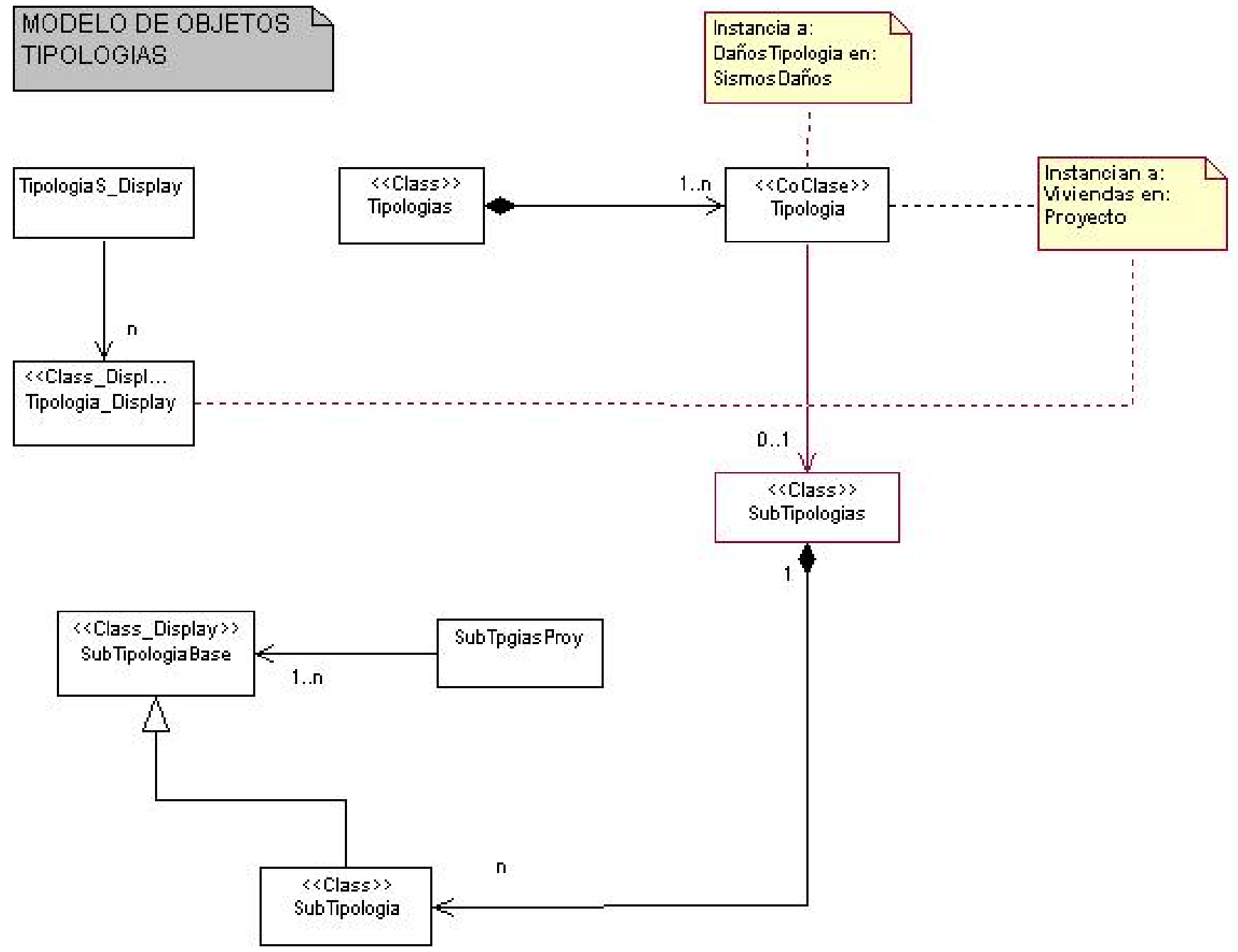

Figura 186. Diagrama de lógica del negocio, sub-diagrama de clases TIPOLOGIAS

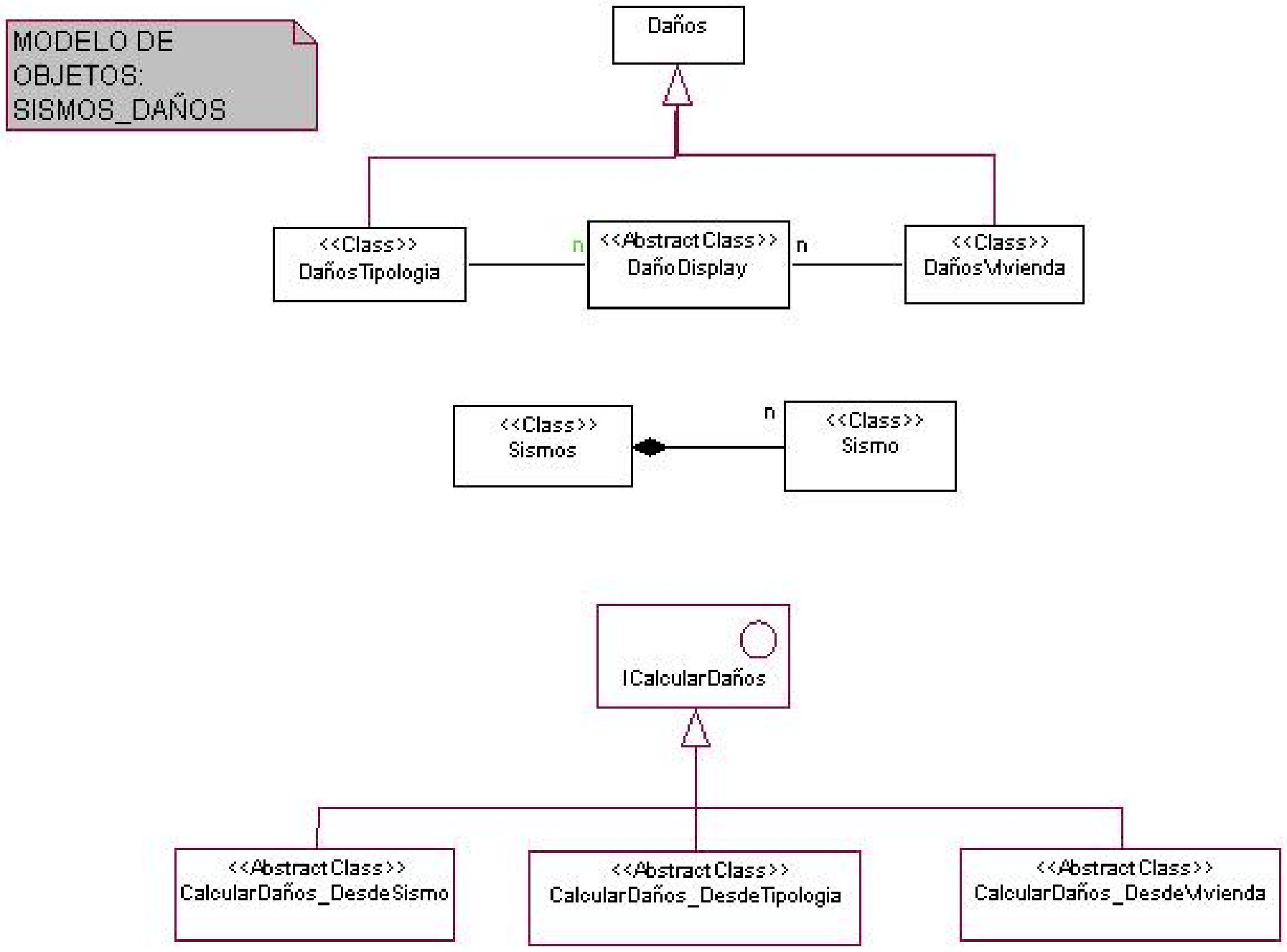

Figura 187. Diagrama de lógica del negocio, sub-diagrama de clases SISMOS_DAÑOS

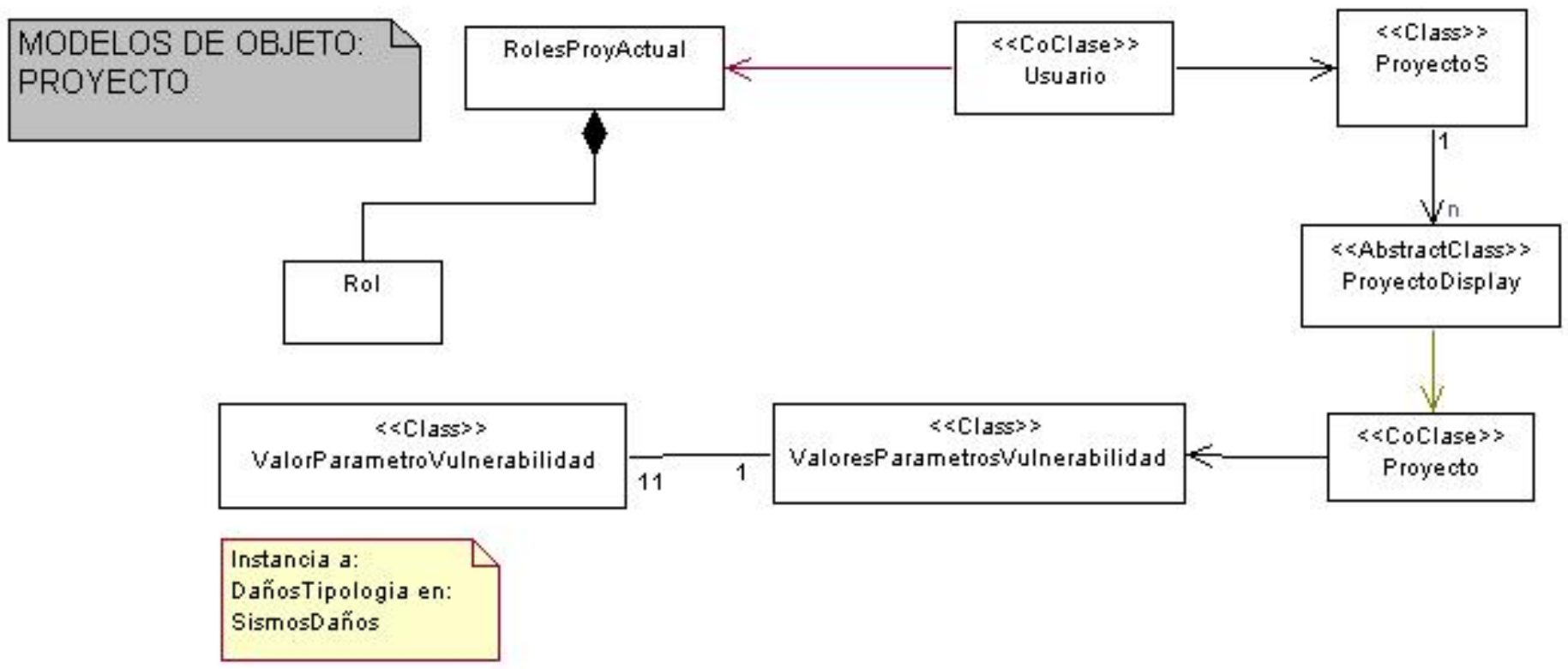

Figura 188. Diagrama de lógica del negocio, sub-diagrama de clases PROYECTO

## A.7 Diagramas de Estado

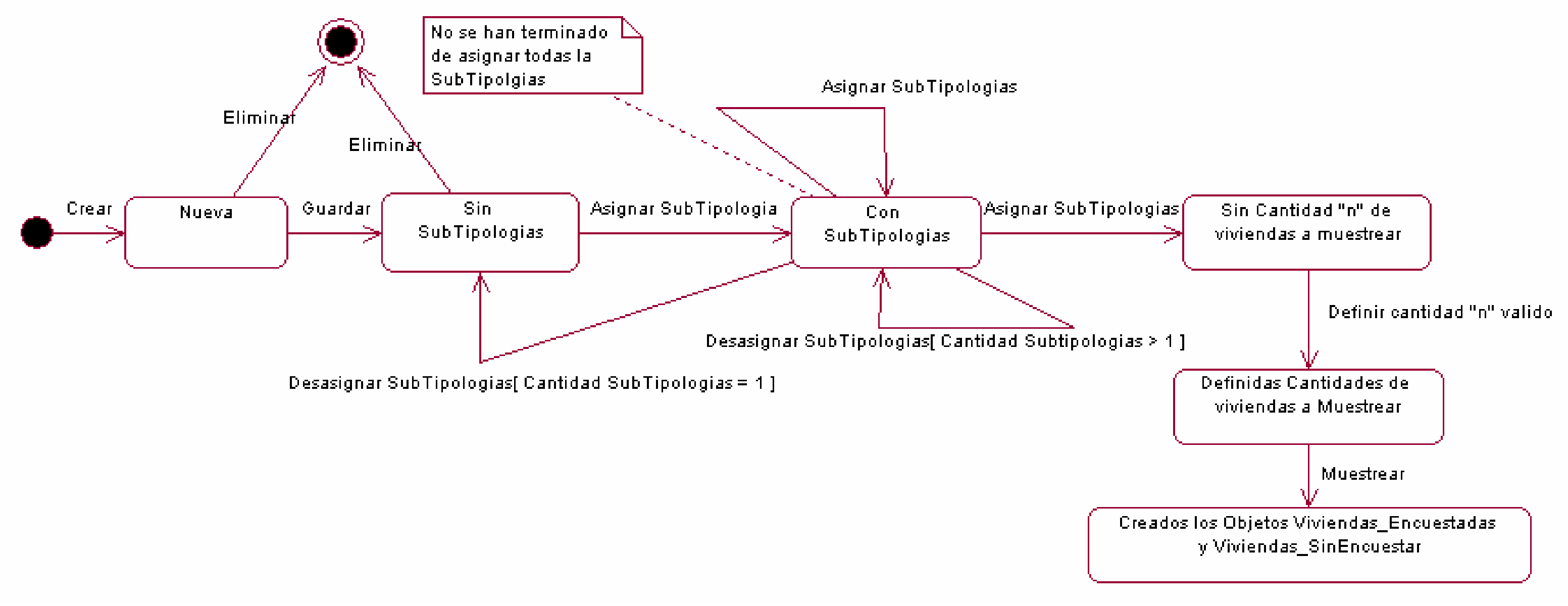

Figura 189. Diagrama de estado para la clase Tipología.

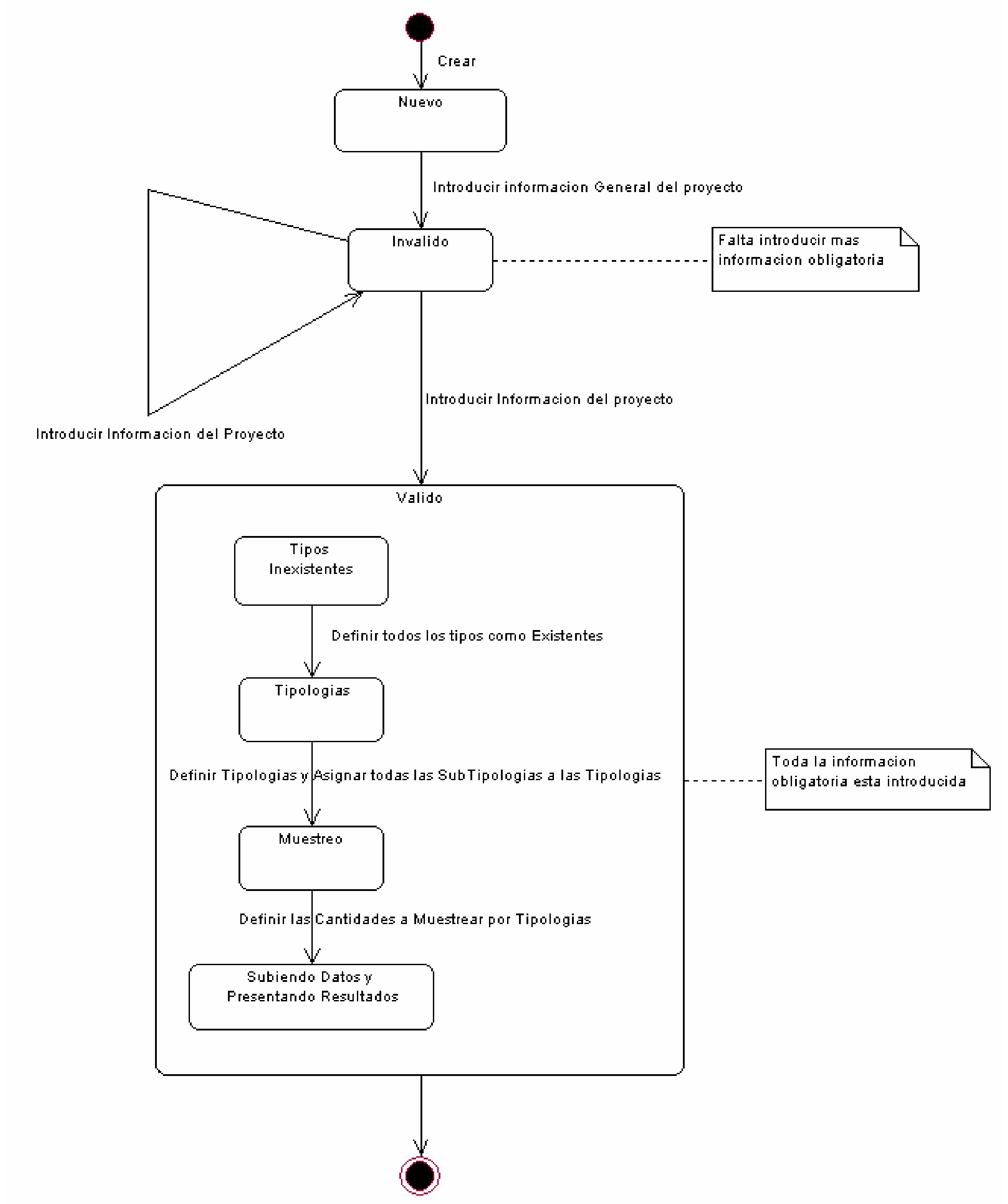

Figura 190. Diagrama de estado para la clase Proyecto.

## A.8 Diagramas de actividad

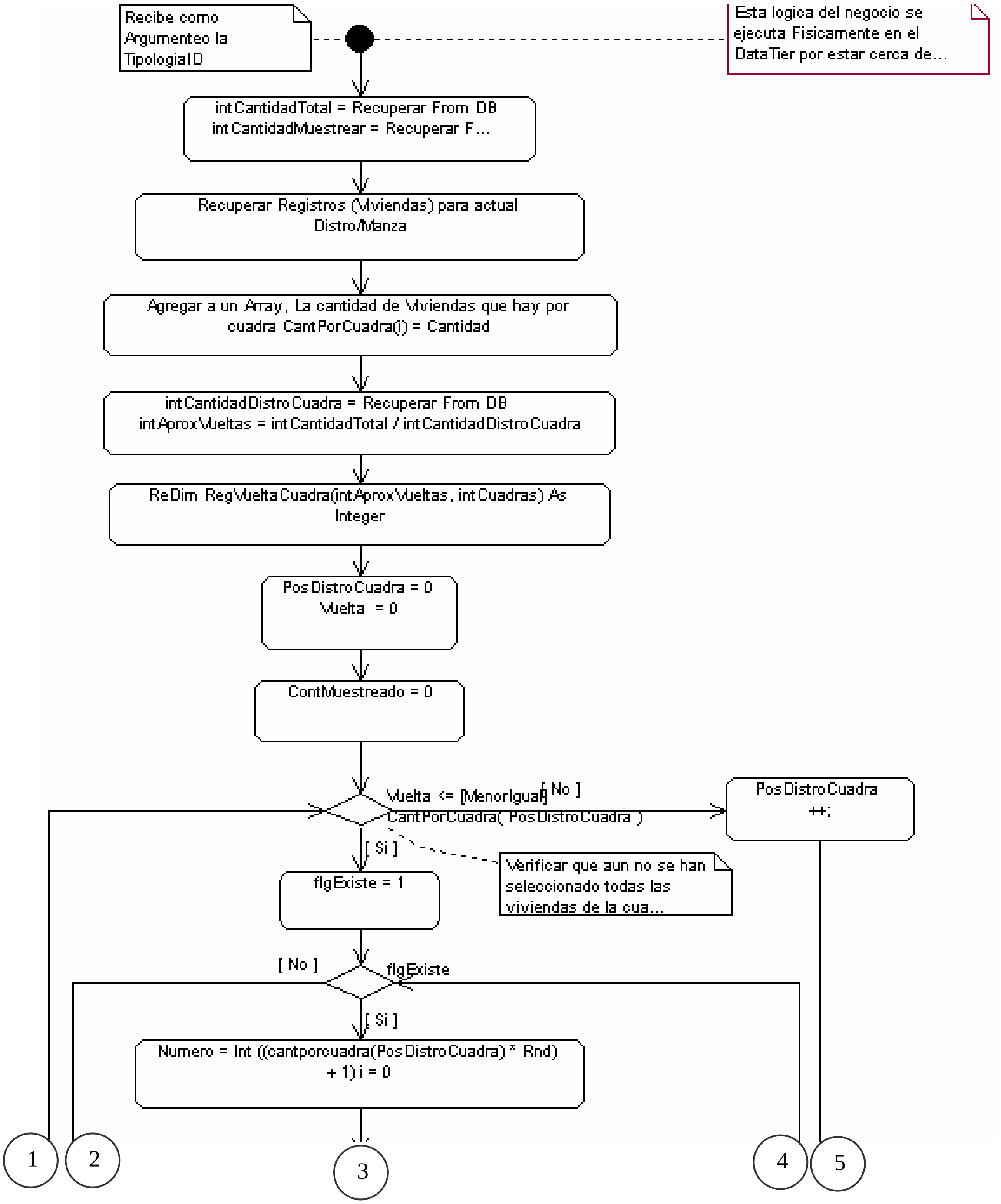

Figura 191. Diagrama de actividad para el método muestrear de la clase Tipología. _Parte superior del diagrama (1/2)

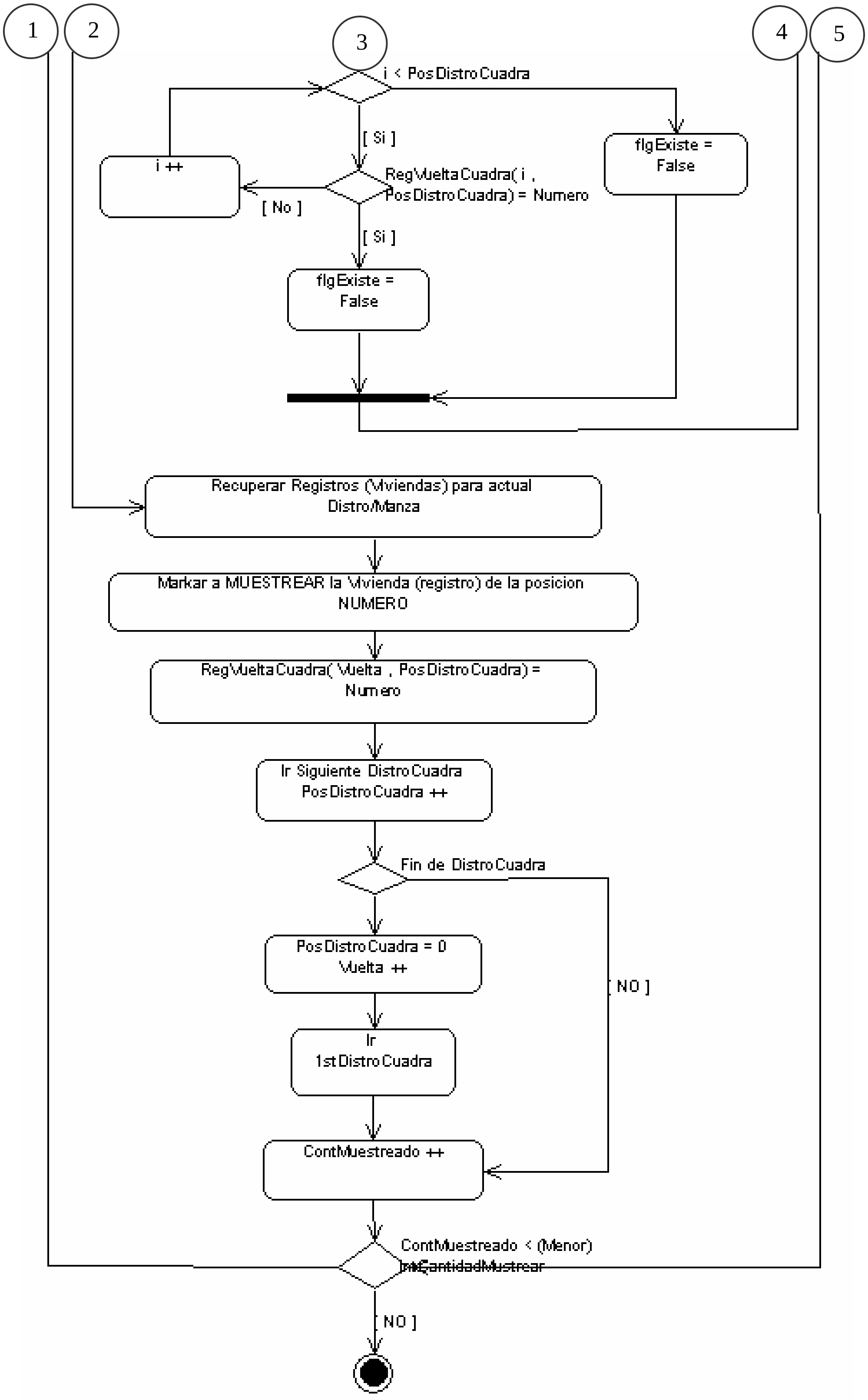

Figura 192. Diagrama de actividad para el método muestrear de la clase Tipología. _Parte inferior del diagrama (2/2)

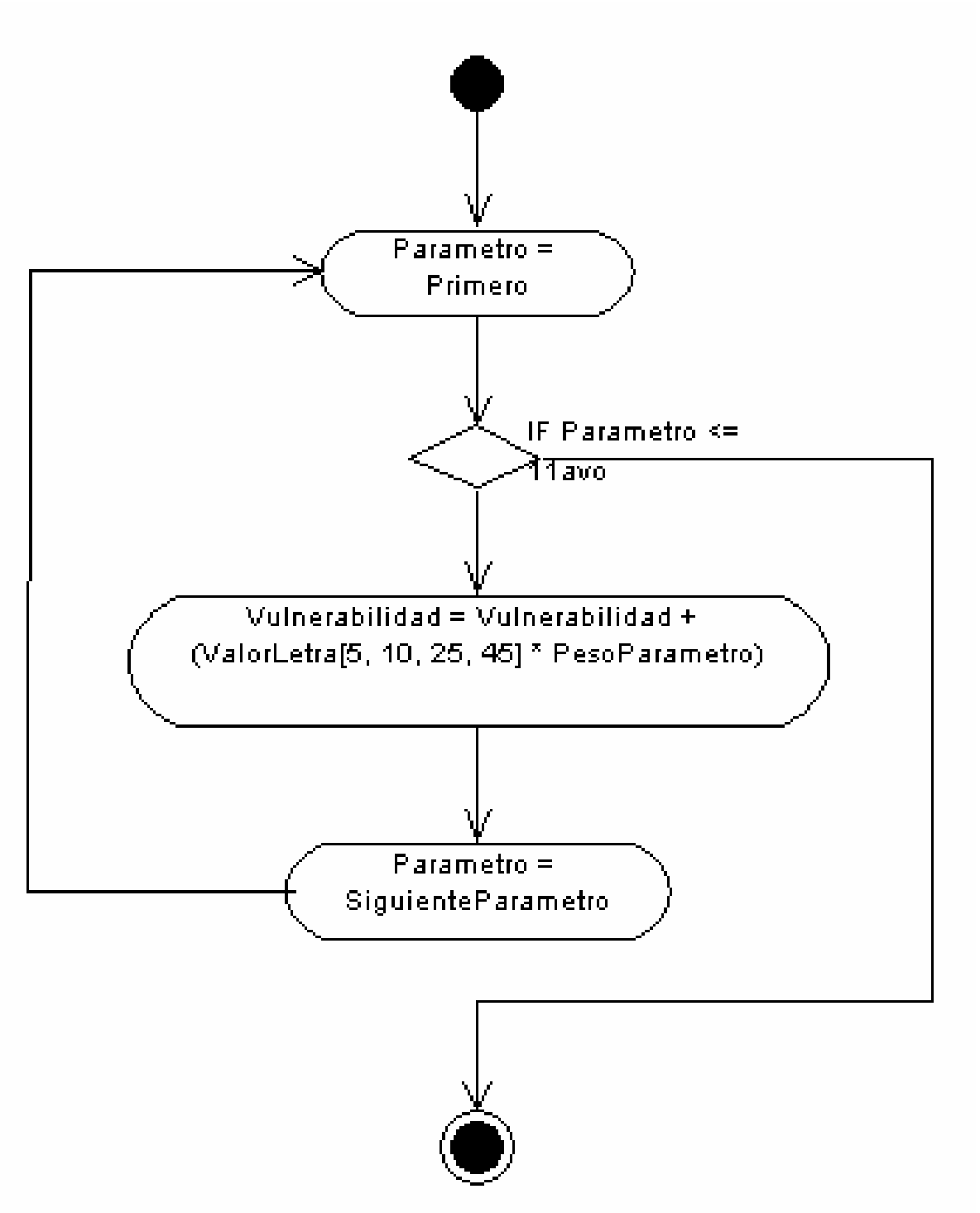

Figura 193. Diagrama de actividad para el método CalcularIndiceVulnerabilidad de la Clase ViviendaCatastralVulnerabilidad

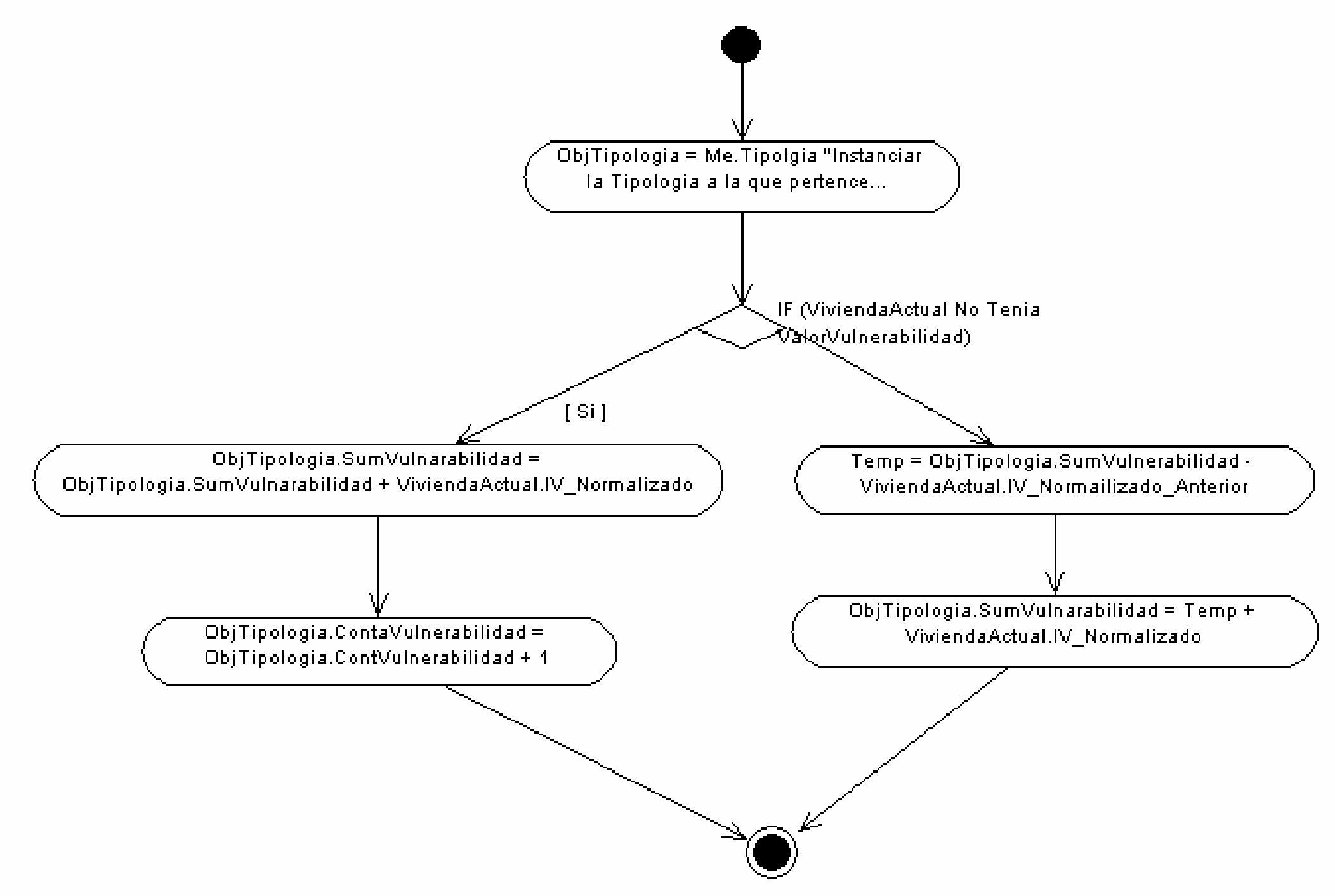

Figura 194. Diagrama de actividad para el método ActualizarIndiceVulEnTipologias de la clase ViviendaCatastralVulnerabilidad

## A.9 Diagramas de componentes

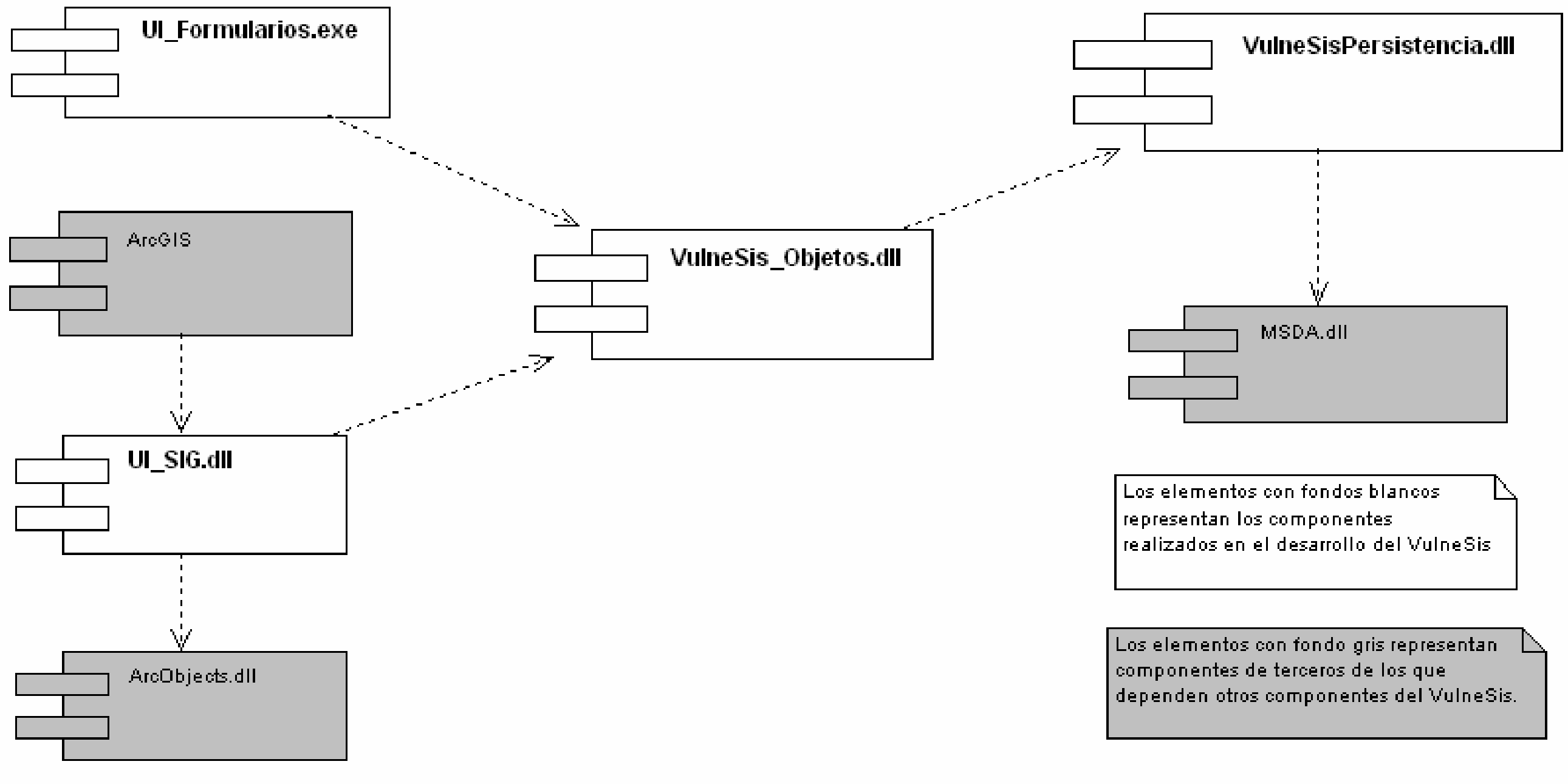

Figura 195. Diagrama de componentes del VulneSis

## A.10 Diagramas de despliegue

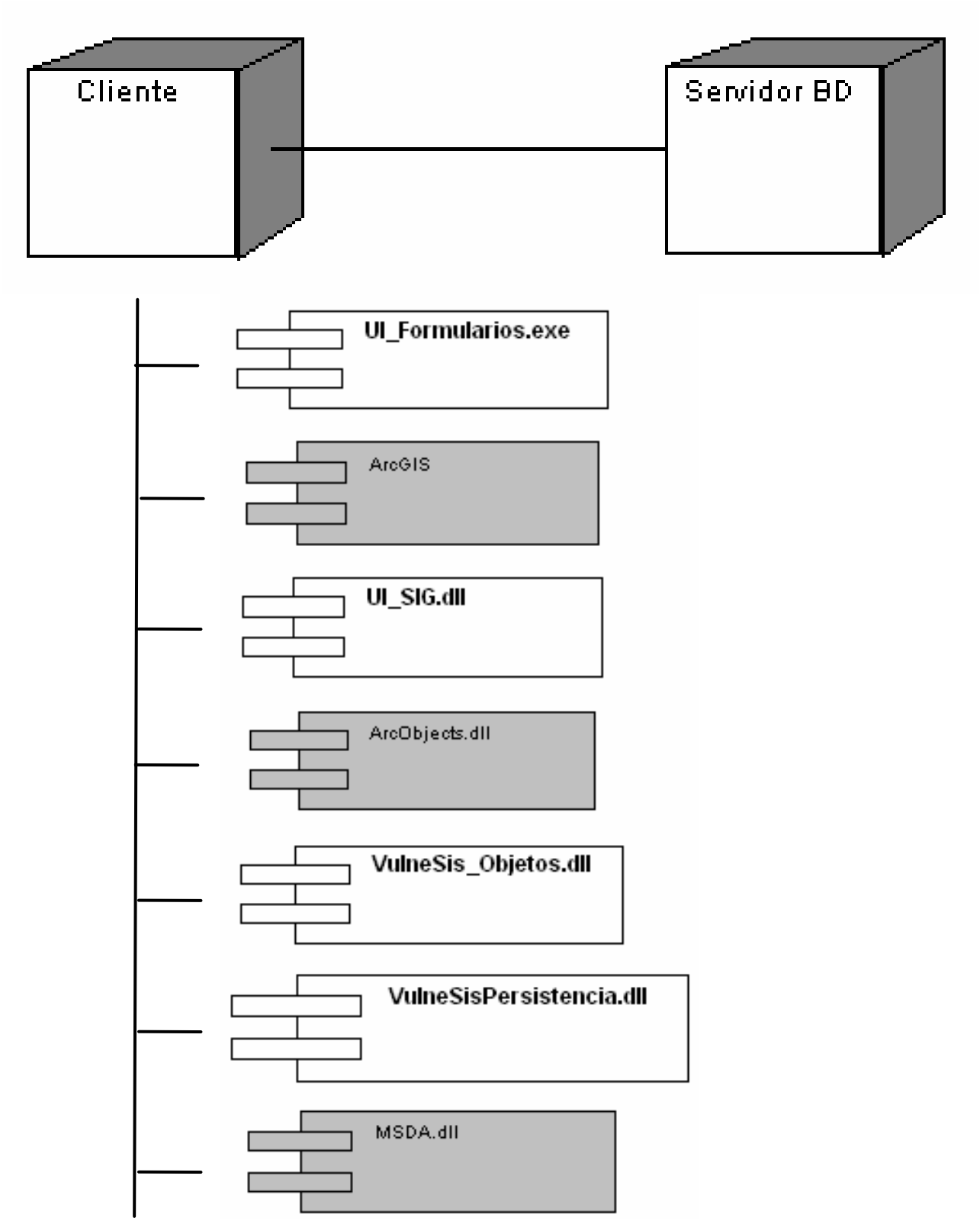

Figura 196. Diagrama de despliegue usado para el VulneSis con un arquitectura física en 2 capas y cliente gordo

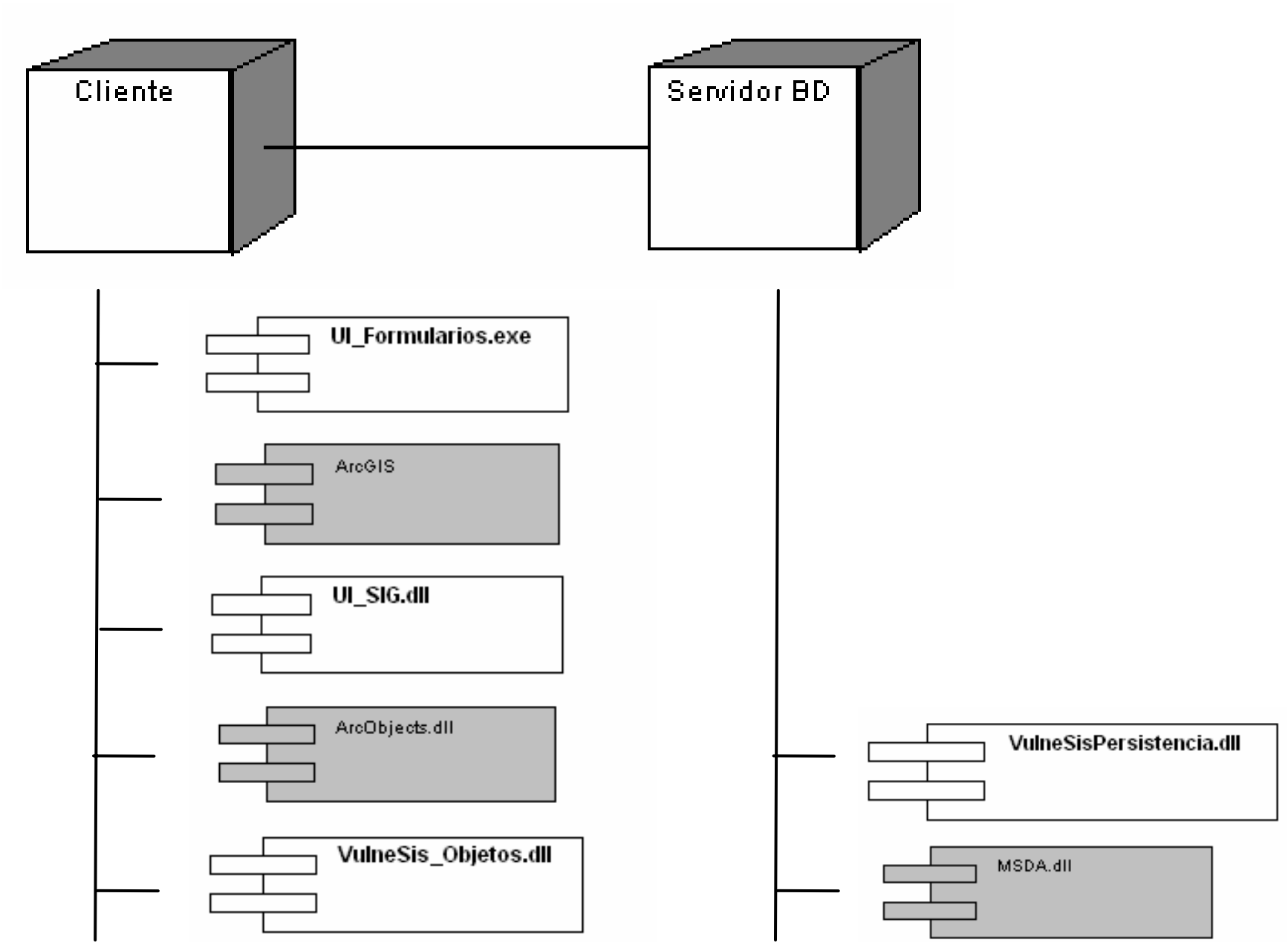

Figura 197. Diagrama de despliegue para el VulneSis con un arquitectura física en 2 capas, delegando al servidor de BD el trabajo de la capa lógica de persistencia (Esta delegación responde bien cuando el Servidor de BD tiene suficiente poder como para hacer sus labores de BD y correr una capa lógica de una aplicación)

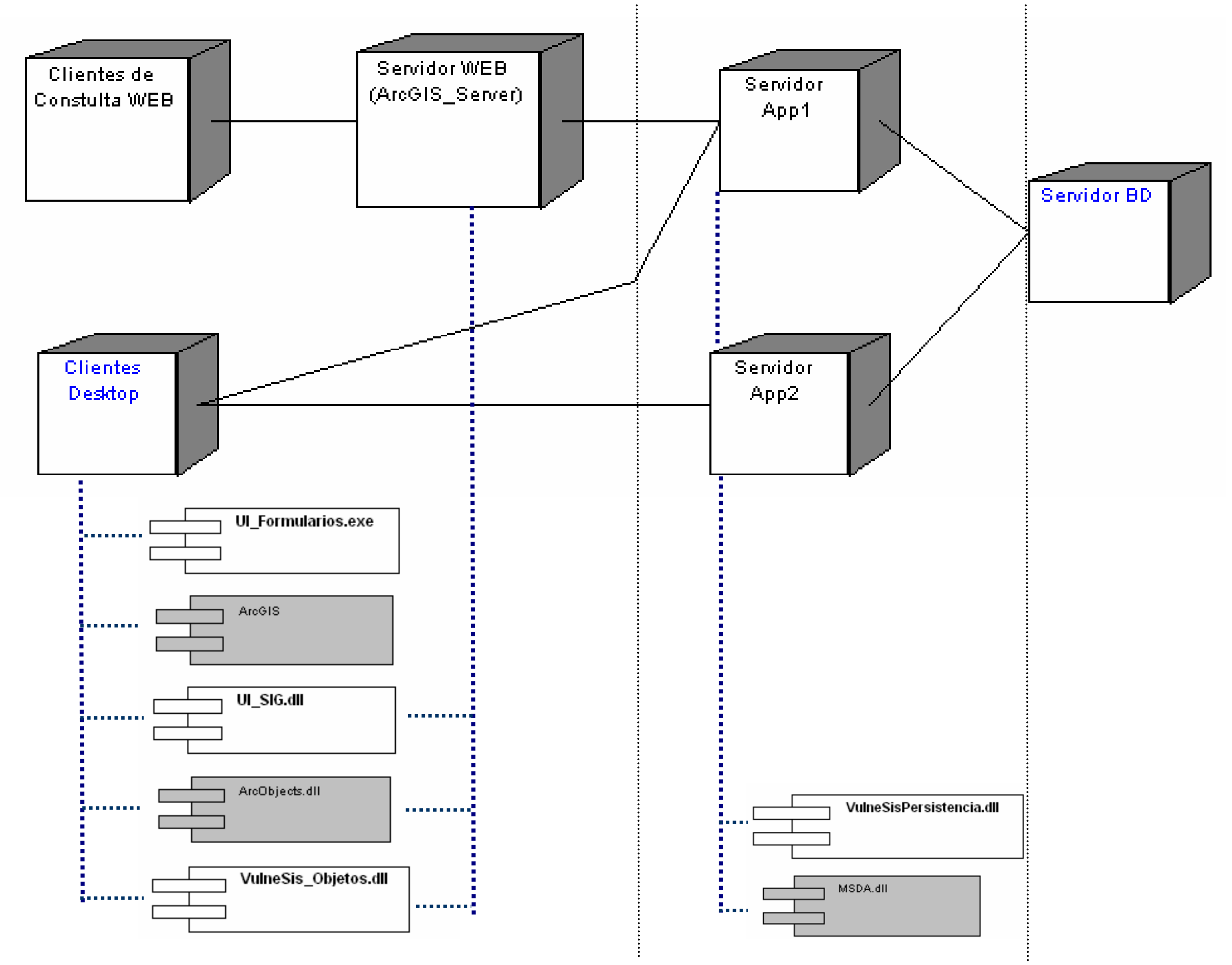

Figura 198. Diagrama de despliegue para el VulneSis
con un arquitectura física en 3 capas, donde a través de un servidor Web y ArcGIS Server se permiten clientes Web del VulneSis
este seria el caso para un ambiente con muchos usuarios por lo que la intervención de Servidores Apps permitirían balancear la carga del Servidor de BD con el que se pudieran conectar a través de un BackBone